\documentclass{ws-ijmpa}
\usepackage[super,compress]{cite}
\usepackage{graphicx}
\begin{document}
\markboth{R.M. Albuquerque et al.}
{Critical analysis of the X(5568) $\dots$}

\def\beq{\begin{equation}}
\def\eeq{\end{equation}}
\def\bea{\begin{eqnarray}}
\def\eea{\end{eqnarray}}
\def\bq{\begin{quote}}
\def\eq{\end{quote}}
\def\ve{\vert}
\def\nnb{\nonumber}
\def\ga{\left(}
\def\dr{\right)}
\def\aga{\left\{}
\def\adr{\right\}}
\def\rar{\rightarrow}
\def\lrar{\Longrightarrow}
\def\llar{\Longleftarrow}		
\def\nnb{\nonumber}
\def\la{\langle}
\def\ra{\rangle}
\def\nin{\noindent}
\def\ba{\vspace*{-0.2cm}\begin{array}}
\def\ea{\end{array}\vspace*{-0.2cm}}
\def\bm{\overline{m}}
\def\ind{\indexentry}
\def\c{\clubsuit}
\def\s{\spadesuit}
\def\b{$\bullet~$}
\def\als{\alpha_s}
\def\as{\ga\frac{\bar{\alpha_s}}{\pi}\dr}
\def\asr{\ga\frac{{\alpha_s}}{\pi}\dr}
\def\gg2{ \la\alpha_s G^2 \ra}
\def\gg3{g^3f_{abc}\la G^aG^bG^c \ra}
\def\ggg4{\la\als^2G^4\ra}
\def\lnu{\log{-\frac{q^2}{\nu^2}}}
\def\calD{ {\cal D} }
\def\ftilde{\tilde f}
\def\ftildeRe{ \Re e \ftilde}
\def\ftildeIm{ \Im m \ftilde}
\def\therho{\theta\rho}
\def\os{{\rm OS}}
\newcommand{\epem}{\mbox{$e^+e^-$}}
\newcommand{\dgr}{^{\rm o}}
\newcommand{\pipi}{\mbox{$\pi\pi$}}
\newcommand{\kkbar}{\mbox{$K\bar K$}}

\def\beq{\begin{equation}}
\def\enq{\end{equation}}
\def\beqa{\begin{eqnarray}}
\def\enqa{\end{eqnarray}}
\def\nnb{\nonumber}
\def\rar{\rightarrow}
\def\MeV{\nobreak\,\mbox{MeV}}
\def\GeV{\nobreak\,\mbox{GeV}}
\def\keV{\nobreak\,\mbox{keV}}
\def\fm{\nobreak\,\mbox{fm}}
\def\Tr{\mbox{ Tr }}
\def\qq{\lag\bar{q}q\rag}
\def\uu{\lag\bar{u}u\rag}
\def\dd{\lag\bar{d}d\rag}
\def\ss{\lag\bar{s}s\rag}
\def\qqs{\lag\bar{s}s\rag}
\def\mix{\lag\bar{q}g\si Gq\rag}
\def\mixs{\lag\bar{s}g\si Gs\rag}
\def\Gd{\lag G^2 \rag}
\def\gG{\lag g^2 G^2 \rag}
\def\GGG{\lag g^3G^3\rag}
\def\kab{\left[(\al+\be)m_c^2-\al\be s\right]}
\def\pli{p^\prime}
\def\ka{\kappa}
\def\lam{\lambda}
\def\La{\Lambda}
\def\gam{\gamma}
\def\Ga{\Gamma}
\def\om{\omega}
\def\rh{\rho}
\def\si{\sigma}
\def\ps{\psi}
\def\ph{\phi}
\def\de{\delta}
\def\al{\alpha}
\def\be{\beta}
\def\alma{\alpha_{max}}
\def\almi{\alpha_{min}}
\def\bemi{\beta_{min}}
\def\lb{\label}
\def\nn{\nonumber}
\def\bd{\boldmath}
\def\MS{\overline{MS}}
\def\Fab{m_Q^2(\al+\be)-\al\be q^2}
\def\Ha{m_Q^2-\al(1-\al) q^2}
\def\Fabs{m_Q^2(\al+\be)-\al\be s}
\def\Has{m_Q^2-\al(1-\al) s}
\newcommand{\rag}{\rangle}
\newcommand{\lag}{\langle}
\newcommand{\bph}{\mbox{\bf $\phi$}}
\newcommand{\rf}{\ref}
\newcommand{\ct}{\cite}
\newcommand{\LogF}[1]{\mbox{$\:{F}_{#1} $}}
\newcommand{\LogH}[1]{\mbox{$\:{H}_{#1} $}}
\newcommand{\ImF}[1]{\mbox{$\:{\cal F}_{#1} $}}
\newcommand{\ImH}[1]{\mbox{$\:{\cal H}_{#1} $}}
\def\pr{\partial}
\def\msb{\overline{MS}}
\def\Bbg{\Bigg{[}}
\def\Bbd{\Bigg{]}}
\def\Bg{\Big{[}}
\def\Bd{\Big{]}}
\def\Lrar{\Longrightarrow}
\def\llrar{\longleftrightarrow}
\def\orar{\overrightarrow}
\def\lrar{\leftrightarrow}

\newcommand{\lgm}{{\,\rm ln }}
\newcommand{\Break}{ \right. \nonumber \\ &{}& \left. }




\title{ Nature of the $X(5568)$ : a critical Laplace sum rule analysis    at N2LO}
\author{R. Albuquerque}
\address{Faculty of Technology,Rio de Janeiro State University (FAT,UERJ), Brazil\\
Email address:  raphael.albuquerque@uerj.br}
\author{S. Narison\footnote{In writing this paper, I learn the sudden death of my collaborator and friend Gerard Mennessier. I dedicate my contribution in this paper for his memory.}}
\address{Laboratoire
Univers et Particules de Montpellier, CNRS-IN2P3, \\
Case 070, Place Eug\`ene
Bataillon, 34095 - Montpellier, France.\\
Email address: snarison@yahoo.fr}

\author{A. Rabemananjara  and   D. Rabetiarivony}

\address{Institut of High-Energy Physics of Madagascar (iHEPMAD)\\
University of Ambohitsaina,
Antananarivo 101, Madagascar}


\maketitle

\begin{history}
\end{history}

\begin{abstract}
\nin
We scrutinize recent QCD spectral sum rules (QSSR) results to lowest order (LO) predicting the masses of the $ BK$ molecule and $(su)(\overline{bd})$
four-quark states.
We improve these results by adding NLO and N2LO corrections to the PT contributions giving a more precise meaning on the $b$-quark mass definition 
used in the analysis. We extract our optimal predictions using Laplace sum rule (LSR) within the {\it standard stability criteria} versus the changes of the external free parameters ($\tau$-sum rule variable, $t_c$ continuum threshold and  subtraction constant $\mu$). The smallness of the higher order PT corrections justifies (a posteriori) the LO order results $\oplus$ the uses of the ambiguous heavy quark mass to that order. However,  our predicted spectra in the range $(5173\sim 5226)$ MeV, summarized  in Table\,\ref{tab:result}, for exotic hadrons built with four different flavours $(buds)$,
do not support some previous    interpretations of the   $D0$ candidate\,\cite{D0}, $X(5568)$, as a pure molecule or a four-quark state.  If experimentally confirmed, it could
result from their mixing with an angle:  sin$2\theta\approx 0.15$. One can also scan the region $(2327\sim 2444)$ MeV (where the $D^*_{s0}(2317)$ might be a good candidate) and the one $(5173\sim 5226)$ MeV for detecting these $(cuds)$ and $(buds)$ unmixed exotic hadrons (if any) via, eventually, their radiative  or $\pi+hadrons$ decays.
 
\ccode{Pac numbers: 11.55.Hx, 12.38.Lg, 13.20-v}
\end{abstract}


\section{Introduction and Experimental Facts}
\nin
Stimulated by the recent observation of the $D0$ collaboration for a narrow $X(5568)$ state ($\Gamma_X=21.9\pm 6.4^{+5.0}_{-2.5}$ MeV) decaying sequencely into $B^0_s\pi^\pm:~B^0_s\to J/\psi\,\phi,~J/\psi\to\mu^+\mu-,~\phi\to K^+K^-$,  where a $J^{P}=0^{+}$ is favoured, 
many papers have been proposed in the literature for explaining its nature. If confirmed\,\footnote{Note that a recent analysis of the LHCb collaboration does not confirm this $D0$ result\,\cite{LHCb}.}, this is the first observation of an hadron bounded with four-different flavours $(buds)$ which cannot be accomodated by the usual quark model but instead by a four-quark or a molecule or some other exotic mechanism. Among different proposals, we select the ones from QCD spectral sum rules (QSSR)\,\cite{SVZa,SVZb}\,\footnote{For reviews where complete references can be found, see e.g: \cite{SNB4a,SNB4b,SNB1,SNB2,SNB3a,SNB3b,SNB3c,RRY,CK}.}, which interpret the $X(5568)$ state either as a $ BK$ molecule\,\cite{TURC} or as a four-quark $(\overline{bu})(ds)$ \cite{TURC1,NIELSEN, STEELE,WANG} exotic bound state.
\section{Interpolating currents}
\hspace*{0.5cm}
\b In the following, we shall analyze different assumptions on the nature of the $X(5568)$ which 
can be specified by the form of the minimal QCD interpolating currents given in Table \ref{tab:current}.
{\scriptsize
\begin{table}[hbt]
 \tbl{Minimal interpolating currents $J_X$ describing the $X(5568)$ }  
    {\small
 {\begin{tabular}{@{}lll@{}} \toprule
&\\
\hline
\hline
\bf Nature&\bf$J^{P}$&\bf Current   \\
\hline
{\it Molecule} &&\\
$ BK$&$0^{+}$&$(\bar b\, i\gamma_5\ u)(\bar d \, i\gamma_5 \, s)$\\
$ B_s\pi$&$0^{+}$&$(\bar b\, i\gamma_5\, s)(\bar d\, i\gamma_5\, u)$\\
$ B^*K$&$1^{+}$&$(\bar b \,i\gamma_\mu\, u)(\bar d \,i\gamma_5\, s)$\\
$ B^*_s\pi$&$1^{+}$&$(\bar b\, i\gamma_\mu s)(\bar d i\gamma_5\, u)$\\
{\it Four-quark} &&\\
&$1^{-}$&$(s^T
C\gamma_5\,u)(\bar b\,\gamma_\mu\gamma_5C\,\bar d^T)+k(s^T
C\,u)(\bar b\,\gamma_\mu\, C\,\bar d^T)$\\
&$1^{+}$&$(s^T
C\gamma_5\,u)(\bar b\,\gamma_\mu C\,\bar d^T)+k(s^T
C\,u)(\bar b\,\gamma_\mu\gamma_5 C\,\bar d^T)$\\
\hline\hline
\end{tabular}}
\label{tab:current}
}
\end{table}
} 
$C$ is the charge conjugation matrix;  $u,d,s,b$ are the quark fields and the summation over colour indices is understood; $k$ is a free mixing parameter. 

\b The corresponding scalar two-point correlator is defined as:
\beq
\psi_X(q^2)=i\int d^4x~e^{-iqx}\la 0\vert J_X(x)J_X^\dagger (0) \vert 0\ra~,
\eeq
The vector or axial-vector correlator reads:
\bea
\Pi^{\mu\nu}_X(q^2)&=&i\int d^4x~e^{-iqx}\la 0\vert J^{\mu}_X(x)\ga J^{\nu}_X(0)\dr^\dagger\vert 0\ra\nnb\\
&\equiv&-\ga g^{\mu\nu}-  \frac {q^\mu q^\nu}{q^2} \dr \Pi^{(1)}_X(q^2) + \frac{q^\mu q^\nu}{q^2} \Pi_X^{(0)}(q^2)~,
\eea
where the longitudinal part : $\Pi_X^{(0)}(q^2)$ is related to the (pseudo)scalar correlator $\psi_X(q^2)$ via a well-known Ward identity. We shall be concerned here with the transverse part $ \Pi^{(1)}_X(q^2)$ which has the quantum number : $J^{P}=1^{+}$ or $1^{-}$ and its longitudinal part:  $ \Pi^{(0)}_X(q^2)$ with the quantum number: $J^{P}=0^{+}$ or $0^-$. 

\b The previous  correlators obey the K\"allen-Lehmann representation or dispersion relation:
\beq
\Pi_X(q^2)=\int_{({M_b}+m_s)^2}^\infty {dt\over {t-q^2-i\epsilon}}{1\over\pi}{\rm Im}\Pi_X(t)+\cdots~,
\eeq
where $\cdots$ represents subtraction constants which are polynomial in $q^2$; $\Pi_X\equiv \psi_X$ or $\Pi_X^{(0,1)}$; 
$
({1/\pi}){\rm Im}\Pi_X(t) \equiv \rho(t)
$
is the spectral function that can be measured experimentally or calculable in QCD for large values of $t$. 
\section{The inverse Laplace transform sum rule (LSR)}
\hspace*{0.5cm} \b One can improve the previous K\"allen-Lehmann representation by applying to both sides $n$-numbers of derivative in 
$Q^2\equiv-q^2$ and by keeping the ratio $n/Q^2\equiv \tau$ fixed. In this way, it becomes an exponential sum rule:
\beq
{\cal L}(\tau,\mu)\equiv \int_{({M_b}+m_s)^2}^\infty {dt}~e^{-t\tau}\rho(t,\mu)~,
\eeq
\beq\label{eq:ratiolsr}
{\cal R}(\tau,\mu) = \frac{\int_{({M_b}+m_s)^2}^{\infty} dt~t~ e^{-t\tau}\rho(t,\mu)}
{\int_{({M_b}+ms)^2}^{\infty} dt~ e^{-t\tau}  \rho(t,\mu)}~,
\eeq
where $\mu$ is the  subtraction point which appears in the approximate QCD series when radiative corrections are included. The set of variables $(\tau,\mu)$  are, in principle, free external parameters. We shall use stability criteria (if any) for extracting the optimal results. 

\b These sum rules firstly derived by SVZ\,\cite{SVZa,SVZb} have been called Borel sum rule due to the factorial suppression factor of the condensate contributions in the OPE. Their quantum mechanics version have been studied by Bell-Bertlmann in \cite{BELLa,BELLb} through the harmonic oscillator where $\tau$ has the property of an imaginary time, while the derivation of their radiative corrections has been firstly shown by Narison-de Rafael \cite{SNRAF} to have the properties of the inverse Laplace sum rule (LSR).
\section{Parametrization of the spectral function and Stability criteria}
\hspace*{0.5cm} \b The ratio\,\cite{SVZa,SVZb,BELLa,BELLb} and double ratio\,\cite{DRSR}\,\footnote{The double ratio of sum rules have been successfully applied in different channels\,\cite{SNGh1,SNGh3,SNGh5,SNmassb,SNhl,SNmassa,SNFORM2,HBARYON1,HBARYON2,NAVARRA}.} of sum  rules :
\beq
{\cal R}_{S/A} (\tau,\nu)~~~~{\rm and}~~~~{\cal R}_{A,S} (\tau,\nu)\equiv {{\cal R}_{A}\over {\cal R}_{S}}~,
\eeq
are useful, as they are  equal  to the
resonance mass squared $M_{S/A}^2$ and their ratio at the $\tau$-stability region:
\beq
{\cal R}_{S/A}\simeq M_{S/A}^2~~~~{\rm and}~~~~{\cal R}_{A,S}\simeq  {M_A^2\over M_S^2}~,
\eeq 
in the Minimal Duality Ansatz (MDA) parametrization of the spectral function:
\bea
\rho_{S/A}(t)&\simeq& f^2_{S/A}M^8_{S/A}\delta(t-M_{S/A}^2)
  \ + \
  ``\mbox{QCD continuum}" \theta (t-t_c),
\label{eq:duality}
\eea
where $f_{S/A}$ is the leptonic decay constant or coupling defined as:
\beq
\la 0| J_S|S\ra=f_{S}M^4_{S}~~{\rm and}~~\la 0| J^\mu_A|A\ra=\epsilon^\mu f_{A}M^5_{A}~,
\label{eq:coupling}
\eeq
respectively for the scalar (S) and axial-vector (A) mesons, where $\epsilon^\mu$ is the $W$-boson polarization, 
while the ``QCD continuum" comes from the discontinuity of the Feynmann diagrams appearing in the SVZ expansion. Though apparently quite simple, different tests of this MDA from complete hadronic data have shown that it can reproduce with high-precision these complete data\,\cite{SNFB12a,SNFB12b,SNB1,SNB2}, while it has been also successfully tested in the large $N_c$ limit of QCD\,\cite{PERISa,PERISb}.

\b Noting that in the previous definition in Table \ref{tab:current},  the bilinear pseudoscalar current acquires an anomalous dimension due to its normalization, thus the $X$-decay constant runs for $n_f=5$ and to $\alpha_s^2$ as :
\beq
f_S(\mu)=\hat f_S \ga -\beta_1a_s\dr^{4/\beta_1}/r_m^2~,~~~~f_A(\mu)=\hat f_A \ga  -\beta_1a_s\dr^{2/\beta_1}/r_m
\label{eq:fhat}
\eeq
where we have introduced the renormalization group invariant coupling $\hat f_{S/A}$; 
$-\beta_1=(1/2)(11-2n_f/3)$ is the first coefficient of the QCD $\beta$-function for $n_f$ flavours and  $a_s\equiv (\alpha_s/\pi)$.  For $n_f=5$ flavours, the QCD corrections read;
\beq
r_m=1+1.176 \ga {\alpha_s\over\pi}\dr +1.501\ga {\alpha_s\over\pi}\dr^2~.  
\eeq

\b Optimal information on the lowest resonance mass and decay constant can be achieved at the stability regions (if they exist) of the set of external variables $(\tau, ~t_c, ~\mu)$. This requirement is similar to the Principle of Minimal Sensitivity (PMS) used, e.g, in \cite{STEVENSON} for optimized perturbative series.

\section{QCD expressions of the $ B^*K$ and $ BK$ spectral functions}
We give below the QCD expressions of the $ B^*K$ and $ BK$ two-point spectral function at lowest order (LO) of perturbation theory and including the quark and gluon condensates within the SVZ operator product expansion (OPE). The value and normalization of these condensates are given in Table\,\ref{tab:param} and in Section\,\ref{sec:input}. 
\subsection{$ B^*K$ $(1^{+})$ axial-vector molecule}
For the axial-vector $1^{+}$ molecule, the spectral function reads up to dimension 6 condensates:
\bea
\rho^{pert}&=&{\frac{{M_b}^8}{5\ 3\ 2^{15 }\pi ^6}\Bigg{[}\frac{5}{x^4}-\frac{96}{x^3}-\frac{945}{x^2}+\frac{480}{x}-60\left(\frac{9}{x^2}+\frac{16}{x}+3\right)\text{Log}\,x+555+x^2\Bigg{]}},\nnb\\
\rho^{\la \bar qq\ra}&=&-{M_b^5\over 2^8\pi^4}\la\bar qq\ra\Bigg{[}{1\over x^2} + {9\over x} + 6\ga {1\over x} + 1\dr\text{Log}\,x - 9 - x\Bigg{]}-\nnb\\
&&\frac{ m_s M_b^4}{2^{11}\pi ^4}\la\bar qq\ra(2-\kappa)\left(\frac{3}{x^2}-\frac{16}{x}-12\,\text{Log}\,x+12+x^2\right),\nnb\\
\rho^{\la G^2\ra}&=&\frac{M_b^4}{3\ 2^{15}\pi ^6}4\pi\la \alpha_s G^2\ra\Bigg{[}\frac{1}{x^2}-\frac{120}{x}-12\left(\frac{4}{x}+7\right)\text{Log}\,x+108+8x+3x^2\Bigg{]},\nnb\\
\rho^{\la \bar q G q\ra}&=&\frac{3M_b^3}{2^9\pi ^4}\la \bar q G q\ra\left(\frac{1}{x}+2\,\text{Log}\,x-x\right)+\frac{m_s  M_b^2}{3\ 2^9\pi ^4}\la \bar q G q\ra(3+2\kappa)\left(\frac{2}{x}-3+x^2\right),\nnb\\
\rho^{\la \bar qq\ra^2}&=&\frac{M_b^2}{3\ 2^5\pi ^2}{\rho\la \bar qq\ra^2} \kappa\left(\frac{2}{x}-3+x^2\right)+\frac{ m_s M_b }{2^5\pi ^2}\rho\la \bar qq\ra^2(2-\kappa)(1-x),\nnb\\
\rho^{\la G^3\ra}&=&\frac{M_b^2}{3^2\ 2^{17}\pi ^6}\la g^3 G^3\ra\Bigg{[}\frac{9}{x^2}-\frac{160}{x}-12\left(\frac{4}{x}+9\right)\text{Log}\,x+144+7x^2\Bigg{]}.
\eea
where :  
$x={M_b}^2/{t}, ~~{\rm and}~~\kappa\equiv\la \bar ss\ra/\la\bar qq \ra 
\simeq \la \bar sGs\ra/\la\bar qGq \ra$
measures the $SU(3)$ breaking of the quark and mixed quark condensates. The contribution of a class of $d=7$ condensate for $m_s=0$  is:
\bea
\rho^{\la \bar qq\ra\la G^2\ra}&=&-\frac{M_b\la \bar qq\ra}{3\ 2^9\pi ^4} 4\pi\la \alpha_s G^2\ra\left(\frac{1}{x}+3\,\text{Log}\,x+5-6x\right)~.
\label{eq:d7axial}
\eea
The previous expressions compared to the ones in the literature are more convenient to use as they are in an integrated form. 
One should note that:

\b In performing the calculation of the spectral function, the heavy quark is put on-shell which corresponds to the implicit use of the pole mass $M_b$ in the previous QCD expression. This feature does not justify the (a priori)  use of the running mass in the sum rule analysis as done in the existing literature. We shall come back to this point later on.

\b In the literature, dimension condensates larger than $d=6$ have been also included in the OPE in order to improve stability of the results. However, one should note that the included condensates are only a part of more general condensates contributions at a given dimension. There is also a poor quantitative control of these high dimension condensates as has been hotly discussed in the past (cases of charmonium and $\rho$ meson channels)\,\cite{SNB1,SNB2} due to the violation of factorization which is already about a factor 3-4 for the $d=6$ condensates. This feature indicates that the error quoted in the final result which does not take into account such a violation has been largely underestimated. Therefore, we refrain to include these terms in the analysis but only consider them as a source of errors. 

\b We have not included in the OPE the $d=2$ tachyonic gluon mass\,\cite{CNZ1,CNZ2}\,\footnote{For reviews, see e.g. \cite{ZAK1,ZAK2}.} as it has been shown to be dual to the higher order terms of the PT series\,\cite{ZAK3} which we shall estimate using a geometric growth of the PT coefficients.

\subsection{$ BK$ $(0^{+})$ scalar molecule}
It reads up to the dimension 6 condensates:
\bea
\rho^{pert}&=&\frac{M_b^8}{5\ 2^{14}\pi ^6}\Bigg{[}\frac{1}{x^4}-\frac{20}{x^3}-\frac{220}{x^2}+\frac{80}{x}-60\left(\frac{2}{x^2}+\frac{4}{x}+1\right)\text{Log}\,x+155+4x\Bigg{]},\nnb\\
\rho^{\la \bar qq\ra}&=&-{M_b^5\over 2^8\pi^4}\la\bar qq\ra\Bigg{[}{1\over x^2} + {9\over x} + 6\ga {1\over x} + 1\dr\text{Log}\,x - 9 - x\Bigg{]}-\nnb\\
&&\frac{{m_s} {M_b}^4}{2^9\pi ^4}\la\bar qq\ra(2-\kappa)\left(\frac{1}{x^2}-\frac{6}{x}-6\,\text{Log}\,x+3+2x\right),\nnb\\
\rho^{\la G^2\ra}&=&\frac{M_b^4}{3\ 2^{13}\pi ^6}4\pi\la \alpha_s G^2\ra\Bigg{[}\frac{5}{x^2}+6\left(\frac{2}{x}-1\right)\text{Log}\,x-9+4x\Bigg{]},\nnb\\
\rho^{\la \bar qGq\ra}&=&\frac{3M_b^3}{2^8\pi ^4}\la \bar q G q\ra\Bigg{[}\frac{3}{x}+\left(\frac{1}{x}+3\right)\text{Log}\,x-2-x\Bigg{]}+\nnb\\
&&\frac{m_s M_b^2}{2^9\pi
^4}{\la \bar q G q\ra}(3+2\kappa)\left(\frac{1}{x}-2+x\right),\nnb\\
\rho^{\la \bar qq\ra^2}&=&\frac{M_b^2}{2^5\pi ^2}{\rho\la \bar qq\ra^2} \kappa\left(\frac{1}{x}-2+x\right)+\frac{m_sM_b }{2^5\pi ^2}{\rho\la \bar qq\ra^2}(2-\kappa)(1-x),\nnb\\
\rho^{\la G^3\ra}&=&\frac{M_b^{2 }}{3\ 2^{15}\pi ^6}\la g^3 G^3\ra\Bigg{[}\frac{1}{x^2}-\frac{21}{x}-6\left(\frac{1}{x}+3\right)\text{Log}\,x+15+5x\Bigg{]}.
\eea
The contribution of a class of $d=7$ condensate for $m_s=0$  is:
\bea
\rho^{\la \bar qq\ra\la G^2\ra}&=&-\frac{M_b\la \bar qq\ra}{3\ 2^9\pi ^4} 4\pi\la \alpha_s G^2\ra\left(\frac{1}{x}+12\,\text{Log}\,x+14-15x\right)~.
\label{eq:d7scalar}
\eea
\subsection{Higher Order (HO) PT corrections to the Spectral function}
\hspace*{0.5cm} \b We extract the Higher Order (HO) PT corrections by considering that the molecule two-point spectral function is the convolution of the two ones built from 
two quark bilinear currents (factorization) as a consequence of the molecule
definition of the state\,\footnote{A such approach has been used in\,\cite{FENOSOA1,FENOSOA2} for estimating {\it for the first time} the higher order perturbative corrections to the spectral functions of molecule
states.}:
\bea\label{eq:qqcurrent}
J_K(x)&\equiv&\bar d i\gamma_5 s ~~~\leadsto ~~~ {1\over \pi}{\rm Im}\psi_K(t)~, \nnb\\
J_B(x)&\equiv&\bar b i\gamma_5 u ~~~ \leadsto ~~~{1\over \pi}{\rm Im}\psi_B(t)~,\nnb\\
J_{B^*}^\mu(x)&\equiv&\bar b \gamma^\mu u ~~~ \leadsto ~~~{1\over \pi}{\rm Im}\Pi_{B^*}(t)~.
\eea
In this way, we obtain the convolution integral for the axial-vector state\,\cite{PICH} :
\bea
{1\over \pi}{\rm Im} \Pi_A(t)&=& \theta (t-(M_b+m_s)^2)\ga 1\over 4\pi\dr^2 t^2 \int_{M_b^2}^{(\sqrt{t}-m_s)^2}\hspace*{-0.5cm}dt_1\int_{m_s^2}^{(\sqrt{t}-\sqrt{t_1})^2}  \hspace*{-1cm}dt_2~\lambda^{3/2}~\nnb\\
&&
\times ~{1\over \pi}{\rm Im} \Pi_{B^*}(t_1) {1\over \pi} {\rm Im} \psi_K(t_2),
\label{eq:axial}
\eea
and for the scalar state\,\cite{PICH,SNPIVO} :
\bea
{1\over \pi}{\rm Im} \psi_S(t)&=& \theta (t-(M_b+m_s)^2)\ga 1\over 4\pi\dr^2 t^2 \int_{M_b^2}^{(\sqrt{t}-m_s)^2}\hspace*{-0.5cm}dt_1\int_{m_s^2}^{(\sqrt{t}-\sqrt{t_1})^2}  \hspace*{-1cm}dt_2~\lambda^{1/2}~\nnb\\
&&\times~\ga {t_1\over t}+ {t_2\over t}-1\dr^2 \times ~{1\over \pi}{\rm Im} \psi_B(t_1) {1\over \pi} {\rm Im} \psi_K(t_2),
\label{eq:scalar}
\eea
with the phase space factor :
\beq
\lambda=\ga 1-{\ga \sqrt{t_1}- \sqrt{t_2}\dr^2\over t}\dr \ga 1-{\ga \sqrt{t_1}+ \sqrt{t_2}\dr^2\over t}\dr~.
\eeq
$M_b$ is the on-shell(pole) perturbative heavy quark mass, while we shall use the running perturbative mass $\bar m_s$ in the $\overline{MS}$-scheme.

\b The perturbative expressions of the bilinear unequal masses pseudoscalar Im$\psi_{K/B}(t)$ and Im$\Pi_{B^*}(t)$ spectral functions are known in the literature up to N2LO corrections \cite{SNB1,SNB2}. The complete LO expression has been evaluated firstly in \cite{FNR}, the NLO corrections for light quarks by \cite{BECCHI}, the complete NLO by \cite{BROAD} and the N2LO with one massless quark by \cite{RVS}. 
\subsection{Relation between the pole and running heavy quark masses}
\hspace*{0.5cm} \b The PT expression of the spectral function obtained using on-shell renormalization can be transformed into the  $\overline{MS}$-scheme by using the relation between the  $\overline{MS}$ running mass $\overline{m}_Q(\mu)$ and the on-shell mass~$M_Q$ , to order $\alpha_s^2$ \cite{TAR,COQUEa,COQUEb,SNPOLEa,SNPOLEb,BROAD2a,AVDEEV,BROAD2b,CHET2a,CHET2b}:
\bea
M_Q &=& \overline{m}_Q(\mu)\Big{[}
1+{4\over 3} a_s+ (16.2163 -1.0414 n_l)a_s^2\nnb\\
&&+\ln{\ga\mu\over M_Q\dr^2} \ga a_s+(8.8472 -0.3611 n_l) a_s^2\dr\nnb\\
&&+\ln^2{\ga\mu\over M_Q\dr^2} \ga 1.7917 -0.0833 n_l\dr a_s^2...\Big{]},
\label{eq:pole}
\eea
for $n_l$ light flavours.
\section{QCD input parameters}\label{sec:input}
{\scriptsize
\begin{table}[hbt]
 \tbl{QCD input parameters:
the original errors for 
$\la\alpha_s G^2\ra$, $\la g^3  G^3\ra$ and $\rho \la \bar qq\ra^2$ have been multiplied by about a factor 3 for a conservative estimate of the errors (see also the text). }  
    {\small
 {\begin{tabular}{@{}lll@{}} \toprule
&\\
\hline
\hline
Parameters&Values& Ref.    \\
\hline
$\alpha_s(M_\tau)$& $0.325(8)$&\cite{SNTAU,BNPa,BNPb,BETHKE,PDG}\\
$\overline{m}_c(m_c)$&$1261(12)$ MeV &average \cite{SNH10a,SNH10b,SNH10c,PDG,IOFFEa,IOFFEb}\\
$\overline{m}_b(m_b)$&$4177(11)$ MeV&average \cite{SNH10a,SNH10b,SNH10c,PDG}\\
$\hat \mu_q$&$(253\pm 6)$ MeV&\cite{SNB1,SNmassa,SNmassb,SNmass98a,SNmass98b,SNLIGHT}\\
$\hat m_s$&$(0.114\pm0.006)$ GeV &\cite{SNB1,SNmassa,SNmassb,SNmass98a,SNmass98b,SNLIGHT}\\
$\kappa\equiv \la \bar ss\ra/\la\bar dd\ra$& $(0.74^{+0.34}_{- 0.12})$&\cite{HBARYON1,HBARYON2,SNB1}\\
$M_0^2$&$(0.8 \pm 0.2)$ GeV$^2$&\cite{JAMI2a,JAMI2b,JAMI2c,HEIDa,HEIDb,HEIDc,SNhl}\\
$\la\alpha_s G^2\ra$& $(7\pm 3)\times 10^{-2}$ GeV$^4$&
\cite{SNTAU,LNT,SNIa,SNIb,YNDU,BELLa,BELLb,BELLc,SNH10a,SNH10b,SNH10c,SNG1,SNG2,SNGH}\\
$\la g^3  G^3\ra$& $(8.2\pm 2.0)$ GeV$^2\times\la\alpha_s G^2\ra$&
\cite{SNH10a,SNH10b,SNH10c}\\
$\rho \alpha_s\la \bar qq\ra^2$&$(5.8\pm 1.8)\times 10^{-4}$ GeV$^6$&\cite{SNTAU,LNT,JAMI2a,JAMI2b,JAMI2c}\\
\hline\hline
\end{tabular}}
\label{tab:param}
}
\end{table}
} 
{\scriptsize
\begin{table}[hbt]
 \tbl{
$\alpha_s(\mu)$ and correlated values of the running mass $\overline{m}_Q(\mu)$ used in the analysis for different values of the  subtraction scale $\mu$. The error in $\overline{m}_Q(\mu)$ has been induced by the one of $\alpha_s(\mu)$ to which one has added the error on their determination given in Table\,\ref{tab:param}. }
    {\small
{\begin{tabular}{@{}llll@{}} \toprule
&\\
\hline

\hline
Input for $BK,~B^*K,~(\overline{bu})(ds)$: $n_f=5$\\
\hline
$\mu$[GeV]&$\alpha_s(\mu)$& $\overline{m}_b(\mu)$[GeV]&\\
3&0.2590(26)&4.474(4)&\\
3.5&0.2460(20)&4.328(2)&\\
Input: $\overline{m}_b(m_b)$&0.2320(20)&4.177&\\
4.5&0.2267(20)&4.119(1)&\\
5.0&0.2197(18)&4.040(1)&\\
5.5&0.2137(17)&3.973(2)&\\
6.0&0.2085(16)&3.914(2)&\\
6.5&0.2040(15)&3.862(2)&\\
7.0&0.2000(15)&3.816(3)&\\
\hline
Input for $ DK,~D^*K,~(\overline{cu})(ds)$ : $n_f=4$\\
\hline
$\mu$[GeV]&$\alpha_s(\mu)$& $\overline{m}_c(\mu)$[GeV]&\\
Input: $\overline{m}_c(m_c)$&0.4084(144)&1.26&\\
1.5&0.3649(110)&1.176(5)&\\
2&0.3120(77)&1.069(9)&\\
2.5&0.2812(61)&1.005(10)&\\
3.0&0.2606(51)&0.961(10)&\\
3.5&0.2455(45)&0.929(11)&\\
4.0&0.2339(41)&0.903(11)& \\
4.5&0.2246(37)&0.882(11)&\\
5.0&0.2169(35)&0.865(11)&\\
5.5&0.2104(33)&0.851(12)&\\
6.0&0.2049(30)&0.838(12)&\\
\hline
\hline
\end{tabular}}
}
\label{tab:alfa}
\end{table}
} 
\nin
\hspace*{0.5cm} \b
The QCD parameters which shall appear in the following analysis will be the charm and bottom quark masses $m_{c,b}$ (we shall neglect  the light quark masses $q\equiv u,d$),
the light quark condensate $\qq$,  the gluon condensates $ \lag
\alpha_sG^2\rag
\equiv \la \alpha_s G^a_{\mu\nu}G_a^{\mu\nu}\ra$ 
and $ \la g^3G^3\ra
\equiv \la g^3f_{abc}G^a_{\mu\nu}G^b_{\nu\rho}G^c_{\rho\mu}\ra$, 
the mixed condensate $\la\bar qGq\ra
\equiv {\la\bar qg\sigma^{\mu\nu} (\lambda_a/2) G^a_{\mu\nu}q\ra}=M_0^2\la \bar qq\ra$ 
and the four-quark 
 condensate $\rho\alpha_s\la\bar qq\ra^2$, where
 $\rho\simeq 3-4$ indicates the deviation from the four-quark vacuum 
saturation. Their values are given in Table \ref{tab:param}. 

\b We shall work with the running
light quark condensates and masses. 
They read:
\beq
{\la\bar qq\ra}(\tau)=-{\hat \mu_q^3  \ga-\beta_1a_s\dr^{2/{
\beta_1}}},~~~~~~~~~
{\la\bar q Gq\ra}(\tau)=-{M_0^2{\hat \mu_q^3} \ga-\beta_1a_s\dr^{1/{3\beta_1}}}~,
\label{d4g}
\eeq
$\hat\mu_q$ is the spontaneous RGI light quark condensate \cite{FNR}. 
We shall use:
\beq   
\alpha_s(M_\tau)=0.325(8) \leadsto  \alpha_s(M_Z)=0.1192(10)
\label{eq:alphas}
\eeq
from $\tau$-decays \cite{SNTAU,BNPa,BNPb} (see also:\,\cite{PICHTAU,DAVIER,BOITOa,BOITOb}), which agree perfectly with the world average 2014 \cite{BETHKE,PDG}: 
\beq
\alpha_s(M_Z)=0.1184(7)~. 
\eeq
The value of the running $\la \bar qq\ra$ condensate is deduced from  the well-known GMOR relation: 
\beq
(m_u+m_d)\la \bar uu+\bar dd\ra=-m_\pi^2f_\pi^2~,
\eeq
where $f_\pi=130.4(2)$ MeV\,\cite{ROSNERb} and the value of $(\overline{m}_u+\overline{m}_d)(2)=(7.9\pm 0.6)$ MeV obtained in  \cite{SNmassa,SNmassb} which agrees with the PDG  in\,\cite{PDG}  and lattice averages in\,\cite{LATT13}. Then, we deduce the RGI light quark spontaneous mass $\hat\mu_q$ given  in Table~\ref{tab:param}. 

\b For the heavy quarks, we shall use the running mass and the corresponding value of $\alpha_s$ evaluated at the scale $\mu$. These sets of correlated parameters are given in Table \ref{tab:alfa} for different values of $\mu$ and for a given number of flavours $n_f$.

\b For the $\la \alpha_s G^2\ra$ condensate, we have enlarged the original error by a factor about 3 in order to have
a conservative result and to recover the original SVZ estimate and the alternative extraction in\,\cite{IOFFEa,IOFFEb} from charmonium sum rules which we consider as the most reliable channel for extracting phenomenologically this condensate.  However, a direct comparison of this  range of values obtained within short QCD series (few terms) with the one from lattice calculations\,\cite{BALIa} obtained within a long QCD series remains to be clarified\,\cite{BALIb}. 

\b Some other estimates of the gluon and four-quark condensates using $\tau$-decay and $e^+e^-\to I=1$ hadrons data can be found in\,\cite{DAVIER,BOITOa,BOITOb,PICHTAU2,FESRa,FESRb}. Due to the large uncertainties induced by the different resummations of the QCD series and by the less-controlled effects of some eventual duality violation, we do not consider explicitly these values in the following analysis. However, we shall see later on that the effects of the gluon and four-quark condensates on the values of the decay constants and masses are quite small though they play an important role in the stability analysis. 
\section{Mass and coupling of the $ B^*K$ $(1^{+})$  axial-vector molecule}
\subsection{$\tau$- and $t_c$-stability criteria at lowest order (LO)}
\begin{figure}[hbt] 
\begin{center}
{\includegraphics[width=6.2cm  ]{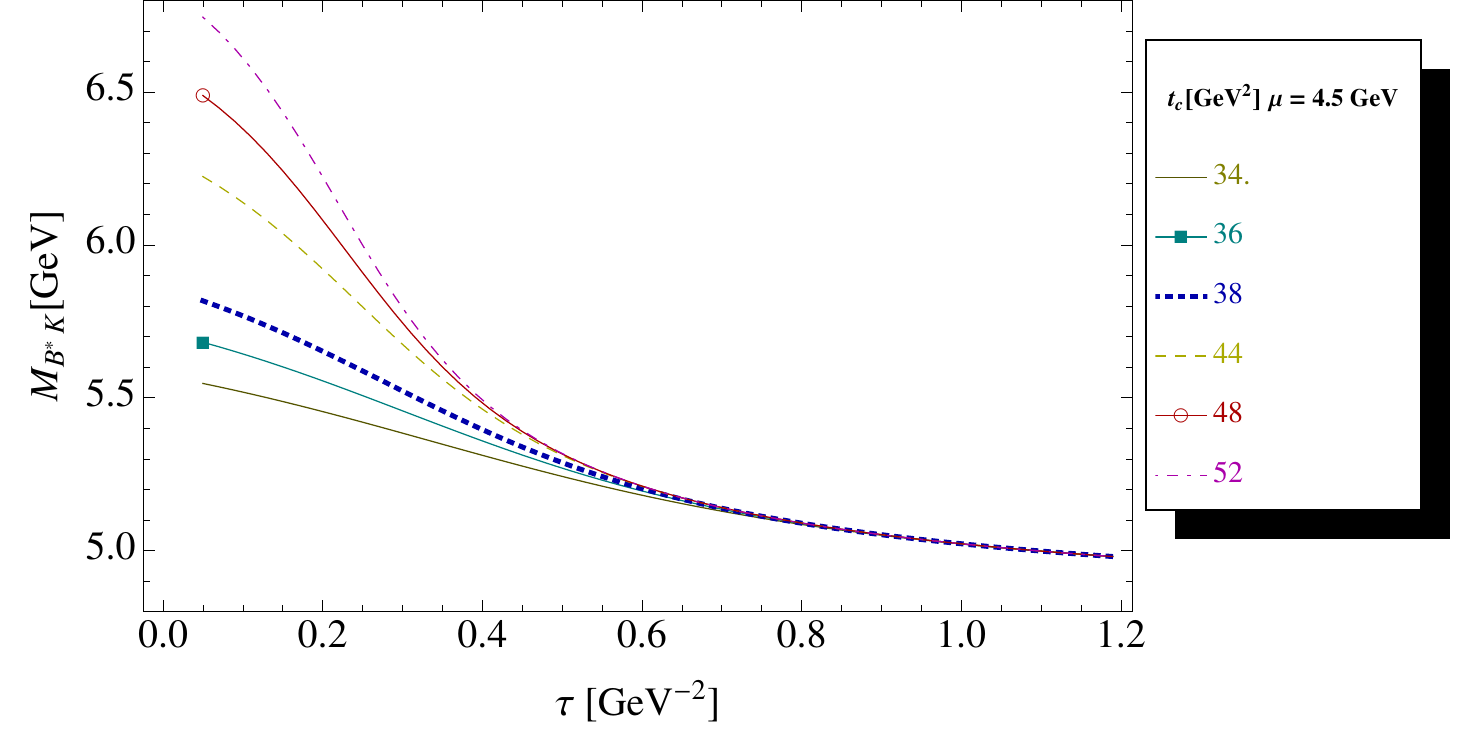}}
{\includegraphics[width=6.2cm  ]{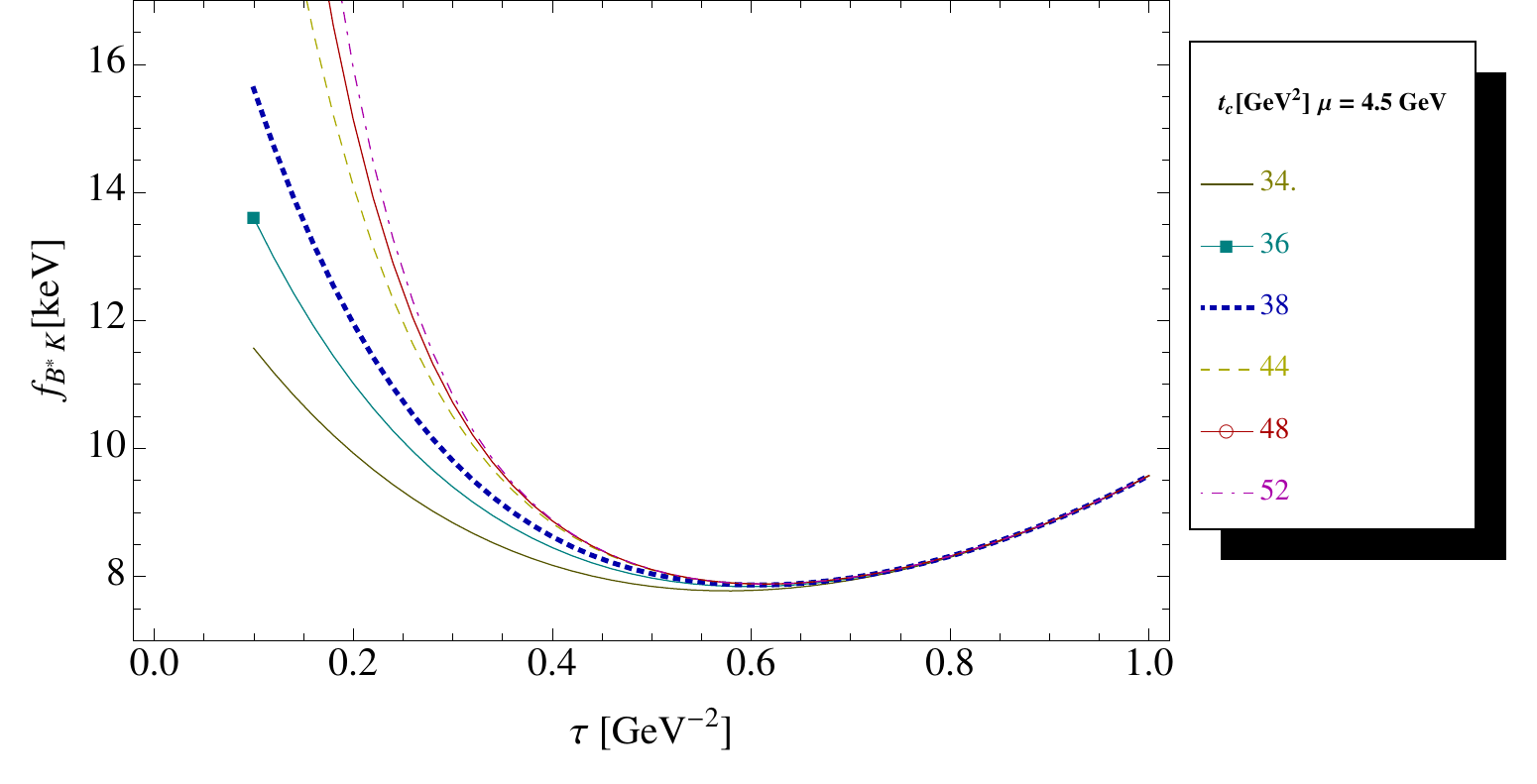}}
\centerline {\hspace*{-3cm} a)\hspace*{6cm} b) }

\caption{
\scriptsize 
{\bf a)} $M_{B^*K}$  at LO as function of $\tau$ for different values of $t_c$, $\mu=4.5$ GeV and for the QCD parameters in Table\,\ref{tab:param} and \ref{tab:alfa}; {\bf b)} The same as a) but for the coupling $f_{B^*K}$.
}
\label{fig:bkstar-lo} 
\end{center}
\end{figure}

\hspace*{0.5cm} \b We show in Fig.\,\ref{fig:bkstar-lo} the $\tau$- and $t_c$-behaviours of the mass and coupling of the $ B^*K$  axial-vector molecule at lowest order (LO) of perturbative QCD. We have used the running $\bar m_b(\mu)$ $b$-quark mass (though not a priori justified), while the OPE is truncated at $d=6$.

\nin

\b One can notice that the stabilities of both curves are reached for $\tau\simeq  (0.45-0.5)$ GeV$^{-2}$. However, an important difference appears here : the coupling is dominated by the $d\leq 6$ non-perturbative condensates (minimum in $\tau$) while the mass is dominated by the perturbative one at the inflexion point. This feature  indicates already the exotic nature of this molecule state compared to ordinary heavy-light $B$-meson\,\cite{SNB4a,SNB4b,SNB1,SNB2,SNFB12a,SNFB12b,SNFB15a,SNFB15b}.

\b We consider as an optimal value from the analysis, the one where the $\tau$-stability about 0.6 GeV$^{-2}$ starts for $t_c\geq 34$ GeV$^2$, while the minimum sensitivity on the change of the continuum threshold $t_c$ is reached  for $t_c\simeq (44-48)$ GeV$^2$. Using $\bar m_b(m_b)$ in Table\,\ref{tab:param}, we consider as a conservative optimal value the results inside the above range:
\bea
M_{B^*K}^{LO}\simeq( 5.19\sim 5.20)~{\rm GeV}~~~~~{\rm and}~~~~~ f_{B^*K}^{LO}\simeq (7.77\sim 7.88)~{\rm keV}
\eea

\b Often added in the literature are the contributions of higher dimension condensates $d\geq 7$
for restoring the $\tau$-stability of the sum rules. However, this procedure is not very helpful as the 
contribution of included high-dimension condensates  come only from one class of diagrams generated by the quark propagator put in an external gluon fields. Moreover, these high-dimension condensates are poorly controlled due to the violation of factorization which already at $d=6$ is expected to be violated by a factor 3-4 (see Table\,\ref{tab:param}). Therefore, we conclude that predictions strongly depending on these high-dimension condensates are not reliable\,\footnote{We shall explicitly check in the case of scalar molecule, which presents the same feature, that the $d=7$ condensates contributions in Eq.\,\ref{eq:d7scalar} are negligible allowing a violation of factorization by a factor 4.}. 
\begin{figure}[hbt] 
\begin{center}
{\includegraphics[width=6.2cm  ]{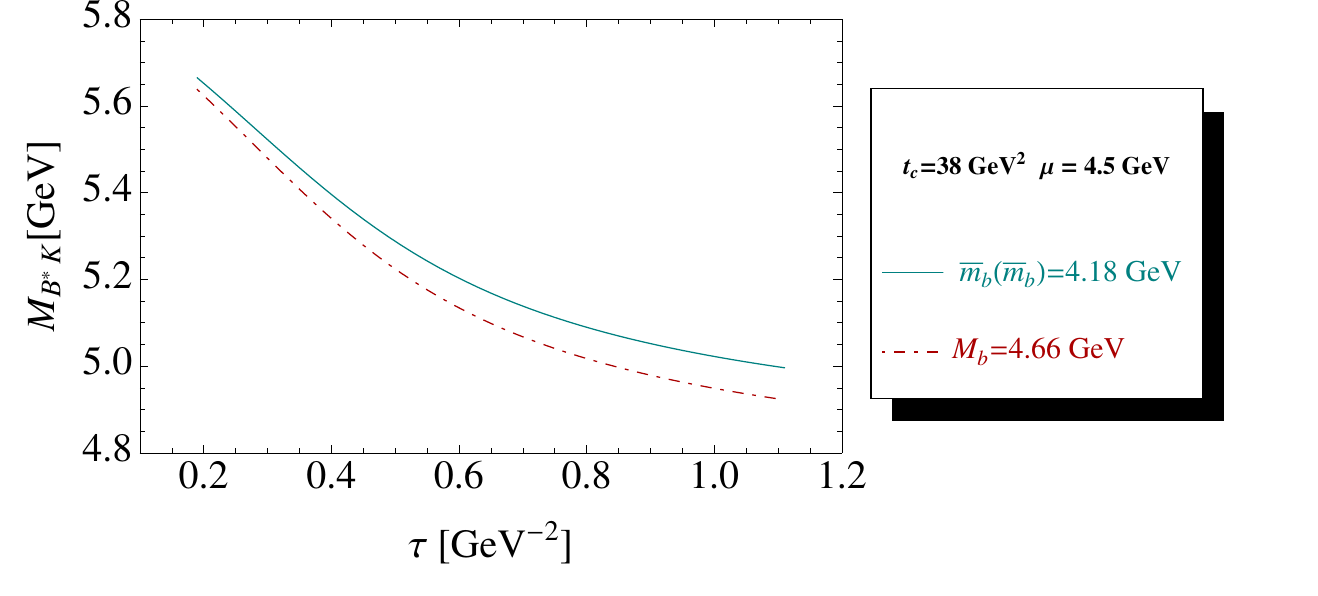}}
{\includegraphics[width=6.2cm  ]{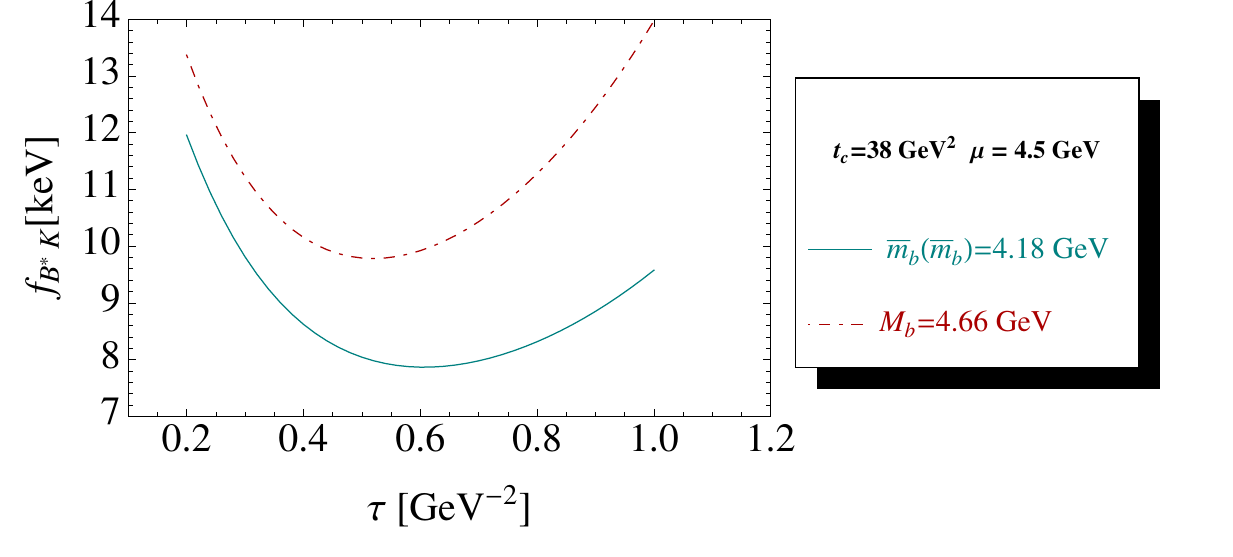}}
\centerline {\hspace*{-3cm} a)\hspace*{6cm} b) }
\caption{
\scriptsize 
{\bf a)} $M_{B^*K}$  at LO as function of $\tau$ for a given value of $t_c=38$ GeV$^2$,  $\mu=4.5$ GeV and for the QCD parameters in Tables\,\ref{tab:param} and \ref{tab:alfa}. The OPE is truncated at $d=6$ and the PT is at lowest order.  We use the on-shell or pole mass $M_b=4.66$ GeV and the running mass $\bar m_b(\bar m_b)=4.18$ GeV; {\bf b)} The same as a) but for the coupling $f_{B^*K}$.
}
\label{fig:bstarkmasspole} 
\end{center}
\end{figure} 
\nin
\subsection{$b$-quark mass ambiguity at lowest order (LO)}
Often used in the existing literature is the value of the running heavy quark mass $\bar m_b(m_b)$ into
the expression of the spectral function evaluated at lowest order (LO) of perturbation 
theory (PT). Though, one should go beyond LO to see the clear selection among the two mass definitions, this is certainly misleading as the spectral function has been evaluated
on-shell such that the mass to be used should be the on-shell or pole quark mass. 
We show in Fig.\,\ref{fig:bstarkmasspole} the comparison of the result when we use the $b$-quark pole mass value of 4.66 GeV\,\cite{PDG} and the running mass $\bar m_b(m_b)=4.18$ GeV in Table\,\ref{tab:param}. This choice
introduces an intrinsic source of error:
\beq
\Delta M_{B^*K}^{LO}\simeq 80 ~{\rm MeV}~~~~~{\rm and}~~~~~ \Delta f_{B^*K}^{LO}\simeq 2~{\rm keV}~,
\eeq
 which is never considered in the existing literature.
\subsection{Higher order perturbative QCD corrections}
For a more reliable prediction, one should improve the previous LO estimate. In so doing, we use the fact that the molecule spectral function is the convolution of the one of two-quark bilinear currents (factorization) which is a consequence of the molecule state definition. In the case of the $B^*K$ molecule, this corresponds to the convolution of the $B^*$ and $K$ mesons spectral functions. The convolution expression is given in Eq.\,\ref{eq:axial}. 

\begin{figure}[hbt] 
\begin{center}
{\includegraphics[width=6.2cm  ]{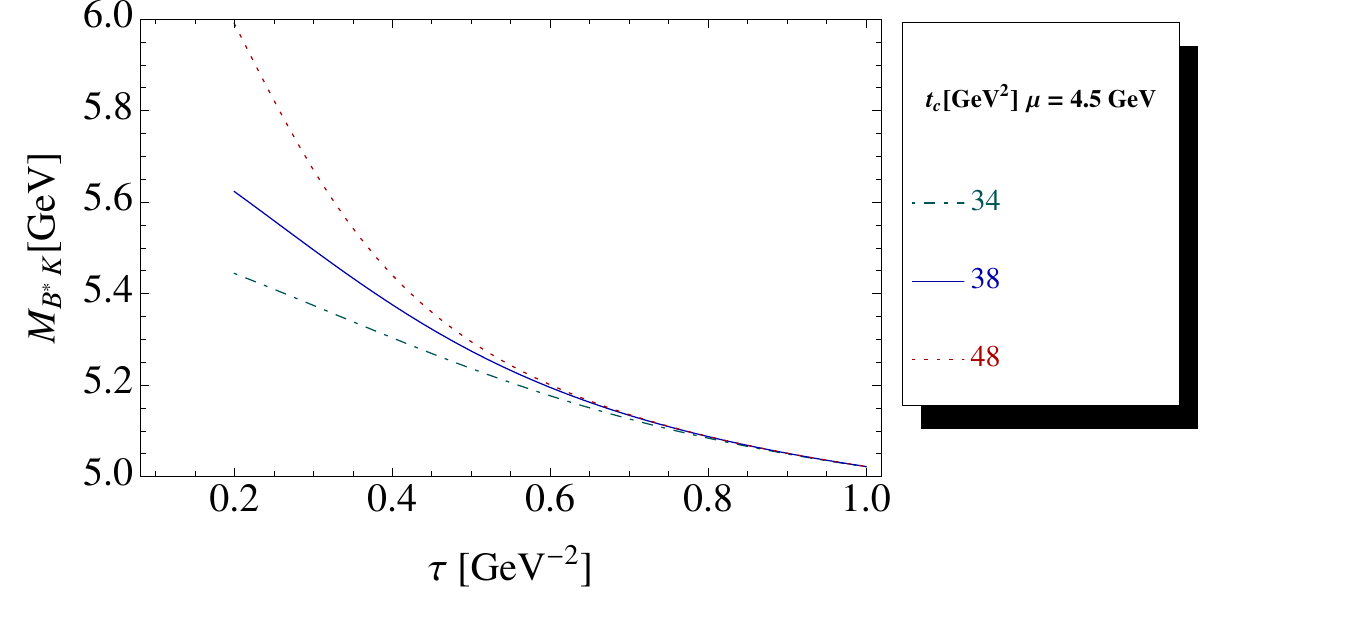}}
{\includegraphics[width=6.2cm  ]{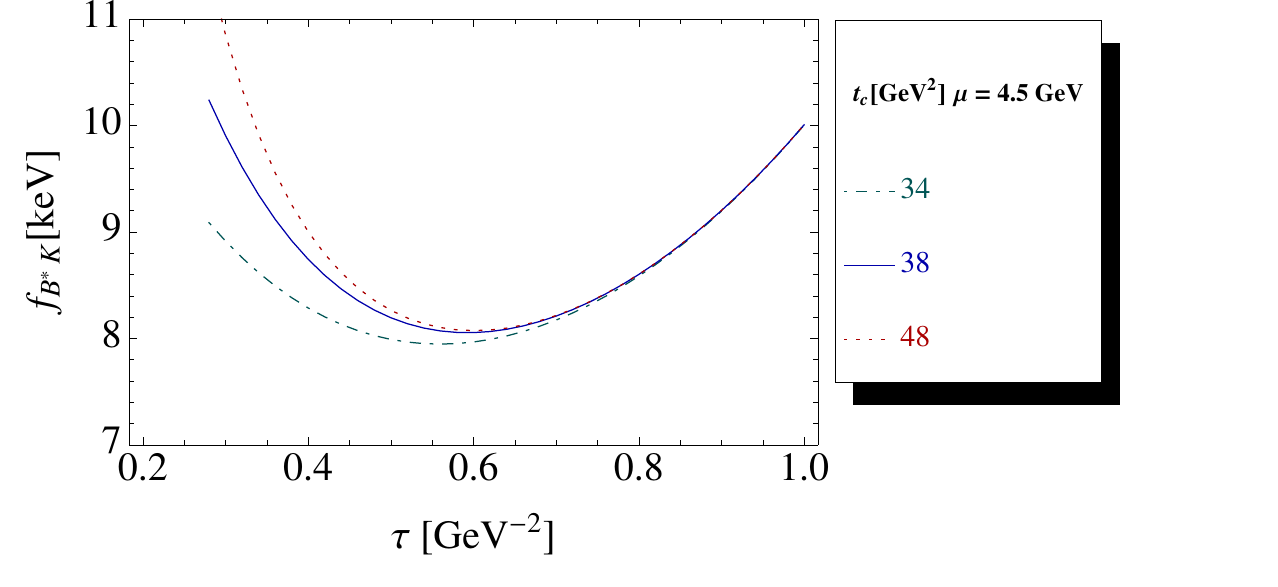}}
\centerline {\hspace*{-3cm} a)\hspace*{6cm} b) }
\caption{
\scriptsize 
{\bf a)} $M_{B^*K}$  at NLO as function of $\tau$ for different values of $t_c$, $\mu=4.5$ GeV and for the QCD parameters in Tables\,\ref{tab:param} and \ref{tab:alfa}; {\bf b)} The same as a) but for the coupling $f_{B^*K}$.
}
\label{fig:bstark-nlo} 
\end{center}
\end{figure}

\b We show the results including the NLO perturbative corrections in Fig.\,\ref{fig:bstark-nlo} versus $\tau$ for three values of $t_c$. 
\nin
The inclusion of the NLO corrections modify the previous LO results for $t_c=(34\sim 48)$ GeV$^2$ and  $\tau\simeq (0.56\sim 0.60)~{\rm GeV}^{-2}$ to:
\beq
M_{B^*K}^{NLO}\simeq( 5200\sim 5201)~{\rm MeV}~~~~{\rm and}~~~~
f_{B^*K}^{NLO}\simeq (7.95\sim 8.07)~{\rm keV}~.
\eeq

\begin{figure}[hbt] 
\begin{center}
{\includegraphics[width=6.2cm  ]{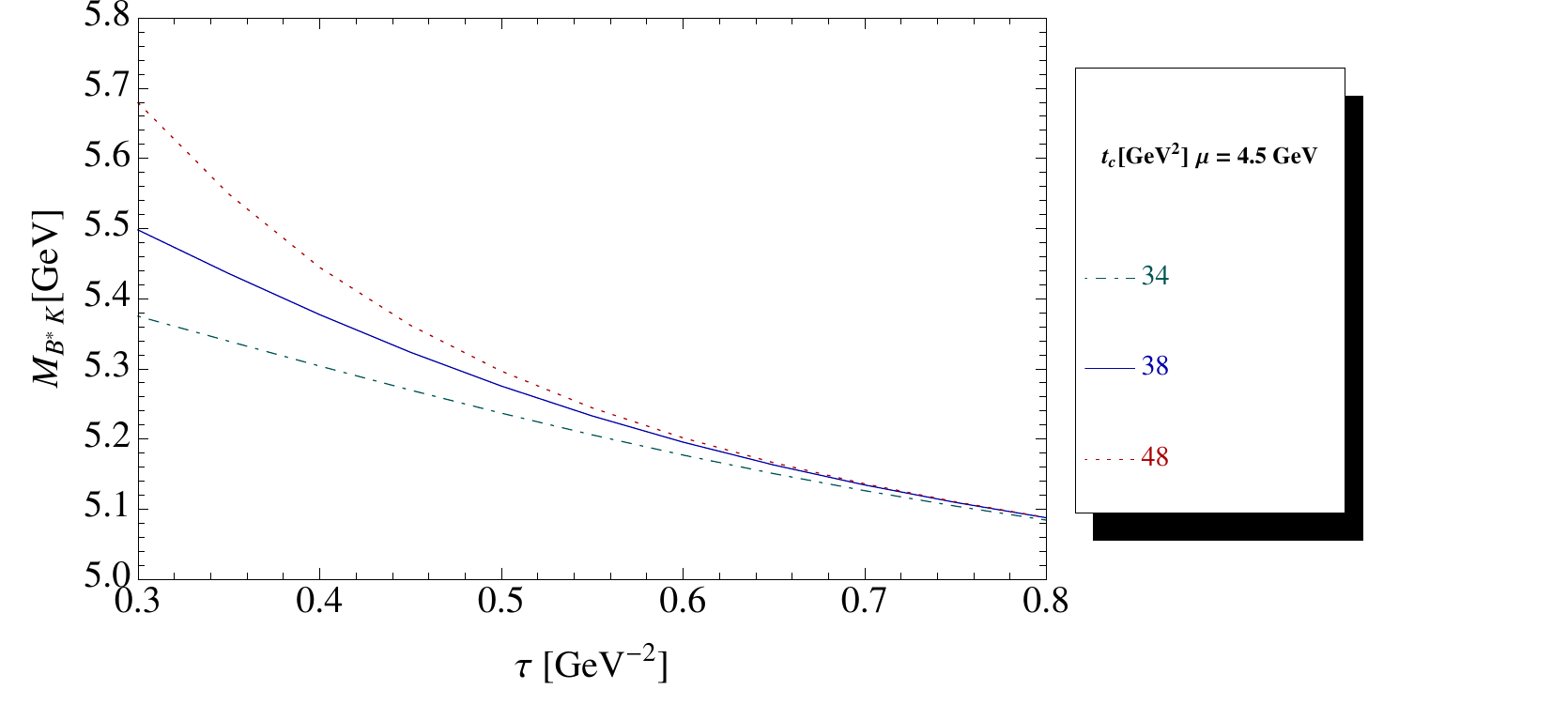}}
{\includegraphics[width=6.2cm  ]{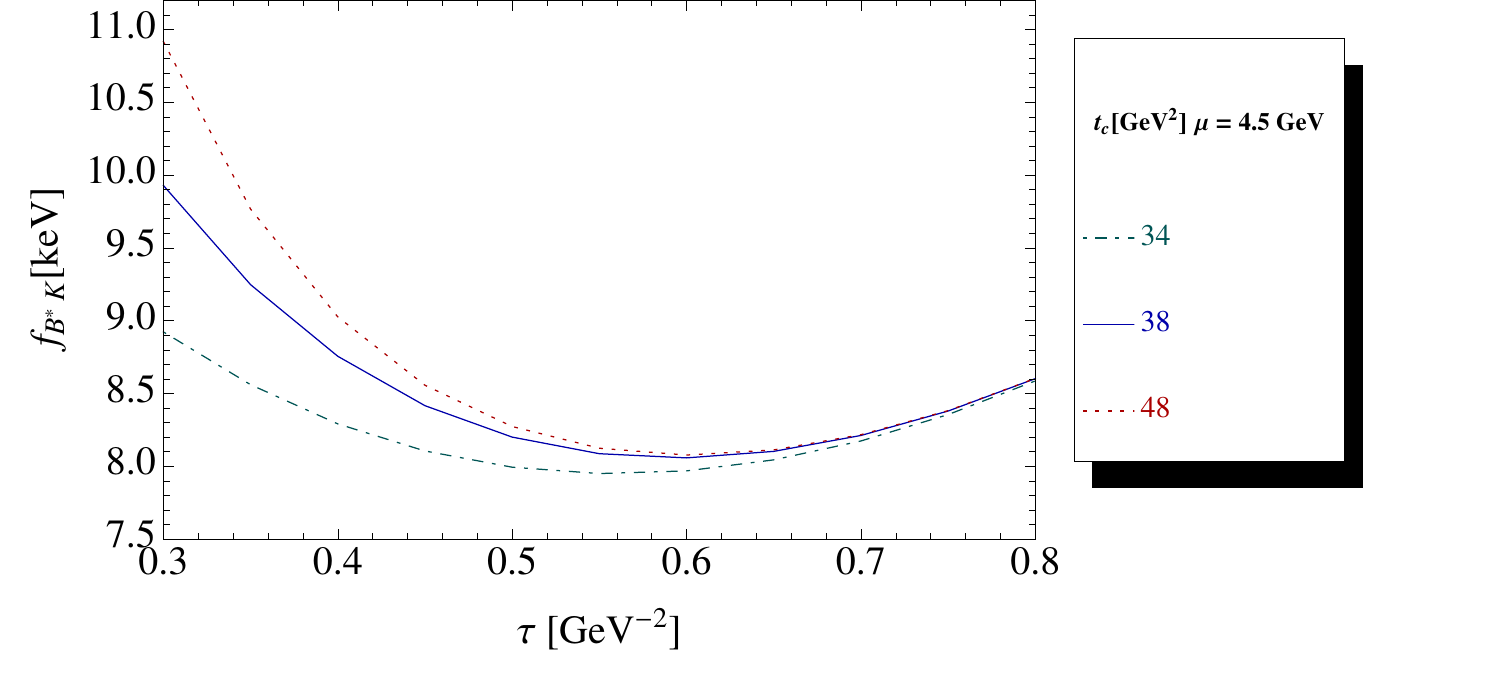}}
\centerline {\hspace*{-3cm} a)\hspace*{6cm} b) }
\caption{
\scriptsize 
{\bf a)} $M_{B^*K}$  at N2LO as function of $\tau$ for  different values of $t_c$, $\mu=4.5$ GeV and for the QCD parameters in Tables\,\ref{tab:param} and \ref{tab:alfa};  {\bf b)} The same as a) but for the coupling $f_{B^*K}$.
}
\label{fig:bstark-n2lo} 
\end{center}
\end{figure}

\b In the same way as in previous analysis, we show the results including the N2LO perturbative corrections in Fig.\,\ref{fig:bstark-n2lo} versus $\tau$ for three values of $t_c$. For $t_c=(34\sim 48)$ GeV$^2$ and  $\tau\simeq (0.58\sim 0.62)~{\rm GeV}^{-2}$, we obtain:
\beq
M_{B^*K}^{N2LO}\simeq( 5185\sim 5187)~{\rm MeV}~~~~{\rm and}~~~~
f_{B^*K}^{N2LO}\simeq (7.95\sim 8.09)~{\rm keV}~.
\eeq

\nin
\begin{figure}[hbt] 
\begin{center}
{\includegraphics[width=6.2cm  ]{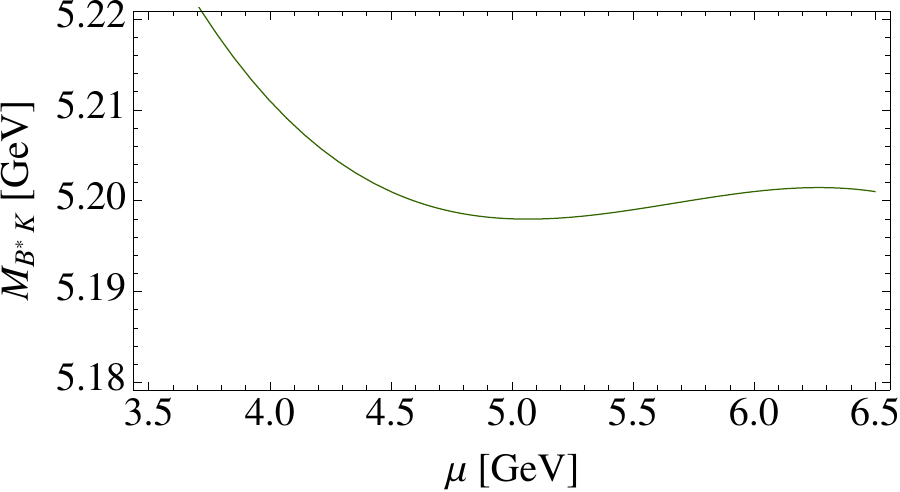}}
{\includegraphics[width=6.2cm  ]{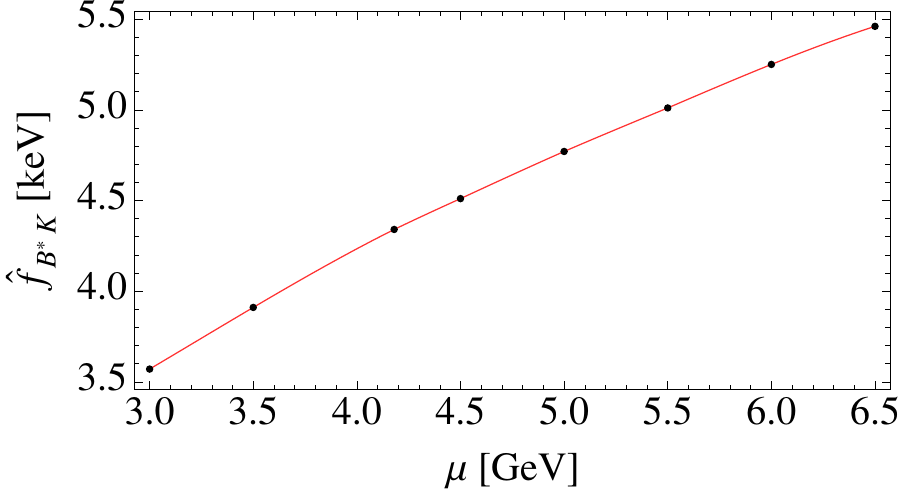}}
\centerline {\hspace*{-3cm} a)\hspace*{6cm} b) }
\caption{
\scriptsize 
{\bf a)} $M_{B^*K}$ at NLO as function of $\mu$ , for the corresponding $\tau$-stability region, for $t_c\simeq 48$ GeV$^2$ and for the QCD parameters in Tables\,\ref{tab:param} and \ref{tab:alfa};  {\bf b)} The same as a) but for the renormalization group invariant coupling $\hat{f}_{B^*K}$.
}
\label{fig:bstark-mu} 
\end{center}
\end{figure} 
\nin
\subsection{$\mu$-subtraction point stability }
We show in Fig.\,\ref{fig:bstark-mu} the dependence of the results obtained at NLO of PT series on the choice of the subtraction constant $\mu$. One can notice that the position of the $\tau$-stability moves slightly with $\mu$ and is in the range of 0.50 to 0.66 GeV$^{-2}$. For each corresponding value of $\tau$-stability, a minimum in $\mu$ is obtained around 5 GeV  for the mass and a slight inflexion point at $\mu\simeq (4.5\sim 5.0)$ GeV for the renormalization group invariant coupling defined in Eq.\,\ref{eq:fhat}, at  which we extract the optimal estimate:
\beq
M_{B^*K}^{NLO}\simeq (5201\sim 5198)~{\rm MeV}~~~~{\rm and}~~~~
\hat{f}_{B^*K}^{NLO}\simeq (4.51\sim 4.77)~{\rm keV}~.
\eeq
\subsection{Test of the convergence of the  PT series}
\begin{figure}[hbt] 
\begin{center}
{\includegraphics[width=6.2cm  ]{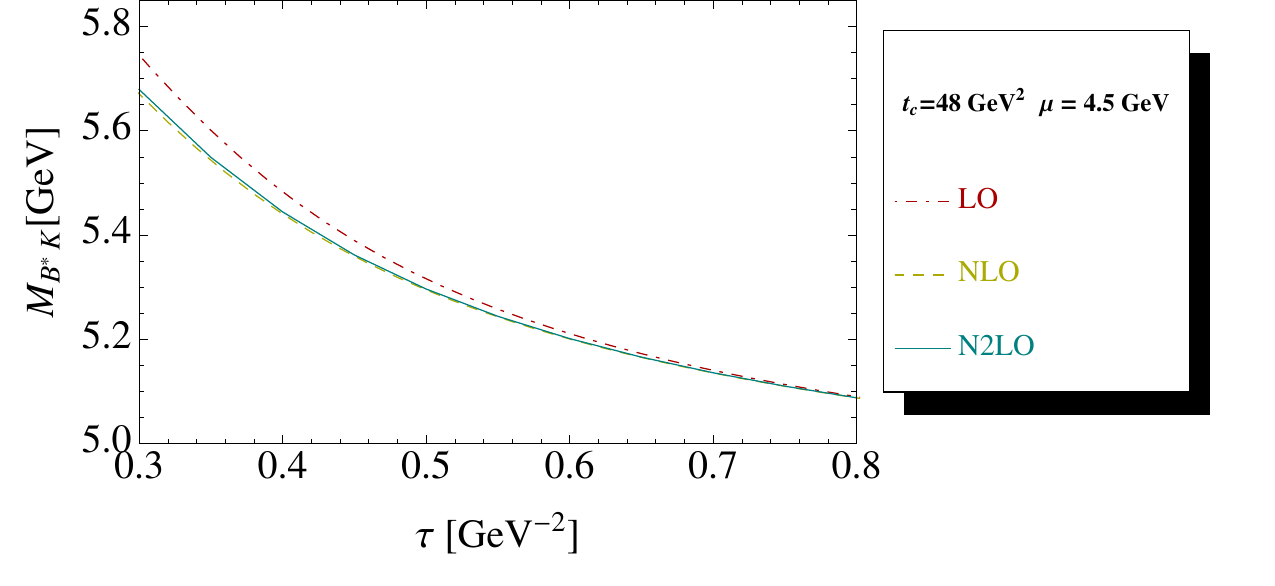}}
{\includegraphics[width=6.2cm  ]{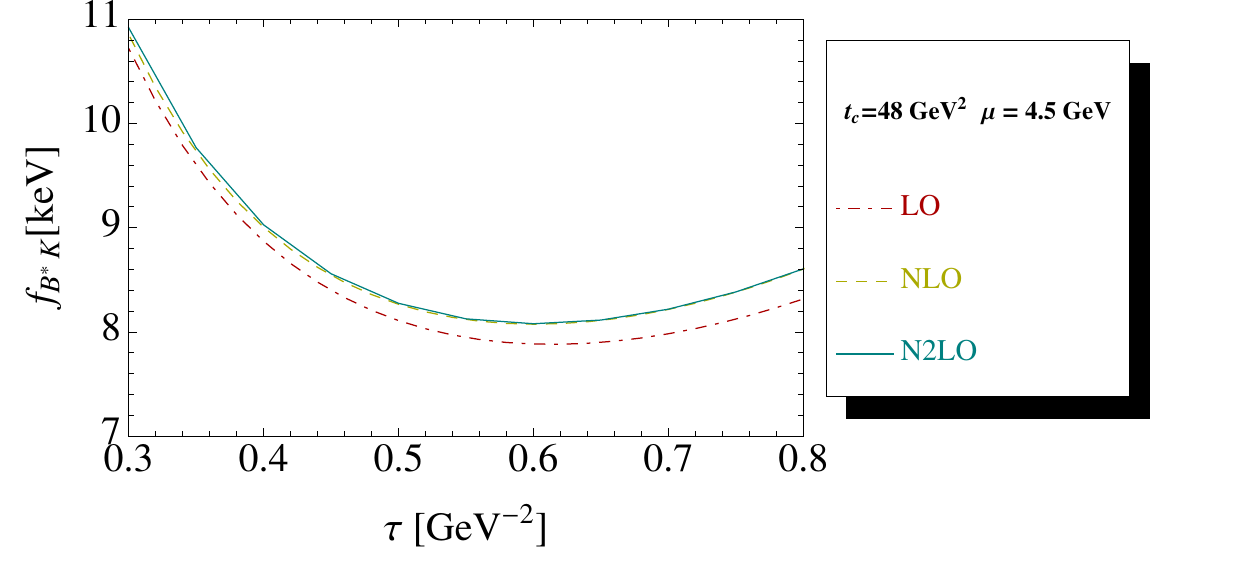}}
\centerline {\hspace*{-3cm} a)\hspace*{6cm} b) }
\caption{
\scriptsize 
{\bf a)} $M_{B^*K}$  as function of $\tau$ for different truncation of the PT series at a given value of $t_c$=48 GeV$^2$, $\mu=4.5$ GeV and for the QCD parameters in Tables\,\ref{tab:param} and \ref{tab:alfa};  {\bf b)} The same as a) but for the coupling $f_{B^*K}$.
}
\label{fig:bstark-lo-n2lo} 
\end{center}
\end{figure}

\hspace*{0.5cm} \b We show,  in Fig.\,\ref{fig:bstark-lo-n2lo}a and \,\ref{fig:bstark-lo-n2lo}b, the behaviour of the results for a given value of $t_c$ for different truncation of the PT series. One can notice small PT corrections from LO to N2LO for the mass  predictions as these corrections tend to compensate in the ratio of sum rules. There is also  a good convergence of the PT series for the coupling prediction. 

\b The good convergence of the PT series indicate that the HO corrections dual to the tachyonic gluon one are negligible.  

\b It also validates {\it (a posteriori)} the results
obtained at LO using the running $b$-quark mass as input as usually done in the existing literature. 
\nin
\subsection{Test of the convergence of the  OPE}
\begin{figure}[hbt] 
\begin{center}
{\includegraphics[width=6.2cm  ]{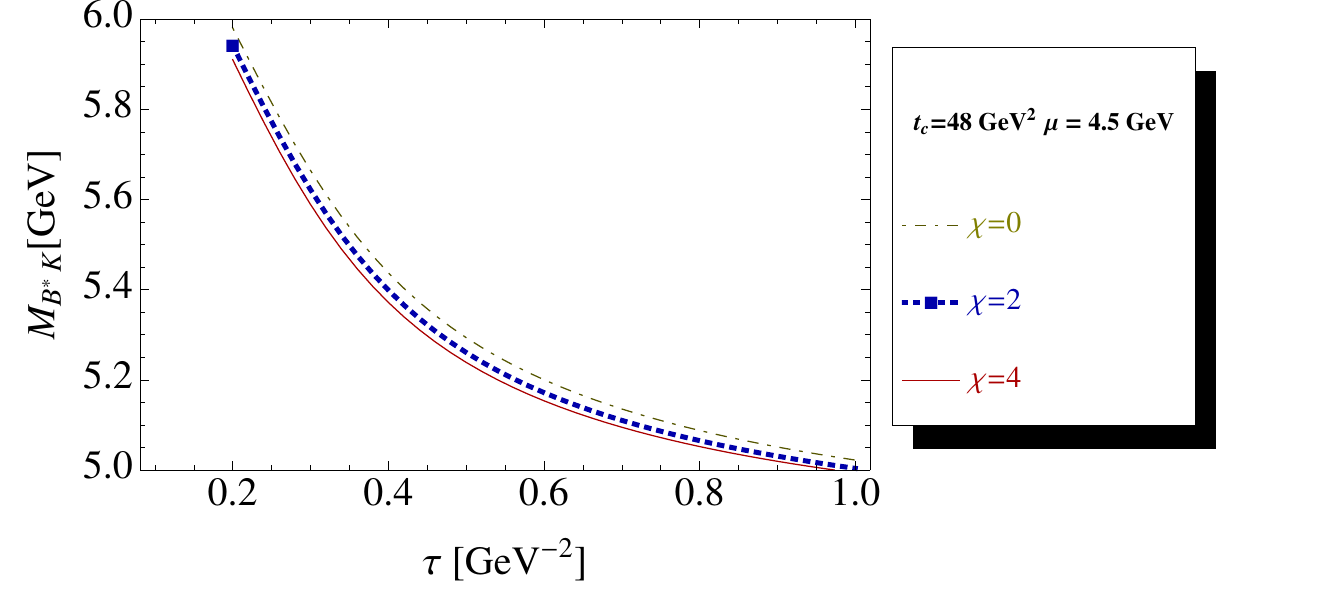}}
{\includegraphics[width=6.2cm  ]{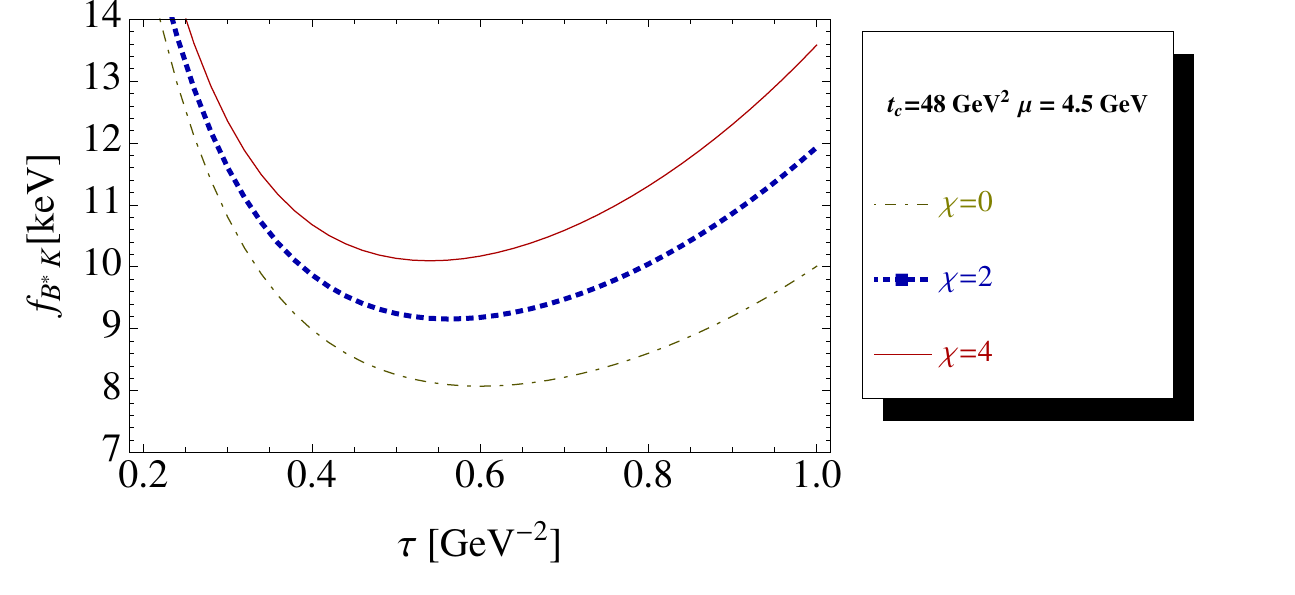}}
\centerline {\hspace*{-3cm} a)\hspace*{6cm} b) }
\caption{
\scriptsize 
{\bf a)} $M_{B^*K}$  as function of $\tau$, for different values of the $d=7$ condensate contribution ($\chi$ measures the violation of factorization),  at a given value of $t_c$=48 GeV$^2$, $\mu=4.5$ GeV and for the QCD parameters in Tables\,\ref{tab:param} and \ref{tab:alfa}; {\bf b)} The same as a) but for the coupling $f_{B^*K}$.
}
\label{fig:bstark-d7} 
\end{center}
\end{figure}

We test the convergence of the OPE by adding the contribution of the $d=7$ condensate given in Eq.\,\ref{eq:d7axial} and allowing a violation of factorization up to a factor $\chi= 4$. We see its effect in Fig.\,\ref{fig:bstark-d7} where the coupling is largely affected while the one on the mass is almost negligible. Taking the maximal error corresponding to $\chi=4$ and $t_c\simeq (34\sim 48)$ GeV$^2$, we estimate the error due to the truncation of the OPE as:
\beq
 \Delta f_{B^*K}^{OPE}\simeq  \pm 1.98 ~{\rm keV}~,~~~~~~~~~~~~~~\Delta M_{B^*K}^{OPE}\simeq \pm 3 ~{\rm MeV}~. 
 \eeq
\nin
\subsection{Final results and error estimates}
We show in Table\,\ref{tab:error} our estimate of the different sources of errors. The main errors from the mass 
come from the localisation of the $\tau$ and $\mu$ stabilities and to lesser extent from the QCD input parameters: $\alpha_s,~\la \bar qq\ra^2$ and the $SU(3)$ breaking parameter $\kappa\equiv \la \bar ss\ra/\la\bar qq \ra$. 
Adding quadratically different
 sources of errors, 
we consider as final estimate to order $\alpha_s^2$ or at  N2LO of the perturbative series and for $\mu=4.5$ GeV:
\bea
M_{B^*K}&\simeq&( 5186\pm 13)~{\rm MeV}~,\nnb\\
\hat f_{B^*K}&\simeq& (4.48\pm 1.45)~{\rm keV}~~\Lrar~~f_{B^*K}(4.5)\simeq (8.02\pm 2.60)~{\rm keV}~.
\label{eq:bstarkn2lo}
\eea
One can notice that the $B^*K$ molecule mass (if any) is expected to be much below the physical $B^*K$ threshold of 5818 MeV. 
{\scriptsize
\begin{table}[hbt]
 \tbl{Different sources of errors for the estimate of the molecule masses (in units of MeV) and couplings (in units of keV) in the $b$-quark channel. }  
    {\scriptsize
 {\begin{tabular}{@{}llllllllll@{}} \toprule
&\\
\hline
\hline
\bf Inputs $[GeV]^d$&$\Delta M_{B^*K}$&$\Delta f_{B^*K}$&$\Delta M_{BK}$&$\Delta f_{BK}$
&$\Delta M_{B^*_s\pi}$&$\Delta f_{B^*_s\pi}$&$\Delta M_{B_s\pi}$&$\Delta f_{B_s\pi}$\\
\hline 
{\it LSR parameters}&\\
$t_c=(34\sim 48)$&1  &0.01& 1& 0.06&    2  &0.09& 0.001& 0.08\\
$\mu=(4.5\sim 5.0)$&2 &0.40&1&0.60&   3 &0.30&3&0.8\\
$\tau=(\tau_{min}\pm 0.02)$&10&0.06&12&0.01&    15&0.01&23&0.06\\
{\it QCD inputs}&\\
$\bar m_b$& 0.83&0.023&0.83&0.023&      0.85&0.03&0.83&0.03\\
$\bar m_s$& 0.17&0.022&0.30&0.020&      0.30&0.03&0.30&0.03\\
$\alpha_s$&5.51&0.18&5.48&0.19&     5.65&0.22&5.51&0.22\\
$\la\bar qq\ra$&0.38&0.006&0.32&0.007&    0.20&0.002&0.18&0.002\\
$\kappa$&4.68&0.93&5.14&0.98&       0.84&0.009&0.75&0.007\\
$\la\alpha_s G^2\ra$&0.96&0.015&1.07&0.016&      0.77&0.01&0.77&0.01\\
$M_0^2$&1.43&0.052&1.0&0.046&      0.79&0.02&0.54&0.02\\
$\la\bar qq\ra^2$&2.97&1.28&2.76&1.32&        2.78&1.50&1.81&1.53\\
$\la g^3G^3\ra$&0.0&0.0&0.0&0.0&0.0&0.0&0.0&0.0\\
$d\geq7$&3.0&1.98&5.0&1.66&5&1.8&1.0&1.48\\
{\it Total errors}&13.4&2.6&15.4&2.4&            17.7&2.40&24.0&2.3\\
\hline\hline
\end{tabular}}
\label{tab:error}
}
\end{table}
} 
\section{Mass and coupling of the $BK$ $(0^{+})$  scalar molecule}
\begin{figure}[hbt] 
\begin{center}
{\includegraphics[width=6.2cm  ]{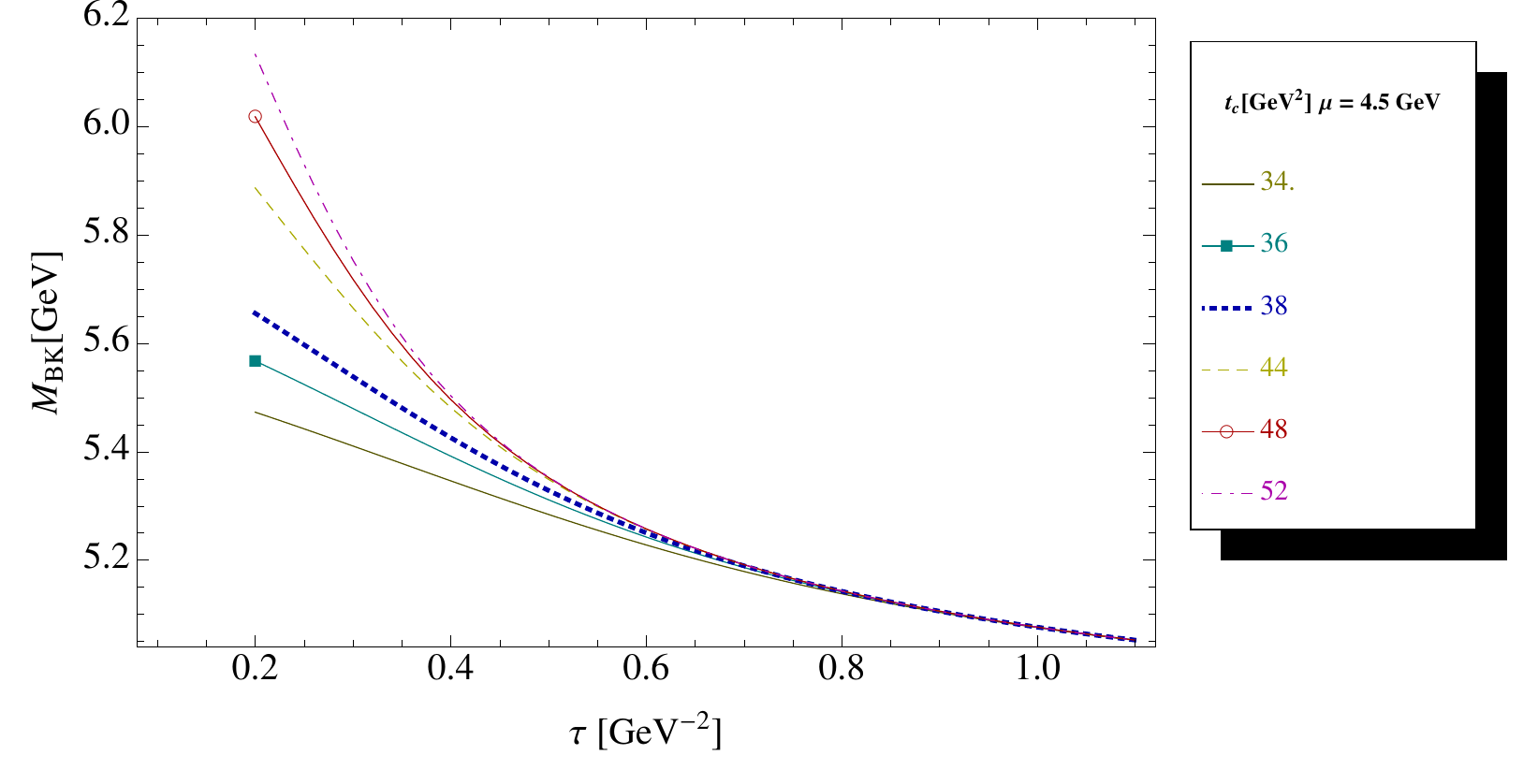}}
{\includegraphics[width=6.2cm  ]{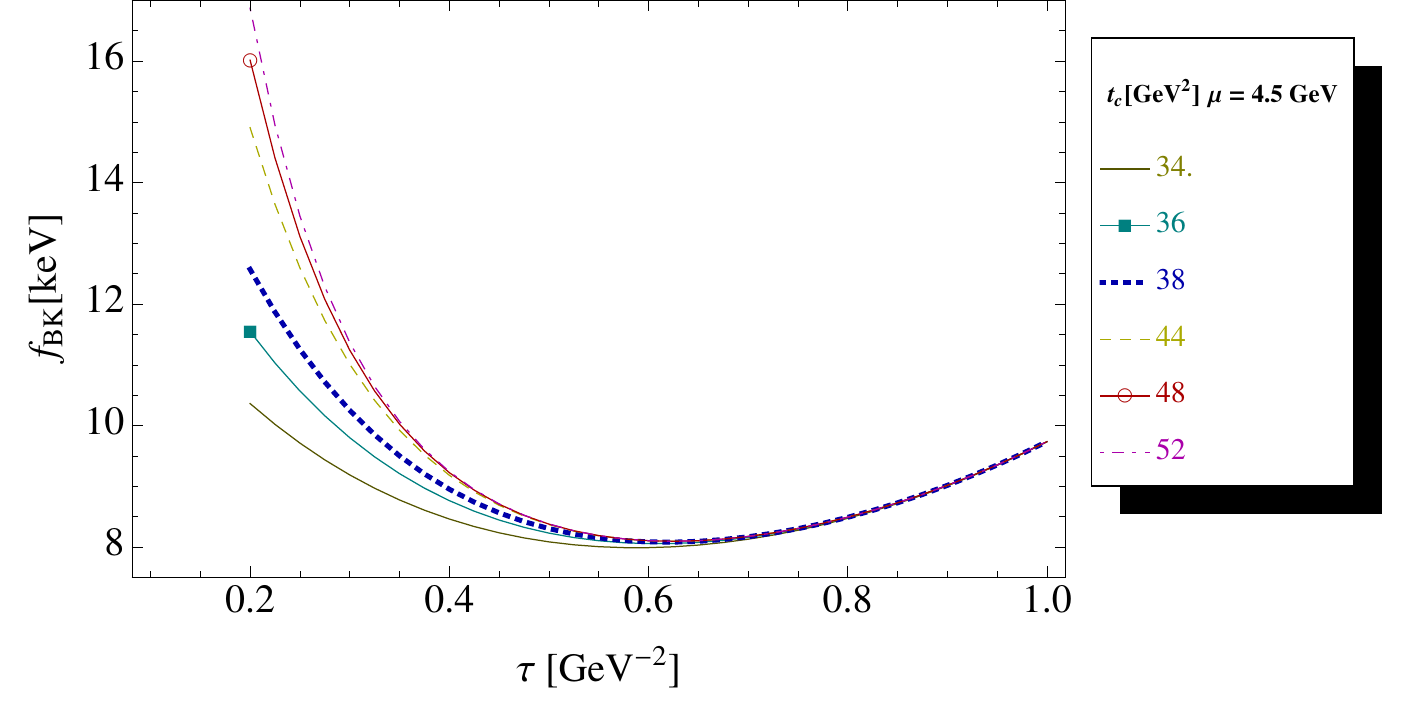}}
\centerline {\hspace*{-3cm} a)\hspace*{6cm} b) }
\caption{
\scriptsize 
{\bf a)} $M_{BK}$  at LO as function of $\tau$ for different values of $t_c$, $\mu=4.5$ GeV and for the QCD parameters in Tables\,\ref{tab:param} and \ref{tab:alfa};  {\bf b)} The same as a) but for the coupling $f_{BK}$.
}
\label{fig:bk-lo} 
\end{center}
\end{figure} 
\nin

\subsection{$\tau$- and $t_c$- stability criteria at lowest order (LO)} 
We redo the previous analysis for the $BK$  scalar molecule. The results are shown in Fig.\,\ref{fig:bk-lo}. For $\tau\simeq  (0.55-0.60) $ GeV$^{-2}$
where the coupling presents a $\tau$-minimum, one obtains for $t_c\simeq (34\sim 48)$ GeV$^2$:
\beq
f_{BK}^{LO}\simeq (8.07\sim 8.21)~{\rm keV}~.
\eeq
For the mass, one obtains an inflexion point for $t_c\simeq (34\sim 48)$ GeV$^2$ and for 
$\tau\simeq 0.60 $ GeV$^{-2}$ which corresponds to: 
\beq
M_{BK}^{LO}\simeq( 5250\sim 5260)~{\rm MeV}~.
\eeq
\begin{figure}[hbt] 
\begin{center}
{\includegraphics[width=6.2cm  ]{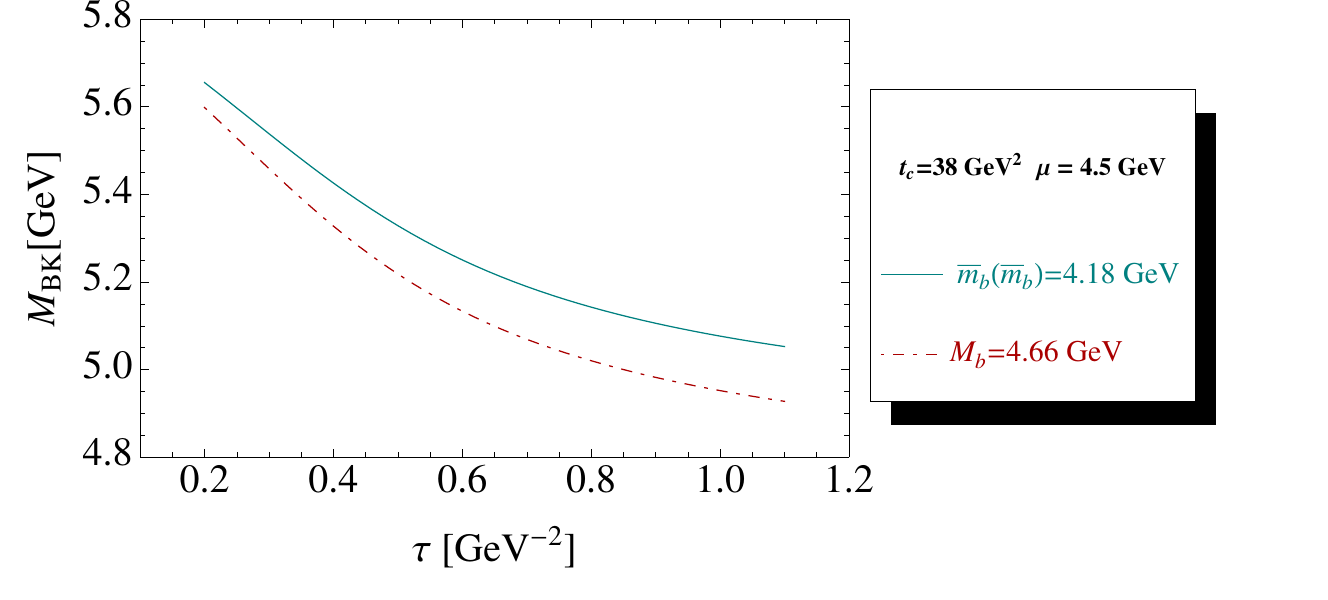}}
{\includegraphics[width=6.2cm  ]{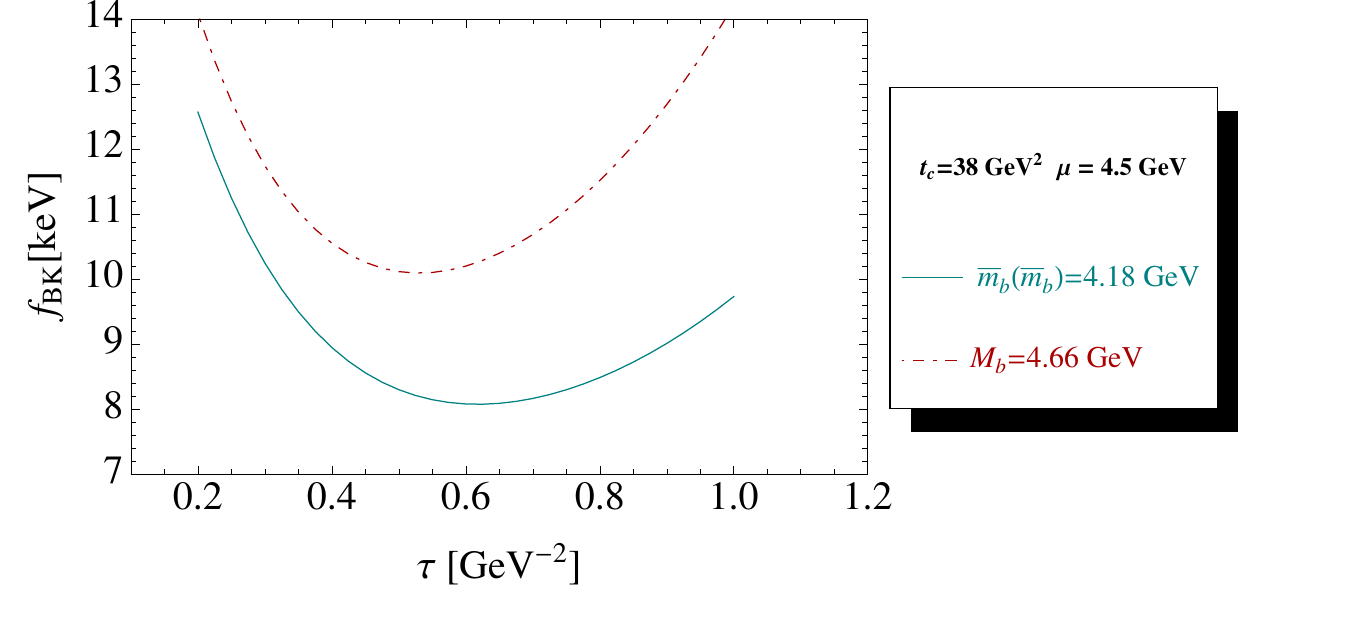}}
\centerline {\hspace*{-3cm} a)\hspace*{6cm} b) }

\caption{
\scriptsize 
{\bf a)} $M_{BK}$  at LO as function of $\tau$ for a given value of $t_c=38$ GeV$^2$,  $\mu=4.5$ GeV and for the QCD parameters in Tables\,\ref{tab:param} and \ref{tab:alfa};  The OPE is truncated at $d=6$.  We compare the effect of  the on-shell or pole mass $M_b=4.66$ GeV and of the running mass $\bar m_b(\bar m_b)=4.18$ GeV; {\bf b)} The same as a) but for the coupling ${f}_{BK}$.
}
\label{fig:bkmasspole} 
\end{center}
\end{figure} 
\nin
\subsection{$b$-quark mass ambiguity at lowest order (LO)}
We show in Fig.\,\ref{fig:bkmasspole} the comparison of the result when we use the $b$-quark pole mass value of 4.66 GeV\,\cite{PDG} and the running mass $\bar m_b(m_b)=4.18$ GeV in Table\,\ref{tab:param}. One can find that this choice
introduces an intrinsic source of error :
\beq
\Delta M_{BK}^{LO}\simeq 120 ~{\rm MeV}~~~~~{\rm and}~~~~~ \Delta f_{BK}^{LO}\simeq 2.1~{\rm keV}~,
\eeq
 which  should be added in the error when one does the LO analysis.
\subsection{Higher order perturbative QCD corrections}
\begin{figure}[hbt] 
\begin{center}
{\includegraphics[width=6.2cm  ]{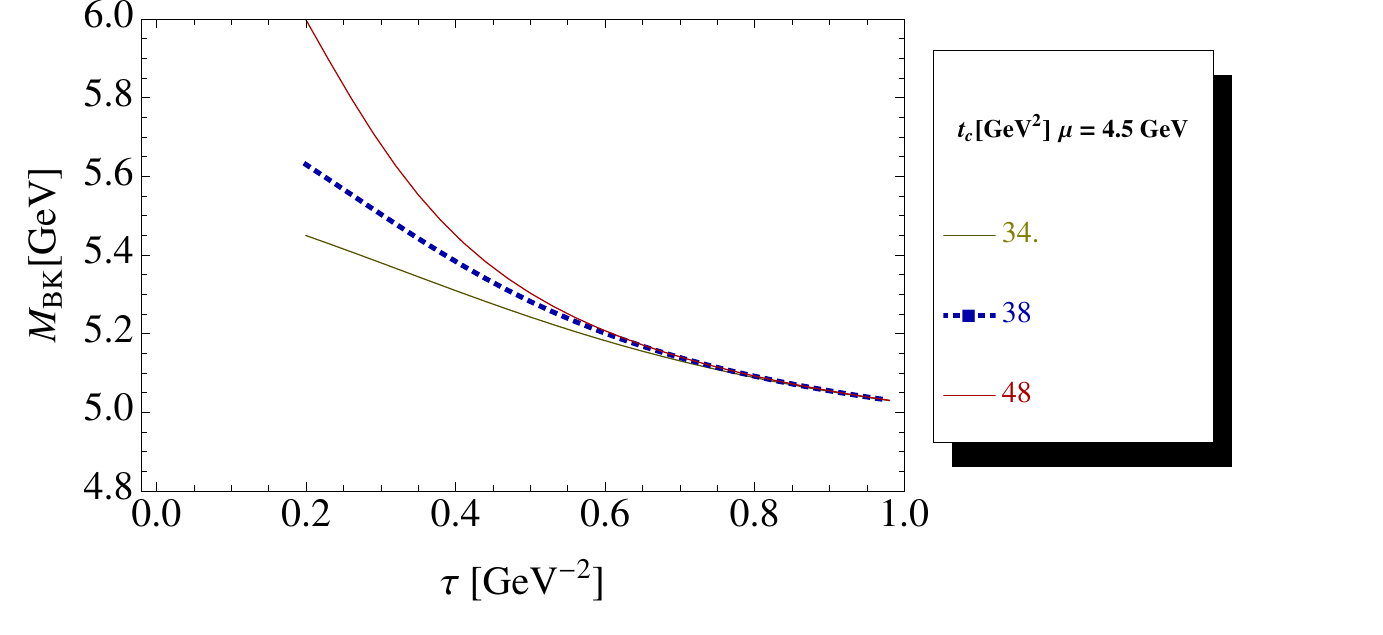}}
{\includegraphics[width=6.2cm  ]{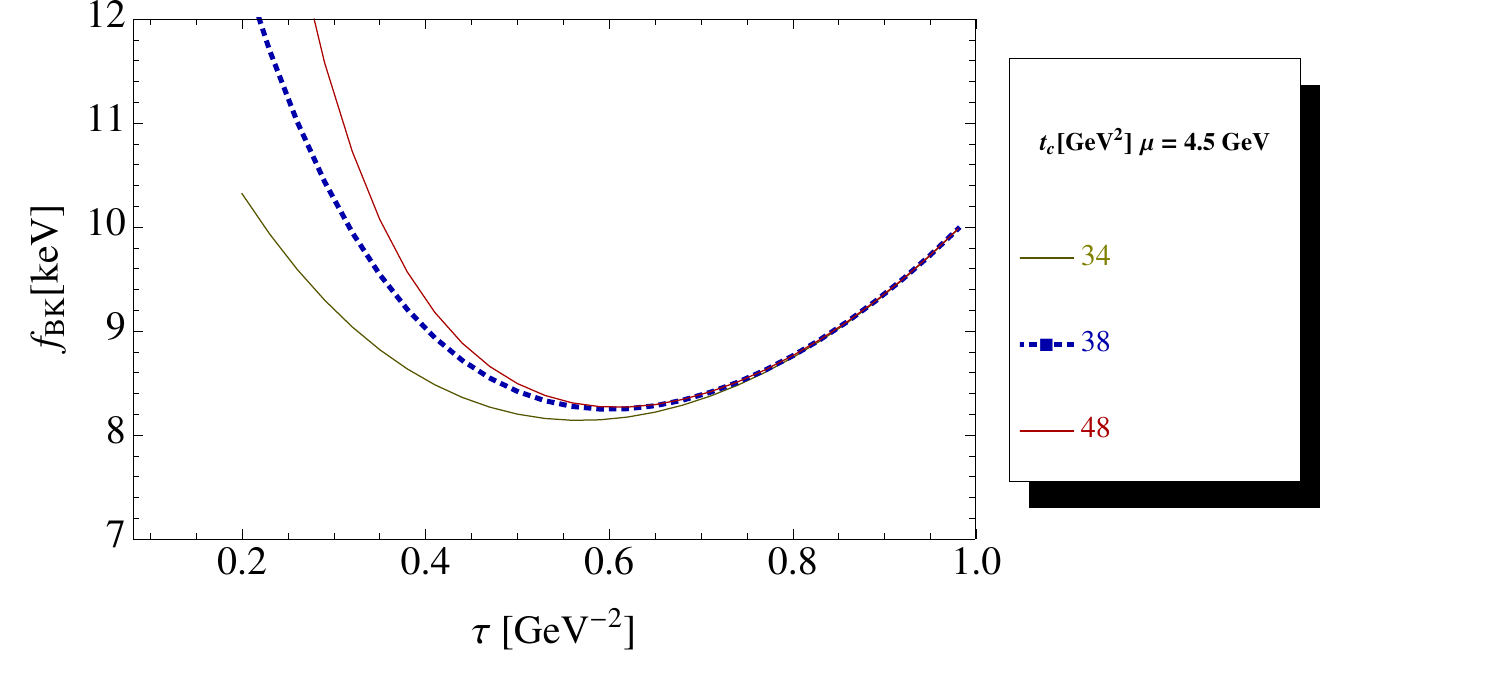}}
\centerline {\hspace*{-3cm} a)\hspace*{6cm} b) }
\caption{
\scriptsize 
{\bf a)} $M_{BK}$  at NLO as function of $\tau$ for different values of $t_c$, for $\mu=4.5$ GeV and for the QCD parameters in Tables\,\ref{tab:param} and \ref{tab:alfa}; {\bf b)} The same as a) but for the coupling ${f}_{BK}$.
}
\label{fig:bk-nlo} 
\end{center}
\end{figure}

\hspace*{0.5cm} \b We improve the previous LO results by including ${\cal O}(\alpha_s)$ (NLO) and ${\cal O}(\alpha^2_s)$ (N2LO) corrections to the LO perturbative expression.
The NLO analysis is shown in Fig.\,\ref{fig:bk-nlo} versus $\tau$, for $\mu=4.5$ GeV  and for different values of $t_c$. We obtain at the optimal stability point $\tau=(0.56\sim 0.62)$ GeV$^{-2}$
and for $t_c\simeq (34\sim 48)$ GeV$^2$:
\beq
 M_{BK}^{NLO}\simeq (5205 \sim 5207)~{\rm MeV}~~~{\rm and}~~~f_{BK}^{NLO}(4.5)\simeq (8.14\sim 8.27)~{\rm keV}~.
  \eeq

\b Adding the N2LO corrections, one obtains similar behaviours as in Fig\,\ref{fig:bk-nlo}. In this case, the results become:
\beq
 M_{BK}^{N2LO}\simeq  5195 ~{\rm GeV}~~~{\rm and}~~~ f_{BK}^{N2LO}(4.5)\simeq (8.18\sim 8.30) ~{\rm keV}~,
\eeq
where one can notice tiny corrections from NLO to N2LO implying also a negligible correction  tachyonic gluon mass contribution. 
\nin
\subsection{$\mu$- subtraction point stability }
\begin{figure}[hbt] 
\begin{center}
{\includegraphics[width=6.2cm  ]{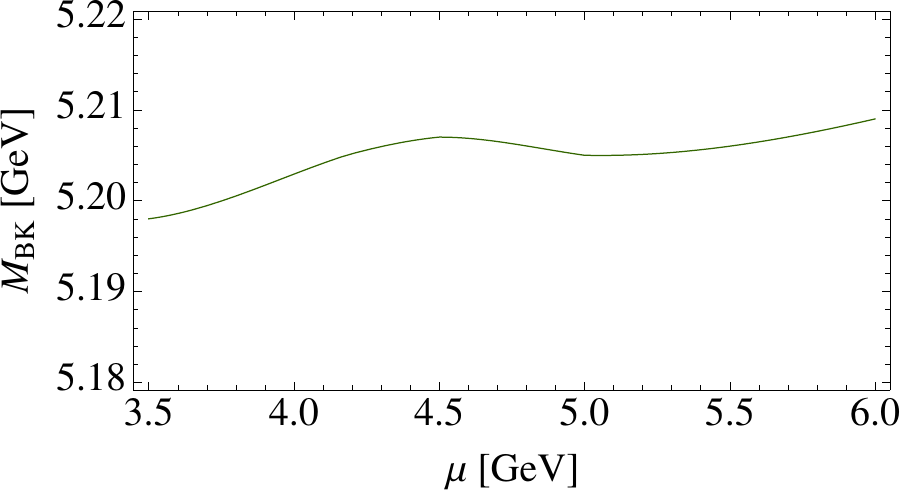}}
{\includegraphics[width=6.2cm  ]{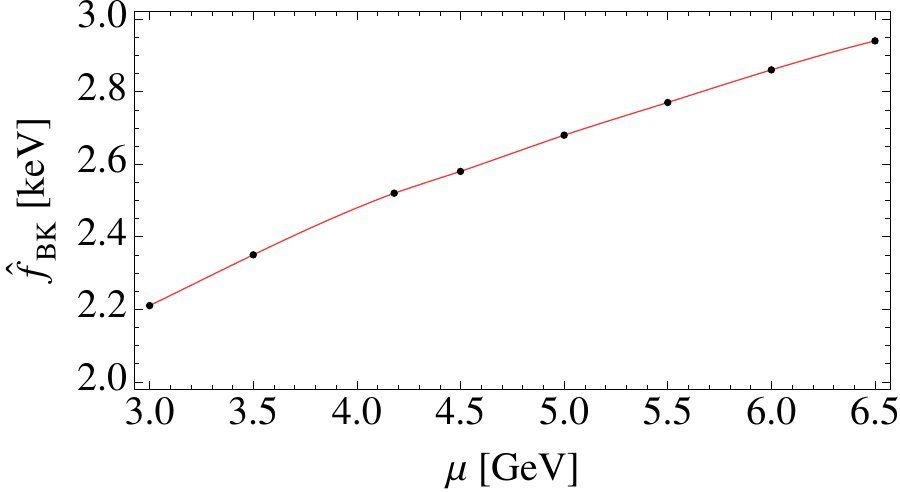}}
\centerline {\hspace*{-3cm} a)\hspace*{6cm} b) }
\caption{
\scriptsize 
{\bf a)} $M_{BK}$ at NLO as function of $\mu$, for the corresponding $\tau$-stability region, for $t_c\simeq 48$ GeV$^2$ and for the QCD parameters in Tables\,\ref{tab:param} and \ref{tab:alfa}; {\bf b)} The same as a) but for the  renormalization group invariant coupling $\hat{f}_{BK}$.
}
\label{fig:bk-mu} 
\end{center}
\end{figure}

We show in Fig.\,\ref{fig:bk-mu} the dependence of $M_{BK}$ and of the renormalization group invariant coupling $\hat{f}_{BK}$ obtained at NLO of PT series on the choice of the  subtraction constant $\mu$.  We consider as optimal  values the ones obtained for $\mu\simeq (4.5-5.0)$ GeV where we have almost a plateau for the mass and a slight inflexion point for $\hat{f}_{BK}$. We deduce:
\beq
M_{BK}^{NLO}(\mu)\simeq (5207\sim 5205)~{\rm MeV}~~~~{\rm and}~~~~
\hat{f}_{BK}^{NLO}(\mu)\simeq (2.58\sim 2.68)~{\rm keV}~.
\eeq
\nin
\begin{figure}[hbt] 
\begin{center}
{\includegraphics[width=6.2cm  ]{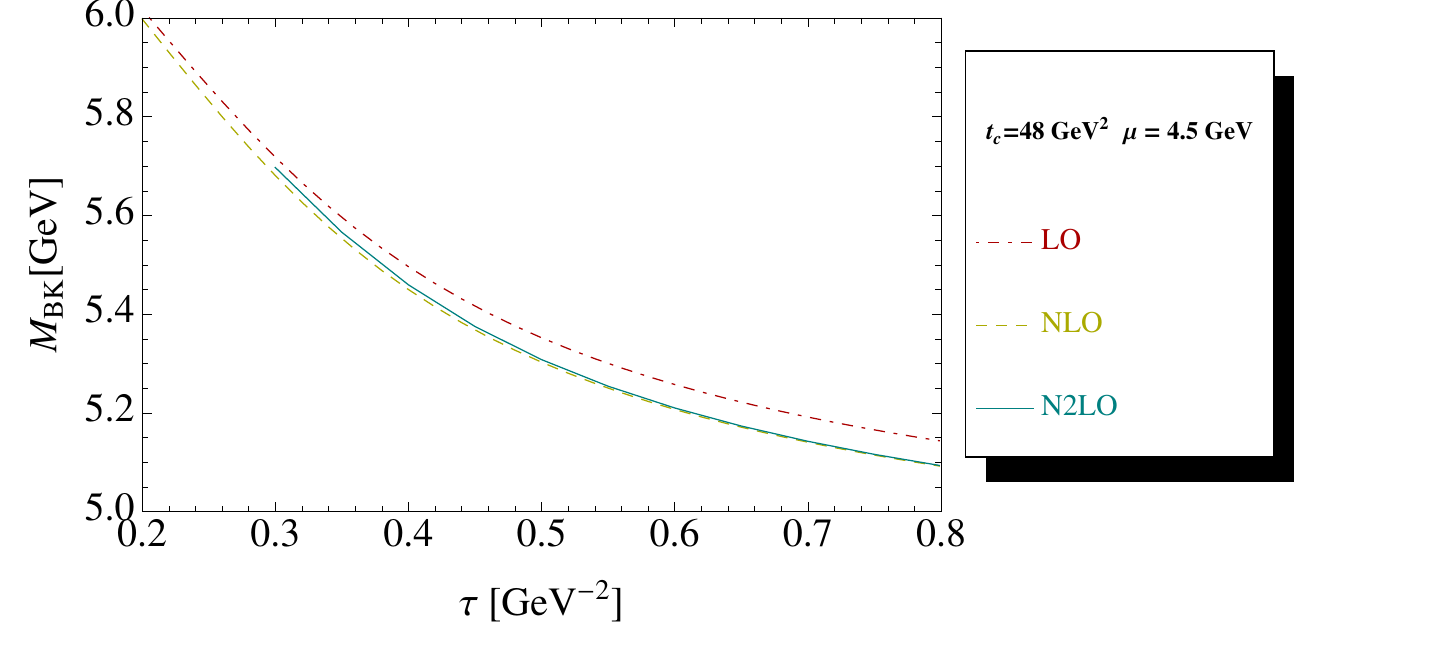}}
{\includegraphics[width=6.2cm  ]{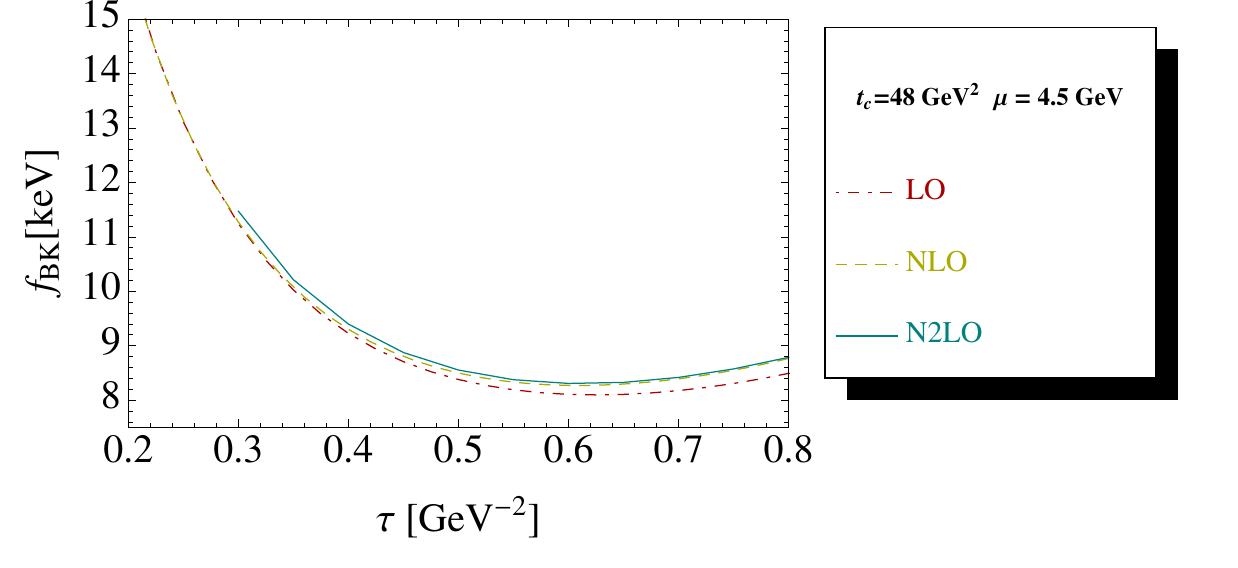}}
\centerline {\hspace*{-3cm} a)\hspace*{6cm} b) }
\caption{
\scriptsize 
{\bf a)} $M_{BK}$  as function of $\tau$ for different truncation of the PT series at a given value of $t_c$=48 GeV$^2$, $\mu=4.5$ GeV and for the QCD parameters in Tables\,\ref{tab:param} and \ref{tab:alfa}; {\bf b)} The same as a) but for the coupling $f_{BK}$.
}
\label{fig:bk-lo-n2lo} 
\end{center}
\end{figure} 
\nin
\subsection{Test of the convergence of the  PT series}
We show in Fig.\,\ref{fig:bk-lo-n2lo} the behaviour of the results for a given value of $t_c$ and of $\mu$ and for different truncation of the PT series. One can notice small PT corrections for the mass predictions because these corrections tend to compensate in the ratio of sum rules. For the coupling, the correction is large from LO to NLO.  For both observables,  one can notice a good convergence of the PT series from NLO to N2LO.  
\subsection{Test of the convergence of the  OPE}
\begin{figure}[hbt] 
\begin{center}
{\includegraphics[width=6.2cm  ]{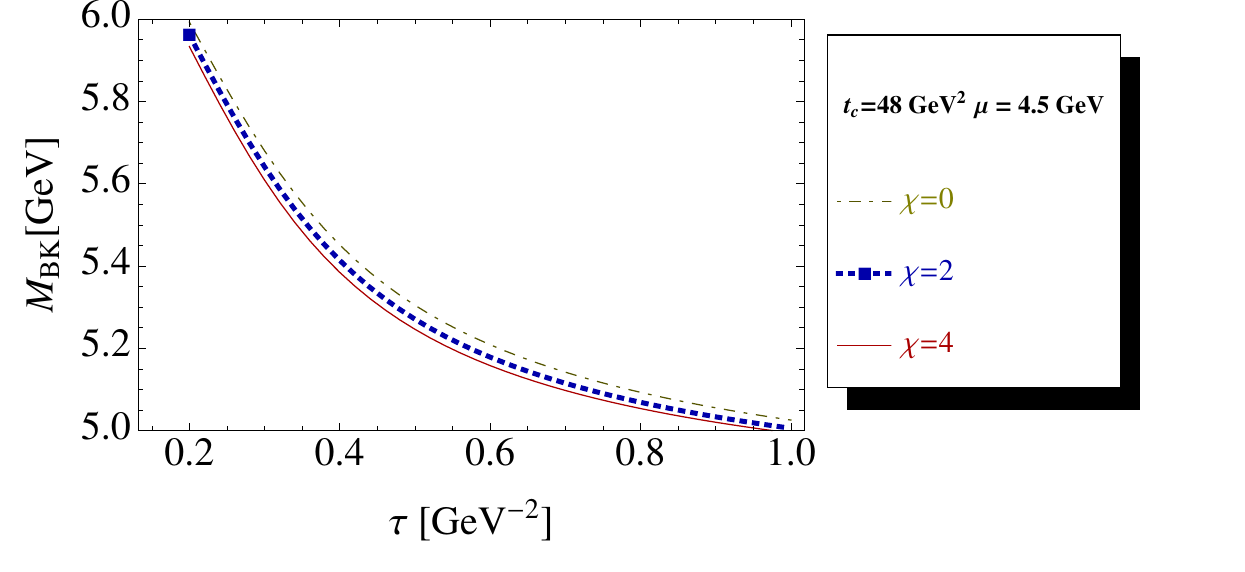}}
{\includegraphics[width=6.2cm  ]{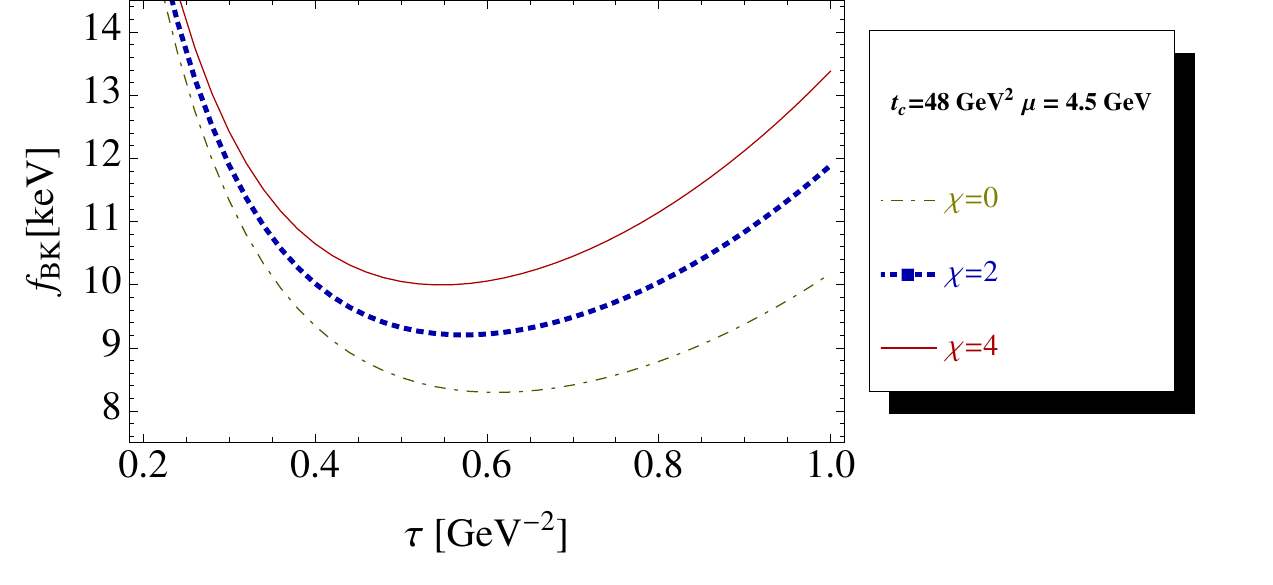}}
\centerline {\hspace*{-3cm} a)\hspace*{6cm} b) }
\caption{
\scriptsize 
{\bf a)} $M_{BK}$  as function of $\tau$, for different values of the $d=7$ condensate contribution ($\chi$ measures the violation of factorization),  at a given value of $t_c$=48 GeV$^2$, $\mu=4.5$ GeV and for the QCD parameters in Tables\,\ref{tab:param} and \ref{tab:alfa}; {\bf b)} The same as a) but for the coupling $f_{BK}$.
}
\label{fig:bk-d7} 
\end{center}
\end{figure}

We show  the effect of a class of the $d=7$ condensate for different values of $\chi$ in Fig.\,\ref{fig:bk-d7}.
Taking the maximal error corresponding to $\chi=4$, we estimate the error due to the truncation of the OPE as:
\beq
 \Delta f_{BK}^{OPE}\simeq  \pm 1.66 ~{\rm keV}~,~~~~~~~~~~~~~~\Delta M_{BK}^{OPE}\simeq \pm 5 ~{\rm MeV}~. 
\eeq
\nin
\subsection{Final results and comparison with previous estimates}
\hspace*{0.5cm}\b We show in Table\,\ref{tab:error} our estimate of the different sources of errors. Like in the case of $B^*K$, the main errors from the mass 
come from the localisation of the $\tau$-stability, $\alpha_s$, $\kappa\equiv \la\bar ss\ra/\la\bar qq\ra$, $\la\bar qq\ra^2$ and $M_0^2$. 
Adding quadratically different
 sources of errors, 
we consider as a final estimate to order $\alpha_s^2$ or at  N2LO of the perturbative series and including an estimate of the $d\geq 7$ dimension condensates:
\bea
 \hat f_{BK}&\simeq& (2.57\pm 0.75) ~{\rm keV}~\Lrar~ 
f_{BK}(4.5)\simeq (8.26\pm 2.40)~{\rm keV}~,\nnb\\
M_{BK}&\simeq&( 5195\pm 15)~{\rm MeV}~.
\label{eq:bkn2lo}
\eea
for $\mu=4.5$ GeV.

\b One can notice that the $BK$ molecule mass (if any) is about 11 MeV heavier than the $B^*K$ molecule and  below the physical $BK$ threshold of 5773 MeV. 

\b Comparing numerically the QCD expression of our spectral function with the one in \cite{TURC}, 
one finds that the perturbative (PT) and $\la \bar qq\ra$ expressions have wrong signs\,\footnote{The negative sign of the PT contribution violates the general positive property of the spectral function. The same wrong sign occurs in Ref.\,\cite{TURC1}.}a good agreement except for the $\la \bar qq\ra^2$ condensate contributions which coefficient is too small compared to ours. 

\b As mentioned previously, we refrain to add the condensate contributions of dimension higher than $d=6$ due to the large uncertainties on the values of these high-dimension condensates (violation of factorization) and to the fact that the contributions of these condensates correspond only to one class of diagrams of the OPE but not to the complete contributions. However, we have seen previously that the effect of the $d=7$ condensates is negligible in the present channel. 

\b Our result differs from the LO one in \cite{TURC} who found the mass $M_{BK}= 5584(137)$ MeV. One can understand the source of the discrepancy as the result of Ref.\,\cite{TURC} has been obtained for the sum rule variable $M^2= (3\sim 6)$ GeV$^2$ or equivalently for $\tau\equiv 1/M^2\simeq  (0.17\sim 0.33)$ GeV$^{-2}$ which is outside the $\tau$-stability region of about 0.6 GeV$^{-2}$ (inflexion point for the mass and minimum for the coupling) as clearly shown in Fig.\,\ref{fig:bk-lo}. In addition to the sensitivity on the variation  of $\tau$\,\footnote{The apparent stability shown in \cite{TURC} comes from the scale chosen for the figure frame.}, one can notice that, in this small $\tau$-region, the result is also very sensitive to the choice of the continuum threshold $t_c$ such that the prediction becomes unreliable. 
\section{$B^*K/BK $ ratio from Double Ratio of Sum Rule (DRSR)}
\begin{figure}[hbt] 
\begin{center}
{\includegraphics[width=6.2cm  ]{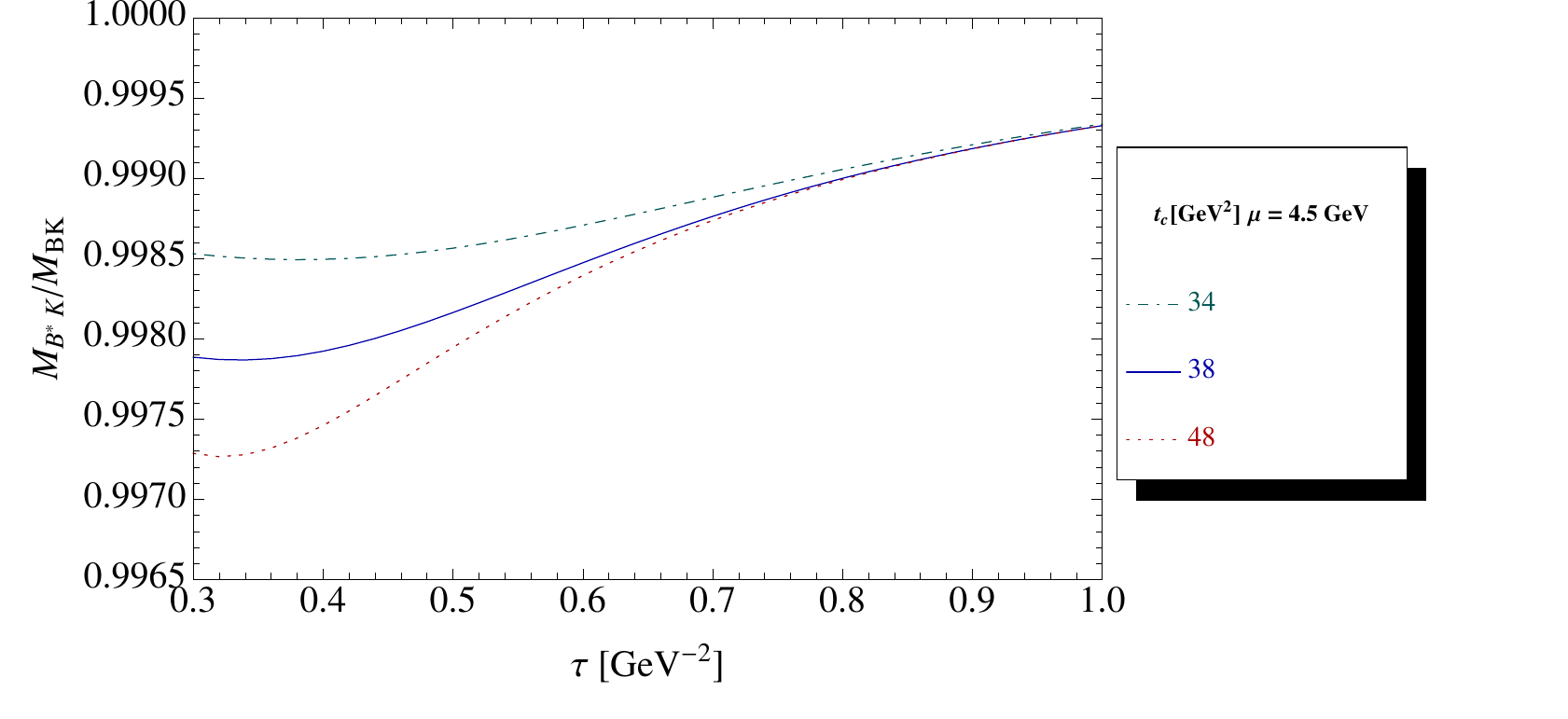}}
{\includegraphics[width=6.2cm  ]{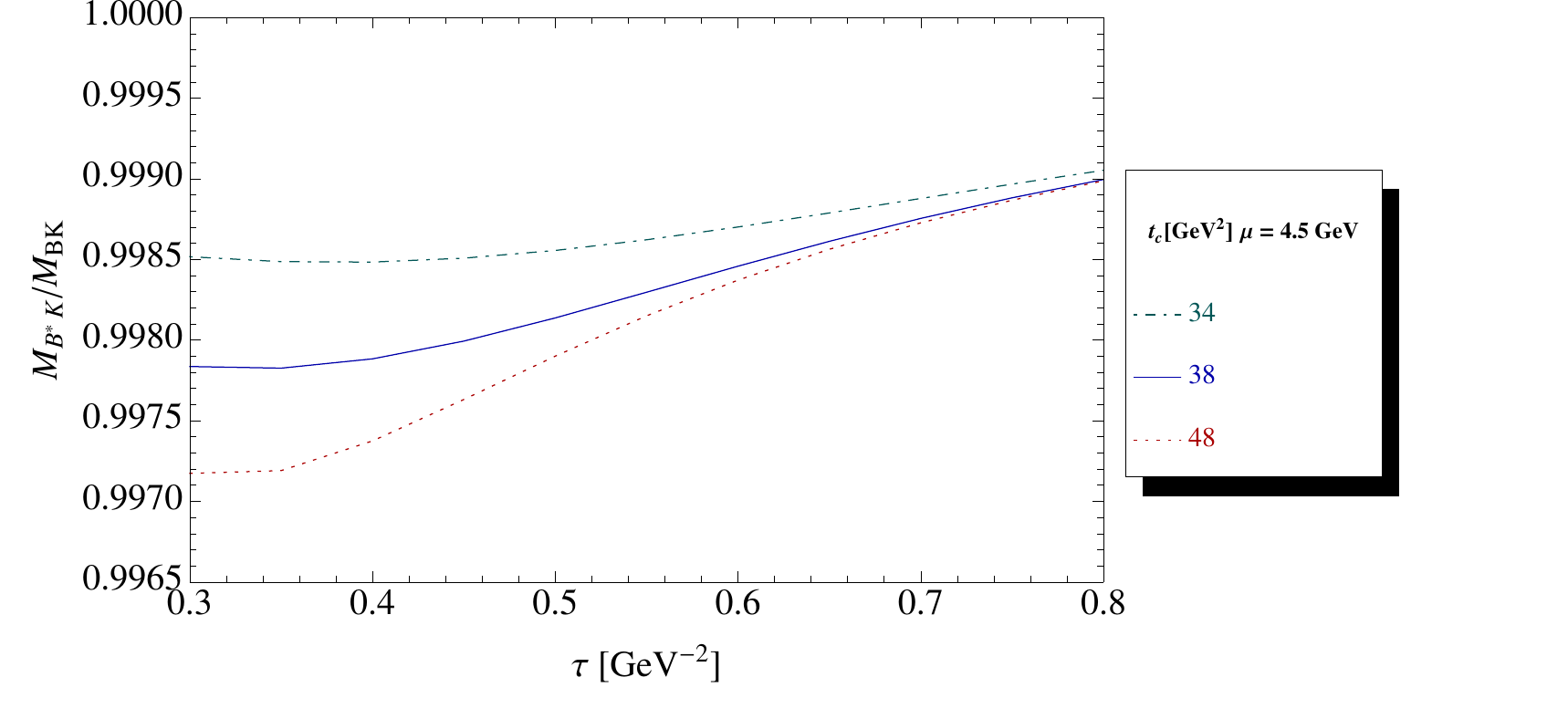}}
\centerline {\hspace*{-3cm} a)\hspace*{6cm} b) }
\caption{
\scriptsize 
{\bf a)} $M_{B^*K}/M_{BK}$ at LO as function of $\tau$ for different values of $t_c$, for $\mu=4.5$ GeV and for the QCD parameters in Tables\,\ref{tab:param} and \ref{tab:alfa}; {\bf b)} The same as a) but at N2LO.
}
\label{fig:rap-mass} 
\end{center}
\end{figure} 
\nin
\begin{figure}[hbt] 
\begin{center}
{\includegraphics[width=6.2cm  ]{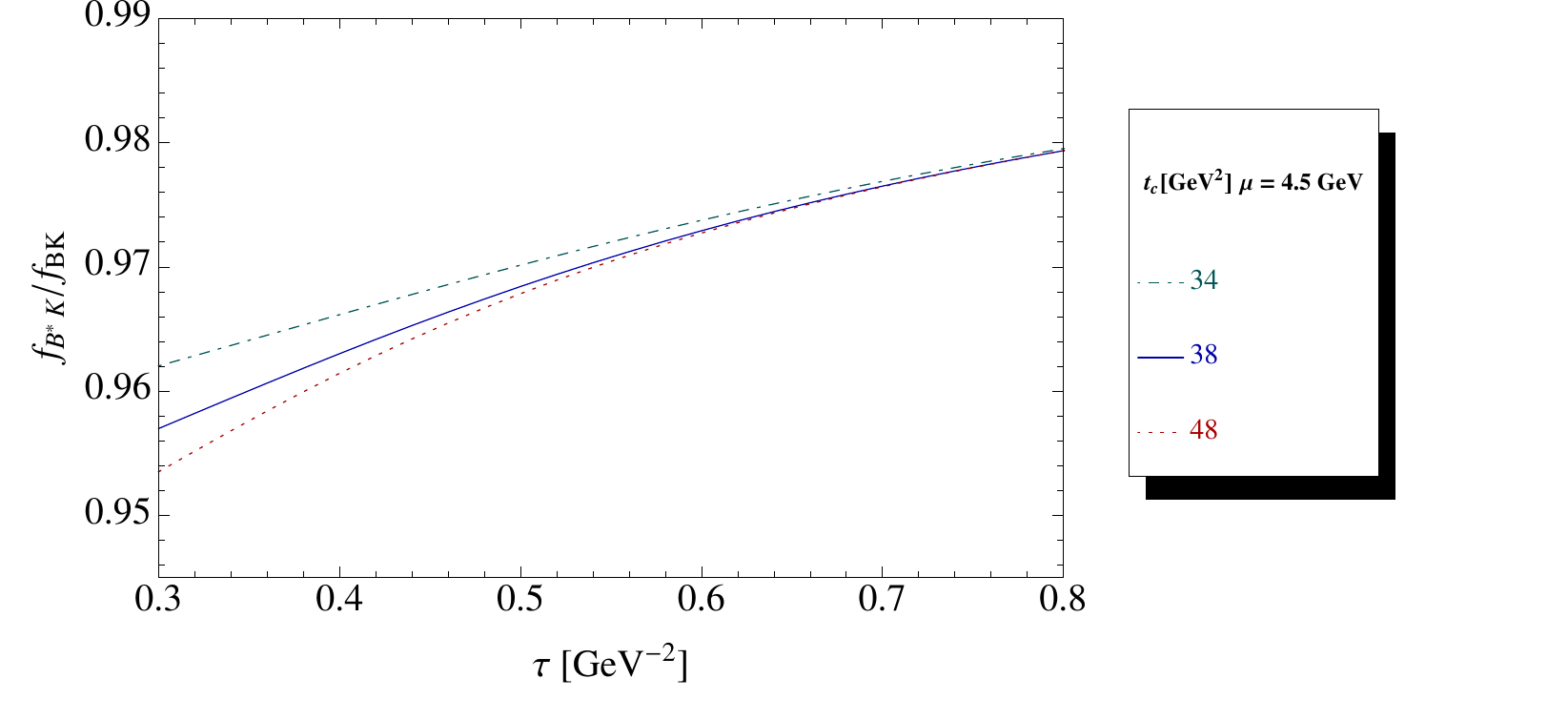}}
{\includegraphics[width=6.2cm  ]{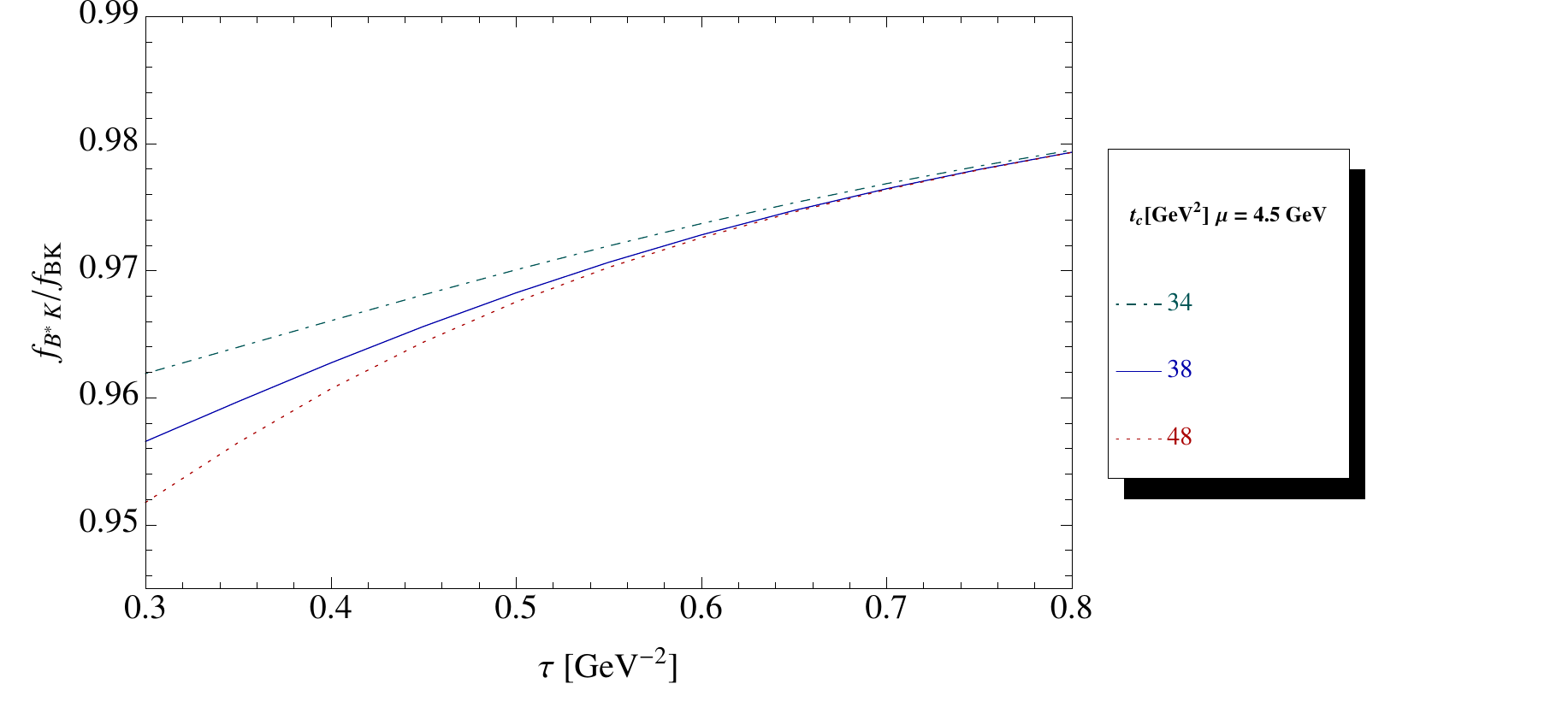}}
\centerline {\hspace*{-3cm} a)\hspace*{6cm} b) }
\caption{
\scriptsize 
{\bf a)} $f_{B^*K}/f_{BK}$ at LO as function of $\tau$ for different values of $t_c$, for $\mu=4.5$ GeV and for the QCD parameters in Tables\,\ref{tab:param} and \ref{tab:alfa};  {\bf b)} The same as a) but at N2LO.
}
\label{fig:rap-fb} 
\end{center}
\end{figure} 
\nin
We cross-check the previous result by a direct estimate of the $B^*K/BK $ mass and coupling ratios using the Double Ratio of Sum Rule (DRSR)\,\cite{DRSR}. As explicitly seen in\,
\cite{SNGh1,SNGh3,SNGh5,SNmassb,SNhl,SNmassa,SNFORM2,HBARYON1,HBARYON2,NAVARRA},
DRSR has the advantage to be free from systematic uncertainties, to be less sensitive to the PT and non-perturbative (NPT) corrections if they act in the same signs in the two channels. The results of the analysis are shown in Figs.\,\ref{fig:rap-mass} and \ref{fig:rap-fb}.  One can see that there is not a good stability in $\tau$ though the results change slightly with it. Taking $\tau\simeq 0.6$ GeV$^{-2}$ where the ratio of sum rules predicting the absolute value of the coupling and masses stabilizes, one deduces at N2LO for $t_c\simeq (34\sim 48)$ GeV$^2$:
\beq
M_{B^*K}/M_{BK}\simeq (0.999\sim 0.998)~,~~~{\rm and}~~~~f_{B^*K}/f_{BK}\simeq (0.973\sim 0.972)~.
\eeq
Using the value of $M_{BK}$ and $f_{BK}$ in Eq.\,\ref{eq:bkn2lo}, one can deduce:
\beq
M_{B^*K}\simeq (5187\pm 15)~{\rm MeV}~,~~~{\rm and}~~~~f_{B^*K}\simeq 
(8.03\pm 1.71)~{\rm keV}~,
\eeq
which confirms the previous estimates in Eq.\,\ref{eq:bstarkn2lo}.
\section{Spectral functions of the $B^*_s\pi$ and $B_s\pi$ molecules}
\subsection{$ B^*_s\pi$ $(1^{+})$ axial-vector molecule}
We shall also consider the two-point spectral function associated to the molecule $B_s\pi$ state with the current given in Table\,\ref{tab:current}. Its QCD expression reads to lowest order:
\bea
\rho^{pert}&=&\frac{M_b^8}{5\ 2^{14}\ \pi ^6}\Bigg{[}\frac{5}{ x^4}-\frac{96}{ x^3}-\frac{945}{ x^2}+\frac{480}{ x}-60\left(\frac{9}{ x^2}+\frac{16}{ x}+3\right)\text{Log}\  x+555+x^2\Bigg{]}+
\nnb\\
&&
\frac{m_sM_b^7}{2^{12}\ \pi ^6}\Bigg{[}\frac{1}{ x^3}+\frac{28}{ x^2}+12\left(\frac{1}{ x^2}+\frac{3}{ x}+1\right)\text{Log}\ x-28- x\Bigg{]}~,
\nnb\\
\rho^{\langle\bar{q}q\rangle}&=&-\frac{M_b^5}{2^8\ \pi ^4}\kappa\la\bar qq\ra \Bigg{[}\frac{1}{ x^2}+\frac{9}{ x}+6\ \left(\frac{1}{ x}+1\right)\ \text{Log}\  x-9- x\Bigg{]}+\nnb\\
&&\frac{m_sM_b^4}{2^{11}\ \pi
^4}\kappa\la\bar qq\ra \left(\frac{3}{ x^2}-\frac{16}{ x}-12\ \text{Log}\  x+12+  x^2\right);\nnb\\
\rho^{\langle G^2\rangle}&=&\frac{M_b^4}{3\ 2^{15}\ \pi ^6}4\pi\la\alpha_s G^2\ra\Bigg{[}\frac{1}{ x^2}-{120\over x}-12\left(\frac{4}{ x}+7\right)\text{Log}\  x+108+8 x+3x^2\Bigg{]}~,
\nnb
\eea
\bea
\rho^{\langle\bar{q}Gq\rangle}&=&\frac{3M_b^3 }{2^9\ \pi ^4}\kappa\la\bar q Gq\ra\left(\frac{1}{ x}+2\ \text{Log}\ x- x\right)-
\frac{m_sM_b^2}{2^{10}\ \pi
^4}\kappa\la\bar q Gq\ra \left(\frac{4}{ x}+6\ \text{Log}\ x-1-4\ x+x^2\right)~,
\nnb\\
\rho^{\langle\bar{q}q\rangle^2}&=&\frac{M_b^2 }{3\ 2^5\ \pi ^2}\rho\la\bar qq\ra^2\left(\frac{2}{ x}-3+ x^2\right)+\frac{m_sM_b }{2^4\ \pi
^2}\rho\la\bar qq\ra^2(1- x)
\nnb\\
\rho^{\langle G^3\rangle}&=&\frac{M_b^{2 }}{3^2\ 2^{17}\ \pi ^6}\la g^3G^3\ra\Bigg{[}\frac{9}{ x^2}-\frac{160}{ x}-12\left(\frac{4}{ x}+9\right)\text{Log}\ x+144+7 x^2\Bigg{]}
\eea
The contribution of a class of $d=7$ condensate for $m_s=0$  is:
\bea
\rho^{\la \bar qq\ra\la G^2\ra}&=&-\frac{M_b\kappa\la \bar qq\ra}{3\ 2^9\pi ^4} 4\pi\la \alpha_s G^2\ra\left(\frac{1}{x}+3\,\text{Log}\,x+5-6x\right)~.
\label{eq:d7bstarspi}
\eea
\subsection{$B_s\pi$ scalar molecule}
The expression of the   $B_s\pi$ scalar two-point spectral function associated to the molecule state with the current given in Table\,\ref{tab:current} reads to lowest order:
\bea
\rho^{pert}&=&\frac{M_b^8}{5\ 2^{14}\ \pi ^6}\Bigg{[}\frac{1}{ x^4}-\frac{20}{ x^3}-\frac{220}{ x^2}+\frac{80}{ x}-60\left(\frac{2}{ x^2}+\frac{4}{ x}+1\right)\text{Log}\  x+155+4 x\Bigg{]}+
\nnb\\
&&
\frac{m_sM_b^7}{2^{12}\ \pi ^6}\Bigg{[}\frac{1}{ x^3}+\frac{28}{ x^2}+12\left(\frac{1}{ x^2}+\frac{3}{ x}+1\right)\text{Log}\ x-28- x\Bigg{]}~,
\nnb\\
\rho^{\langle\bar{q}q\rangle}&=&-\frac{M_b^5}{2^8\ \pi ^4}\kappa\la\bar qq\ra \Bigg{[}\frac{1}{ x^2}+\frac{9}{ x}+6\ \left(\frac{1}{ x}+1\right)\ \text{Log}\  x-9- x\Bigg{]}+\nnb\\
&&\frac{m_sM_b^4}{2^9\ \pi
^4}\kappa\la\bar qq\ra\left(\frac{1}{ x^2}-\frac{6}{ x}-6\ \text{Log}\  x+3+2\  x\right);\nnb\\
\rho^{\langle G^2\rangle}&=&\frac{M_b^4 }{3\ 2^{13}\ \pi ^6}4\pi\la\alpha_s G^2\ra\Bigg{[}\frac{5}{ x^2}+6\left(\frac{2}{ x}-1\right)\text{Log}\  x-9+4 x\Bigg{]}~,
\nnb\\
\rho^{\langle\bar{q}Gq\rangle}&=&\frac{3M_b^3 }{2^8\ \pi ^4} \kappa\la\bar q Gq\ra\Bigg{[}\frac{3}{ x}+\left(\frac{1}{ x}+3\right)\text{Log}\ x-2- x\Bigg{]}+\nnb\\
&&\frac{m_sM_b^2}{2^8\ \pi
^4} \kappa\la\bar q Gq\ra \left(\frac{1}{ x}+3\ \text{Log}\ x+1-2\ x\right)~,
\nnb\\
\rho^{\langle\bar{q}q\rangle^2}&=&\frac{M_b^2 }{2^5\ \pi ^2}\rho\la\bar qq\ra^2\left(\frac{1}{ x}-2+ x\right)+\frac{m_sM_b }{2^4\ \pi
^2}\rho\la\bar qq\ra^2(1- x)
\nnb\\
\rho^{\langle G^3\rangle}&=&\frac{M_b^{2 }}{3\ 2^{15}\ \pi ^6}\la g^3G^3\ra\Bigg{[}\frac{1}{ x^2}-\frac{21}{ x}-6\left(\frac{1}{ x}+3\right)\text{Log}\ x+15+5 x\Bigg{]}
\eea
The contribution of a class of $d=7$ condensate for $m_s=0$  is:
\bea
\rho^{\la \bar qq\ra\la G^2\ra}&=&-\frac{M_b\kappa\la \bar qq\ra}{3\ 2^9\pi ^4} 4\pi\la \alpha_s G^2\ra\left(\frac{1}{x}+12\,\text{Log}\,x+14-15x\right)~.
\label{eq:d7bspi}
\eea

\section{Mass and coupling of the $ B^*_s\pi$ $(1^+)$ axial-vector  molecule}
\begin{figure}[hbt] 
\begin{center}
{\includegraphics[width=6.2cm  ]{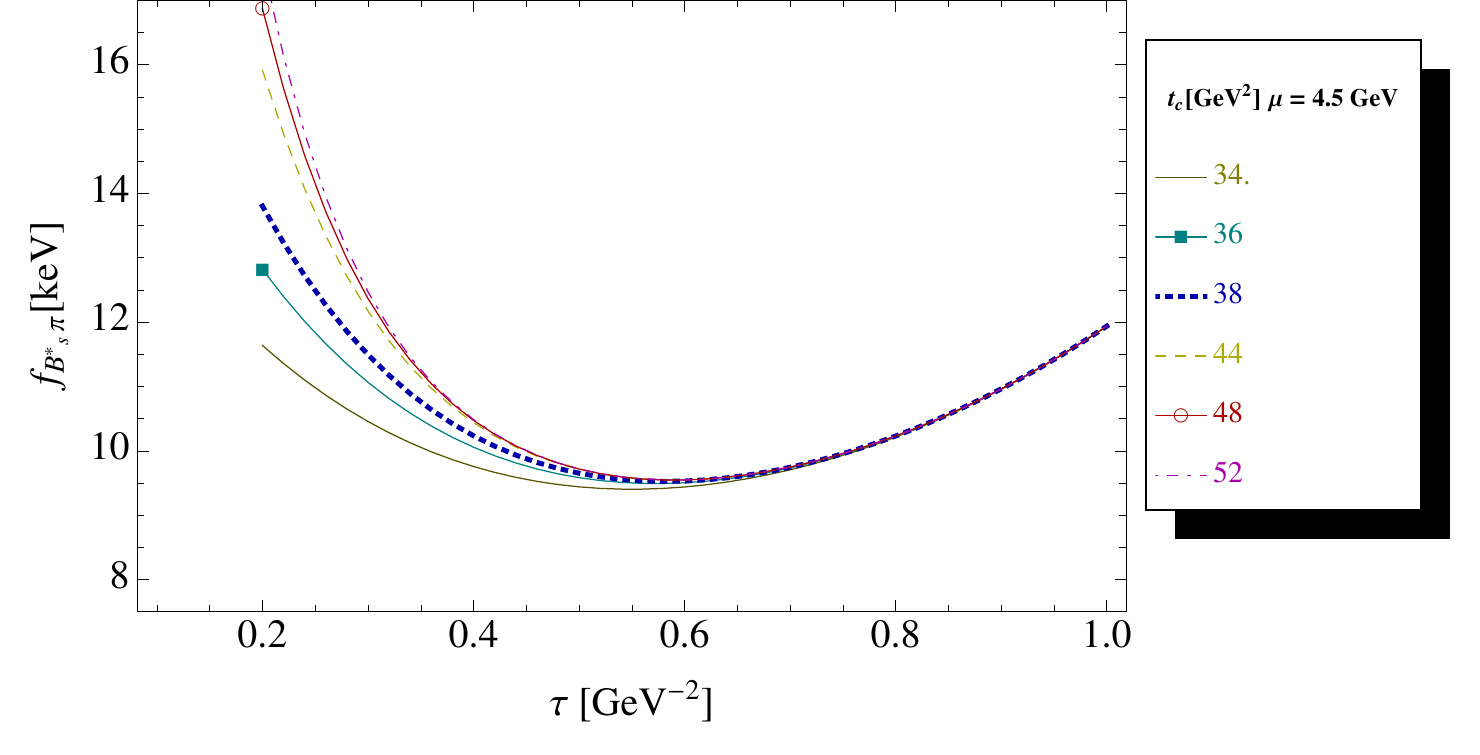}}
{\includegraphics[width=6.2cm  ]{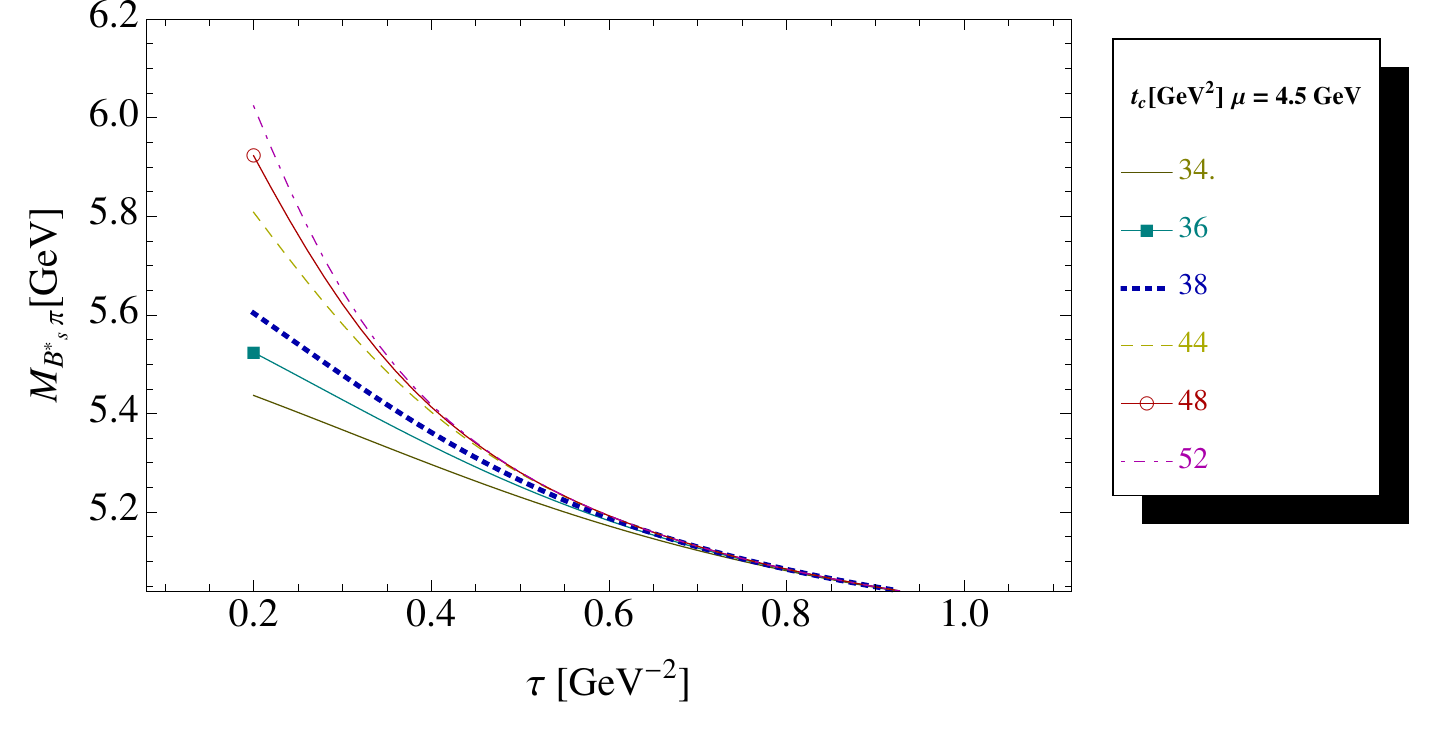}}
\centerline {\hspace*{-3cm} a)\hspace*{6cm} b) }
\caption{
\scriptsize 
{\bf a)} $f_{B^*_s\pi}$  at LO as function of $\tau$ for different values of $t_c$, $\mu=4.5$ GeV and for the QCD parameters in Tables\,\ref{tab:param} and \ref{tab:alfa};  {\bf b)} The same as a) but for the mass.
}
\label{fig:bstarspi-lo} 
\end{center}
\end{figure} 
\nin
\subsection{$\tau$- and $t_c$-stability criteria at lowest order (LO)} 
The results of the analysis are shown in Fig.\,\ref{fig:bstarspi-lo}. For $\tau\simeq  (0.58-0.62) $ GeV$^{-2}$
where the coupling presents a $\tau$-minimum and the mass an inflexion point, one obtains for $t_c\simeq (34\sim 48)$ GeV$^2$:
\beq
f_{B^*_s\pi}^{LO}\simeq (9.41\sim 9.55)~{\rm keV}~~~~{\rm and}~~~~
M_{B^*_s\pi}^{LO}\simeq 5208~{\rm MeV}~.
\eeq
\subsection{$b$-quark mass ambiguity at lowest order (LO)}
\begin{figure}[hbt] 
\begin{center}
{\includegraphics[width=6.2cm  ]{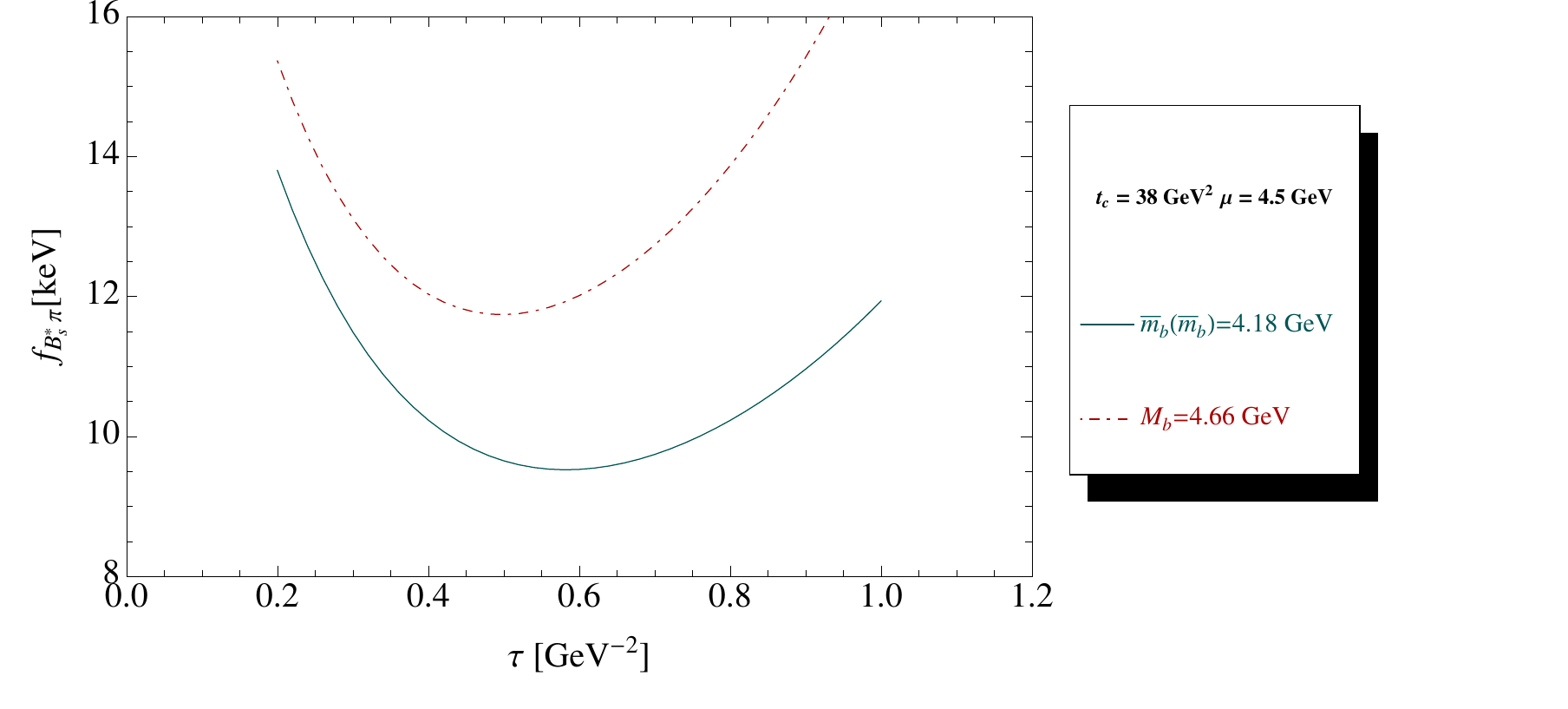}}
{\includegraphics[width=6.2cm  ]{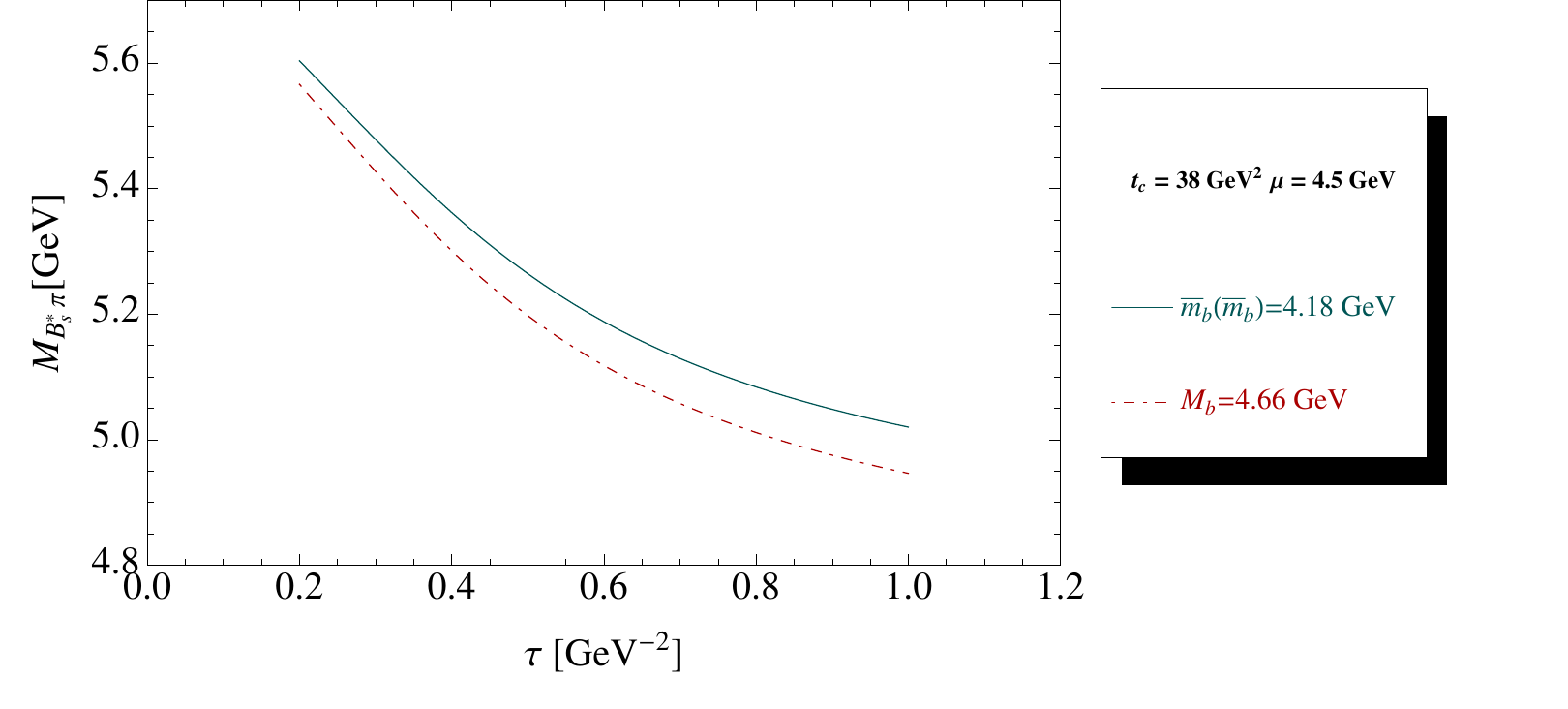}}
\centerline {\hspace*{-3cm} a)\hspace*{6cm} b) }

\caption{
\scriptsize 
{\bf a)} $f_{B^*_s\pi}$  at LO as function of $\tau$ for a given value of $t_c=38$ GeV$^2$,  $\mu=4.5$ GeV and for the QCD parameters in Tables\,\ref{tab:param} and \ref{tab:alfa};  The OPE is truncated at $d=6$.  We compare the effect of  the on-shell or pole mass $M_b=4.66$ GeV and of the running mass $\bar m_b(\bar m_b)=4.18$ GeV; {\bf b)} The same as a) but for the mass.
}
\label{fig:bstarspi-constituent} 
\end{center}
\end{figure} 
\nin
In Fig.\,\ref{fig:bstarspi-constituent}, we show the influence on the choice of the quark mass value on the coupling and mass at lowest order. One can see
an effect of about 2.2 keV (23\%) for the coupling and about 4 MeV for the mass  $M_{B^*_s\pi}$.
\subsection{$\mu$-dependence of the result at NLO}
\begin{figure}[hbt] 
\begin{center}
{\includegraphics[width=6.2cm  ]{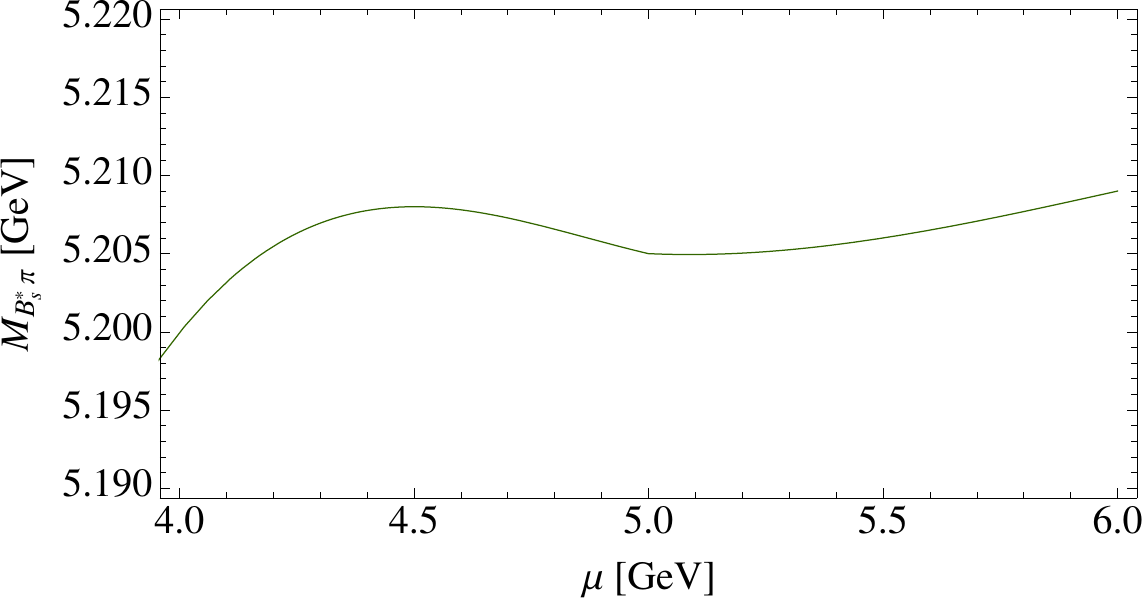}}
{\includegraphics[width=6.2cm  ]{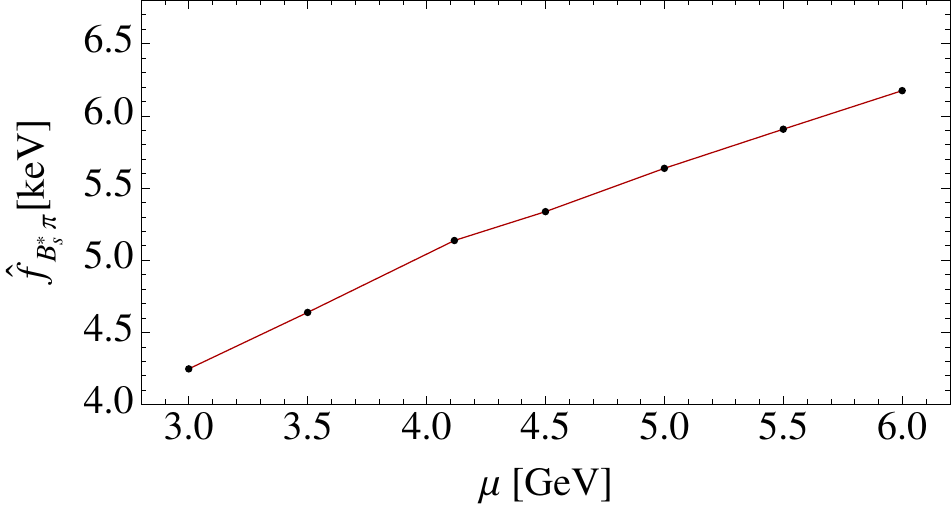}}
\centerline {\hspace*{-3cm} a)\hspace*{6cm} b) }
\caption{
\scriptsize 
{\bf a)} $M_{B^*_s\pi}$ at NLO as function of $\mu$, for the corresponding $\tau$-stability region, for $t_c\simeq 48$ GeV$^2$ and for the QCD parameters in Tables\,\ref{tab:param} and \ref{tab:alfa}; {\bf b)} The same as a) but for the  renormalization group invariant coupling $\hat{f}_{B^*_s\pi}$.
}
\label{fig:bstarspi-mu} 
\end{center}
\end{figure} 
\nin
We study the $\mu$-dependence of the result including $\alpha_s$-corrections. 
\subsection{Test of the convergence of the  PT series}
\begin{figure}[hbt] 
\begin{center}
{\includegraphics[width=6.2cm  ]{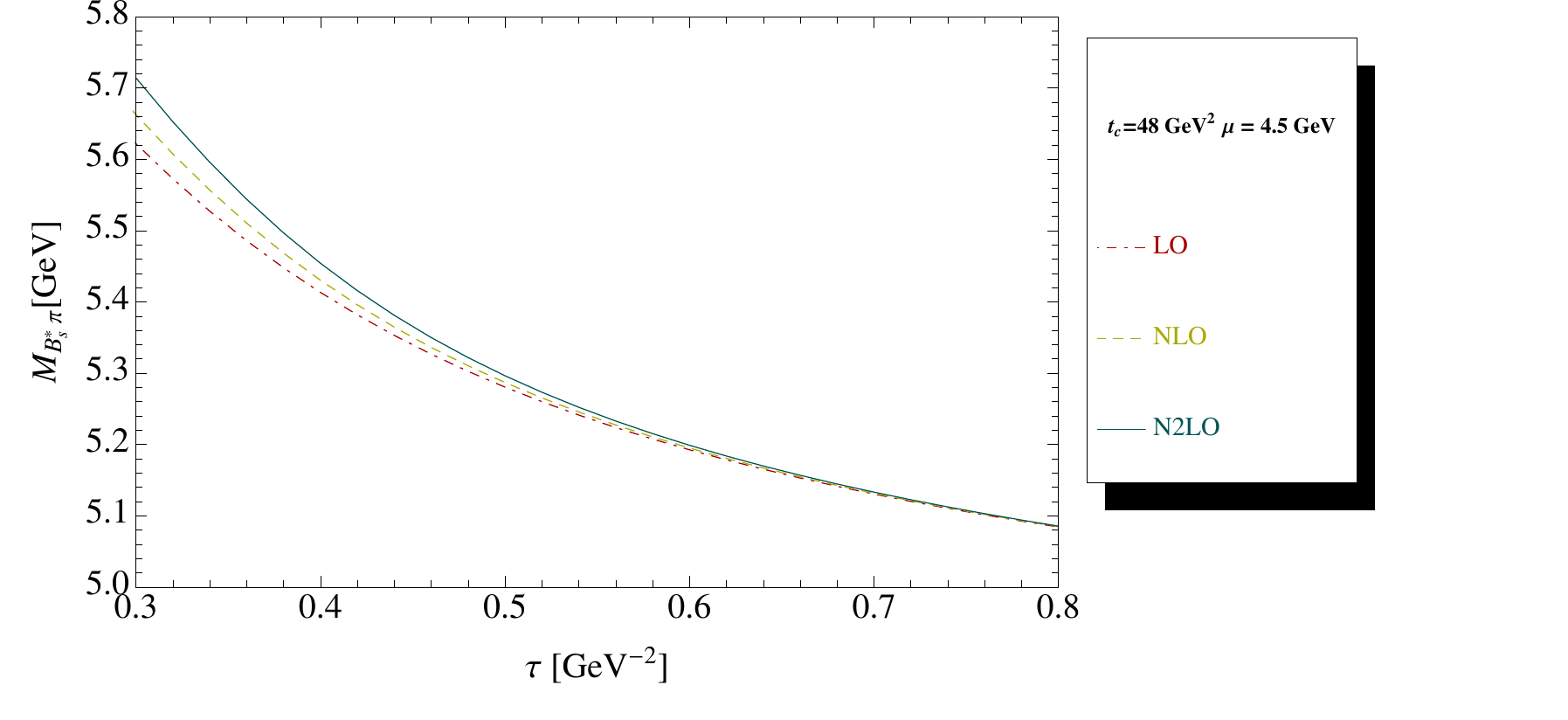}}
{\includegraphics[width=6.2cm  ]{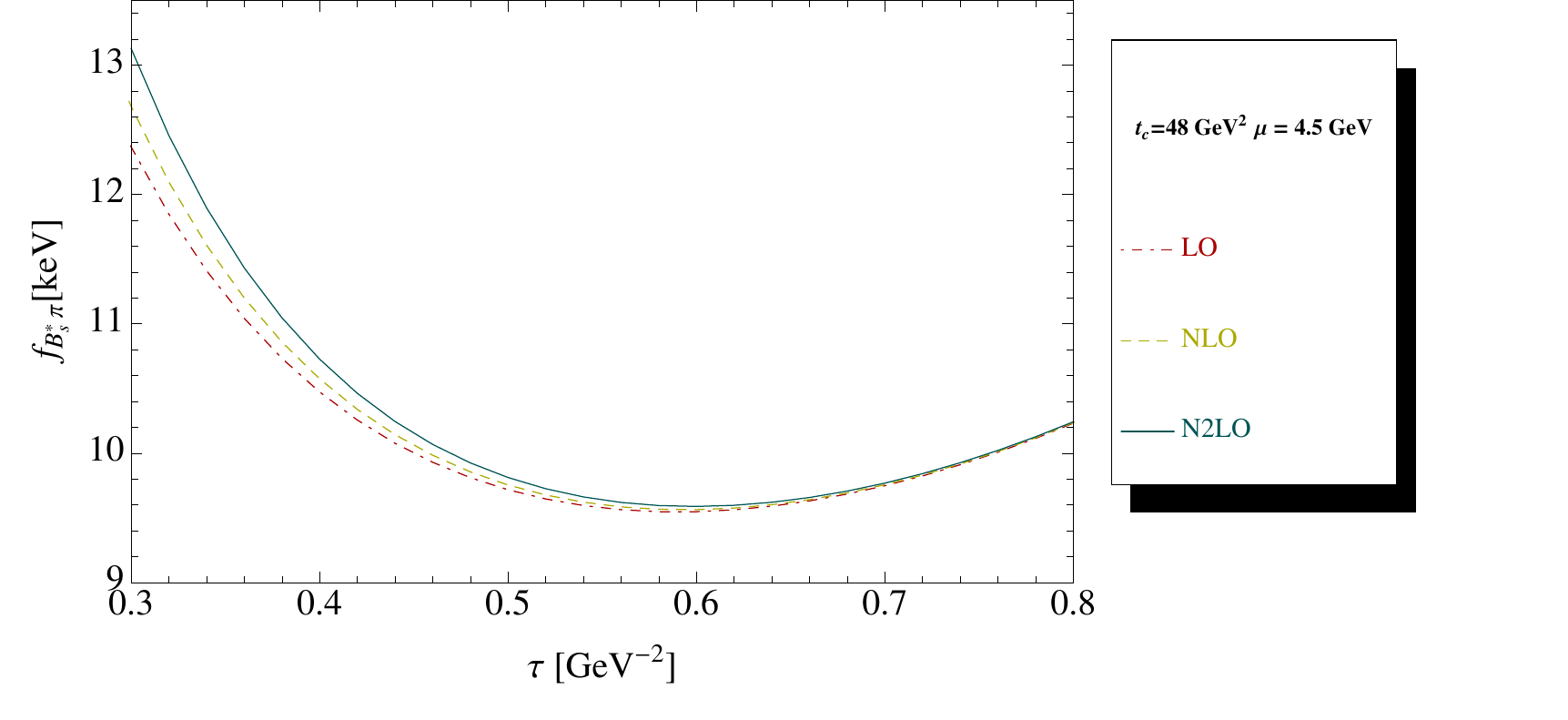}}
\centerline {\hspace*{-3cm} a)\hspace*{6cm} b) }
\caption{
\scriptsize 
{\bf a)} $M_{B^*_s\pi}$  as function of $\tau$ for different truncation of the PT series at a given value of $t_c$=48 GeV$^2$, $\mu=4.5$ GeV and for the QCD parameters in Tables\,\ref{tab:param} and \ref{tab:alfa}; {\bf b)} The same as a) but for the coupling $f_{B^*_s\pi}$.
}
\label{fig:bstarspi-lo-n2lo} 
\end{center}
\end{figure}

We show in Fig.\,\ref{fig:bstarspi-lo-n2lo} the behaviour of the results for a given value of $t_c$ and of $\mu$ and for different truncation of the PT series. One can notice small PT corrections for the mass and coupling predictions.  
\nin
\subsection{Error induced by the OPE}
\begin{figure}[hbt] 
\begin{center}
{\includegraphics[width=6.2cm  ]{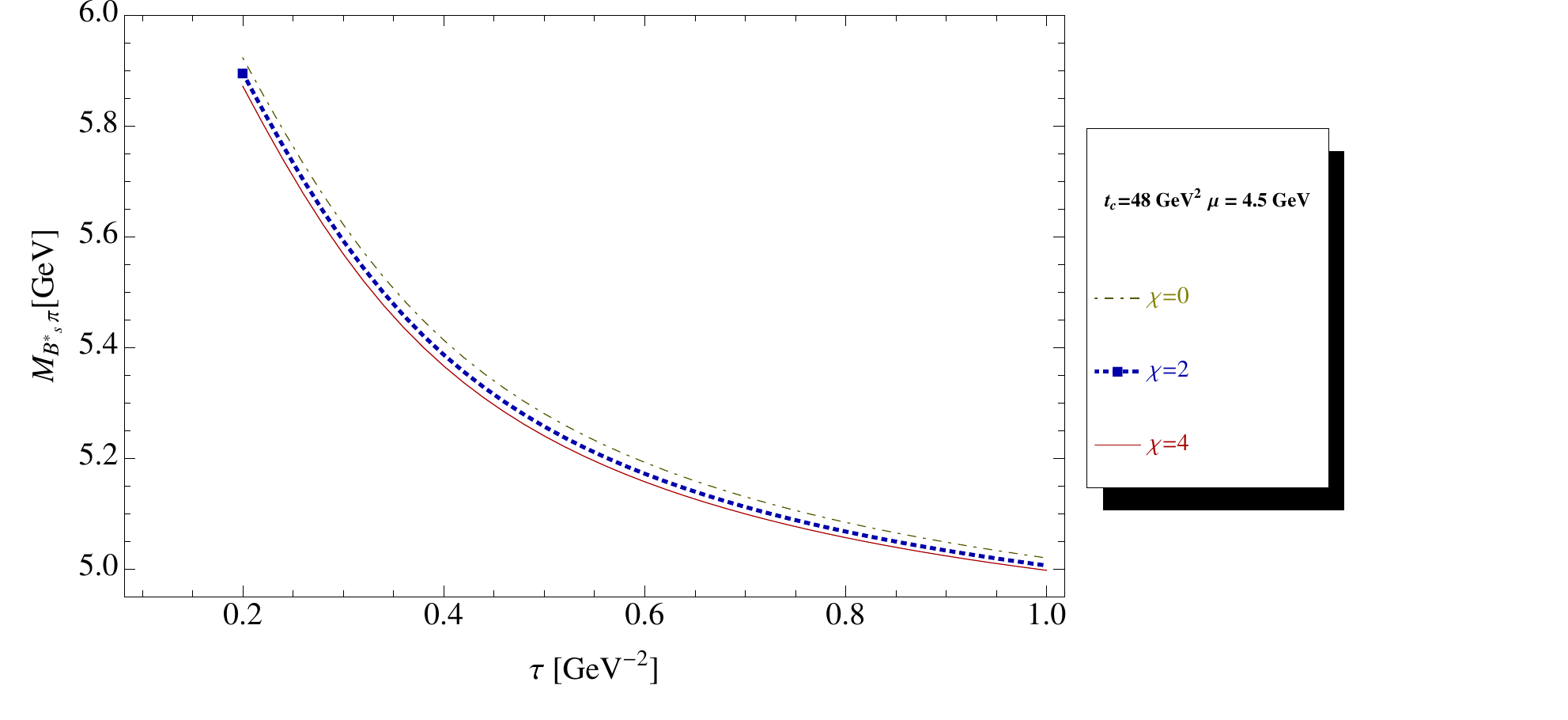}}
{\includegraphics[width=6.2cm  ]{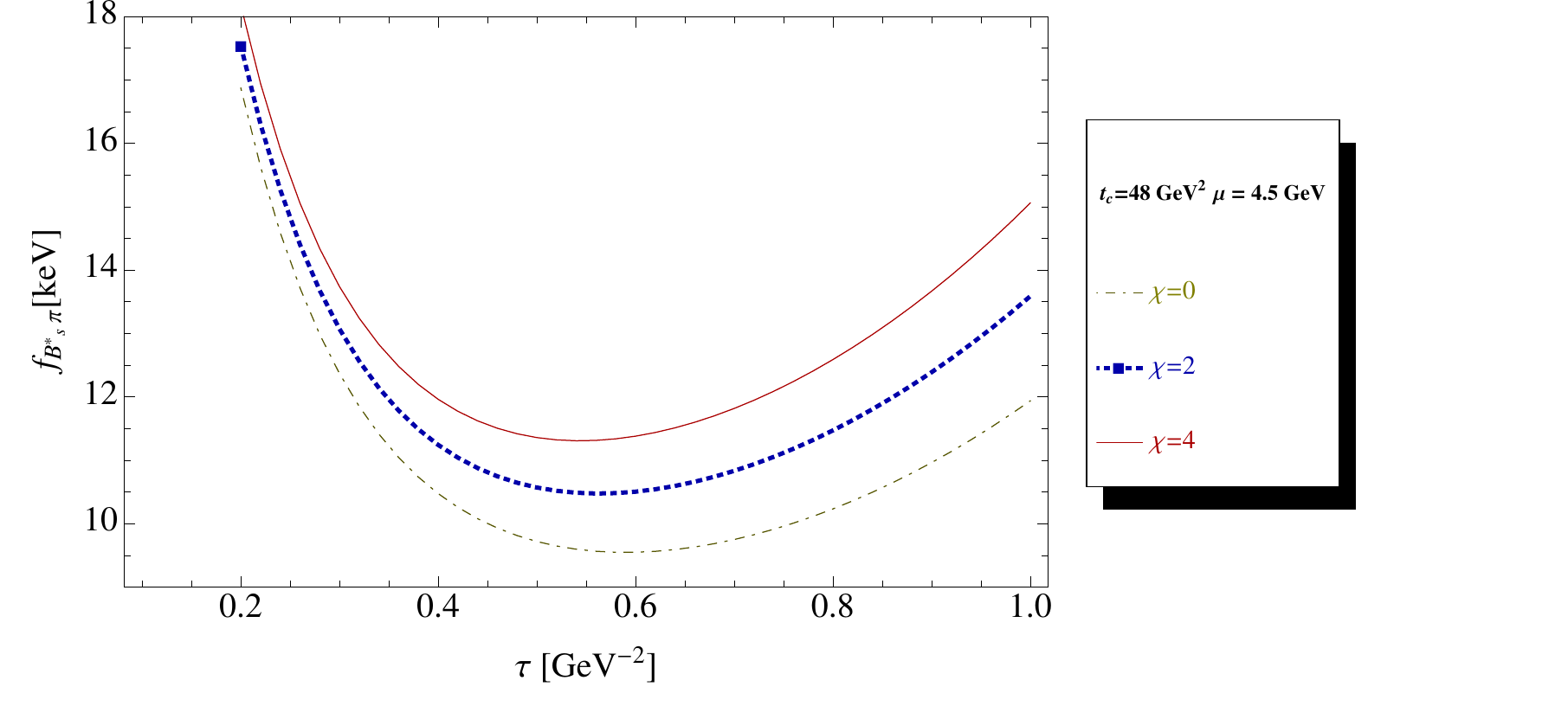}}
\centerline {\hspace*{-3cm} a)\hspace*{6cm} b) }
\caption{
\scriptsize 
{\bf a)} $M_{B^*_s\pi}$  as function of $\tau$, for different values of the $d=7$ condensate contribution ($\chi$ measures the violation of factorization),  at a given value of $t_c$=48 GeV$^2$, $\mu=4.5$ GeV and for the QCD parameters in Tables\,\ref{tab:param} and \ref{tab:alfa}; {\bf b)} The same as a) but for the coupling $f_{B^*_s\pi}$.
}
\label{fig:bstarspi-d7} 
\end{center}
\end{figure}

Like in previous section, we estimate this effect by taking a violation of about a factor 4 of the factorization assumption
for the $d=7$ condensates. The analysis is shown in Fig.\,\ref{fig:bstarspi-d7} from which we deduce:
\beq
 \Delta f_{B^*_s\pi}^{OPE}\simeq  \pm 1.80 ~{\rm keV}~,~~~~~~~~~~~~~~\Delta M_{B^*_s\pi}^{OPE}\simeq \pm 5 ~{\rm MeV}~. 
\eeq
\nin
\subsection{Final results at N2LO}
We conclude from the previous analysis the results at N2LO and for $\mu=$5 GeV:
\bea
M_{B^*_s\pi}&\simeq&( 5200\pm 18 )~{\rm MeV}~,
\nnb\\
 \hat f_{B^*_s\pi}&\simeq& (5.61\pm 1.32) ~{\rm keV}~~\Lrar~~ 
f_{B^*_s\pi}(4.5)\simeq (10.23\pm 2.40)~{\rm keV}~.
\label{eq:bstarspi-n2lo}
\eea
\section{Mass and coupling of the $ B_s\pi$ ($0^+$) scalar  molecule}
\subsection{$\tau$- and $t_c$-stability criteria at lowest order (LO)} 
\begin{figure}[hbt] 
\begin{center}
{\includegraphics[width=6.2cm  ]{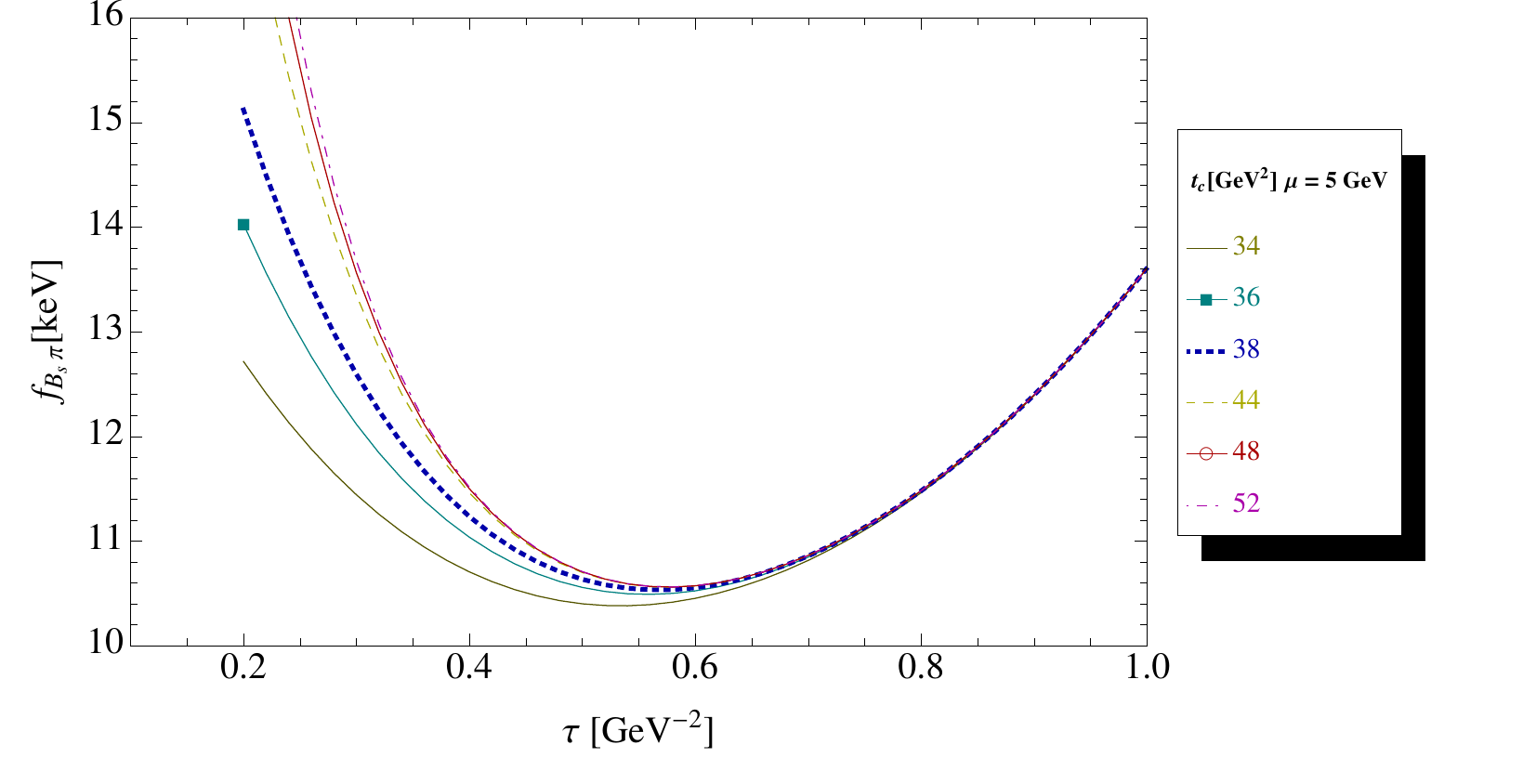}}
{\includegraphics[width=6.2cm  ]{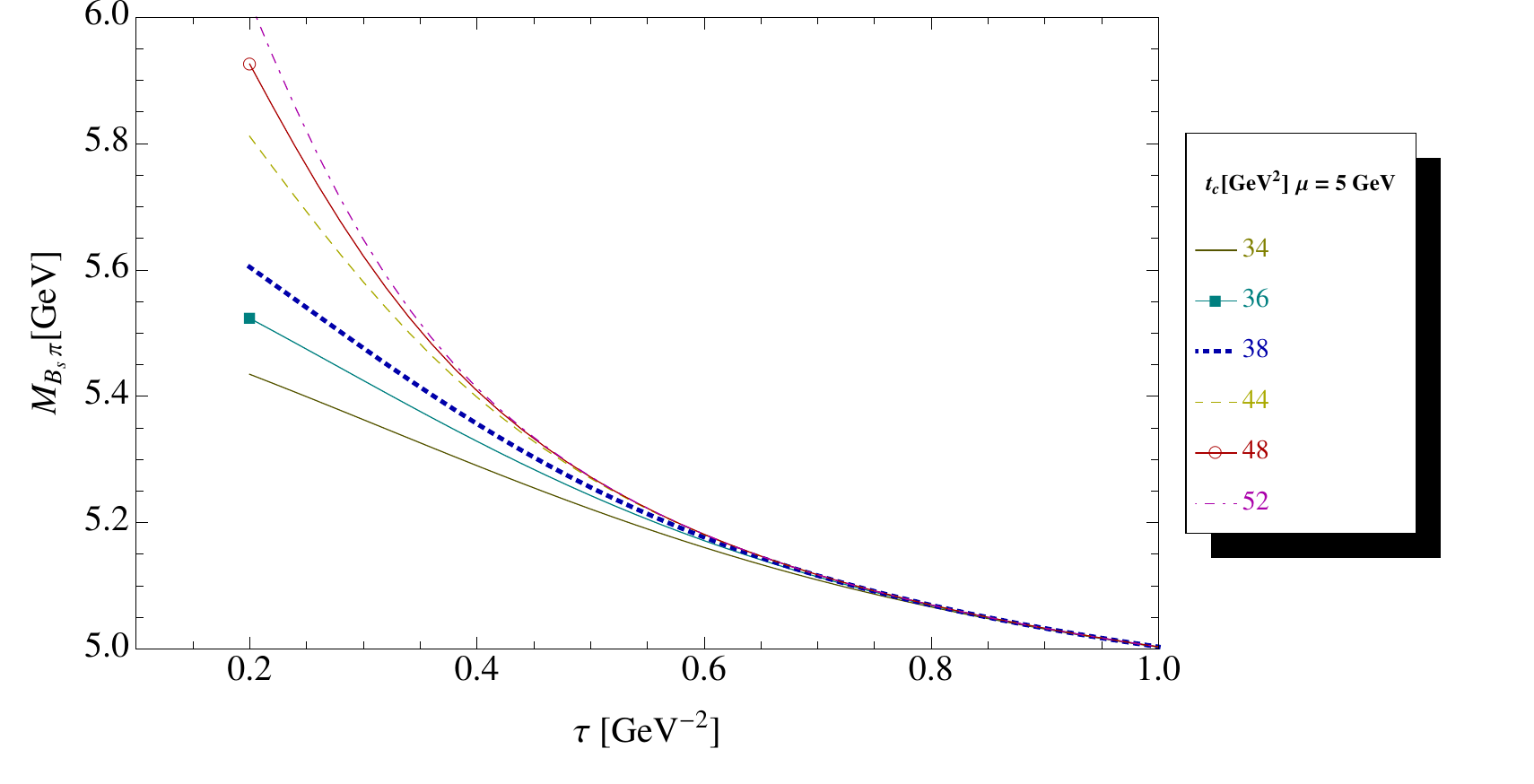}}
\centerline {\hspace*{-3cm} a)\hspace*{6cm} b) }
\caption{
\scriptsize 
{\bf a)} $f_{B_s\pi}$  at LO as function of $\tau$ for different values of $t_c$, $\mu=4.5$ GeV and for the QCD parameters in Tables\,\ref{tab:param} and \ref{tab:alfa};  {\bf b)} The same as a) but for the mass  $M_{B_s\pi}$.
}
\label{fig:bspi-lo} 
\end{center}
\end{figure} 
\nin
The results of the analysis are shown in Fig.\,\ref{fig:bspi-lo}. For $\tau\simeq  (0.58-0.62) $ GeV$^{-2}$
where the coupling presents a $\tau$-minimum and the mass an inflexion point, one obtains for $t_c\simeq (34\sim 48)$ GeV$^2$:
\beq
f_{B_s\pi}^{LO}\simeq (9.65\sim 9.80)~{\rm keV}~~~~{\rm and}~~~~
M_{B_s\pi}^{LO}\simeq( 5234\sim 5235)~{\rm MeV}
~.
\eeq
\subsection{$b$-quark mass ambiguity at lowest order (LO)}
\begin{figure}[hbt] 
\begin{center}
{\includegraphics[width=6.2cm  ]{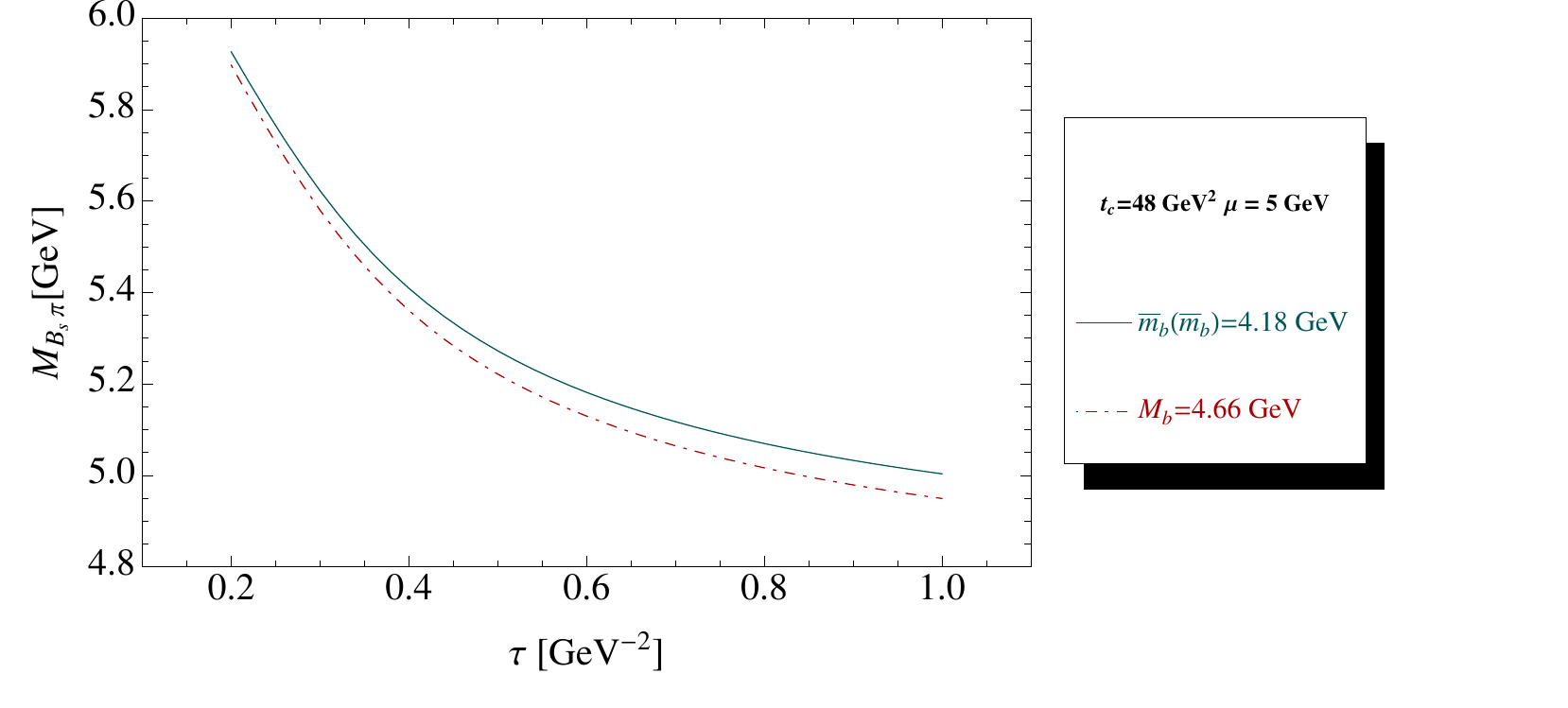}}
{\includegraphics[width=6.2cm  ]{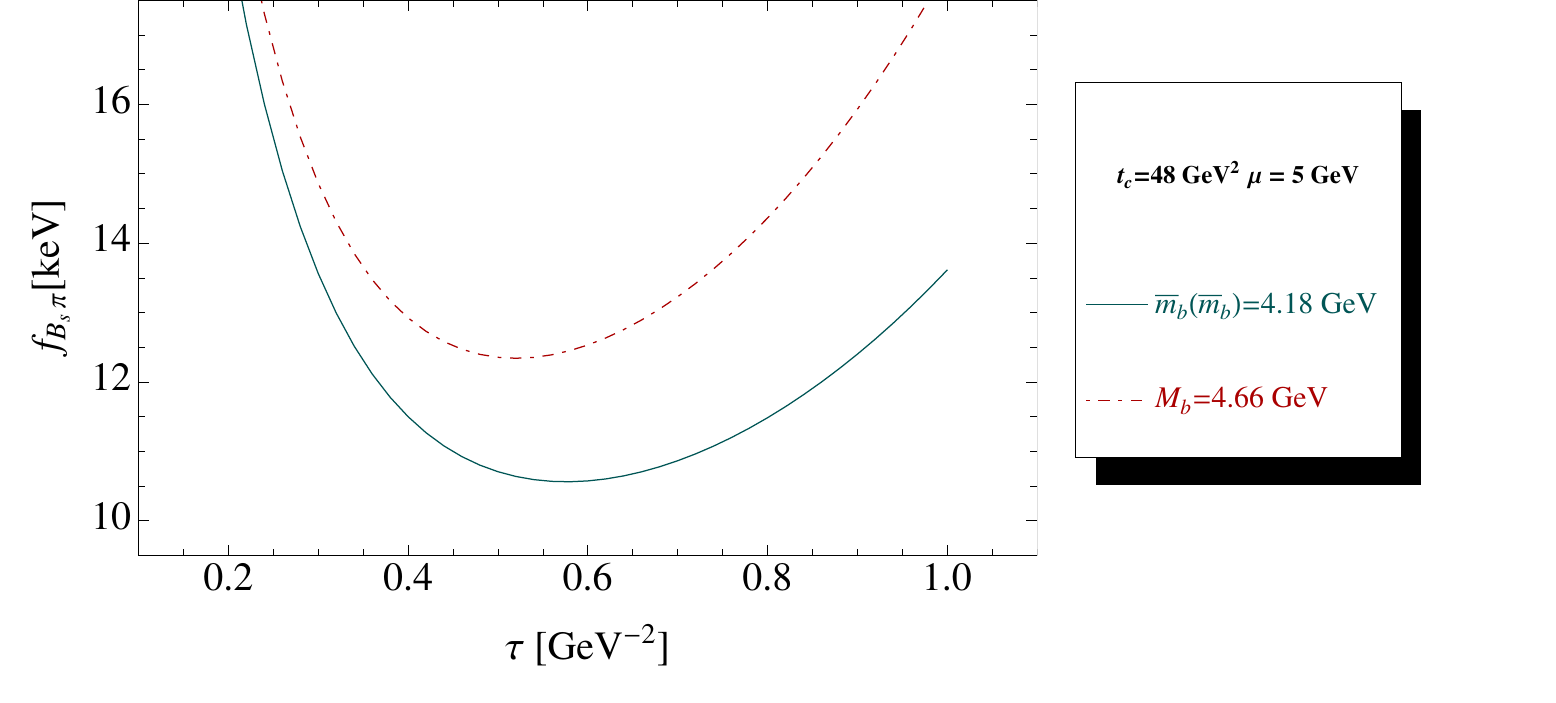}}
\centerline {\hspace*{-3cm} a)\hspace*{6cm} b) }

\caption{
\scriptsize 
{\bf a)} $f_{B_s\pi}$  at LO as function of $\tau$ for a given value of $t_c=48$ GeV$^2$,  $\mu=4.5$ GeV and for the QCD parameters in Tables\,\ref{tab:param} and \ref{tab:alfa};  The OPE is truncated at $d=6$.  We compare the effect of  the on-shell or pole mass $M_b=4.66$ GeV and of the running mass $\bar m_b(\bar m_b)=4.18$ GeV; {\bf b)} The same as a) but for the mass $M_{B_s\pi}$.
}
\label{fig:bspi-constituent} 
\end{center}
\end{figure} 
\nin
In Fig.\,\ref{fig:bspi-constituent}, we show the influence on the choice of the quark mass value on the coupling and mass at lowest order. One can see
an effect of about 27\% for the coupling and about 5\% for the mass.
\subsection{$\mu$-subtraction point stability }
\begin{figure}[hbt] 
\begin{center}
{\includegraphics[width=6.2cm  ]{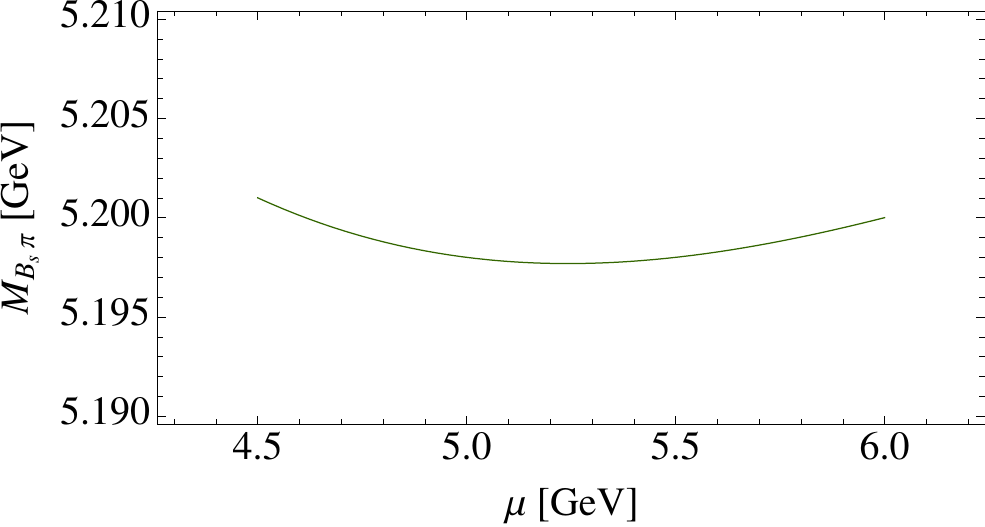}}
{\includegraphics[width=6.2cm  ]{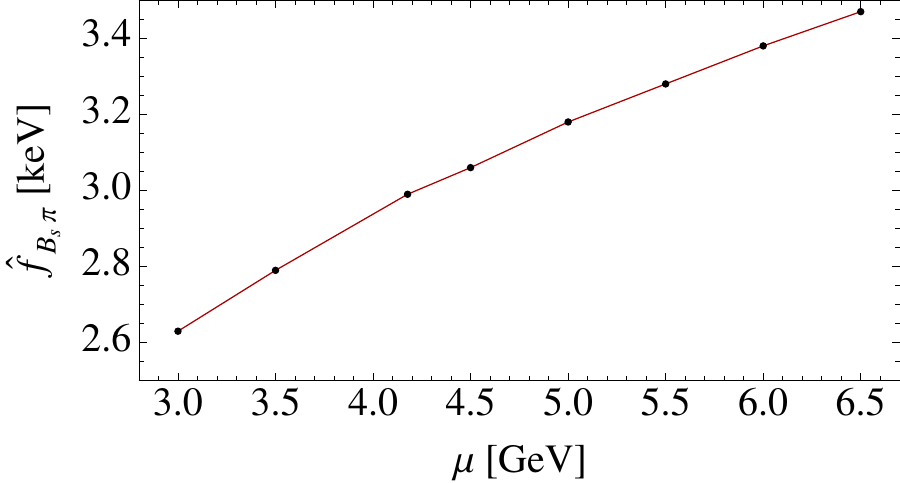}}
\centerline {\hspace*{-3cm} a)\hspace*{6cm} b) }
\caption{
\scriptsize 
{\bf a)} $M_{B_s\pi}$ at NLO as function of $\mu$, for the corresponding $\tau$-stability region, for $t_c\simeq 48$ GeV$^2$ and for the QCD parameters in Tables\,\ref{tab:param} and \ref{tab:alfa}; {\bf b)} The same as a) but for the  renormalization group invariant coupling $\hat{f}_{B_s\pi}$.
}
\label{fig:bspi-mu} 
\end{center}
\end{figure} 
\nin
We show in Fig.\,\ref{fig:bspi-mu} the dependence of $M_{B_s\pi}$ and of the renormalization group invariant coupling $\hat{f}_{{B_s\pi}}$ obtained at NLO of PT series on the choice of the  subtraction constant $\mu$.  We consider as an optimal  values the ones obtained for $\mu\simeq 5.0$ GeV where we have a minimum for the mass and a slight inflexion point for $\hat{f}_{B_s\pi}$. We deduce for $t_c\simeq (34\sim 48)$ GeV$^2$:
\beq
M_{B_s\pi}^{NLO}(\mu)\simeq (5196\sim 5198)~{\rm MeV}~~~~{\rm and}~~~~
\hat{f}_{B_s\pi}^{NLO}(\mu)\simeq ( 3.12\sim 3.18)~{\rm keV}~.
\eeq
\begin{figure}[hbt] 
\begin{center}
{\includegraphics[width=6.2cm  ]{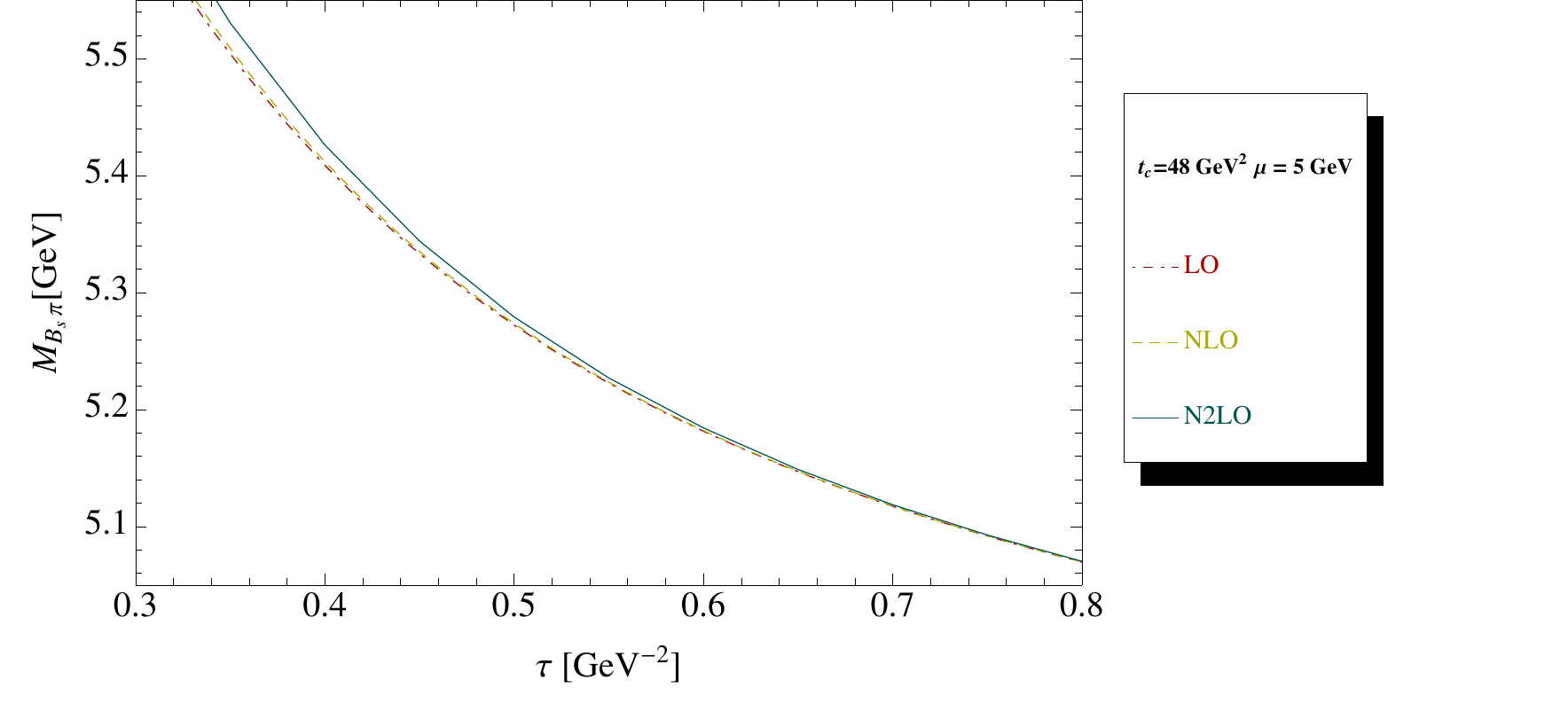}}
{\includegraphics[width=6.2cm  ]{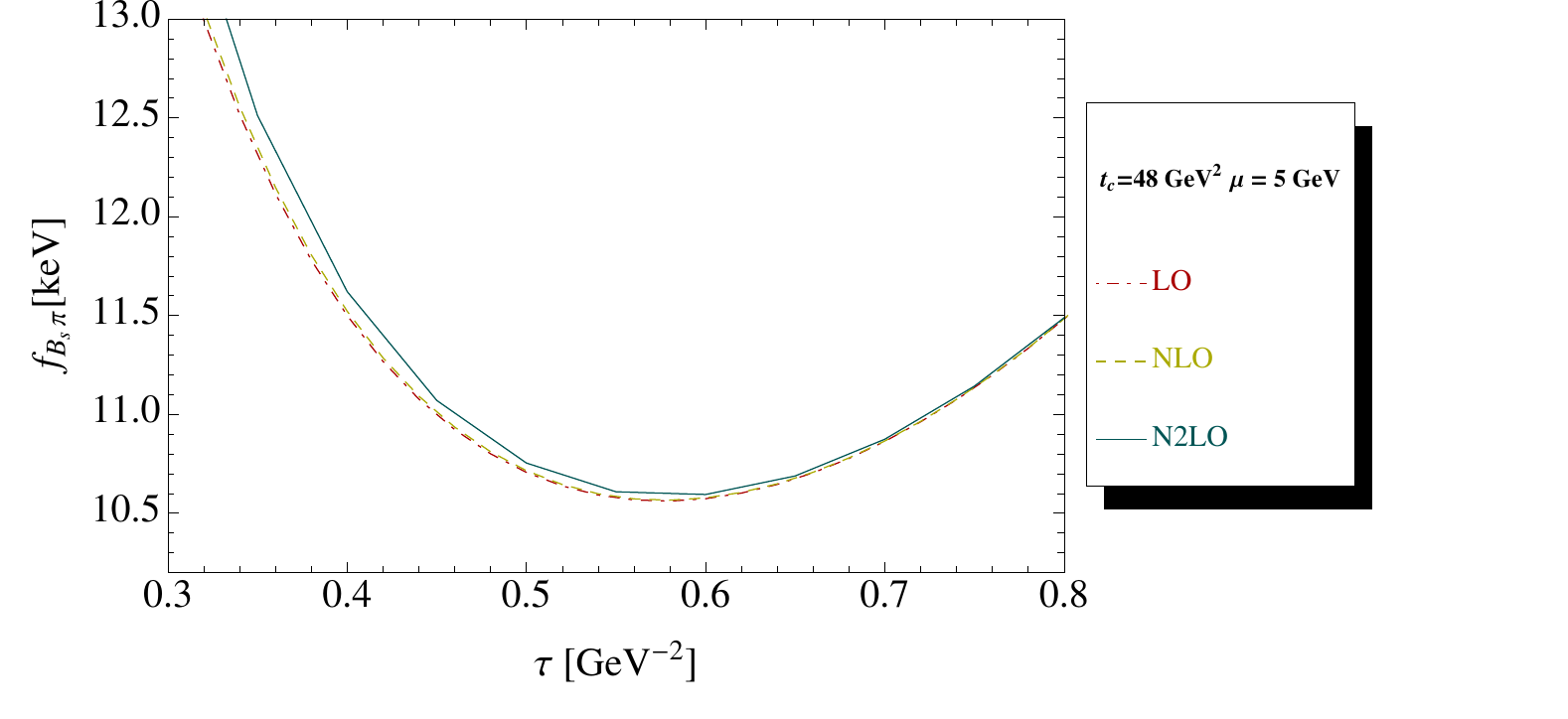}}
\centerline {\hspace*{-3cm} a)\hspace*{6cm} b) }
\caption{
\scriptsize 
{\bf a)} $M_{B_s\pi}$  as function of $\tau$ for different truncation of the PT series at a given value of $t_c$=48 GeV$^2$, $\mu=4.5$ GeV and for the QCD parameters in Tables\,\ref{tab:param} and \ref{tab:alfa}; {\bf b)} The same as a) but for the coupling $f_{B_s\pi}$.
}
\label{fig:bspi-lo-n2lo} 
\end{center}
\end{figure} 
\nin
\subsection{Test of the convergence of the  PT series}
We show in Fig.\,\ref{fig:bspi-lo-n2lo} the behaviour of the results for a given value of $t_c$  and for different truncation of the PT series. One can notice small PT corrections for the mass predictions because these corrections tend to compensate in the ratio of sum rules. For the coupling, the correction is large from LO to NLO.  For both observables,  one can notice a good convergence of the PT series from LO to N2LO.  
\subsection{Error induced by the OPE}
\begin{figure}[hbt] 
\begin{center}
{\includegraphics[width=6.2cm  ]{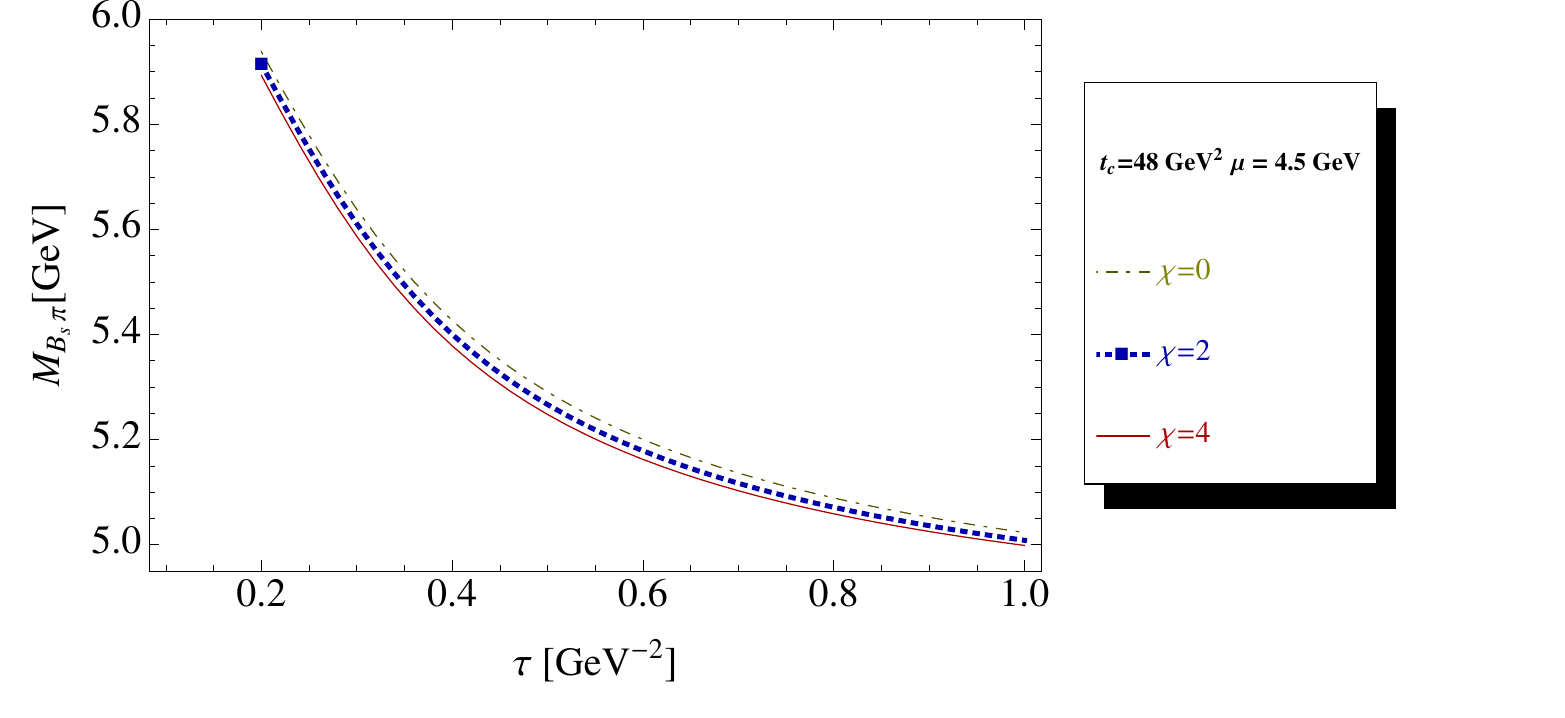}}
{\includegraphics[width=6.2cm  ]{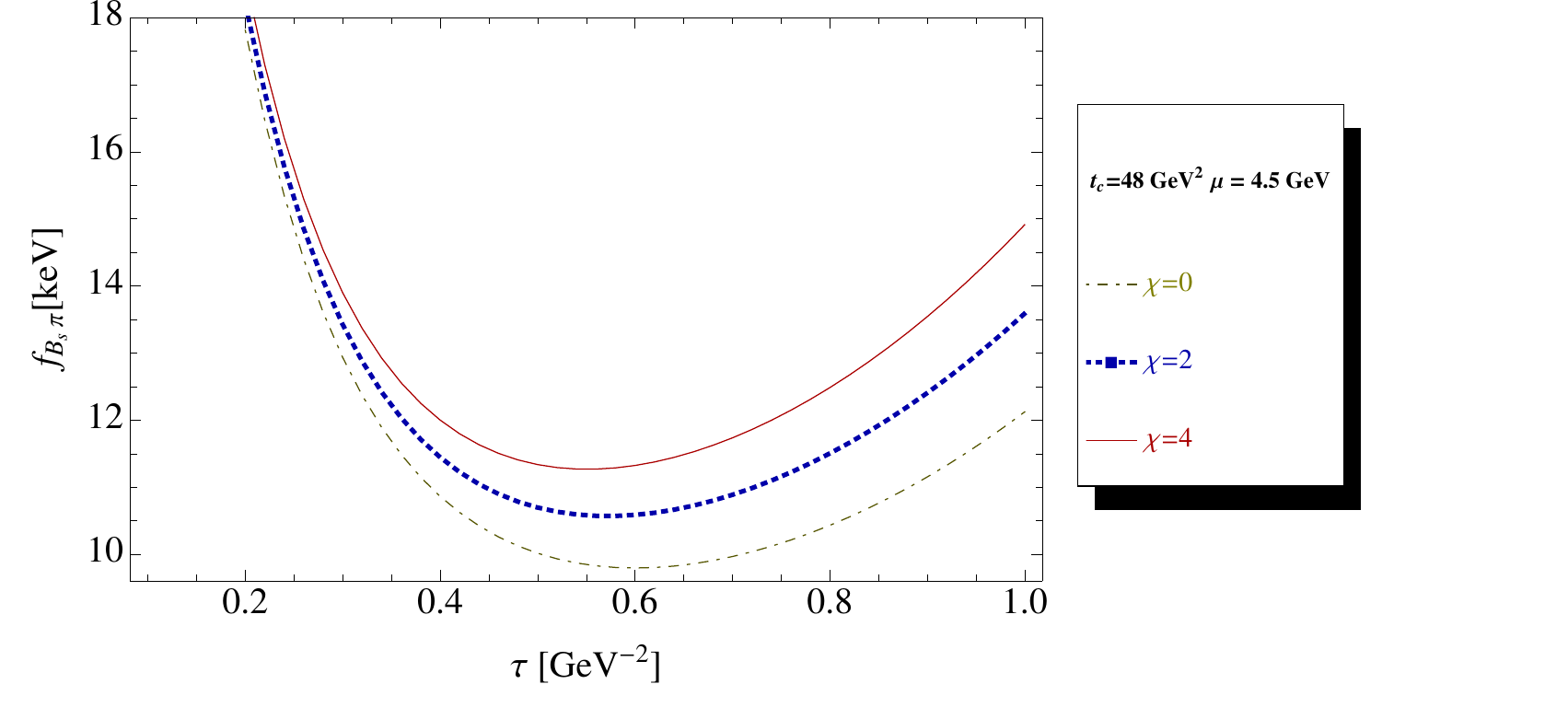}}
\centerline {\hspace*{-3cm} a)\hspace*{6cm} b) }
\caption{
\scriptsize 
{\bf a)} $M_{B_s\pi}$  as function of $\tau$, for different values of the $d=7$ condensate contribution ($\chi$ measures the violation of factorization),  at a given value of $t_c$=48 GeV$^2$, $\mu=4.5$ GeV and for the QCD parameters in Tables\,\ref{tab:param} and \ref{tab:alfa}; {\bf b)} The same as a) but for the coupling $f_{B_s\pi}$.
}
\label{fig:bspi-d7} 
\end{center}
\end{figure}

Like in previous section, we estimate this effect by taking a violation of about a factor 4 of the factorization assumption
for the $d=7$ condensates. The analysis is shown in Fig.\,\ref{fig:bspi-d7} from which we deduce for $t_c\simeq (34\sim 48)$ GeV$^2$:
\beq
 \Delta f_{B_s\pi}^{OPE}\simeq  \pm 1.48 ~{\rm keV}~,~~~~~~~~~~~~~~\Delta M_{B_s\pi}^{OPE}\simeq \pm 1~{\rm MeV}~. 
\eeq
\nin
\subsection{Final results}
We show in Table\,\ref{tab:error} our estimate of the different sources of errors. 
Adding quadratically the different
 sources of errors, 
we consider as a final estimate for $\mu$ 5 GeV and to order $\alpha_s^2$ or at  N2LO of the perturbative series:
\bea
M_{B_s\pi}&\simeq&( 5199\pm 24)~{\rm MeV}~,\nnb\\
\hat f_{B_s\pi}&\simeq& (3.15\pm 0.70)~{\rm keV}~~~~\Lrar~~~~f_{B_s\pi}\simeq (10.5\pm 2.3)~{\rm keV} ,
\label{eq:bspin2lo}
\eea
which we list in Table\,\ref{tab:result}.
One can notice that the $ B_s\pi$ molecule mass (if any) is also expected to be below the physical $ B_s\pi$ threshold of 5500 MeV. 
\section{QCD expression of the $0^+$  scalar four-quark $(su)( \overline{bd})$ states}
Using the expression of the vector current in Table\,\ref{tab:current}, we deduce the expression of the
scalar spectral function from the longitudinal part of the correlator. 
\bea
\rho^{pert}&=&\frac{(1 +k^{2}){M_{b}}^8}{5\, 3\, 2^{13}\pi^{4}}\Bigg{[}\frac{1}{x^{4}}-\frac{16}{x^3}-\frac{65}{x^2}+\frac{160}{x}-60\left(\frac{1}{x^2}-1\right){\rm Log}(x)
-65-16x+x^2\Bigg{]},\nnb\\
\rho^{\langle\bar{q}q\rangle}&=&-\frac{(1-k^2)M_{b}^{5}}{3\, 2^6\,\pi^4}\langle\bar{q}q\rangle\Bigg{[} \frac{1}{x^2}+\frac{9}{x}+6\left( \frac{1}{x}+1\right){\rm Log}(x)-9-x\Bigg{]}
\nnb\\&&
-\frac{m_s m_{b}^{4}}{3\, 2^9\,\pi^4}\langle\bar{q}q\rangle\Bigg{[} 2(1-k^2)-(1+k^2)\kappa\Bigg{]}
\Bigg{[} \frac{1}{x^2}+12\,{\rm Log}(x)+12-16x+3x^2\Bigg{]},\nnb\\
\rho^{\langle G^2\rangle}&=&\frac{(1+k^2)M_{b}^{4}}{3^2\,2^{13}\,\pi^6}4\pi\langle \alpha_{s}G^2\rangle\Bigg{[}\frac{5}{x^2}+\frac{40}{x}+24\left( \frac{1}{x}+2\right){\rm Log}(x)-18-32x+5x^2\Bigg{]},\nnb\\
\rho^{\langle\bar{q}Gq\rangle}&=&\frac{(1-k^2)m_{b}^{3}}{2^7\,\pi^4}\langle\bar{q}Gq\rangle\left[ \frac{1}{x}+2\,{\rm Log}(x)-x\right]
\nnb\\&&
+\frac{m_s M_{b}^{2}}{3\, 2^8\,\pi^4}\langle\bar{q}Gq\rangle\left( 6(1-k^2)+(1+k^2)\kappa\right)(1-x)^2,\nnb\\
\rho^{\langle\bar{q}q\rangle^2}&=&\frac{(1-k^2)M_{b}^{2}}{3\,2^3\,\pi^2}\rho\langle\bar{q}q\rangle\kappa(1-x)^2
+\frac{m_s M_b}{3\,2^3\,\pi^2}\rho\langle\bar{q}q\rangle^2\Bigg{[}2(1+k^2)-(1-k^2)\kappa \Bigg{]}(1-x)~,\nnb\\
\rho^{\langle G^3\rangle}&=&\frac{(1+k^2)M_{b}^{2}}{3^2\,2^{15}\,\pi^6}\langle g_{s}^{3}G^3\rangle\left[ \frac{1}{x^2}+\frac{8}{x}+36\, {\rm Log}(x)+24-40x+7x^2\right]~.
\eea
The contribution of a class of $d=7$ condensate for $m_s=0$  is:
\bea
\rho^{\la \bar qq\ra\la G^2\ra}&=&-(1-k^2)\frac{M_b\la \bar qq\ra}{3^2\ 2^8\pi ^4} 4\pi\la \alpha_s G^2\ra\left(\frac{2}{x}+6\,\text{Log}\,x+7-9x\right)~.
\label{eq:d7scalar4}
\eea
\begin{figure}[hbt] 
\begin{center}
{\includegraphics[width=6.2cm  ]{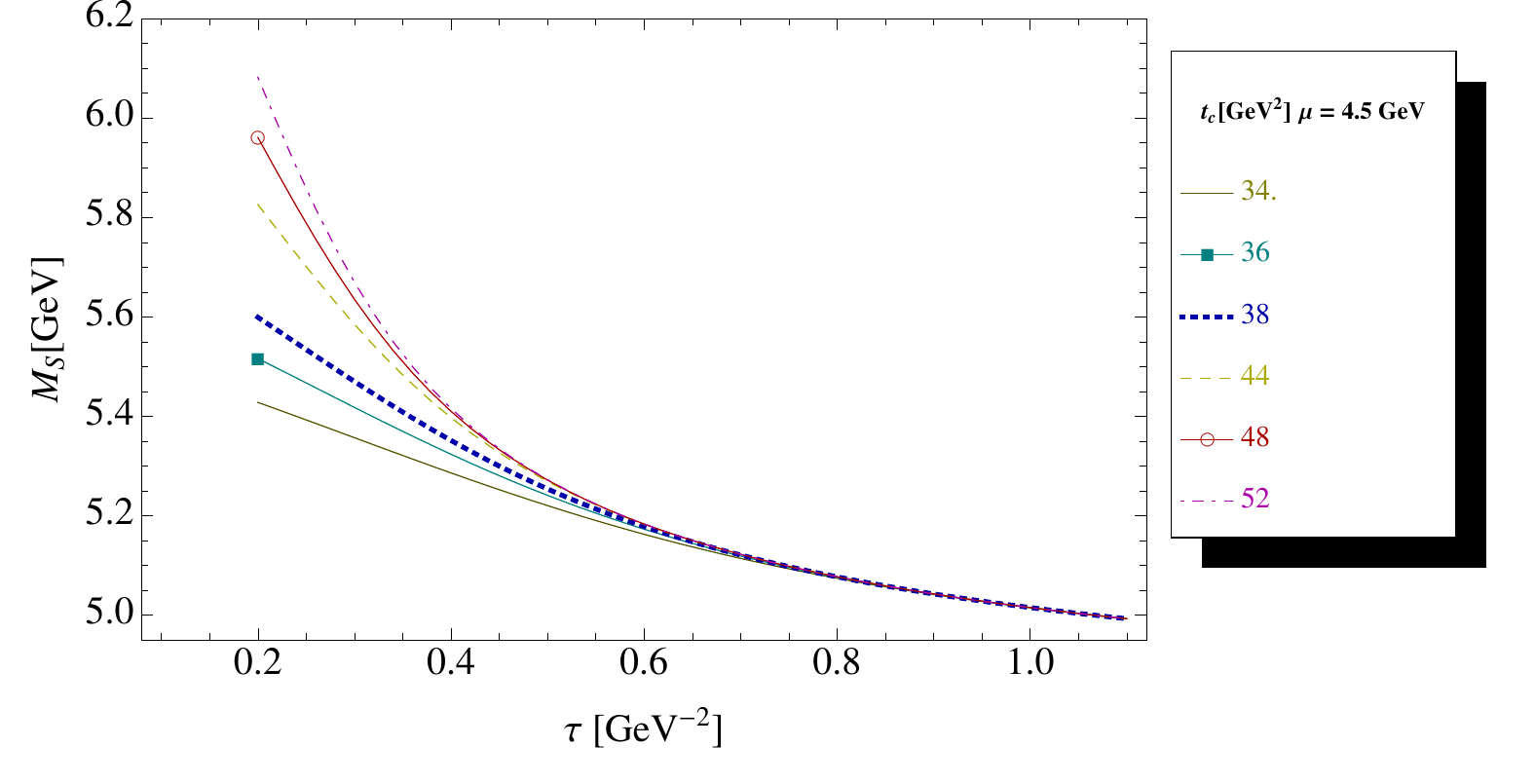}}
{\includegraphics[width=6.2cm  ]{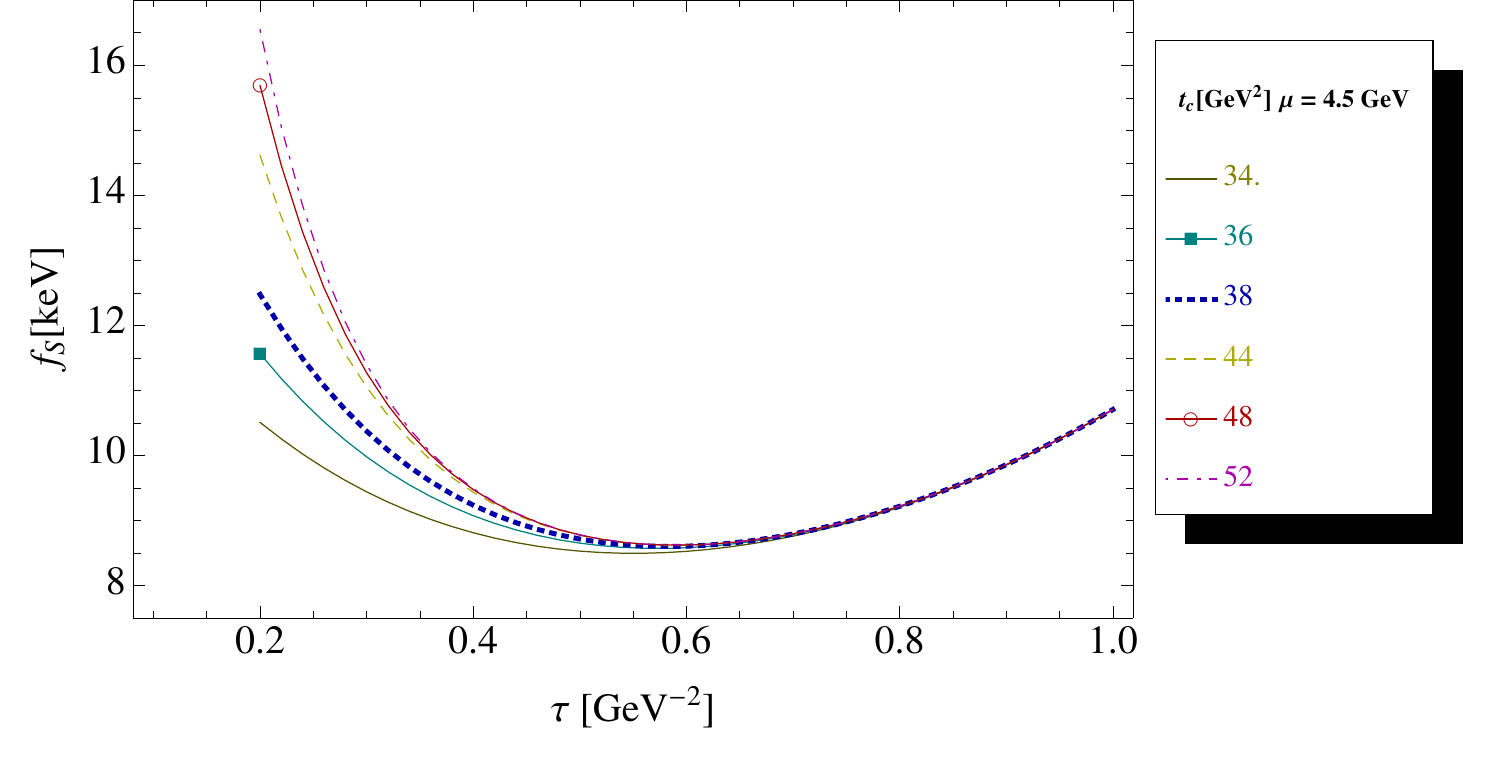}}
\centerline {\hspace*{-3cm} a)\hspace*{6cm} b) }
\caption{
\scriptsize 
{\bf a)} $M_{S_b}$  at LO as function of $\tau$ for different values of $t_c$, $\mu=4.5$ GeV, mixing of currents $k=0$ and for the QCD parameters in Tables\,\ref{tab:param}  and \ref{tab:alfa};   {\bf b)} The same as a) but for the coupling $f_S$.
}
\label{fig:s-lo} 
\end{center}
\end{figure} 
\nin
\begin{figure}[hbt] 
\begin{center}
{\includegraphics[width=6.2cm  ]{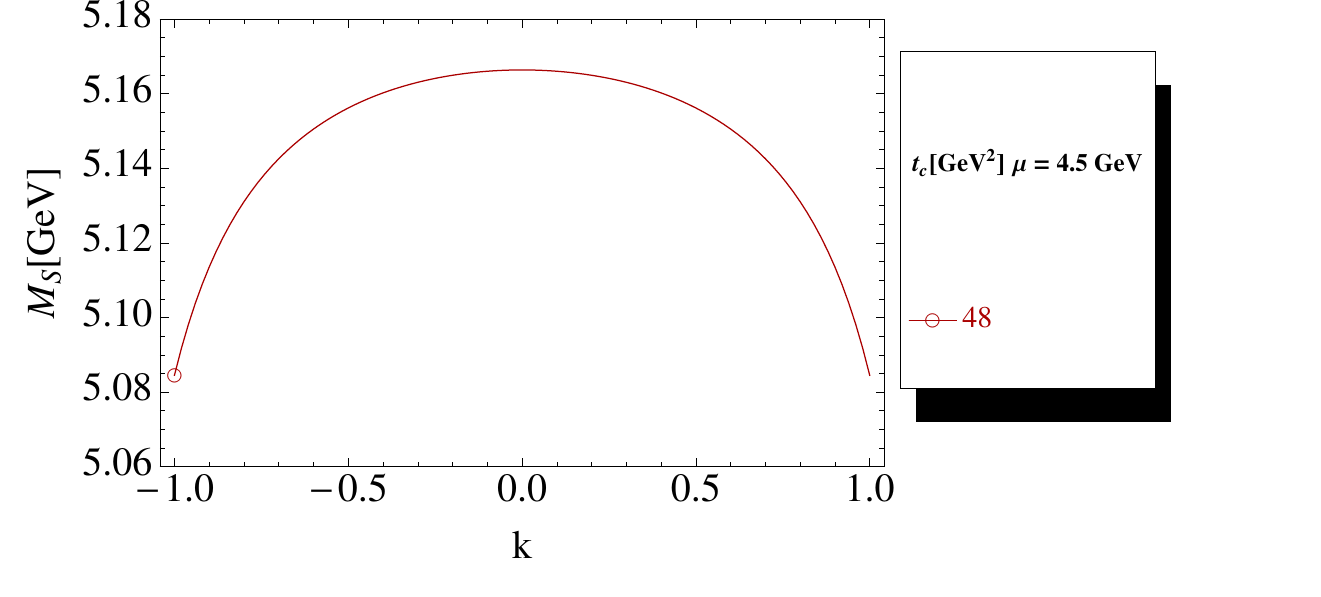}}
{\includegraphics[width=6.2cm  ]{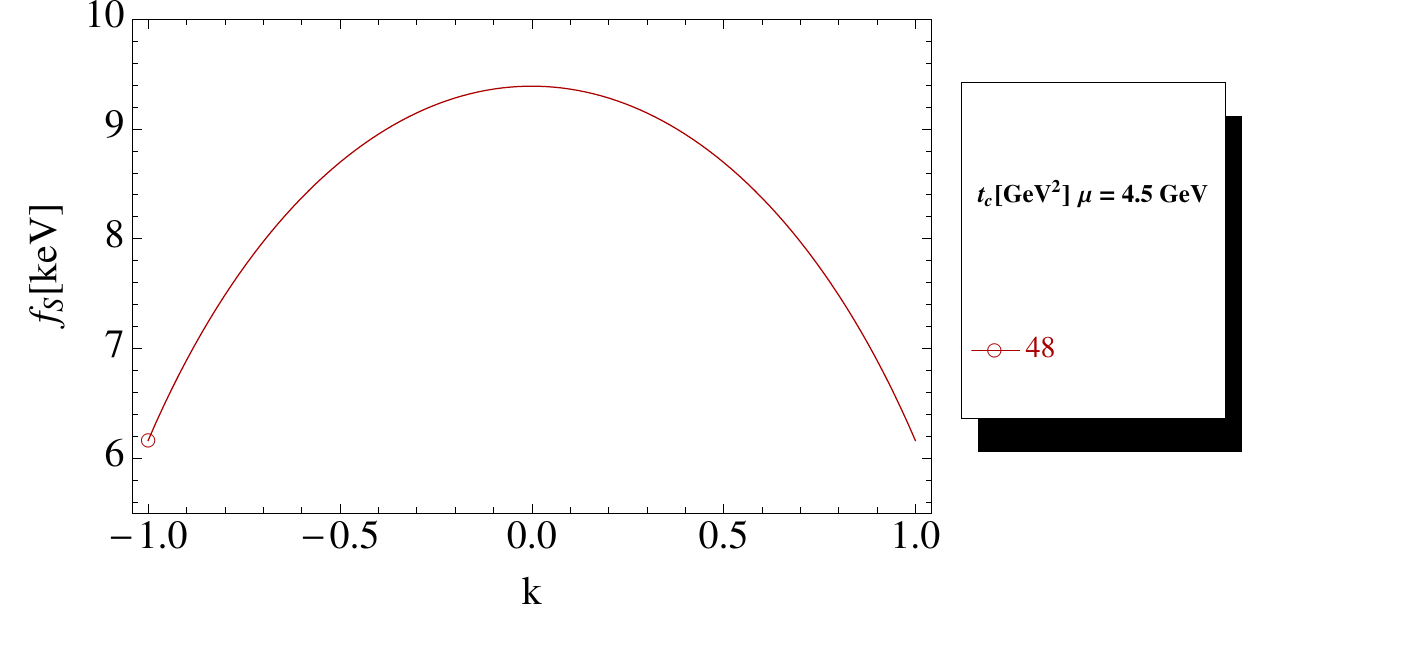}}
\centerline {\hspace*{-3cm} a)\hspace*{6cm} b) }
\caption{
\scriptsize 
{\bf a)} $M_{S_b}$  at LO as function of the mixing of currents $k$, for $\tau=0.6$ GeV$^{-2}, ~\mu=4.5$ GeV, $t_c$=48 GeV$^2$  and for the QCD parameters in Tables\,\ref{tab:param} and \ref{tab:alfa};  {\bf b)} The same as a) but for the coupling $f_{S_b}$.
}
\label{fig:s-lo-k} 
\end{center}
\end{figure} 
\nin
\section{Mass and coupling of the $0^{+}$ scalar $S_b$ four-quark state}
\subsection{$\tau$- and $t_c$-stabilities}
We study the $\tau$- and $t_c$-stabilities of the mass and coupling predictions in Fig.\,\ref{fig:s-lo} by fixing the  subtraction point $\mu$=4.5 GeV and the mixing of current $k$ defined in Table\,\ref{tab:current} to be equal to zero. We have an inflexion point for the 
mass and a minimum for the coupling for $\tau\simeq (0.56\sim 0.6) $ GeV$^{-2}$. This $\tau$-stability is reached from $t_c\simeq 34$ GeV$^2$ 
while $t_c$-stability is obtained around $t_c=48$ GeV$^2$. 
\subsection{Optimal choice of the four-quark currents}
We show in Fig.\,\ref{fig:s-lo-k} the behaviour of the mass and coupling predictions versus the mixing of current $k$ defined in Table\,\ref{tab:current} given the value $\tau\simeq 0.6$ GeV$^{-2}, ~\mu=4.5$ GeV, $t_c$=48 GeV$^2$  and for the QCD parameters in Table\,\ref{tab:param}. One can notice that the optimal choice is obtained for $k=0$ which can simply be understood analytically by taking the 
zero of the derivative of the spectral function versus k:
\beq
{\partial \rho\over \partial k}=0~~~\Lrar~~~ k=0.
\eeq
{\scriptsize
\begin{table}[hbt]
 \tbl{Different sources of errors for the estimate of the axial-vector $(A_{b,c})$ and scalar four-quark $(S_{b,c})$ masses (in units of MeV) and couplings (in units of keV). }  
    {\scriptsize
 {\begin{tabular}{@{}llllllllll@{}} \toprule
&\\
\hline
\hline
\bf Inputs $[GeV]^d$&$\Delta M_{S_b}$&$\Delta f_{S_b}$&$\Delta M_{A_b}$&$\Delta f_{A_b}$
&$\Delta M_{S_c}$&$\Delta f_{S_c}$&$\Delta M_{A_c}$&$\Delta f_{A_c}$\\
\hline 
{\it LSR parameters}&\\
$t^c_c=(12\sim 18)$&  -- &--& --&--&     0.5  & 2.7&7 & 2\\
$t^b_c=(34\sim 48)$& 4  &0.16& 2& 0.12&  --   & --&-- &-- \\
$\mu^c=(2.0\sim 2.5)$&--  & --&-- & --&     0.5 &11&9.5&15\\
$\mu^b=(4.5\sim 5.0)$& 1 &0.34&6&0.36&     -- &--&--&--\\
$\tau=\tau_{min}\pm 0.02$&15&0.01&12&0.01&       30&0.12&28&0.2\\
{\it QCD inputs}&\\
$\bar m_{b,c}$& 0.85&0.024&0.85&0.026&                      5.0&2.2&5.5&3.2\\
$\bar m_s$& 0.58&0.058&0.13&0.025&            1.62&1.57&1.66&0.67\\
$\alpha_s$& 5.60&0.19&5.64&0.21&                    7.9&4.7&7.86&6.70\\
$\la\bar qq\ra$&0.63&0.004&0.43&0.006&            6.4&0.58&3.7&0.33\\
$\kappa$&1.88&1.11&4.86&1.05&                28.3&21.5&12.9&29.1\\
$\la\alpha_s G^2\ra$& 0.63&0.010&0.48&0.008&           3.35&1.0&1.43&0.74\\
$M_0^2$&1.95&0.056&1.43&0.054&                             4.1&3.43&1.65&3.06\\
$\la\bar qq\ra^2$&4.4&1.48&2.96&1.45&          51&29.6&30.6&37.4\\
$\la g^3G^3\ra$&0.0&0.0&0.0&0.0&0.0&0.0&0.0&0.0\\
$d\geq 7$&1.0&1.4&2.0&1.49&11.0&25.4&10.0&23.3\\
{\it Total errors}&17.4&2.36&16.1&2.37&                68.0&46.5&47.0&55.0\\
\hline\hline
\end{tabular}}
\label{tab:error4}
}
\end{table}
} 
\subsection{Lowest Order (LO) results}
Therefore, to the LO approximation, we deduce from Fig.\,\ref{fig:s-lo} at the stability region:
\beq
M_{S_b}^{LO}\simeq( 5.19\sim 5.18)~{\rm GeV}~~~~{\rm and}~~~~
f_{S_b}^{LO}\simeq (8.49\sim 8.62)~{\rm keV}~.
\eeq
\subsection{$b$-quark mass ambiguity at lowest order (LO)}
\begin{figure}[hbt] 
\begin{center}
{\includegraphics[width=6.2cm  ]{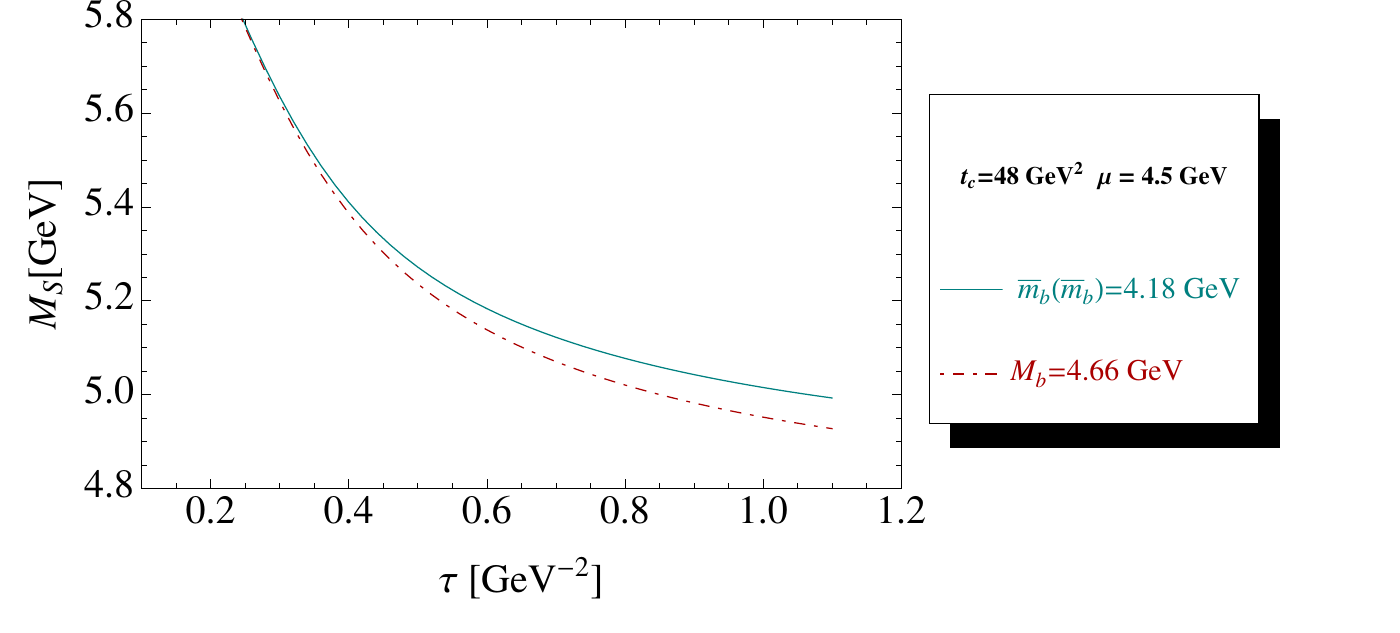}}
{\includegraphics[width=6.2cm  ]{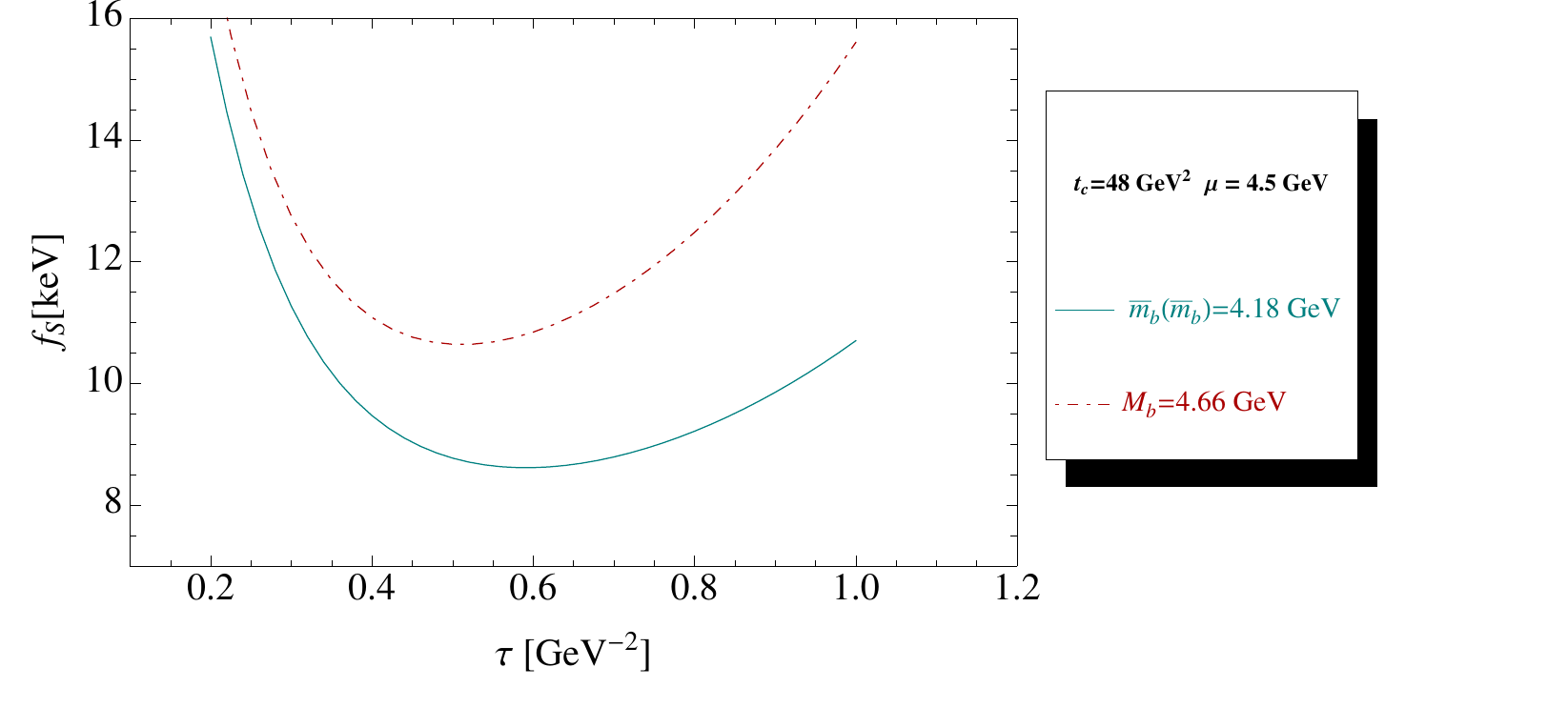}}
\centerline {\hspace*{-3cm} a)\hspace*{6cm} b) }
\caption{
\scriptsize 
{\bf a)} $M_{S_b}$  at LO as function of $\tau$ for a given value of $t_c=48$ GeV$^2$,  $\mu=4.5$ GeV, mixing of currents $k=0$ and for the QCD parameters in Table\,\ref{tab:param}  and \ref{tab:alfa};  The OPE is truncated at $d=6$.  We compare the effect of  the on-shell or pole mass $M_b=4.66$ GeV and of the running mass $\bar m_b(\bar m_b)=4.18$ GeV; {\bf b)} The same as a) but for the coupling $f_{S_b}$.
}
\label{fig:smasspole} 
\end{center}
\end{figure}

We compare in Fig.\,\ref{fig:smasspole}  the result when we use the $b$-quark pole mass value of 4.66 GeV\,\cite{PDG} and the running mass $\bar m_b(m_b)=4.18$ GeV in Table\,\ref{tab:param}. One can find that this ( a priori) choice
introduces an intrinsic source of error :
\beq
\Delta M_{S_b}^{LO}\simeq 43 ~{\rm MeV}~~~~~{\rm and}~~~~~ \Delta f_{S_b}^{LO}\simeq 2.6~{\rm keV}~,
\eeq
 which  should be added into the error when one does the LO analysis.
\nin
\begin{figure}[hbt] 
\begin{center}
{\includegraphics[width=6.2cm  ]{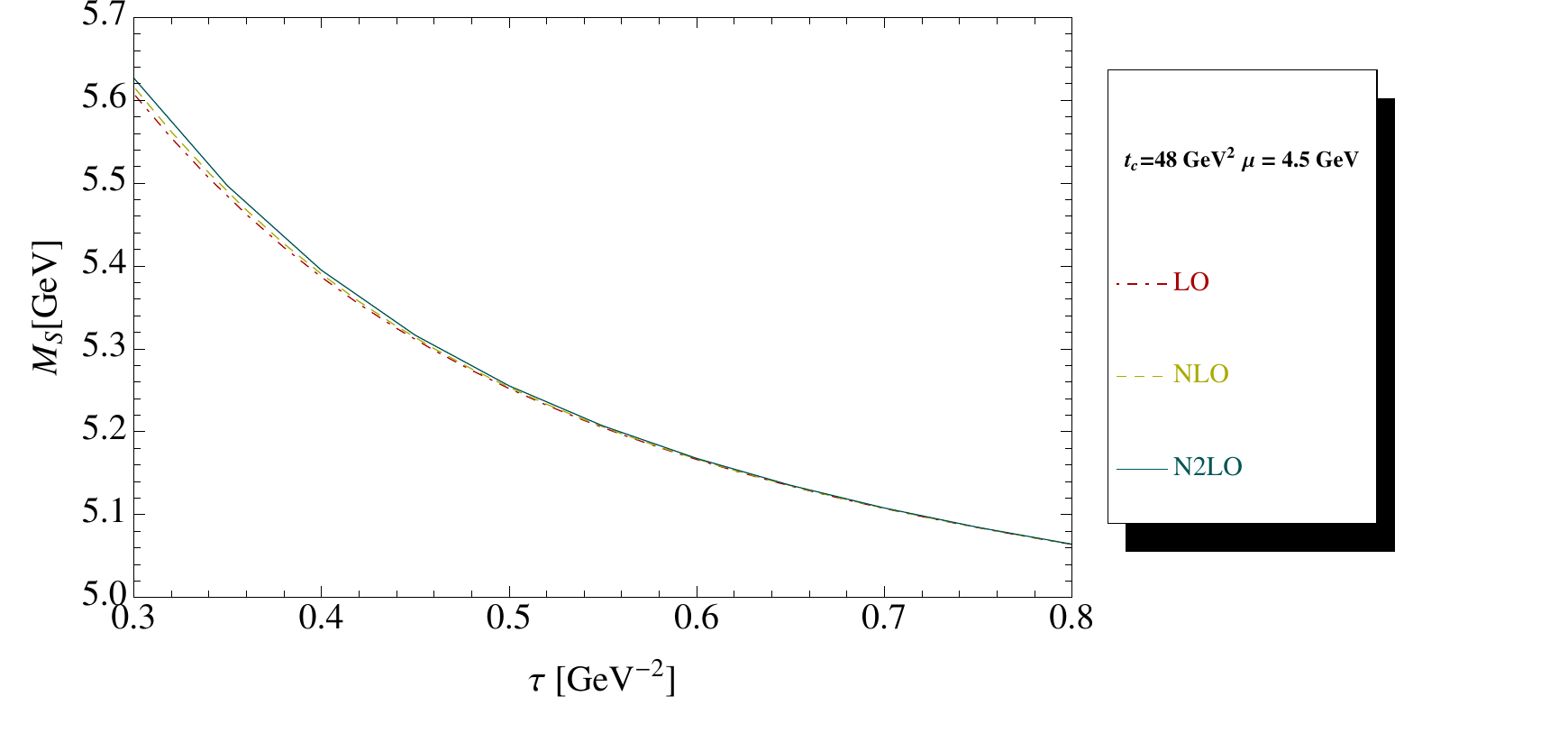}}
{\includegraphics[width=6.2cm  ]{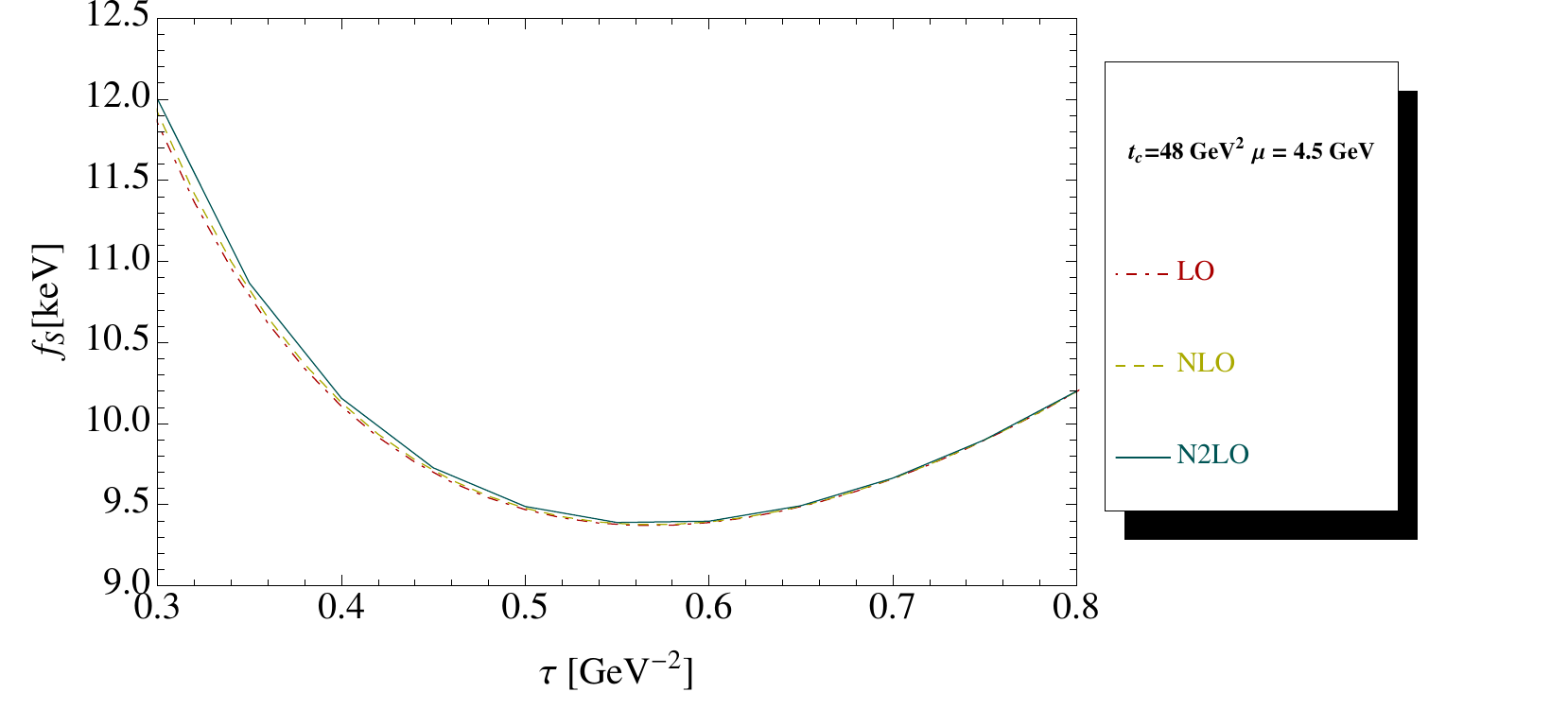}}
\centerline {\hspace*{-3cm} a)\hspace*{6cm} b) }
\caption{
\scriptsize 
{\bf a)} $M_{S_b}$  as function of $\tau$ for different truncation of the PT series at a given value of $t_c$=48 GeV$^2$, $\mu=4.5$ GeV and for the QCD parameters in Tables\,\ref{tab:param}  and \ref{tab:alfa};  {\bf b)} The same as a) but for the coupling $f_{S_b}$.
}
\label{fig:s-lo-n2lo} 
\end{center}
\end{figure} 
\nin
\subsection{Comparison of the LO results with the previous ones}
\hspace*{0.5cm} \b We also compare our QCD expression of the spectral function with the ones in \cite{NIELSEN,STEELE,WANG} where all authors use the four-quark current :
\beq
J_{4q}=(s^T\gamma_5Cu)(\bar b \gamma_5 C\bar d^T)~,
\eeq
which corresponds to the optimal choice $k=0$, while, equivalently, we take the longitudinal part of the correlator associated to the current
defined in Table\,\ref{tab:current}:

\hspace*{0.5cm}-- We do not agree with the QCD expression given in Ref.\,\cite{TURC1} (see footnote on page 18).\\
\hspace*{0.5cm}-- We agree with the perturbative and $\la \alpha_s G^2\ra$ coefficient of \cite{STEELE}, while the \\
\hspace*{1.3cm} coefficient of their  $\la\bar qq\ra$ is higher than ours by about a factor 1.3. We \\
\hspace*{1.3cm} differs with \cite{STEELE} for the four-quark
condensate $\la\bar qq\ra^2$ coefficient which is about\\
\hspace*{1.3cm} a factor 3 smaller in absolute value than the one of \cite{NIELSEN}. 

\hspace*{0.5cm}-- Ref.\,\cite{WANG} does not provide the expressions of his spectral function such that  \\ \hspace*{1.3cm} no comparison can be done.

\b Hopefully, such discrepancies will affect only slightly the numerical analysis below because the contributions of the condensates are small though necessary corrections in the OPE.

\b One can compare our numerical results with the ones in \cite{NIELSEN,STEELE,WANG} where the running mass $\bar m_b(m_b)\simeq (4.18\sim 4.24)$ GeV has been implicitly
used in different papers:

\hspace*{0.5cm}-- The authors in Ref.\,\cite{NIELSEN} choose $\tau \simeq (0.33\sim 0.45)$\,GeV$^{-2}$, $t_c\simeq (35 \sim 37)$ 
\hspace*{1.2cm} GeV$^2$, 
 $\bar m_b(\bar m_b)\simeq 4.24$ GeV and found 
$M_{S_b}= (5.58\pm 0.17)$ GeV. 

\hspace*{0.5cm}-- The authors in Ref.\,\cite{STEELE} take $\tau \simeq (0.14\sim 0.17)$ GeV$^{-2}$, $t_c\simeq 34$ GeV$^2,$ 
\hspace*{1.2cm} $ \bar m_b(\bar m_b)=4.18$ GeV and obtain : 
$M_{S_b}= (5.58\pm 0.14)$ GeV. 

\hspace*{0.5cm}-- The author in Ref.\,\cite{WANG} uses  $\tau \simeq (0.20\sim 0.22)$ GeV$^{-2}$, $t_c\simeq 37$ GeV$^2,$ 
\hspace*{1.2cm} $\bar m_b(\bar m_b)=4.18$ GeV and predicts : 
$M_{S_b}= (5.57\pm 0.12)$ GeV and $f_{S_b}=6.9$\\
\hspace*{1.2cm} \,{\rm keV}. 

\b Using the previous values of the set $(\tau,t_c, \bar m_b(\bar m_b))$ choosen by previous authors, and our set of QCD parameters in Table\,\ref{tab:param}, we deduce from Fig.\,\ref{fig:s-lo}:

\hspace*{0.5cm}-- $M_{S_b}=(5.32\sim 5.46)$ GeV  (Ref.\,\cite{NIELSEN} choice)~,

\hspace*{0.5cm}-- $M_{S_b}=(5.46\sim 5.47)$ GeV  (Ref.\,\cite{STEELE} choice)~,

\hspace*{0.5cm}-- $M_{S_b}=(5.54\sim 5.57)$ GeV  (Ref.\,\cite{WANG} choice)~,
\\
which agrees within the errors with the previous results. 

\b However, one should note that in the low-value of $\tau$ outside the stability regions used by different authors, the mass and the coupling are very sensitive to the 
value of the continuum threshold $t_c$ as can be inspected from Fig.\,\ref{fig:s-lo}. In addition,  the value of the mass is also affected by the change of $\tau$\,\footnote{Like in \cite{TURC}, the apparent stability shown in the figures of \cite{NIELSEN,STEELE,WANG} is only due to the choosen scale of the frame and to the small range of variation of $\tau$ or equivalently $M^2$.} such that these predictions become unreliable. 
\subsection{Improvement of the LO results including HO corrections}
\hspace*{0.5cm} \b
We improve the previous LO results by including higher order (HO) perturbative corrections. In so doing, we assume that these corrections are dominated by the one from factorized diagrams (leading order in 1/$N_c$-expansion where $N_c$ is the colour number). This assumption can be supported by the analysis of the $(\bar bd)(\bar db)$ four-quark correlator controlling the $\bar B^0B^0$ mixing by \cite{SNPIVO} where it has been shown that the non-factorizable diagram gives a small $\alpha_s$-correction of about 10\% of the
factorizable one. Within this assumption, we proceeds like in the case of the molecule by using the four-quark correlator as a convolution of two correlator built from bilinear quark anti-quark fields after Fierz transformations. 

\b We show the effects of these PT corrections in Fig.\,\ref{fig:s-lo-n2lo} for $\mu=4.5$ GeV and $t_c=48$ GeV$^2$, where one can notice that these corrections are small. 
\begin{figure}[hbt] 
\begin{center}
{\includegraphics[width=6.2cm  ]{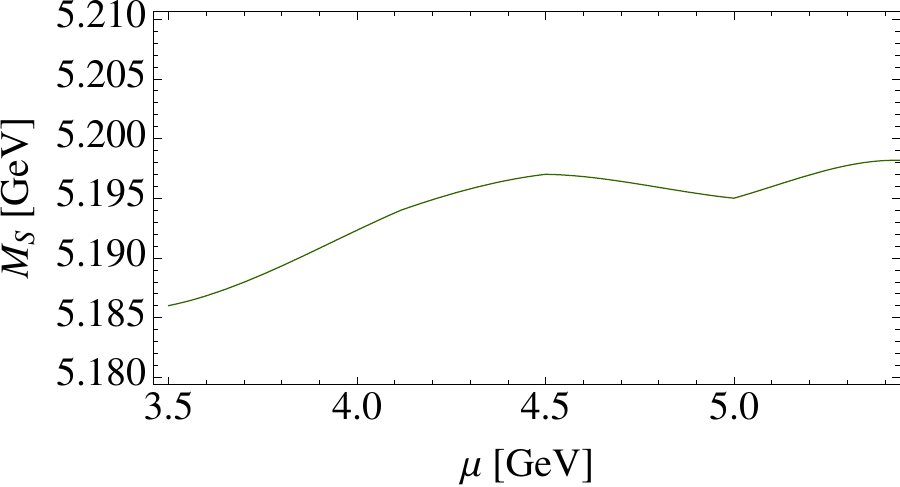}}
{\includegraphics[width=6.2cm  ]{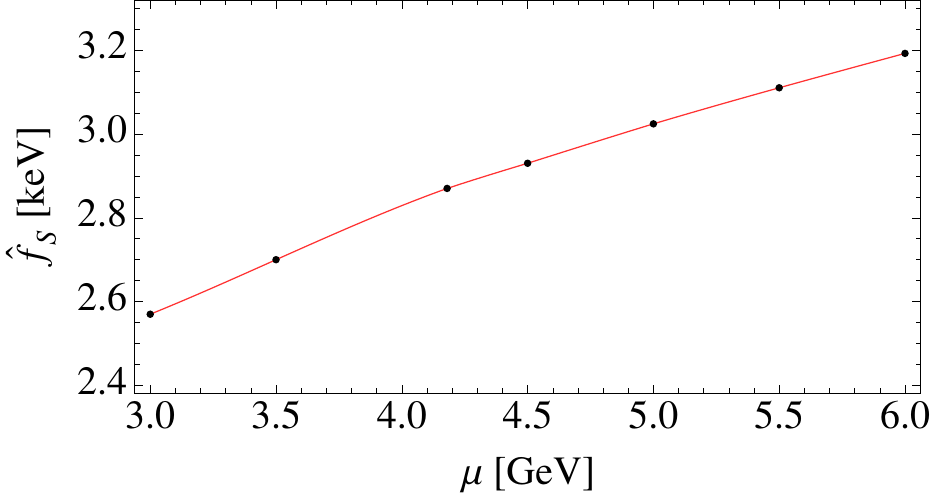}}
\centerline {\hspace*{-3cm} a)\hspace*{6cm} b) }
\caption{
\scriptsize 
{\bf a)} $M_{S_b}$ at NLO as function of $\mu$, for the corresponding $\tau$-stability region, for $t_c\simeq 48$ GeV$^2$ and for the QCD parameters in Tables\,\ref{tab:param}  and \ref{tab:alfa};  {\bf b)} The same as a) but for the renormalization group invariant coupling $\hat{f}_{S_b}$.
}
\label{fig:s-mu} 
\end{center}
\end{figure} 
\nin
\subsection{$\mu$-subtraction point stability }
We show in Fig.\,\ref{fig:s-mu} the dependence of $M_{S_b}$ and of the renormalization group invariant coupling $\hat{f}_{S_b}$ obtained at NLO of PT series on the choice of the  subtraction constant $\mu$.  We consider as an optimal  values the ones obtained for $\mu\simeq (4.5\sim 5)$ GeV where we have a plateau for the mass and a slight inflexion point for $\hat{f}_{BK}$. We deduce:
\bea
M_{S_b}^{NLO}(4.5)&\simeq& (5197\sim 5195)~{\rm MeV}~,\nnb\\
\hat{f}_{S_b}^{NLO}&\simeq& (2.93\sim 3.02)~{\rm keV} ~~~~ \Lrar ~~~~ f_{S_b}^{NLO}(4.5)\simeq (9.38\sim 10.06)~{\rm keV}~.
\label{eq:s-mu}
\eea
\subsection{Error induced by the truncation of the OPE}
\begin{figure}[hbt] 
\begin{center}
{\includegraphics[width=6.2cm  ]{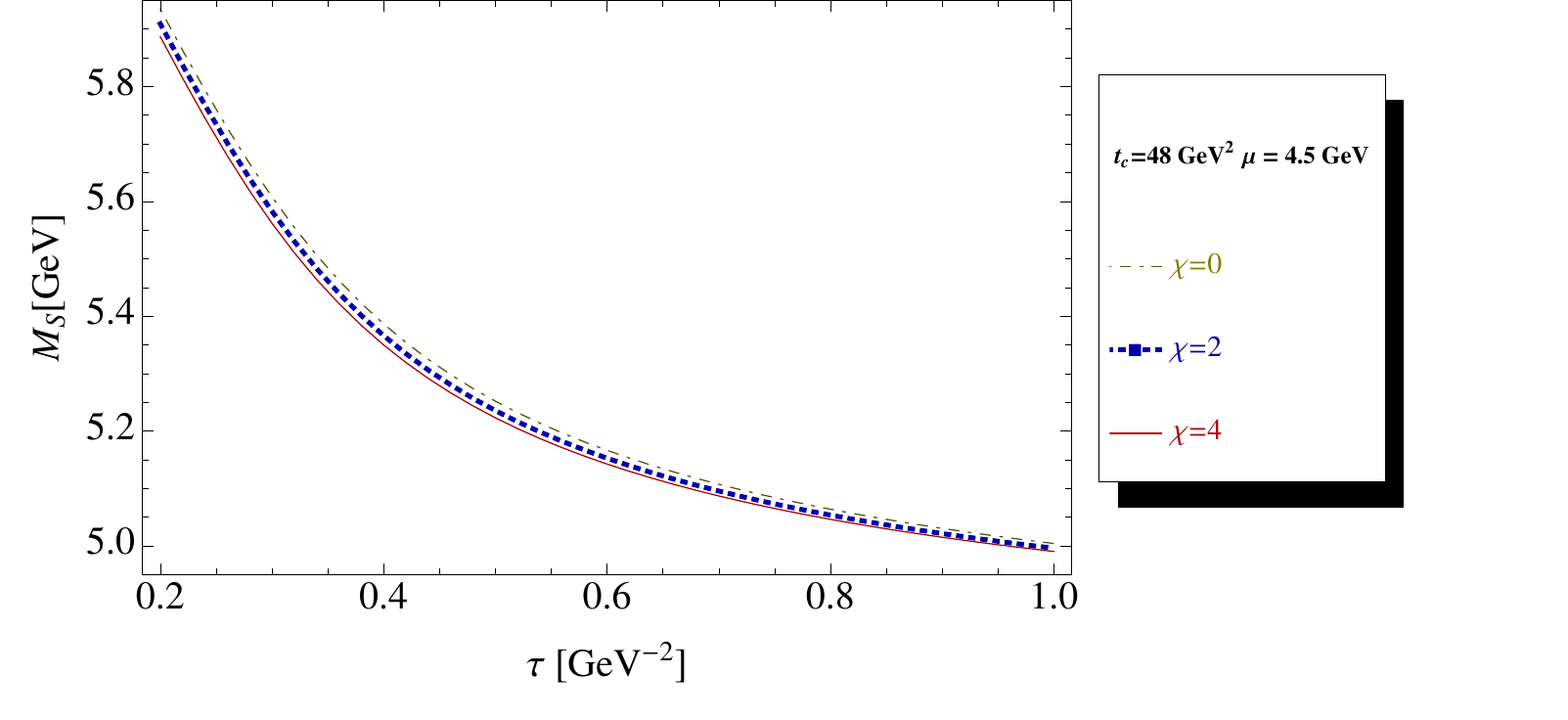}}
{\includegraphics[width=6.2cm  ]{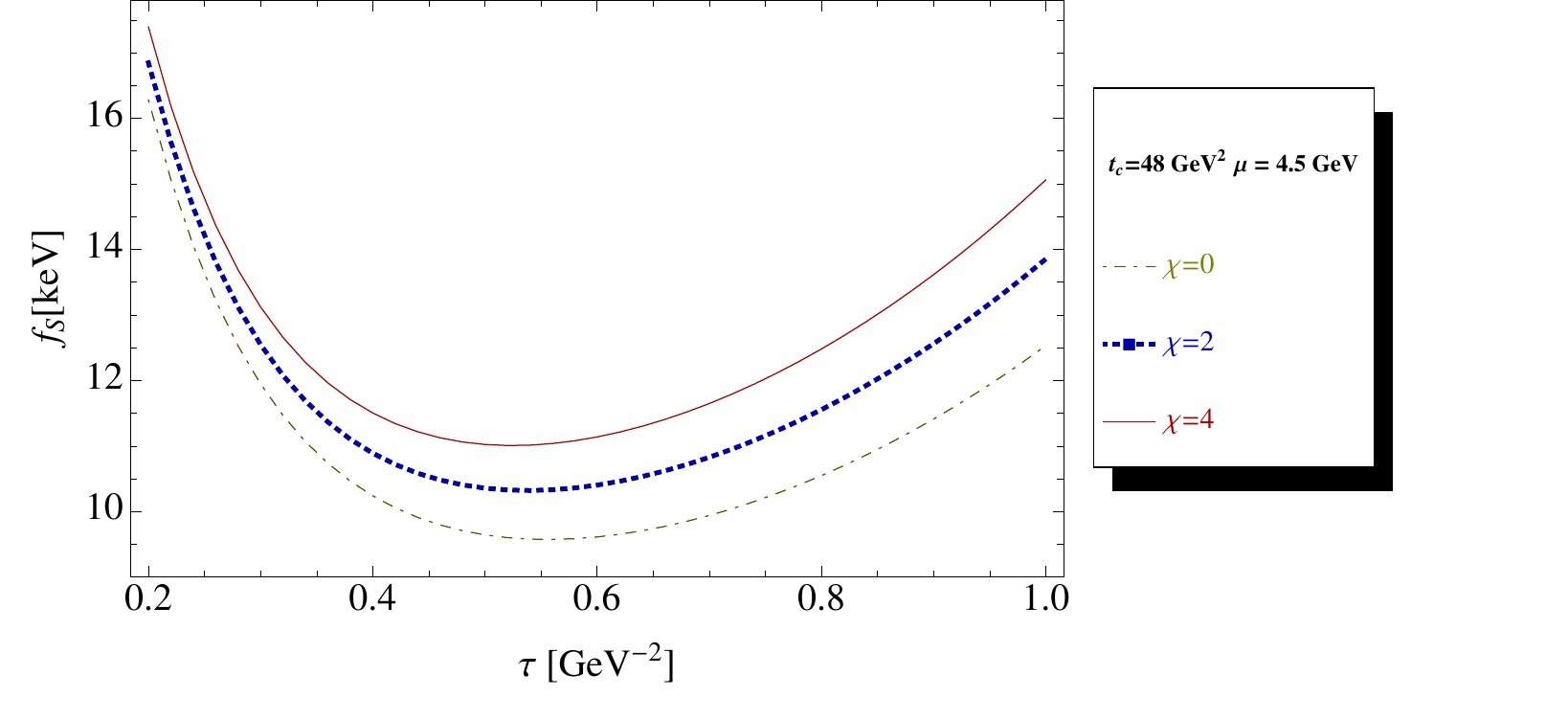}}
\centerline {\hspace*{-3cm} a)\hspace*{6cm} b) }
\caption{
\scriptsize 
{\bf a)} $M_{S_b}$  as function of $\tau$, for different values of the $d=7$ condensate contribution ($\chi$ measures the violation of factorization),  at a given value of $t_c$=48 GeV$^2$, $\mu=4.5$ GeV and for the QCD parameters in Tables\,\ref{tab:param} and \ref{tab:alfa}; {\bf b)} The same as a) but for the coupling $f_{S_b}$.
}
\label{fig:sb-d7} 
\end{center}
\end{figure}

We show in Fig.\,\ref{fig:sb-d7} the effect of a class of $d=7$ condensates for different values of the factorization violation parameter $\chi$.
Our estimate of the error induced by the truncation of the OPE corresponds to the choice $\chi=4$. 
\nin
\subsection{Final results}
We conclude from previous analysis that the mass and coupling of the $0^{+}$ scalar four-quark state to N2LO is:
\bea
M_{S_b}&\simeq& (5196\pm 17)~{\rm MeV}~,\nnb\\
\hat{f}_{S_b}&\simeq& (2.98\pm 0.70)~{\rm keV}~~~~ \Lrar ~~~~f_{S_b}(4.5)\simeq (9.99\pm 2.36)~{\rm keV}~,
\label{eq:s-final}
\eea 
where the errors come from Table\,\ref{tab:error4}. 
Our mass prediction is lower than previous sum rule results\,\cite{NIELSEN,STEELE,WANG} and
the $D0$ candidate $X(5568)$. 
\section{Spectral function of the axial-vector four-quark $(su)( \overline{bd})$ state}
The expression of the axial-vector spectral function comes from the transverse part
of the corresponding correlator built from the $1^{+}$ current. It reads:
\bea
\rho^{pert}&=&\frac{(1+k^2) M_b^8}{5\ 3\ 2^{13} \pi^6} \Bigg{[}{5\over x^4}-\frac{96}{x^3}-\frac{945}{x^2}+\frac{480}{x}-60 \left(\frac{9}{x^2}+\frac{16}{x}+3\right){\rm Log} (x)+555+x^2 \Bigg{]},\nnb\\
\rho^{\la\bar qq\ra}&=&-\frac{(1-k^2) M_b^5}{3\ 2^6\ \pi^4}  \langle\bar qq \rangle \Bigg{[}\frac{1}{x^2}+\frac{9}{x}+6 \left(\frac{1}{x}+1\right) {\rm Log} (x)-9-x \Bigg{]}\nnb\\
&&-\frac{m_s  M_b^4}{3\ 2^9\ \pi^4}  \langle \bar qq \rangle \Bigg{[}2 (1-k^2)-(1+k^2) \kappa \Bigg{]}  \left(\frac{3}{x^2}-\frac{16}{x}-12\, {\rm Log} (x)+12+x^2\right)~,\nnb\\
\rho^{\langle G^2 \rangle}&=&-\frac{ (1+k^2)m_b^4}{3^3\ 2^{13}\ \pi^6}  4\pi\langle \alpha_sG^2 \rangle \Bigg{[}\frac{3}{x^2}+\frac{212}{x}+48 \left(\frac{2}{x}+3\right) {\rm Log} (x)-198-12 x-5 x^2\Bigg{]},\nnb\\
\rho^{\langle \bar qGq \rangle}&=&\frac{(1-k^2) M_b^3 }{2^7\ \pi^4}\langle \bar qGq \rangle \left(\frac{1}{x}+2 \, {\rm Log} (x)-x\right)\nnb\\
&&+\frac{ m_s M_b^2}{3^2\ 2^8\ \pi^4 }\langle \bar qGq \rangle \Bigg{[}6(1-k^2)
  +(1+k^2) \kappa \Bigg{]} \left(\frac{2}{x}-3+x^2\right),\nnb
   \eea
  \bea
\rho^{\langle \bar qq \rangle^2}&=&\frac{(1-k^2) M_b^2} {3^2\ 2^3\ \pi^2} \rho\langle \bar qq \rangle^2 \kappa\left(\frac{2}{x}-3+x^2\right)
\nnb\\&&
+\frac{m_s M_b}{3\ 2^3\ \pi^2} \rho \langle \bar qq \rangle^2  \Bigg{[}2 (1+k^2) -(1-k^2) \kappa \Bigg{]}(1-x)~,\nnb\\
\rho^{\langle G^3 \rangle}&=&\frac{ (1+k^2) M_b^2 }{3^3\ 2^{15}\ \pi^6 }\langle g_s^3G^3 \rangle \Bigg{[}\frac{9}{x^2}-\frac{160}{x}-12 \left(\frac{4}{x}+9\right) {\rm Log} (x)+144+7 x^2\Bigg{]}.
\eea
The contribution of a class of $d=7$ condensate for $m_s=0$  is:
\bea
\rho^{\la \bar qq\ra\la G^2\ra}&=&-(1-k^2)\frac{M_b\la \bar qq\ra}{3^2\ 2^8\pi ^4} 4\pi\la \alpha_s G^2\ra\left(\frac{2}{x}+6\,\text{Log}\,x+7-9x\right)~.
\label{eq:d7axial4}
\eea
\section{Mass and coupling of the $1^{+}$ axial-vector  four-quark state}
\begin{figure}[hbt] 
\begin{center}
{\includegraphics[width=6.2cm  ]{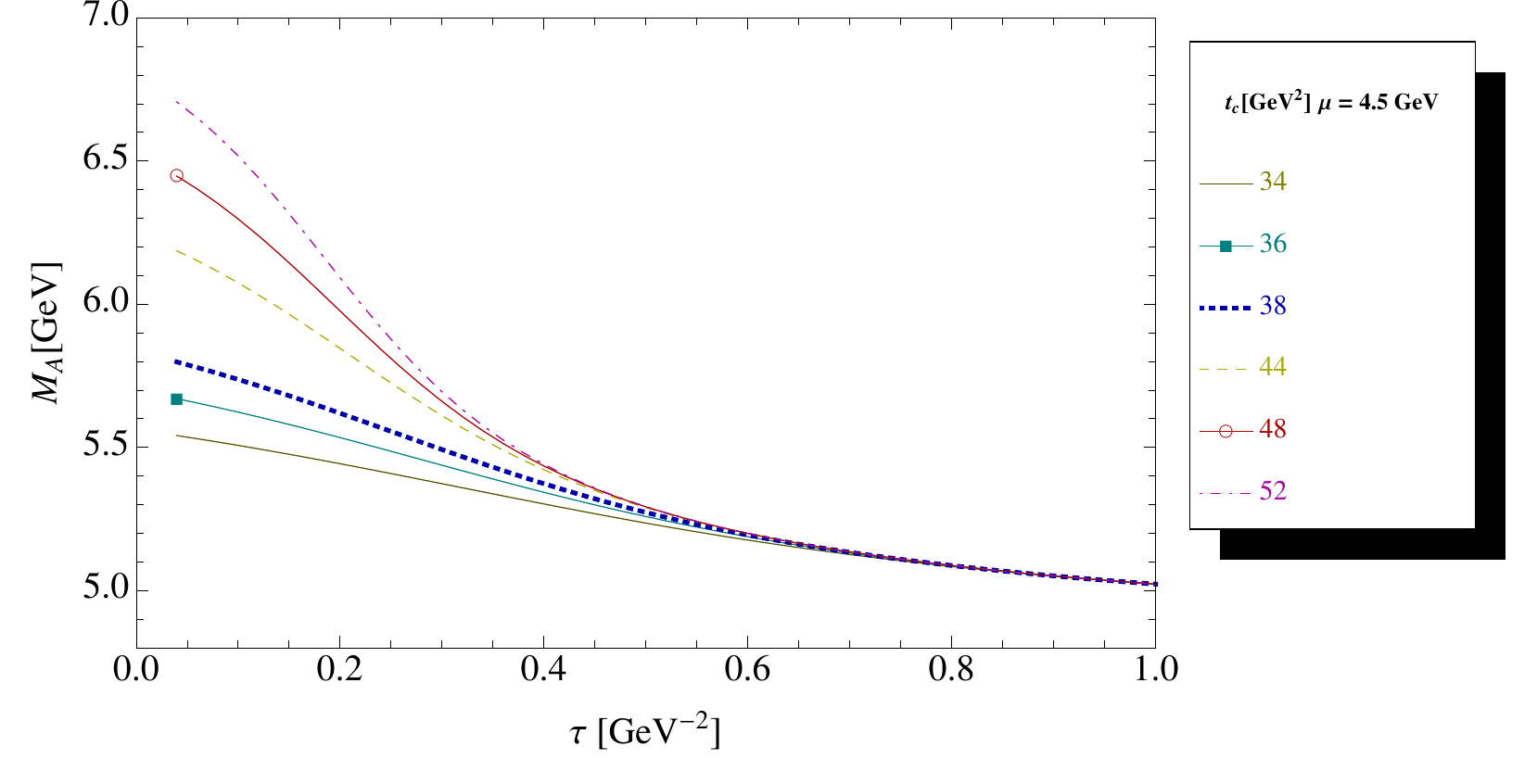}}
{\includegraphics[width=6.2cm  ]{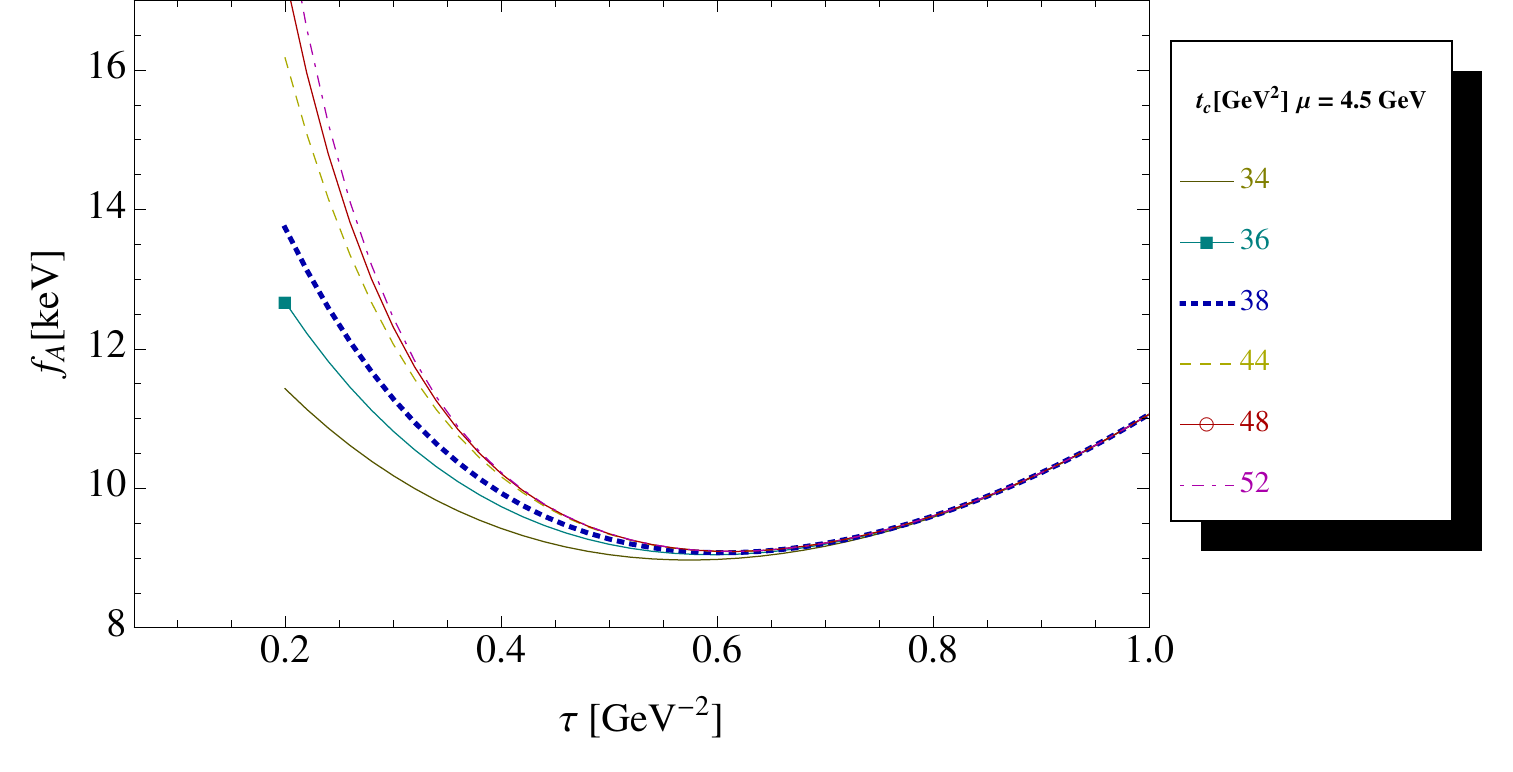}}
\centerline {\hspace*{-3cm} a)\hspace*{6cm} b) }
\caption{
\scriptsize 
{\bf a)} $M_{A_b}$  at LO as function of  $\tau$ and different values of $t_c$. We use $\mu=4.5$ GeV, the mixing parameter $k=0$ and the QCD parameters in Tables\,\ref{tab:param}  and \ref{tab:alfa};  {\bf b)} The same as a) but for the coupling $f_{A_b}$.
}
\label{fig:a-lo} 
\end{center}
\end{figure} 
\nin
\begin{figure}[hbt] 
\begin{center}
{\includegraphics[width=6.2cm  ]{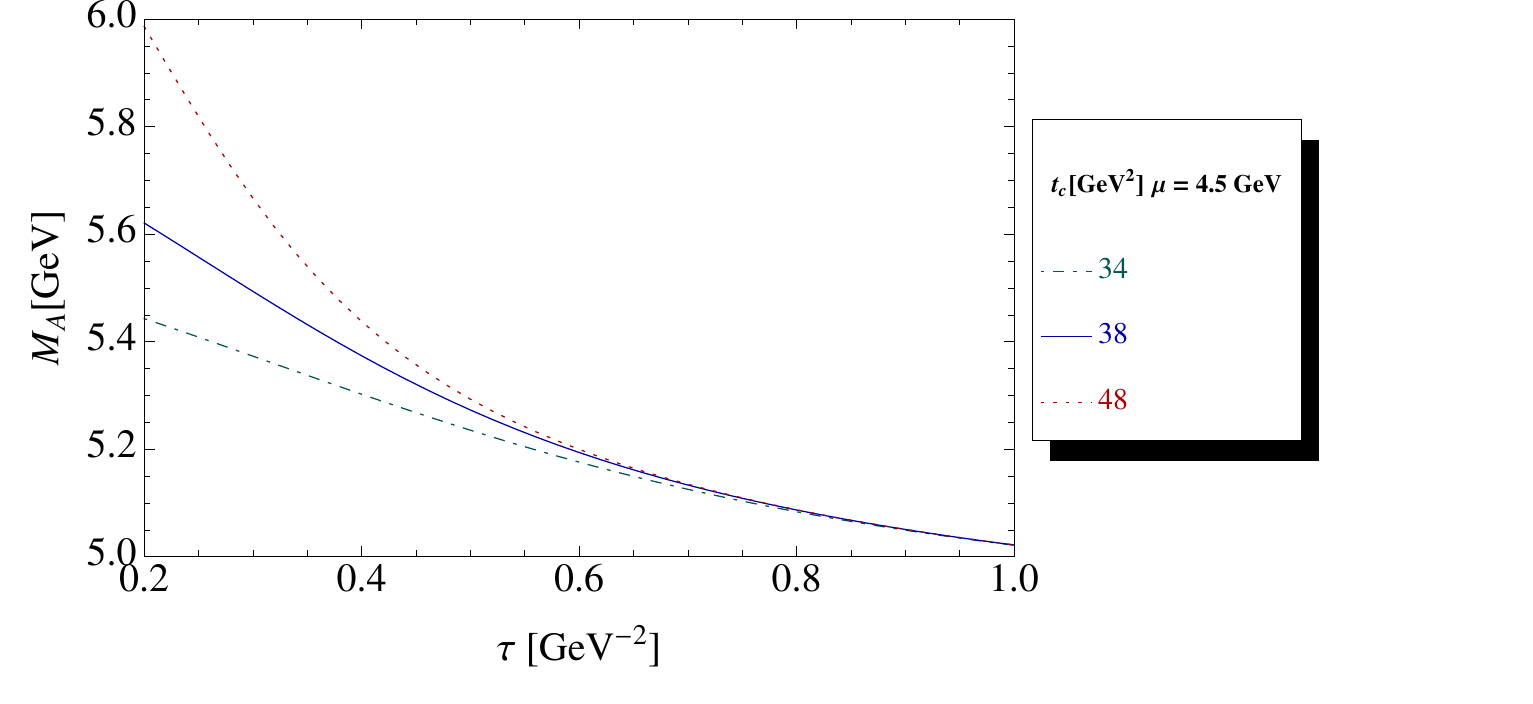}}
{\includegraphics[width=6.2cm  ]{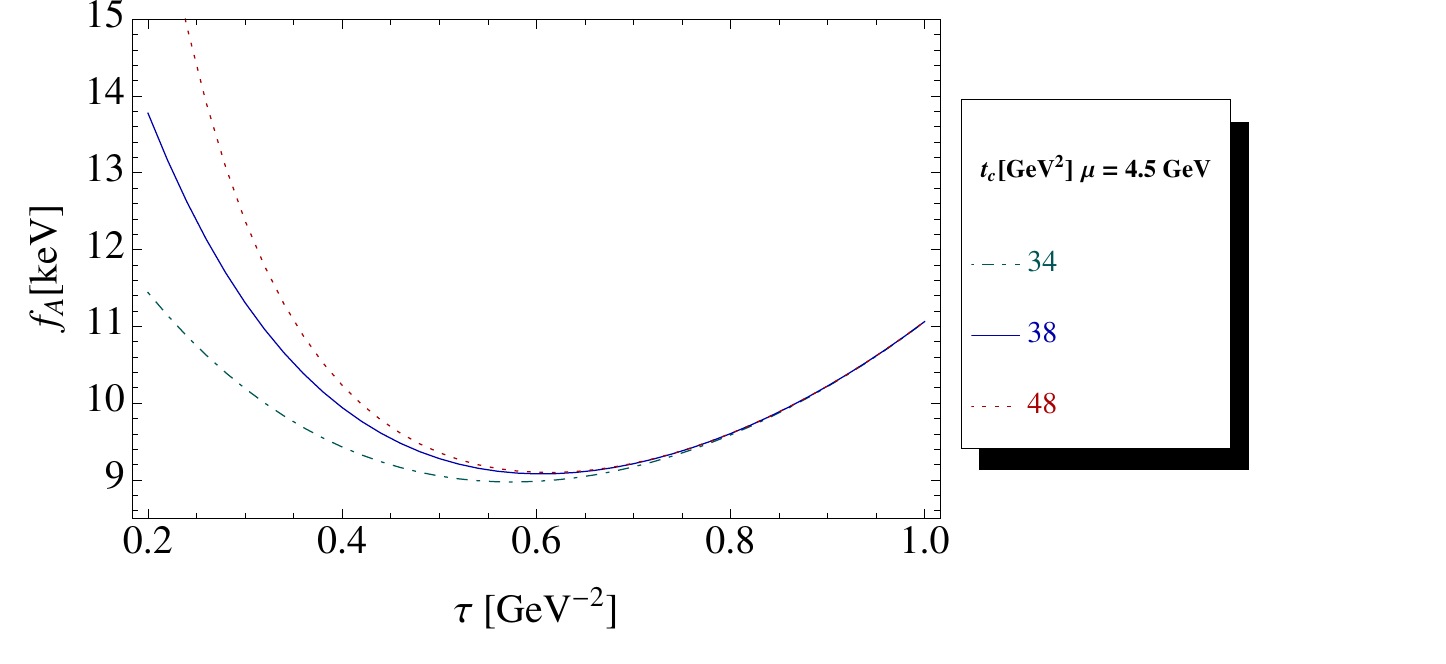}}
\centerline {\hspace*{-3cm} a)\hspace*{6cm} b) }
\caption{
\scriptsize 
The same as Fig.\,\ref{fig:a-lo} but for NLO.
}
\label{fig:a-nlo} 
\end{center}
\end{figure} 
\nin
\begin{figure}[hbt] 
\begin{center}
{\includegraphics[width=6.2cm  ]{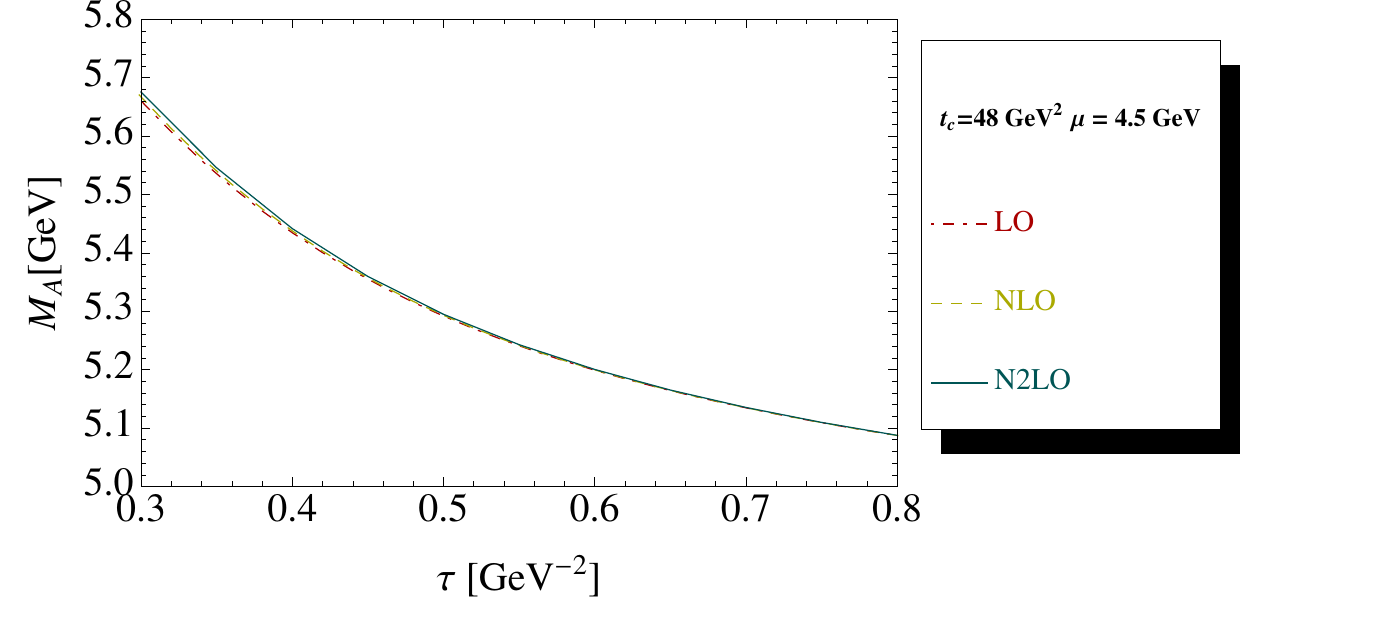}}
{\includegraphics[width=6.2cm  ]{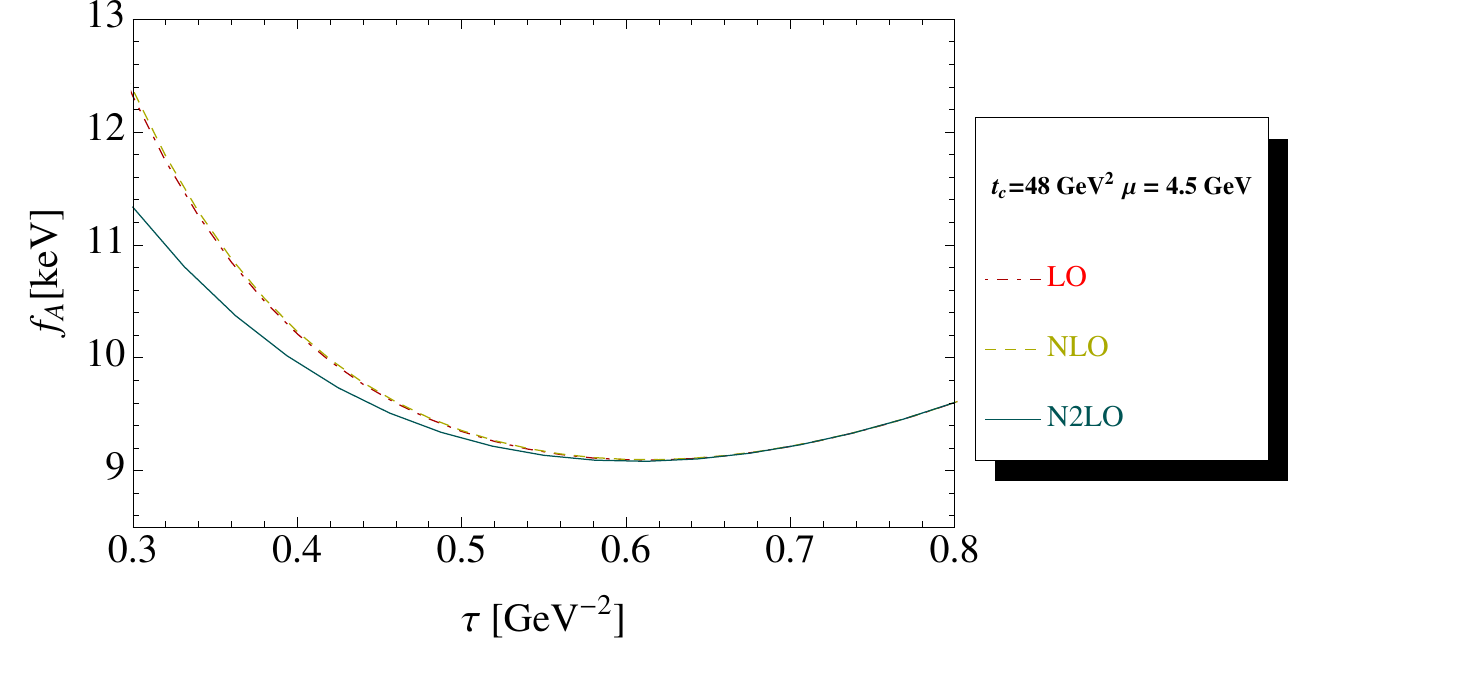}}
\centerline {\hspace*{-3cm} a)\hspace*{6cm} b) }
\caption{
\scriptsize 
{\bf a)} $M_{A_b}$  as function of $\tau$ for different truncation of the PT series at a given value of $t_c$=48 GeV$^2$, $\mu=4.5$ GeV and for the QCD parameters in Tables\,\ref{tab:param}  and \ref{tab:alfa};  {\bf b)} The same as a) but for the coupling $f_{A_b}$.
}
\label{fig:a-lo-n2lo} 
\end{center}
\end{figure} 
\nin
\begin{figure}[hbt] 
\begin{center}
{\includegraphics[width=6.2cm  ]{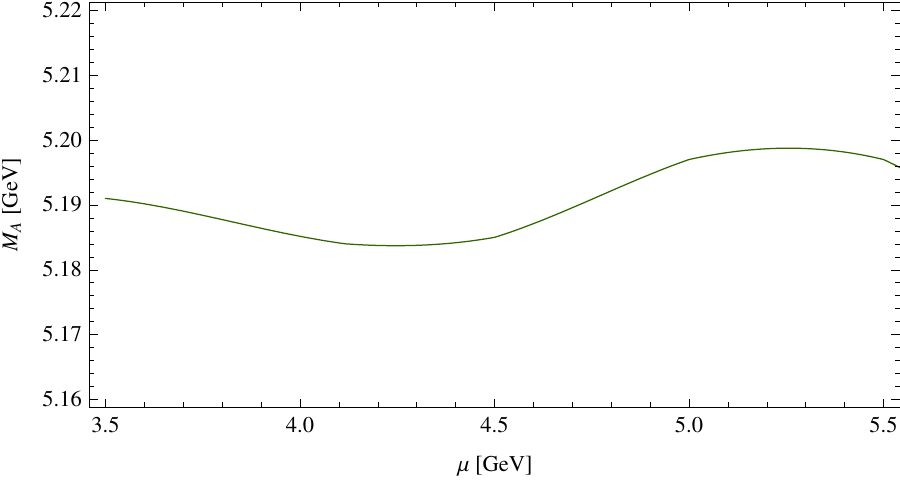}}
{\includegraphics[width=6.2cm  ]{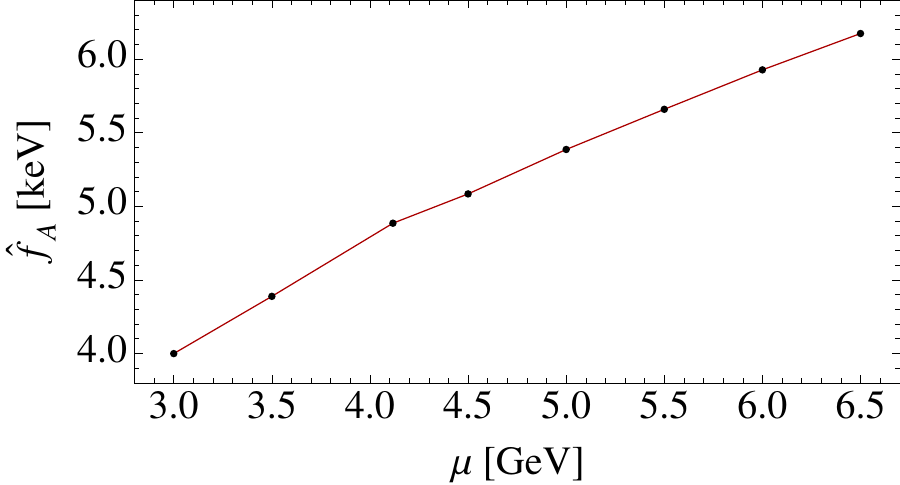}}
\centerline {\hspace*{-3cm} a)\hspace*{6cm} b) }
\caption{
\scriptsize 
{\bf a)} $M_{A_b}$ at NLO as function of $\mu$, for the corresponding $\tau$-stability region, for $t_c\simeq 48$ GeV$^2$ and for the QCD parameters in Tables\,\ref{tab:param}  and \ref{tab:alfa};  {\bf b)} The same as a) but for the renormalization group invariant coupling $\hat{f}_{A_b}$.
}
\label{fig:a-mu} 
\end{center}
\end{figure} 
\nin
\begin{figure}[hbt] 
\begin{center}
{\includegraphics[width=6.2cm  ]{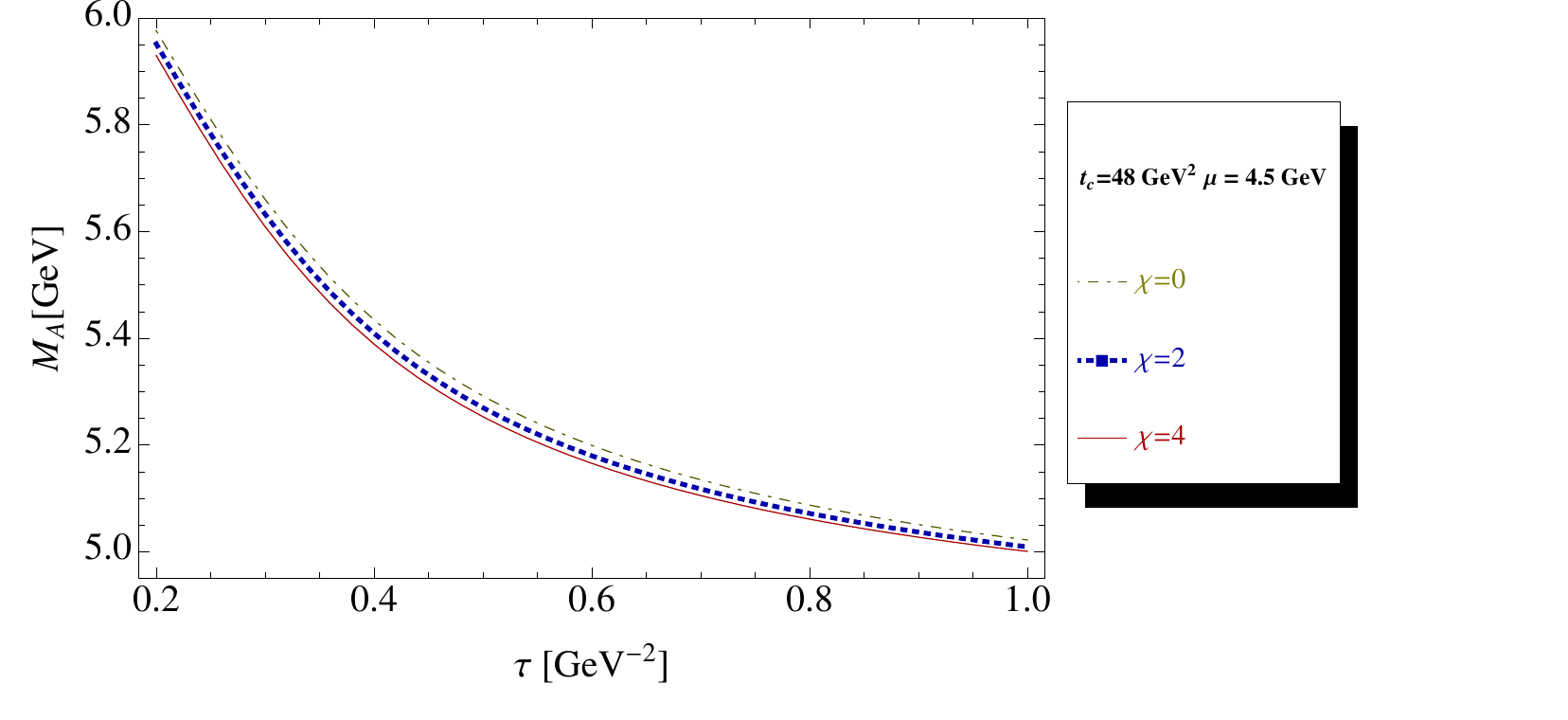}}
{\includegraphics[width=6.2cm  ]{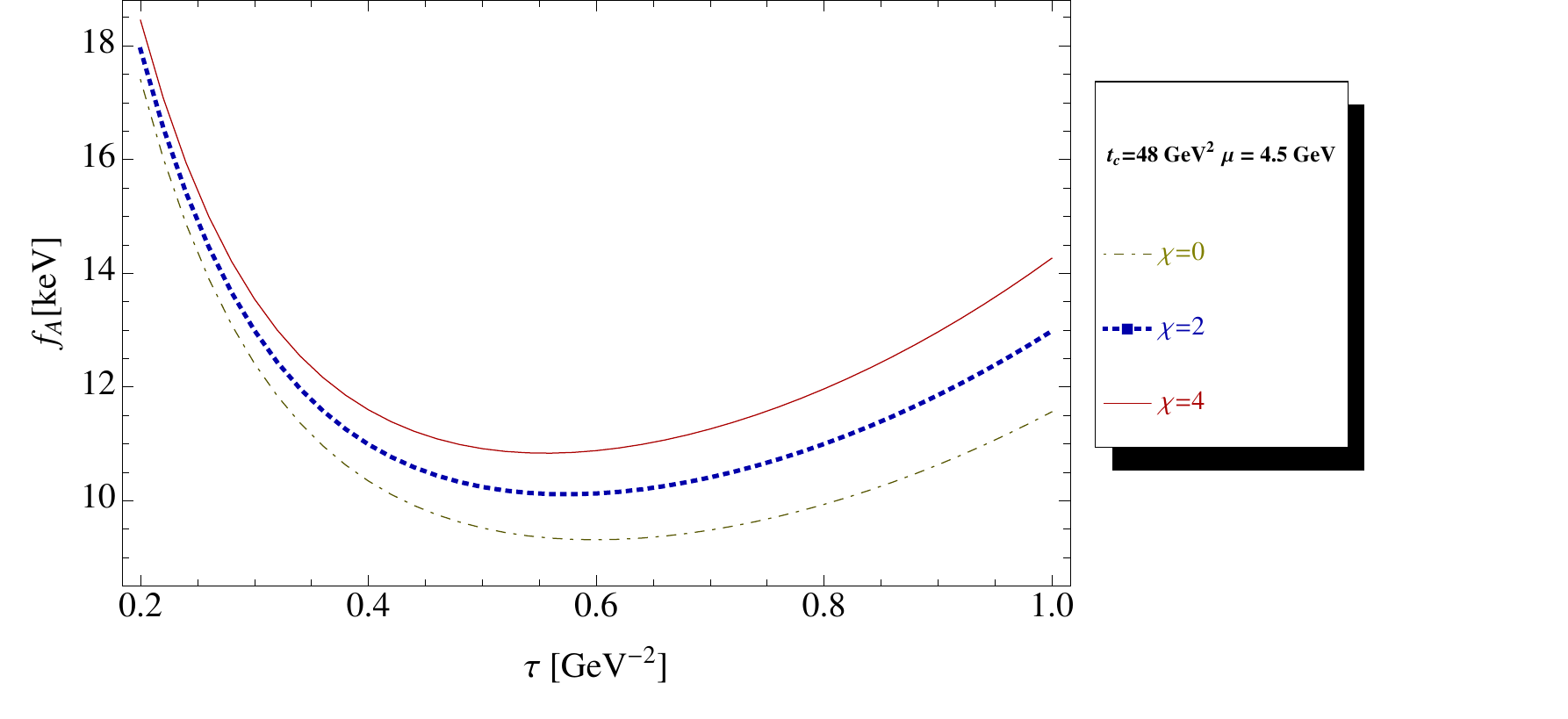}}
\centerline {\hspace*{-3cm} a)\hspace*{6cm} b) }
\caption{
\scriptsize 
{\bf a)} $M_{A_b}$  as function of $\tau$, for different values of the $d=7$ condensate contribution ($\chi$ measures the violation of factorization),  at a given value of $t_c$=48 GeV$^2$, $\mu=4.5$ GeV and for the QCD parameters in Tables\,\ref{tab:param} and \ref{tab:alfa}; {\bf b)} The same as a) but for the coupling $f_{A_b}$.
}
\label{fig:ab-d7} 
\end{center}
\end{figure} 
\nin
The analysis is similar to the one in previous sections. The curves have the same feature as the one of the $0^{+}$ scalar state. The optimal choice of the interpolating current is also obtained for
$k=0$.
\subsection{$\tau$- and $t_c$-stabilities}
\hspace*{0.5cm} \b We show in Fig.\,\ref{fig:a-lo} the $\tau$-behaviour of the mass and coupling at LO  for  different $t_c$ and for $\mu$=4.5 GeV.  
\subsection{NLO result and HO PT corrections}
The NLO result is shown in Fig.\,\ref{fig:a-nlo} and a comparison of the effects of PT corrections is shown in Fig.\,\ref{fig:a-lo-n2lo}.  We have $\tau$-stability for $\tau\simeq (0.52\sim 0.56)$ GeV$^{-2}$ while the corresponding range of $t_c$ values is from 34 to 48 GeV$^2$. 
\subsection{$\mu$- subtraction point stability at NLO}
We study the effect of the  subtraction point in Fig.\,\ref{fig:a-mu} where we have stability for $\mu\simeq 4.5$ GeV.
\subsection{Error induced by the truncation of the OPE}
We show in Fig.\,\ref{fig:ab-d7} the effect of a class of $d=7$ condensates for different values of the factorization violation parameter $\chi$.
Our estimate of the error induced by the truncation of the OPE corresponds to the choice $\chi=4$. 
\subsection{Final results}
From previous analysis, we deduce the final results to N2LO:
\bea
M_{A_b}&\simeq& (5186\pm 16)~{\rm MeV}~,\nnb\\
\hat{f}_{A_b}&\simeq& (5.05\pm 1.32)~{\rm keV}~~~~ \Lrar ~~~~f_{A_b}\simeq (9.04\pm 2.37)~{\rm keV}~,
\label{eq:a-final}
\eea 
where the errors come from the quadratic sum of the ones in Table\,\ref{tab:error4}. 
\section{Extension of the analysis to the charm quark}
One can naturally extend the previous analysis done in the $b$-quark channel to the charm quark.
In so doing, we replace the $b$-quark mass by the $c$-quark one, use $n_f=4$ and the QCD parameters in Tables\,  \ref{tab:param} and \ref{tab:alfa}. 
\subsection{$0^{+}$ scalar $ DK$ molecule}
\hspace*{0.5cm}\b The analysis is very similar to the previous one which we illustrate for the case of a scalar $DK$ molecule. In  Fig.\,\ref{fig:dk-lo}, we show the $\tau$-behaviour of the mass and coupling to lowest order (LO) for different values of $t_c$, for $\mu=2$ GeV and for the input parameters in Tables\,\ref{tab:param} and \ref{tab:alfa}. 
One can see a $\tau$-stability of about 0.6 GeV$^{-2}$ starting from $t_c=12$ GeV$^2$ while $t_c$-stability is reached from $t_c=18$ GeV$^2$. At lowest order, one obtains for $\mu$=2 GeV:
\begin{figure}[hbt] 
\begin{center}
{\includegraphics[width=6.2cm  ]{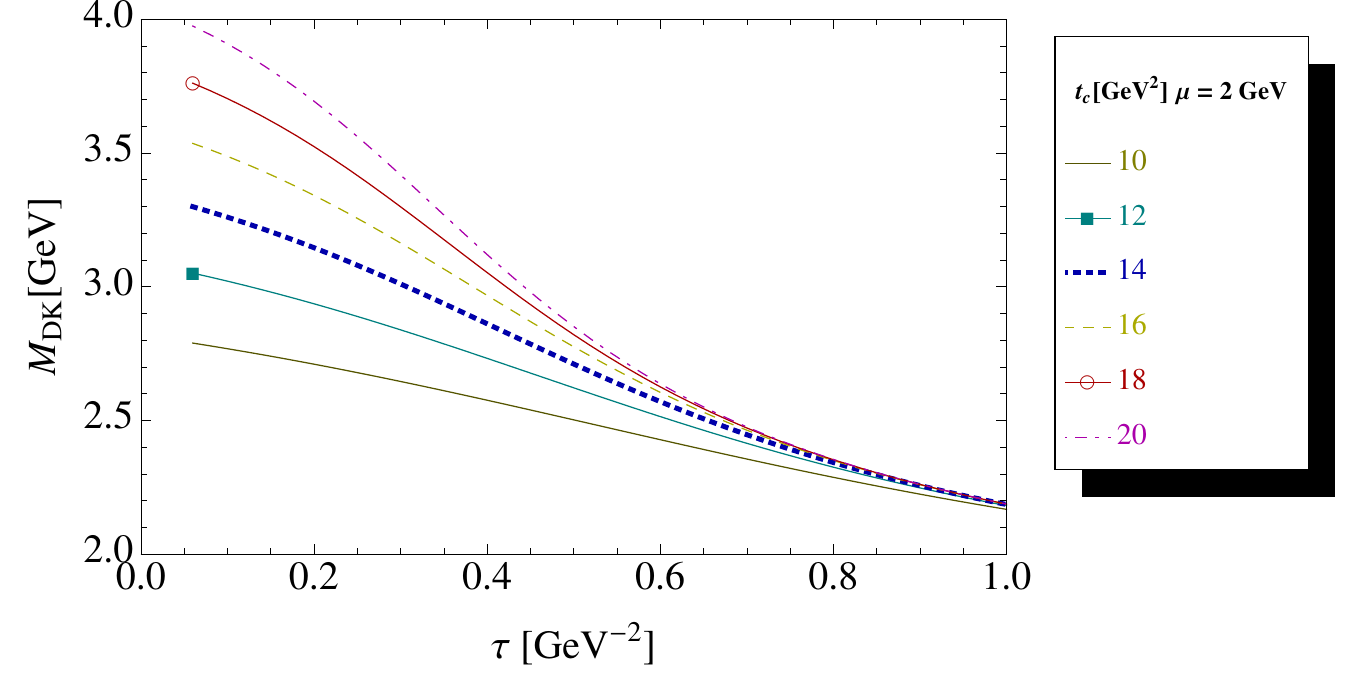}}
{\includegraphics[width=6.2cm  ]{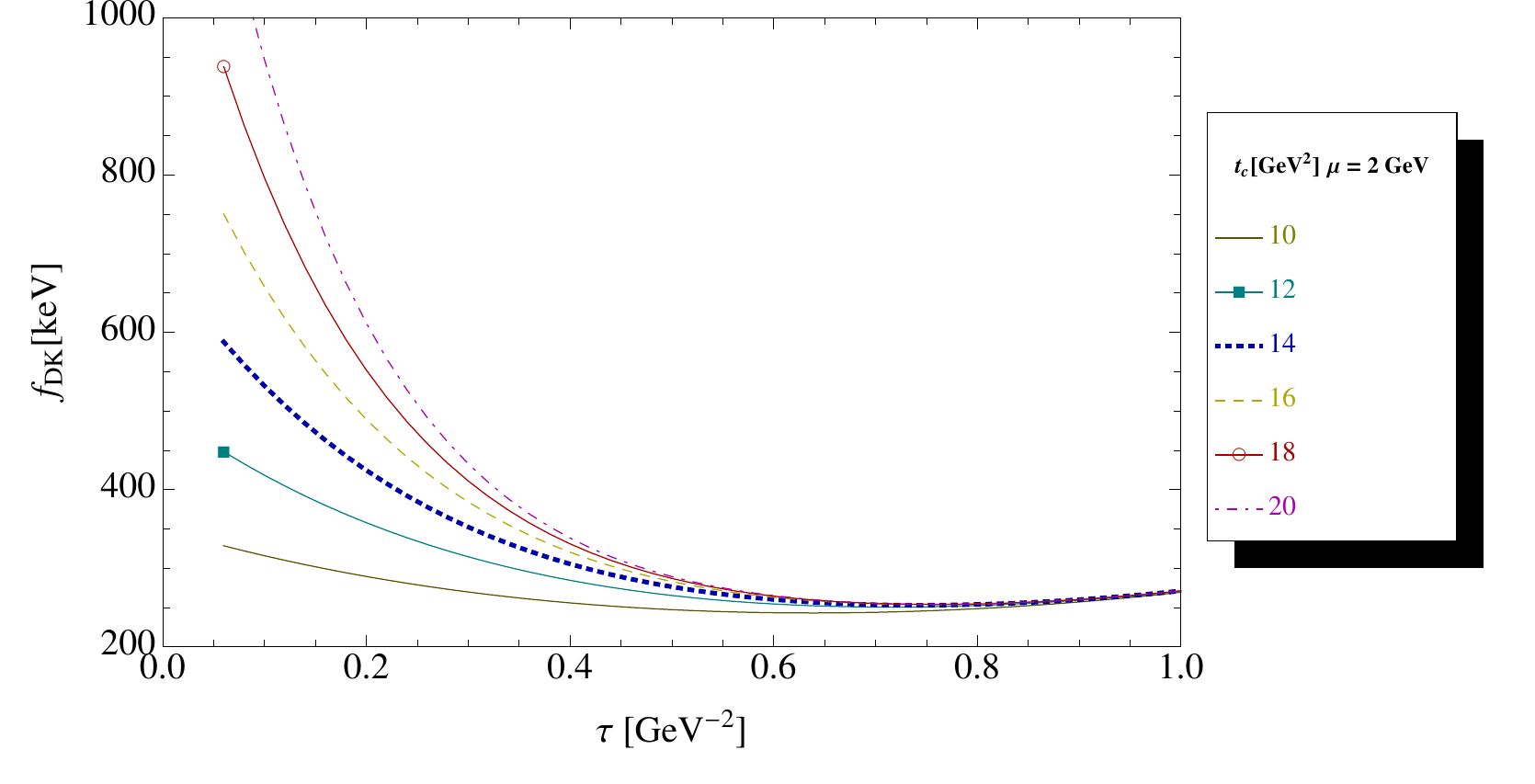}}
\centerline {\hspace*{-3cm} a)\hspace*{6cm} b) }
\caption{
\scriptsize 
{\bf a)} $M_{DK}$  at LO as function of  $\tau$ and different values of $t_c$. We use $\mu=2$ GeV, the mixing parameter $k=0$ and the QCD parameters in Tables\,\ref{tab:param}  and \ref{tab:alfa};  {\bf b)} The same as a) but for the coupling $f_{DK}$.
}
\label{fig:dk-lo} 
\end{center}
\end{figure} 
\nin
\begin{figure}[hbt] 
\begin{center}
{\includegraphics[width=6.2cm  ]{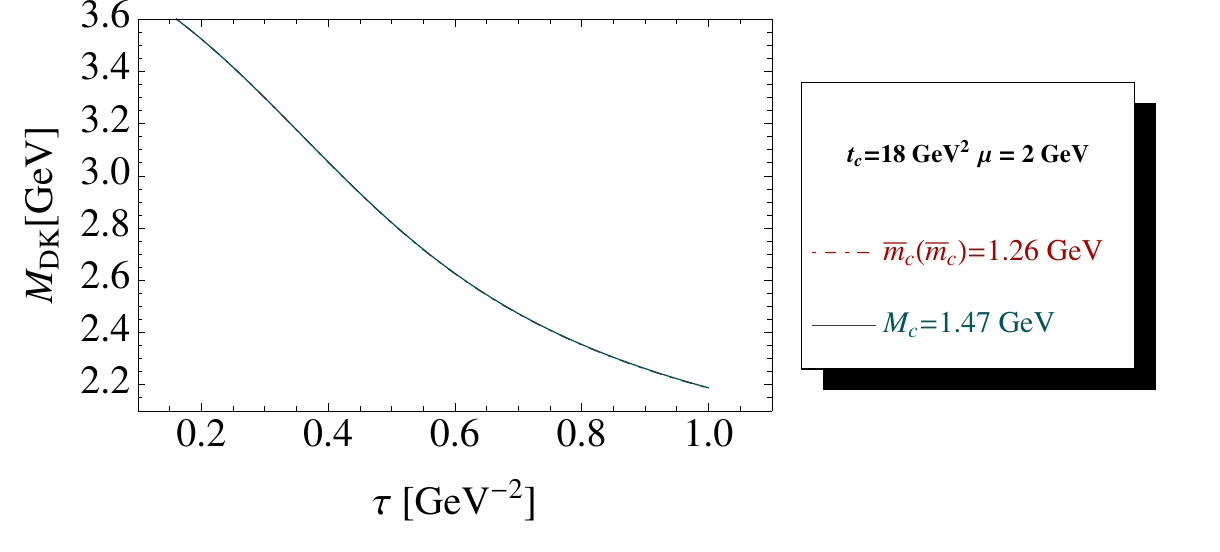}}
{\includegraphics[width=6.2cm  ]{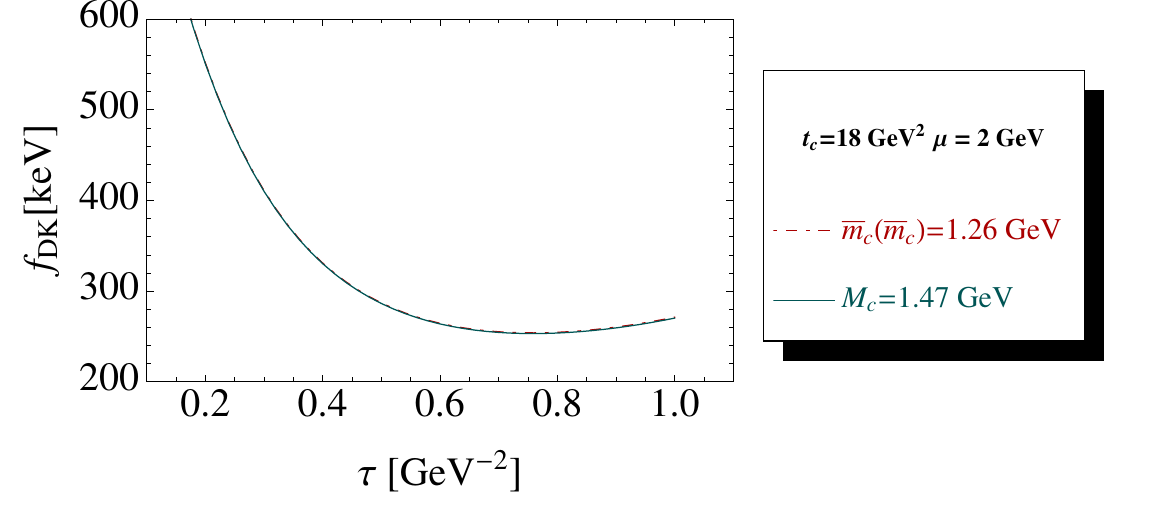}}
\centerline {\hspace*{-3cm} a)\hspace*{6cm} b) }
\caption{
\scriptsize 
{\bf a)} $M_{DK}$  at LO as function of $\tau$ for a given value of $t_c=18$ GeV$^2$,  $\mu=2$ GeV, mixing of currents $k=0$ and for the QCD parameters in Tables\,\ref{tab:param} and \ref{tab:alfa}. The OPE is truncated at $d=6$.  We compare the effect of  the on-shell or pole mass $M_c=1.47$ GeV and of the running mass $\bar m_c(\bar m_c)=1.26$ GeV; {\bf b)} The same as a) but for the coupling $f_{DK}$.
}
\label{fig:dkmasspole} 
\end{center}
\end{figure} 
\nin
\begin{figure}[hbt] 
\begin{center}
{\includegraphics[width=6.2cm  ]{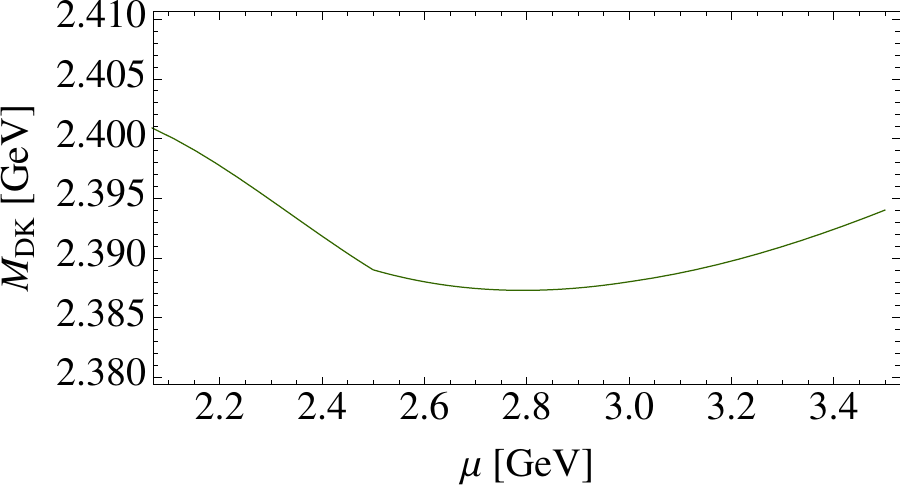}}
{\includegraphics[width=6.2cm  ]{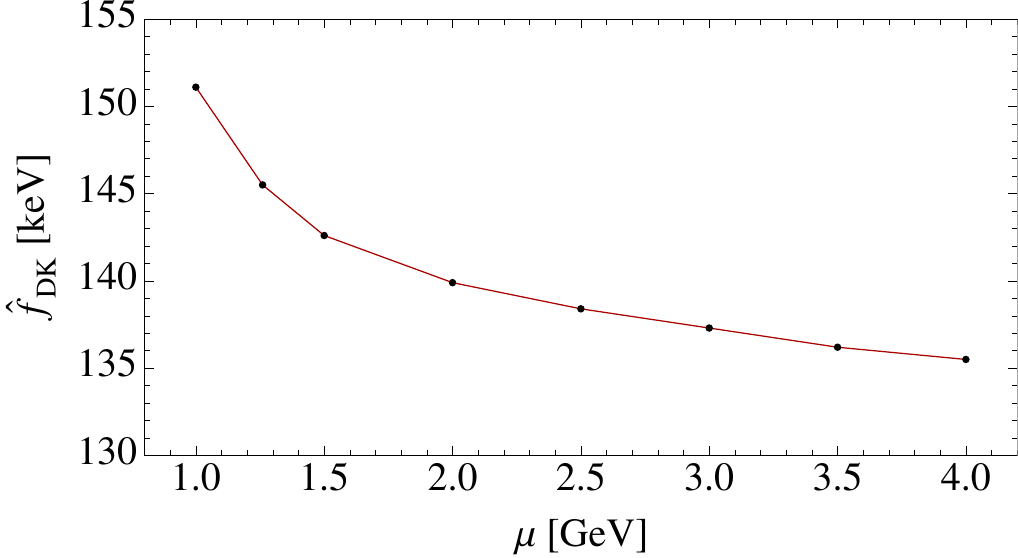}}
\centerline {\hspace*{-3cm} a)\hspace*{6cm} b) }
\caption{
\scriptsize 
{\bf a)} $M_{DK}$ at NLO as function of $\mu$, for the corresponding $\tau$-stability region, for $t_c\simeq 18$ GeV$^2$ and for the QCD parameters in Tables\,\ref{tab:param} and \ref{tab:alfa}. {\bf b)} The same as a) but for the renormalization group invariant coupling $\hat{f}_{DK}$.
}
\label{fig:dk-mu} 
\end{center}
\end{figure} 
\nin
\begin{figure}[hbt] 
\begin{center}
{\includegraphics[width=6.2cm  ]{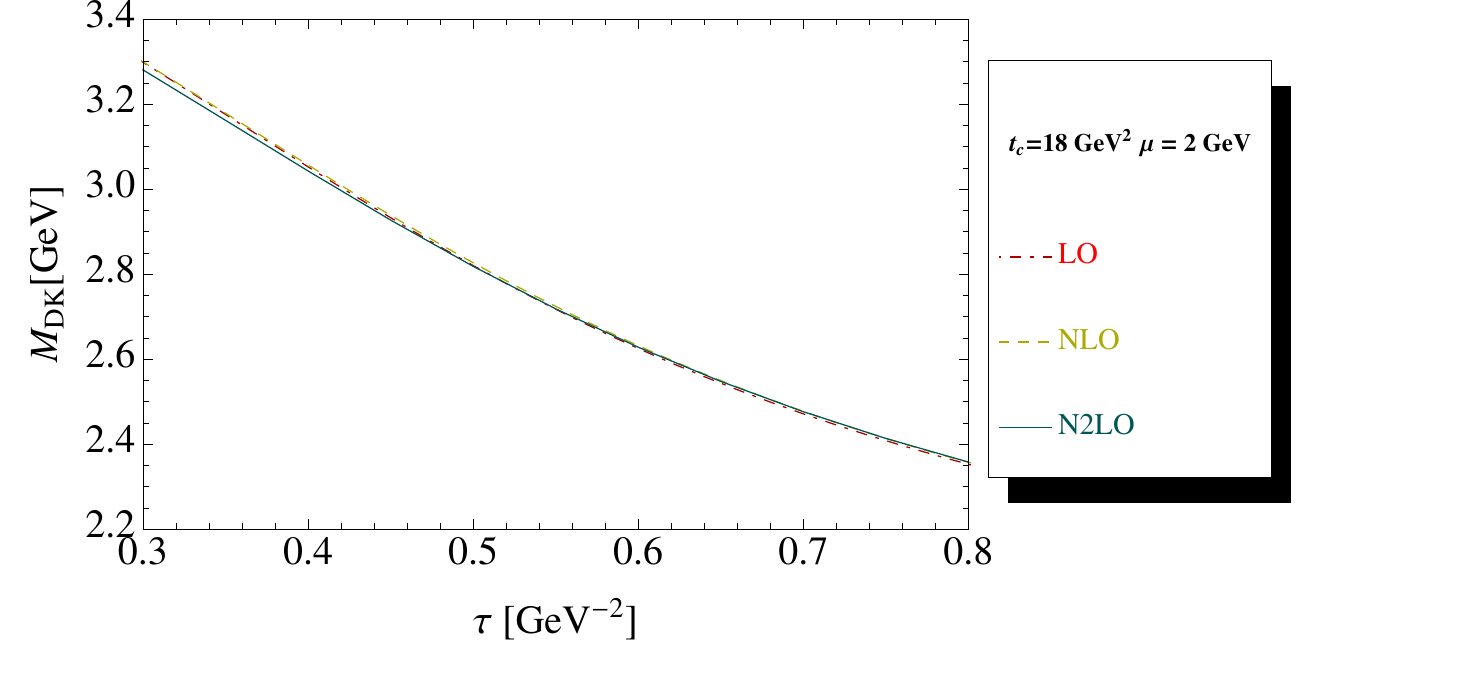}}
{\includegraphics[width=6.2cm  ]{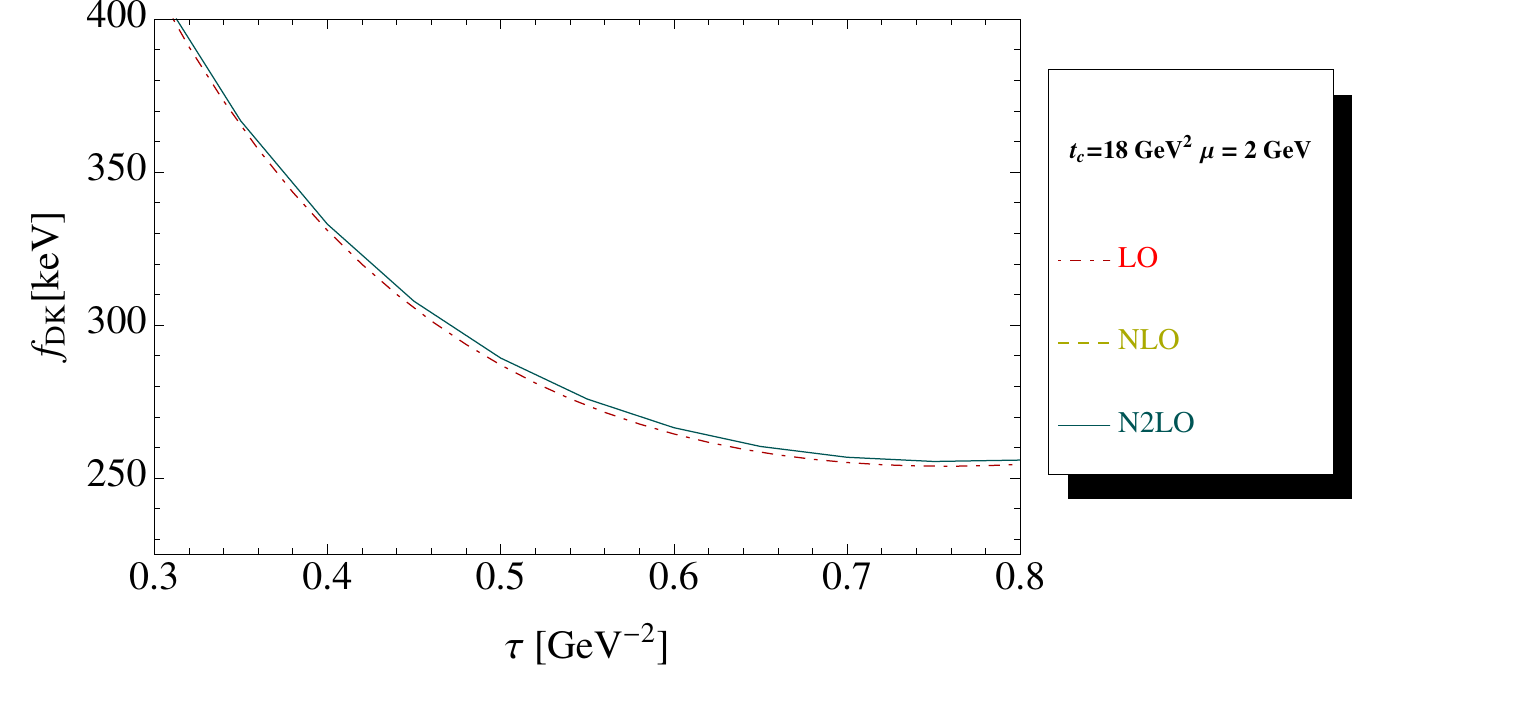}}
\centerline {\hspace*{-3cm} a)\hspace*{6cm} b) }
\caption{
\scriptsize 
{\bf a)} $M_{DK}$  as function of $\tau$ for different truncation of the PT series at a given value of $t_c$=18 GeV$^2$, $\mu=2$ GeV and for the QCD parameters in Tables\,\ref{tab:param} and \ref{tab:alfa}; {\bf b)} The same as a) but for the coupling $f_{DK}$.
}
\label{fig:dk-lo-n2lo} 
\end{center}
\end{figure} 
\nin
\begin{figure}[hbt] 
\begin{center}
{\includegraphics[width=6.2cm  ]{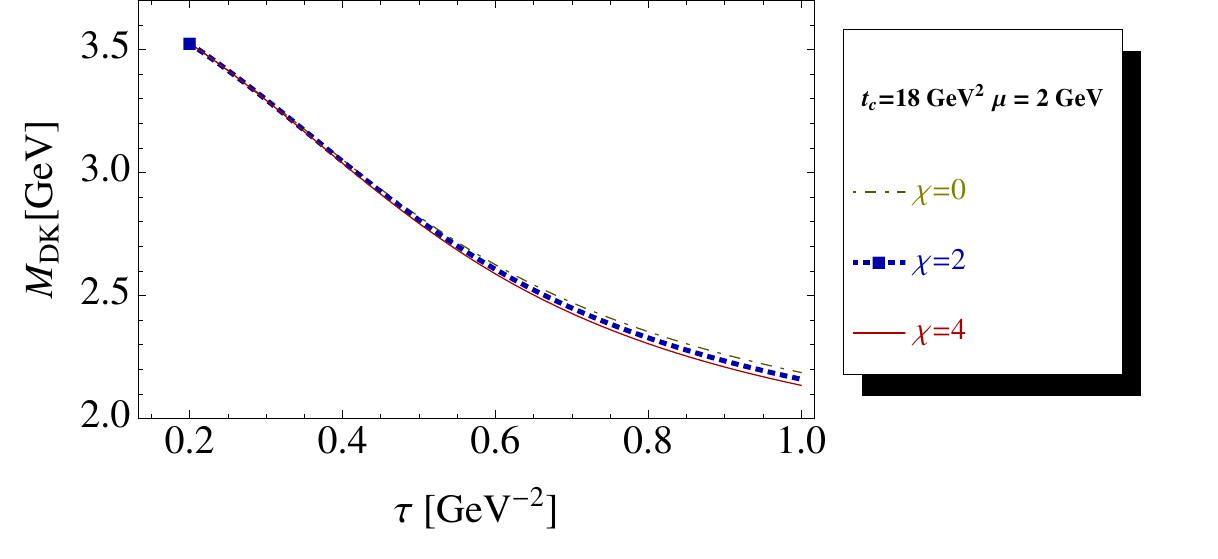}}
{\includegraphics[width=6.2cm  ]{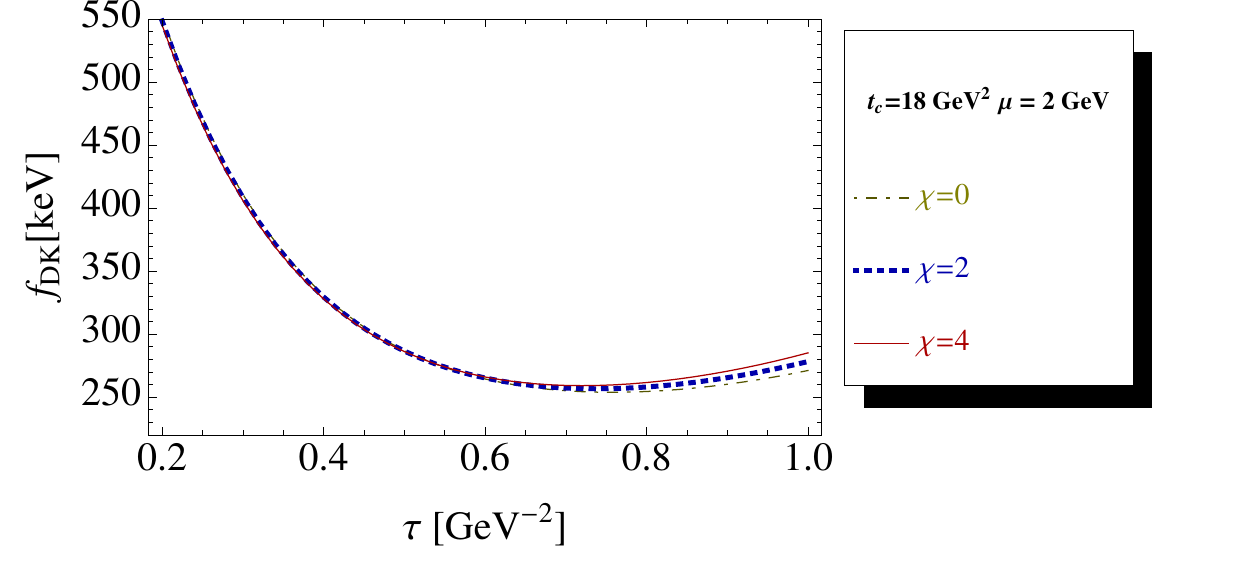}}
\centerline {\hspace*{-3cm} a)\hspace*{6cm} b) }
\caption{
\scriptsize 
{\bf a)} $M_{DK}$  as function of $\tau$, for different values of the $d=7$ condensate contribution ($\chi$ measures the violation of factorization),  at a given value of $t_c$=18 GeV$^2$, $\mu=2$ GeV and for the QCD parameters in Tables\,\ref{tab:param} and \ref{tab:alfa}; {\bf b)} The same as a) but for the coupling $f_{DK}$.
}
\label{fig:dk-d7} 
\end{center}
\end{figure} 
\nin
{\scriptsize
\begin{table}[hbt]
 \tbl{Different sources of errors for the estimate of the molecule masses (in units of MeV) and couplings (in units of keV) in the $c$-quark channel. }  
    {\scriptsize
 {\begin{tabular}{@{}llllllllll@{}} \toprule
&\\
\hline
\hline
\bf Inputs $[GeV]^d$&$\Delta M_{D^*K}$&$\Delta f_{D^*K}$&$\Delta M_{DK}$&$\Delta f_{DK}$
&$\Delta M_{D^*_s\pi}$&$\Delta f_{D^*_s\pi}$&$\Delta M_{D_s\pi}$&$\Delta f_{D_s\pi}$\\
\hline 
{\it LSR parameters}&\\
$t_c=(12\sim 18)$& 2  &1.5& 0.0 & 1.5&        8.5 & 2.5&2.5 & 2\\
$\mu=(2.0\sim 2.5)$&9 &13.5&5&15.5&         4 &14.5&4.5&17\\
$\tau=\tau_{min}\pm 0.02$&24&0.17&24&0.10&        30&0.12&23.2&0.06\\
{\it QCD inputs}&\\
$\bar m_c$& 5.4&2.8&5.5&6.7&                      5.0&3.32&5.58&3.58\\
$\bar m_s$& 0.13&0.55&0.13&0.48&           0.85&0.76&1.0&0.63\\
$\alpha_s$&7.9&5.8&8.1&6.68&                    9.0&6.55&7.92&7.32\\
$\la\bar qq\ra$&3.7&0.24&3.4&0.51&           1.58&0.16&0.36&0.36\\
$\kappa$&11.4&22.5&13.5&27.9&                  6.83&0.7&8.10&1.58\\
$\la\alpha_s G^2\ra$&2.1&1.3&4.7&1.95&            2.16&1.13&0.60&1.83\\
$M_0^2$&2.3&2.8&1.60&1.38&                              0.90&1.28&0.60&0.47\\
$\la\bar qq\ra^2$&32.1&32.5&29.4&34.9&            34&37.37&23.05&41.23\\
$\la g^3G^3\ra$&0&0&0.0&0.0&0.0&0.0&0.0&0.0\\
$d\geq 7$&3&31.4&1&4.7&3.5&27.6&7&2\\
{\it Total errors}&48.1&52.4&42&48&                  48.1&49.4&36.7&46\\
\hline\hline
\end{tabular}}
\label{tab:errorc}
}
\end{table}
} 
\beq
M_{DK}^{LO}\simeq( 2395\sim 2397)~{\rm MeV}~~~~{\rm and}~~~~
f_{DK}^{LO}\simeq (250\sim 254)~{\rm keV}~.
\eeq

\b The analysis of the $\mu$-behaviour in Fig.\,\ref{fig:dk-mu} indicates a $\mu$-stability for 
$
\mu=2~{\rm GeV}~.
$

\b The effects of the truncation of the PT series are shown in Fig.\,\ref{fig:dk-lo-n2lo}. One can notice that the PT corrections are small both for the coupling and
for the mass which justify the LO result and the (a priori) use of the running heavy quark mass. It also shows that the effect of the tachyonic gluon mass is small by duality with HO corrections. 

\b The convergence of the OPE is tested in Fig.\,\ref{fig:dk-d7} by adding the contribution of the $d=7$ condensate. It induces  an error:
\beq
 \Delta f_{DK}^{OPE}\simeq  \pm 4.7 ~{\rm keV}~,~~~~~~~~~~~~~~\Delta M_{DK}^{OPE}\simeq \pm 1 ~{\rm MeV}~. 
\eeq

\b From the previous analysis, we deduce the final result including N2LO PT perturbative $\oplus$ $d\leq 6$ dimension contributions taking $t_c \simeq 12\sim 18$ GeV$^2$ and $\mu$=2~GeV:
\bea
M_{DK}&\simeq&( 2402\pm 42)~{\rm MeV}~,\nnb\\
\hat f_{DK}&\simeq &(139\pm 26)~{\rm keV}~~~~\Lrar ~~~~f_{DK}\simeq (254\pm 48)~{\rm keV}~,
\eea
where the different sources of errors come from Table\,\ref{tab:errorc}.
\subsection{$1^{+}$ axial-vector $ D^*K$ molecule}
\begin{figure}[hbt] 
\begin{center}
{\includegraphics[width=6.2cm  ]{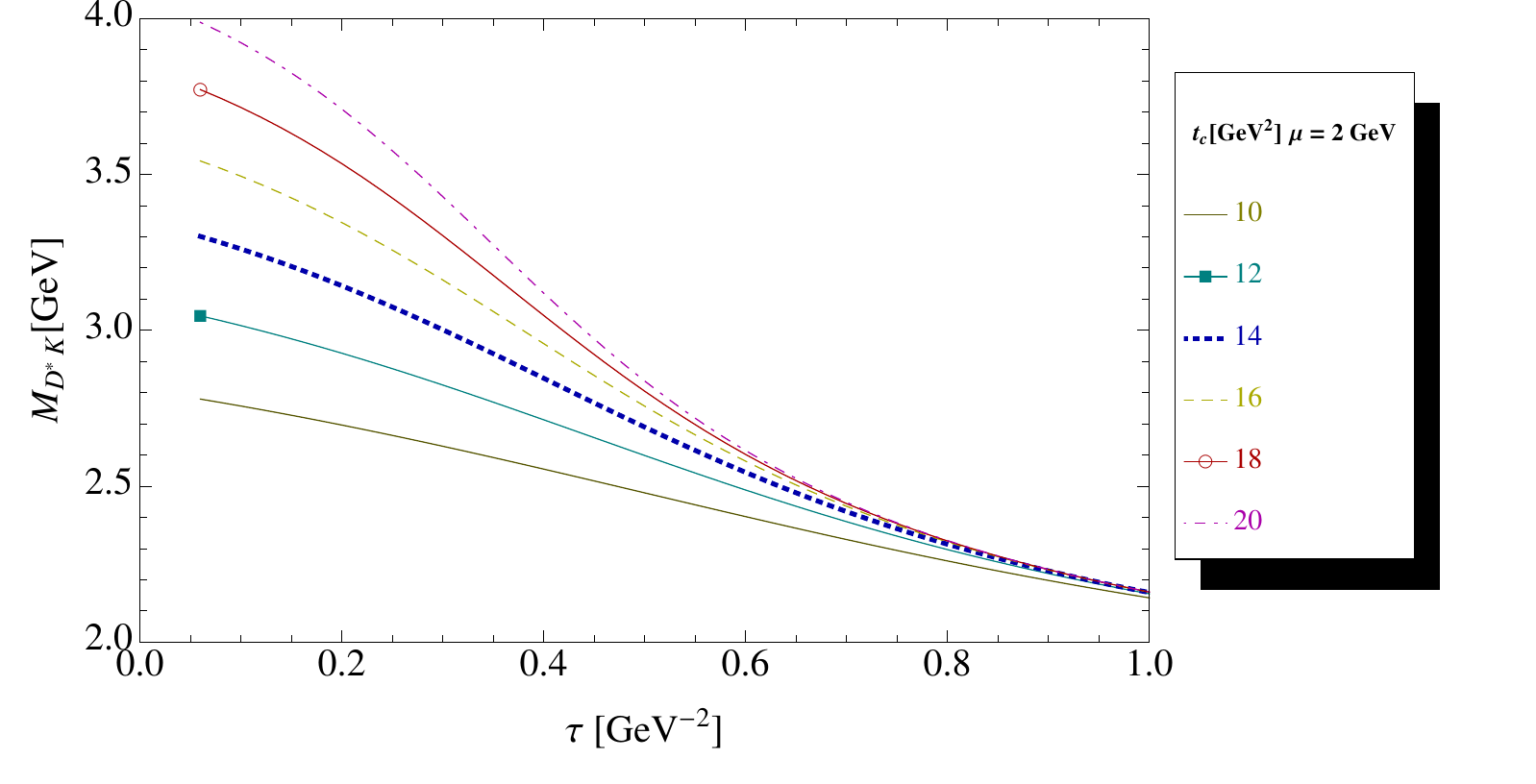}}
{\includegraphics[width=6.2cm  ]{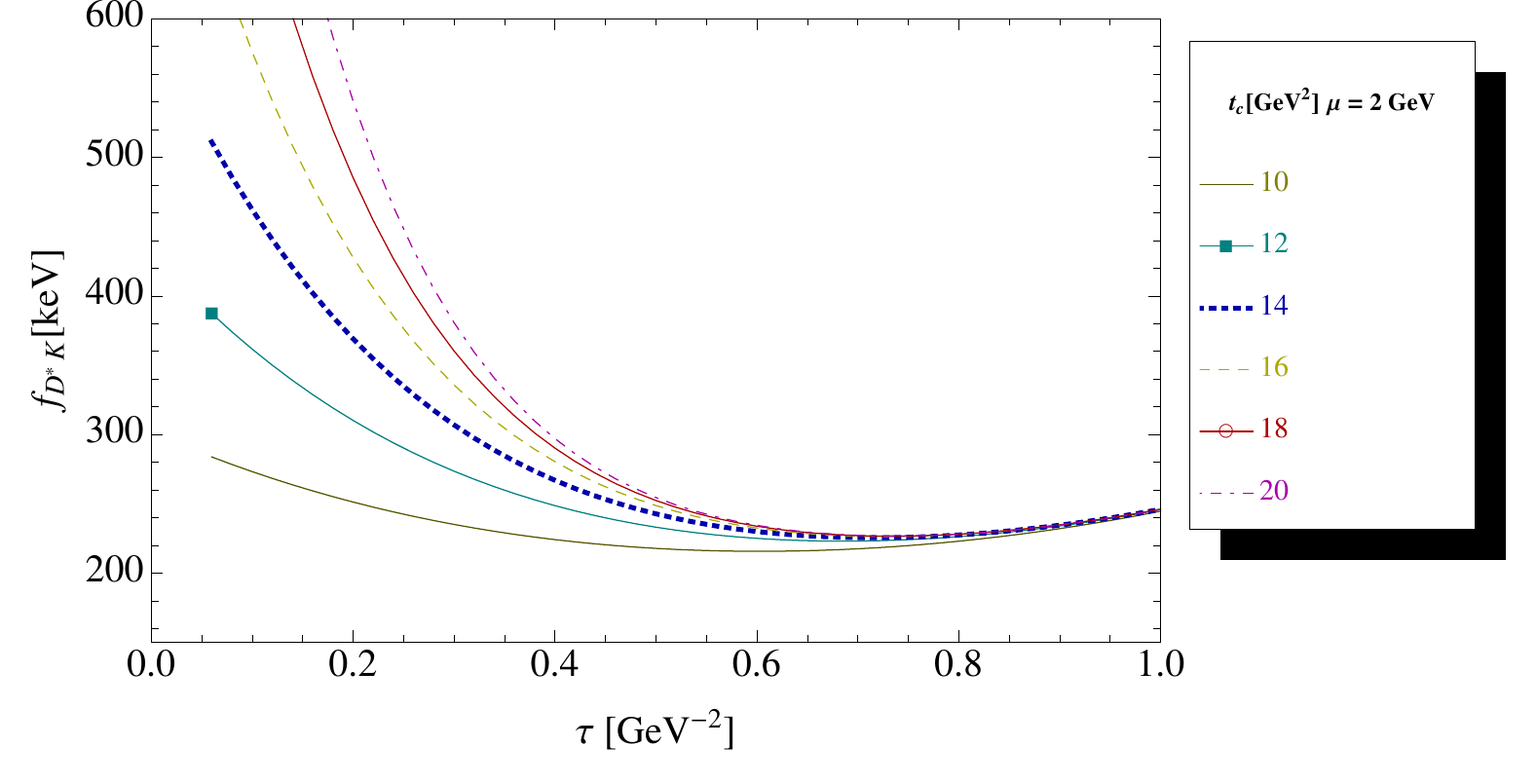}}
\centerline {\hspace*{-3cm} a)\hspace*{6cm} b) }
\caption{
\scriptsize 
{\bf a)} $M_{D^*K}$  at LO as function of  $\tau$ and different values of $t_c$. We use $\mu=2$ GeV, the mixing parameter $k=0$ and the QCD parameters in Tables\,\ref{tab:param}  and \ref{tab:alfa}; {\bf b)} The same as a) but for the coupling $f_{D^*K}$.
}
\label{fig:dstark-lo} 
\end{center}
\end{figure}

\begin{figure}[hbt] 
\begin{center}
{\includegraphics[width=6.2cm  ]{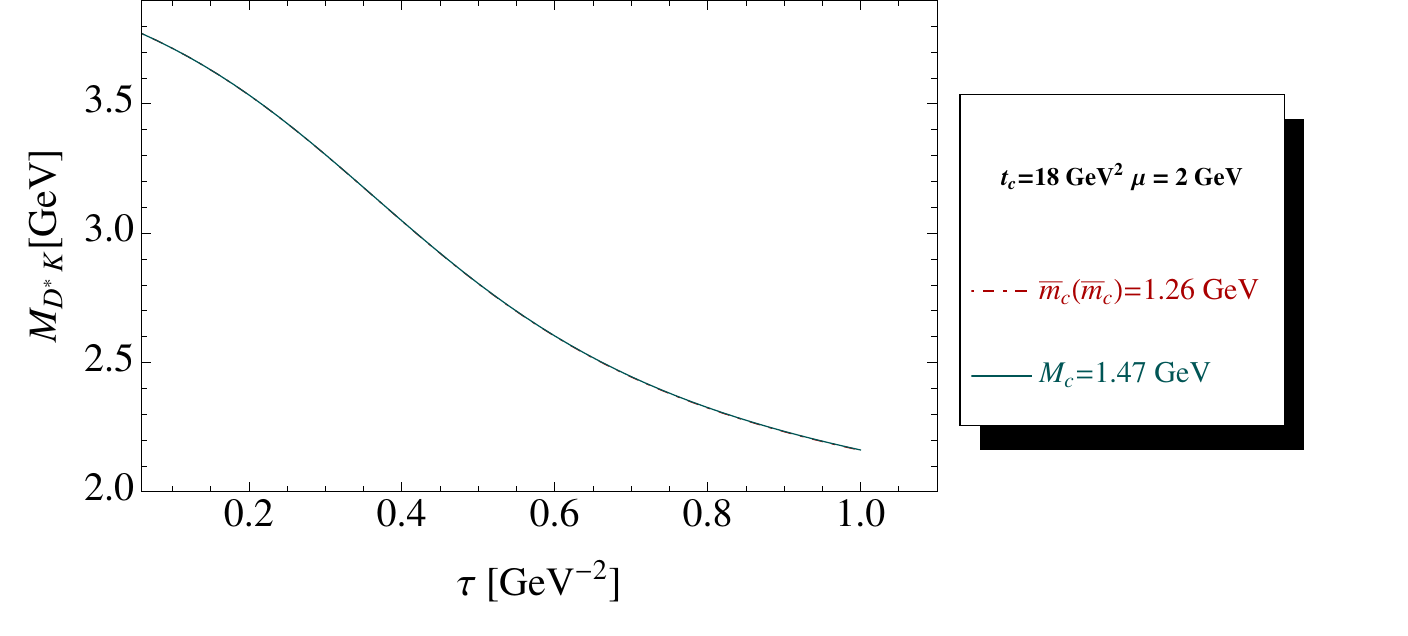}}
{\includegraphics[width=6.2cm  ]{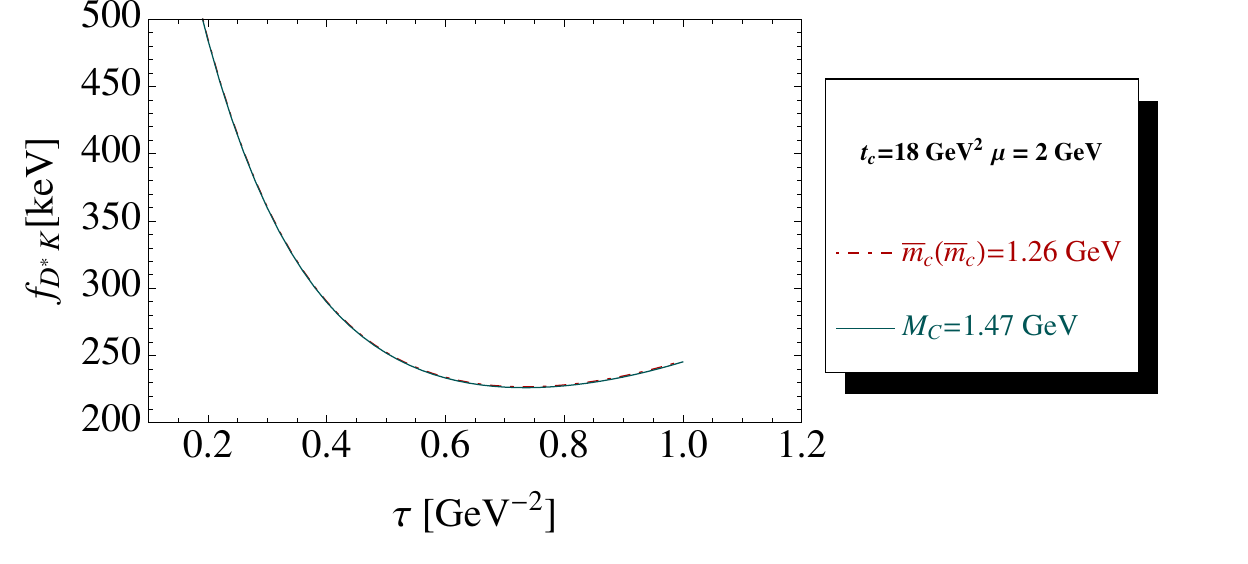}}
\centerline {\hspace*{-3cm} a)\hspace*{6cm} b) }
\caption{
\scriptsize 
{\bf a)} $M_{D^*K}$  at LO as function of $\tau$ for a given value of $t_c=18$ GeV$^2$,  $\mu=2$ GeV, mixing of currents $k=0$ and for the QCD parameters in Tables\,\ref{tab:param} and \ref{tab:alfa}. The OPE is truncated at $d=6$.  We compare the effect of  the on-shell or pole mass $M_c=1.47$ GeV and of the running mass $\bar m_c(\bar m_c)=1.26$ GeV; {\bf b)} The same as a) but for the coupling $f_{D^*K}$.
}
\label{fig:dstarkmasspole} 
\end{center}
\end{figure}

\begin{figure}[hbt] 
\begin{center}
{\includegraphics[width=6.2cm  ]{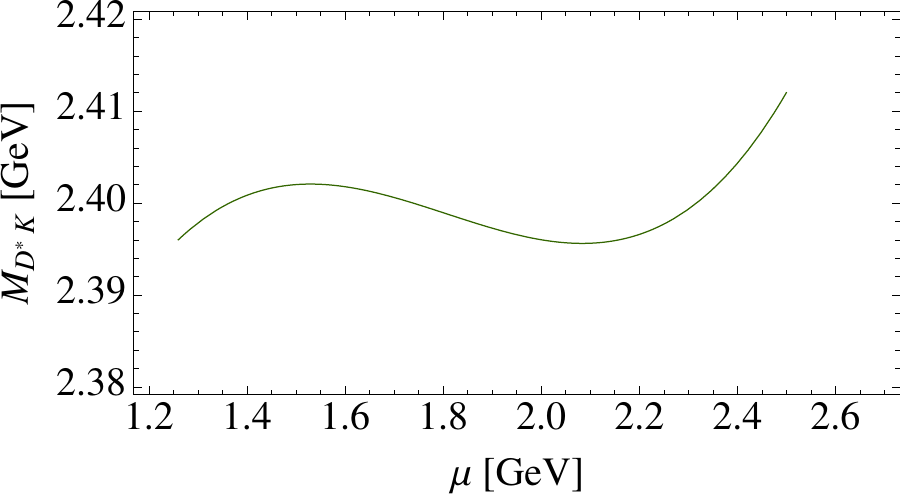}}
{\includegraphics[width=6.2cm  ]{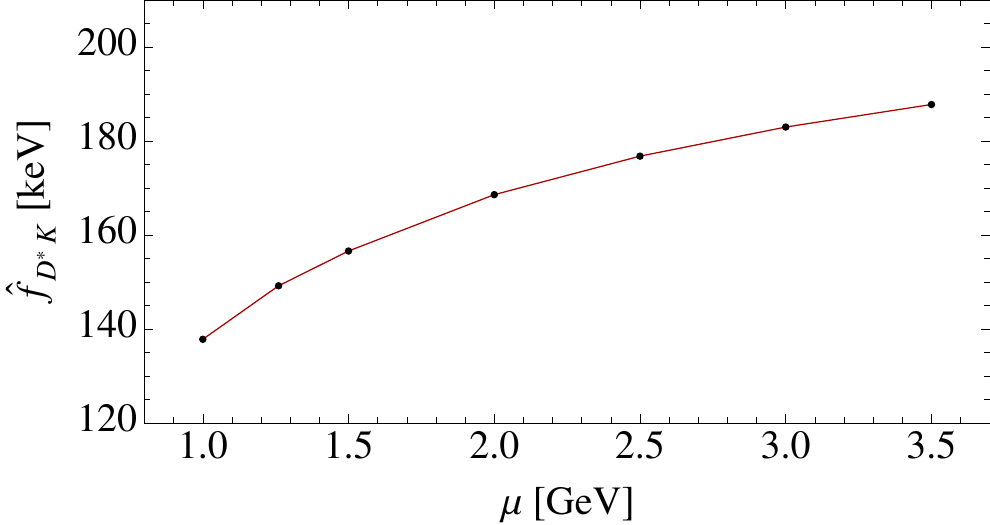}}
\centerline {\hspace*{-3cm} a)\hspace*{6cm} b) }
\caption{
\scriptsize 
{\bf a)} $M_{D^*K}$ at NLO as function of $\mu$, for the corresponding $\tau$-stability region, for $t_c\simeq 18$ GeV$^2$ and for the QCD parameters in Tables\,\ref{tab:param} and \ref{tab:alfa}; {\bf b)} The same as a) but for the renormalization group invariant coupling $\hat{f}_{D^*K}$.
}
\label{fig:dstark-mu} 
\end{center}
\end{figure}

\begin{figure}[hbt] 
\begin{center}
{\includegraphics[width=6.2cm  ]{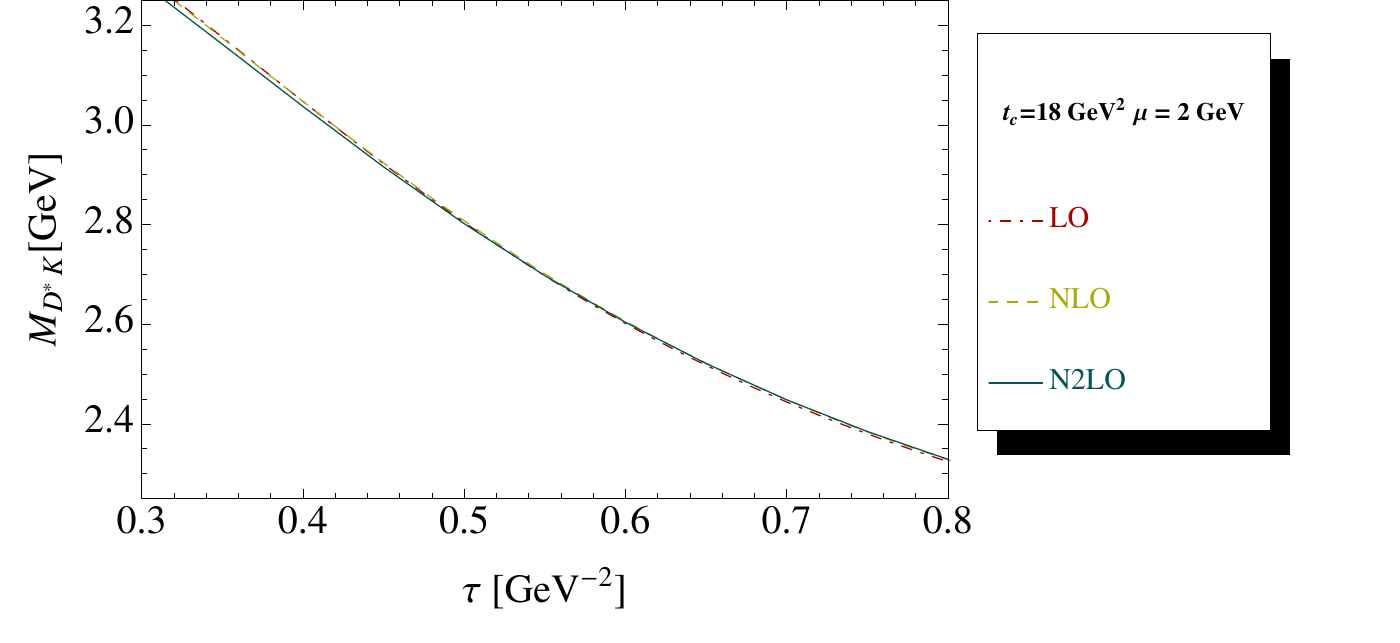}}
{\includegraphics[width=6.2cm  ]{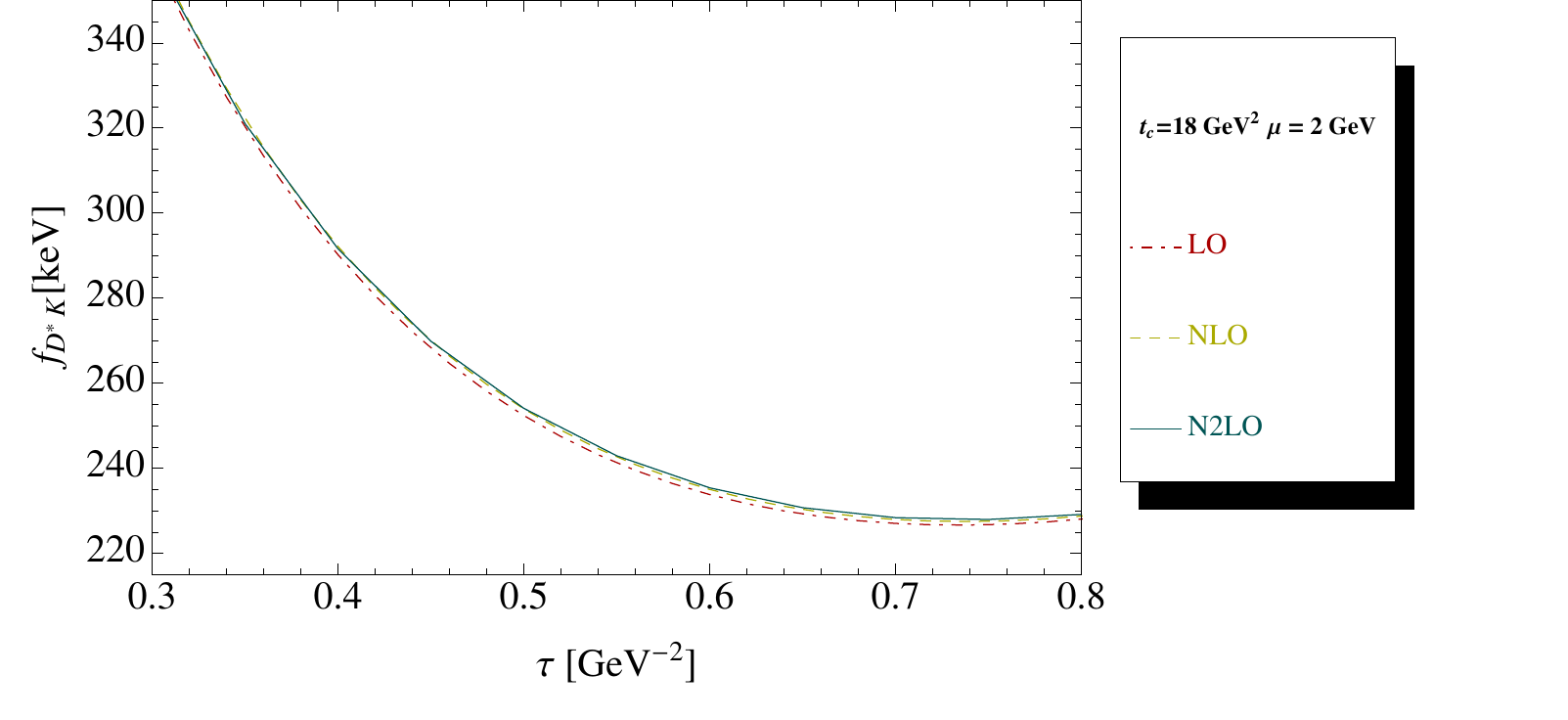}}
\centerline {\hspace*{-3cm} a)\hspace*{6cm} b) }
\caption{
\scriptsize 
{\bf a)} $M_{D^*K}$  as function of $\tau$ for different truncation of the PT series at a given value of $t_c$=18 GeV$^2$, $\mu=2$ GeV and for the QCD parameters in Tables\,\ref{tab:param} and \ref{tab:alfa}; {\bf b)} The same as a) but for the coupling $f_{D^*K}$.
}
\label{fig:dstark-lo-n2lo} 
\end{center}
\end{figure}

\begin{figure}[hbt] 
\begin{center}
{\includegraphics[width=6.2cm  ]{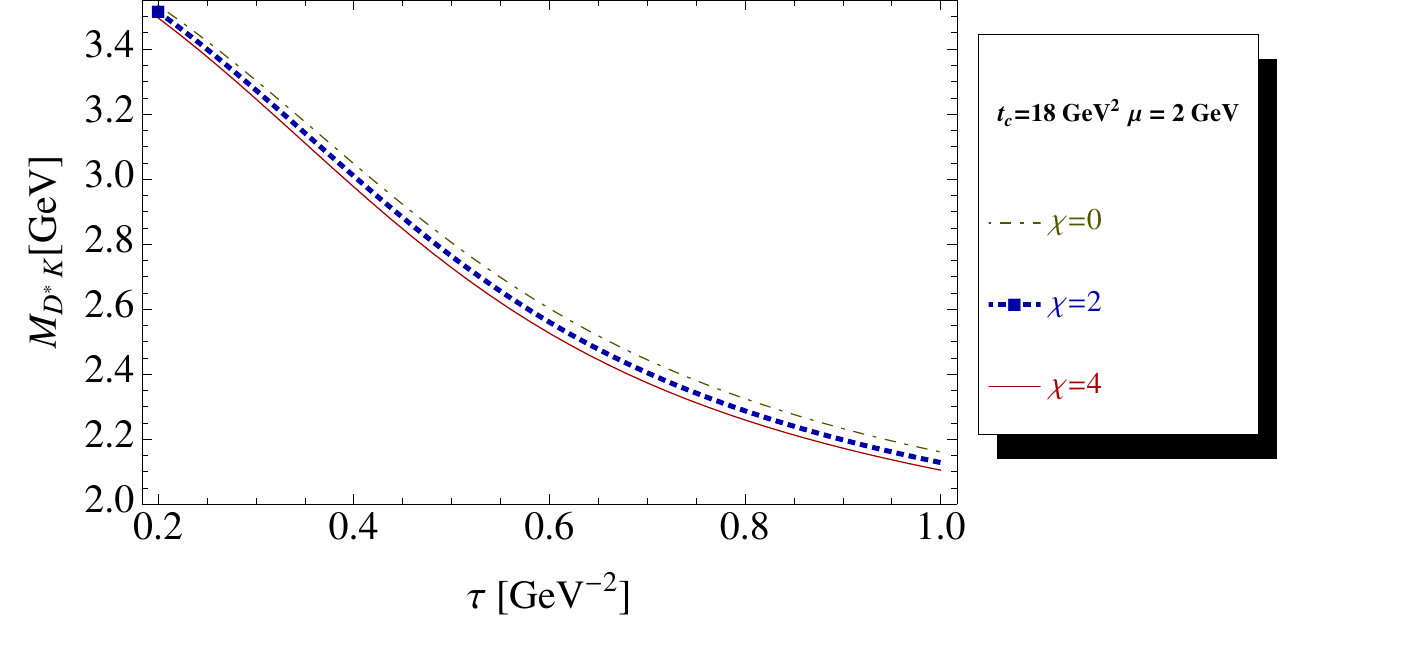}}
{\includegraphics[width=6.2cm  ]{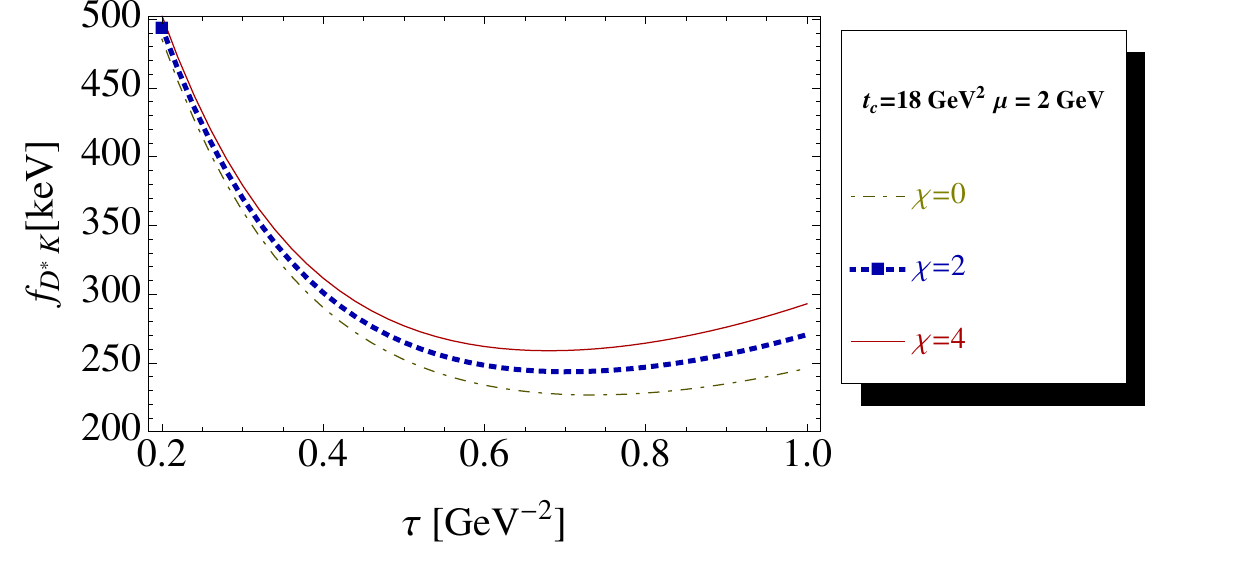}}
\centerline {\hspace*{-3cm} a)\hspace*{6cm} b) }
\caption{
\scriptsize 
{\bf a)} $M_{D^*K}$  as function of $\tau$, for different values of the $d=7$ condensate contribution ($\chi$ measures the violation of factorization),  at a given value of $t_c$=18 GeV$^2$, $\mu=2$ GeV and for the QCD parameters in Tables\,\ref{tab:param} and \ref{tab:alfa}; {\bf b)} The same as a) but for the coupling $f_{D^*K}$.
}
\label{fig:dstark-d7} 
\end{center}
\end{figure}

\hspace*{0.5cm}
\b The analysis of the $1^{+}$ $ D^*K$ molecule exhibits the same feature as the one of the $ DK$ molecule. We show the results of the analysis
in Figs\,\ref{fig:dstark-lo} to \ref{fig:dstark-d7}. The optimal result is obtained for $\mu\simeq$ 2 GeV:
\bea
M_{D^*K}&\simeq&( 2395\pm 48)~{\rm MeV}~,\nnb\\
\hat f_{D^*K}&\simeq& ( 155\pm 36 )~{\rm keV}~~~~\Lrar ~~~~f_{D^*K}\simeq (226\pm 52)~{\rm keV}~,
\eea
where the different sources of errors come from Table\,\ref{tab:errorc}.
\nin
\nin
\nin
\nin
\nin
\subsection{$1^{+}$ axial- vector $ D^*_s\pi$ molecule}
\begin{figure}[hbt] 
\begin{center}
{\includegraphics[width=6.2cm  ]{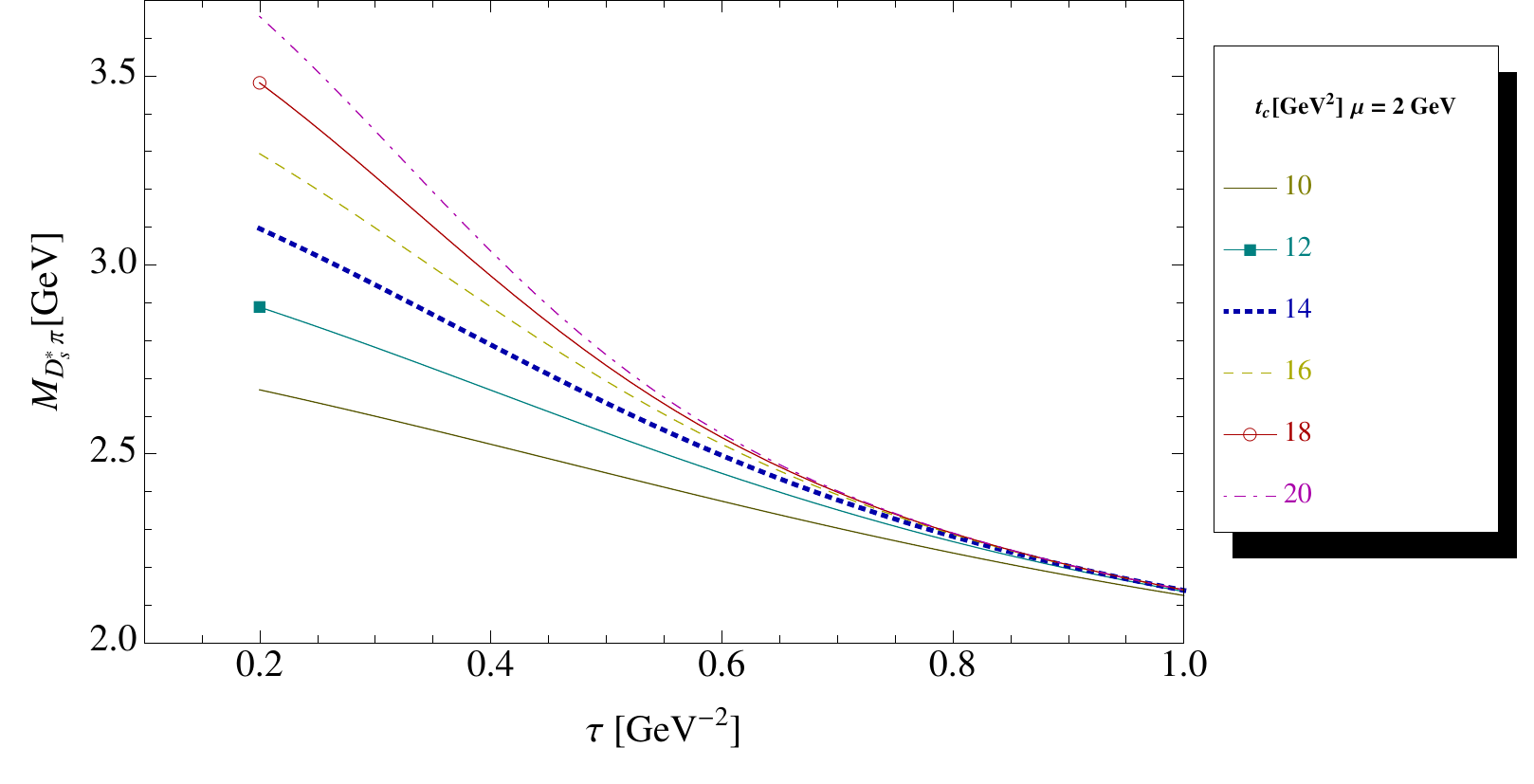}}
{\includegraphics[width=6.2cm  ]{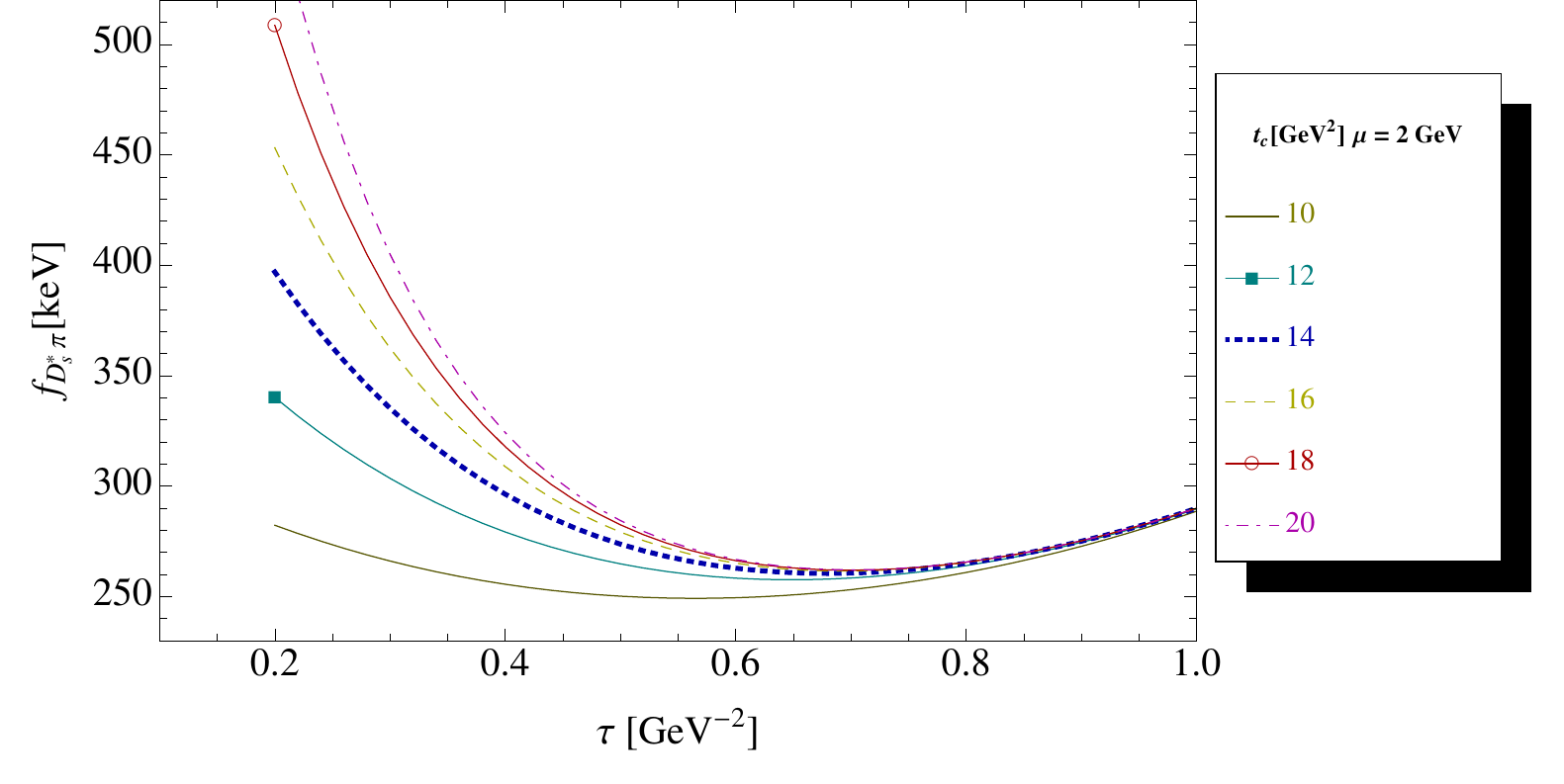}}
\centerline {\hspace*{-3cm} a)\hspace*{6cm} b) }
\caption{
\scriptsize 
{\bf a)} $M_{D^*_s\pi}$  at LO as function of  $\tau$ and different values of $t_c$. We use $\mu=2$ GeV, the mixing parameter $k=0$ and the QCD parameters in Tables\,\ref{tab:param}  and \ref{tab:alfa}; {\bf b)} The same as a) but for the coupling $f_{D^*_s\pi}$.
}
\label{fig:dstarspi-lo} 
\end{center}
\end{figure}

\begin{figure}[hbt] 
\begin{center}
{\includegraphics[width=6.2cm  ]{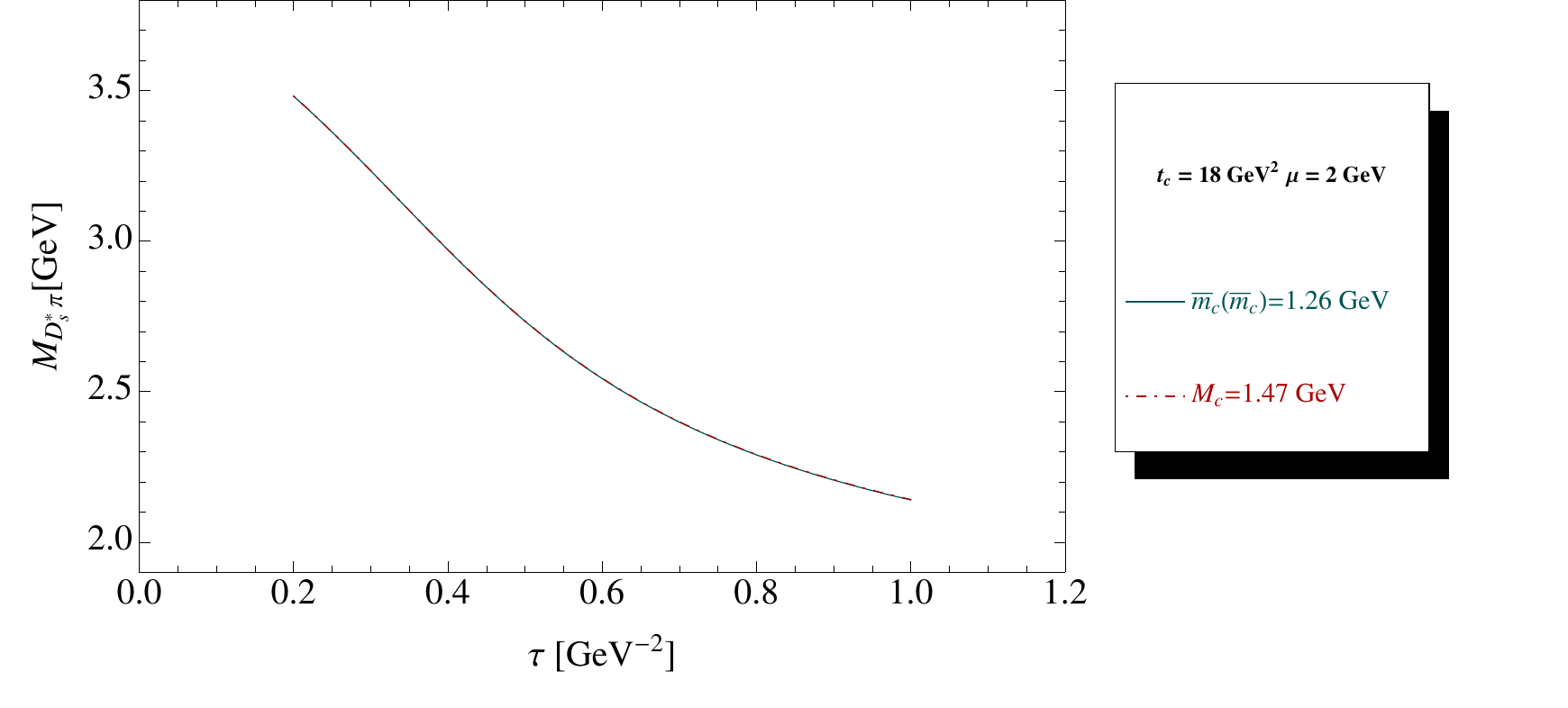}}
{\includegraphics[width=6.2cm  ]{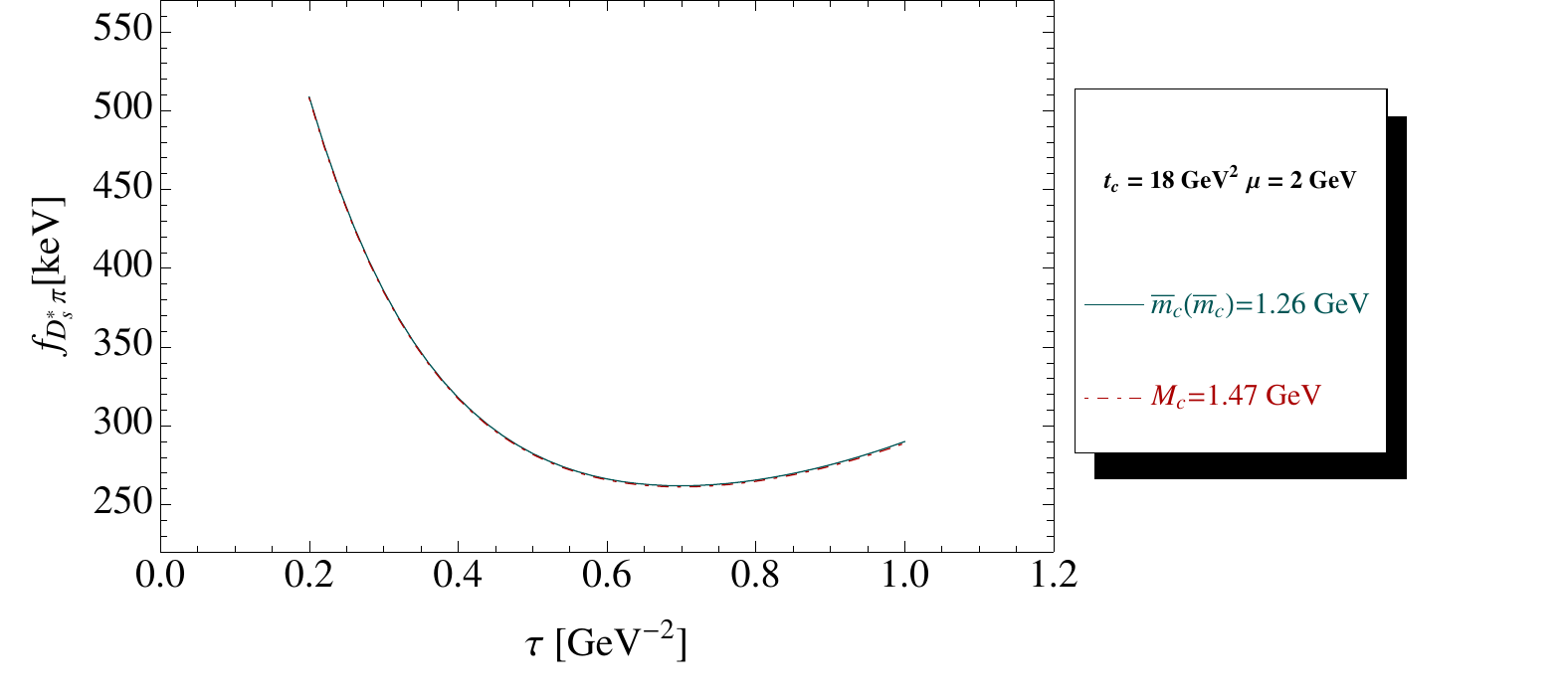}}
\centerline {\hspace*{-3cm} a)\hspace*{6cm} b) }
\caption{
\scriptsize 
{\bf a)} $M_{D^*_s\pi}$  at LO as function of $\tau$ for a given value of $t_c=18$ GeV$^2$,  $\mu=2$ GeV and for the QCD parameters in Tables\,\ref{tab:param} and \ref{tab:alfa}. The OPE is truncated at $d=6$.  We compare the effect of  the on-shell or pole mass $M_c=1.47$ GeV and of the running mass $\bar m_c(\bar m_c)=1.26$ GeV; {\bf b)} The same as a) but for the coupling $f_{D^*_s\pi}$.
}
\label{fig:dstarspimasspole} 
\end{center}
\end{figure}

\begin{figure}[hbt] 
\begin{center}
{\includegraphics[width=6.2cm  ]{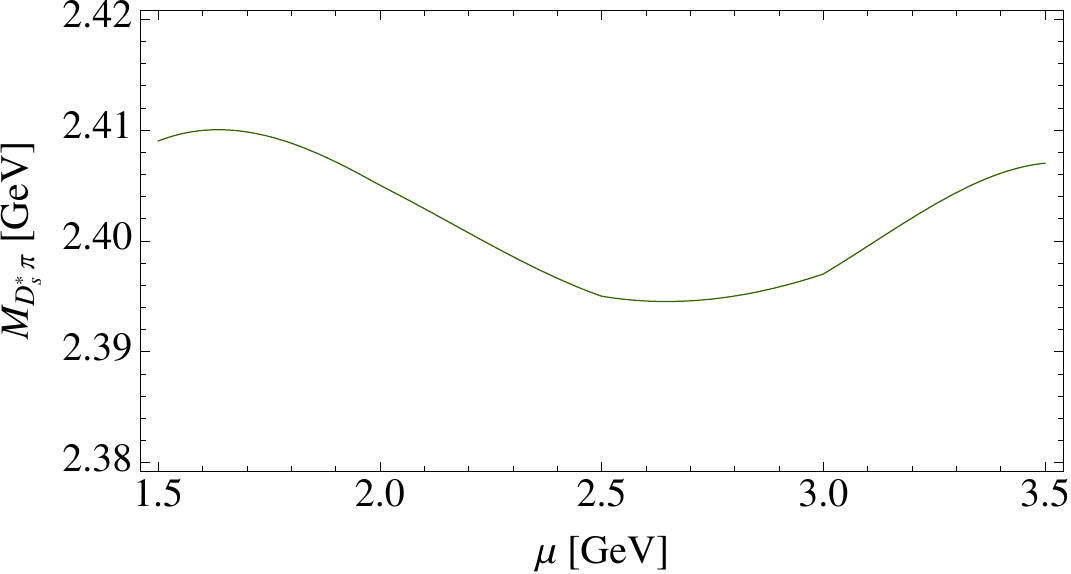}}
{\includegraphics[width=6.2cm  ]{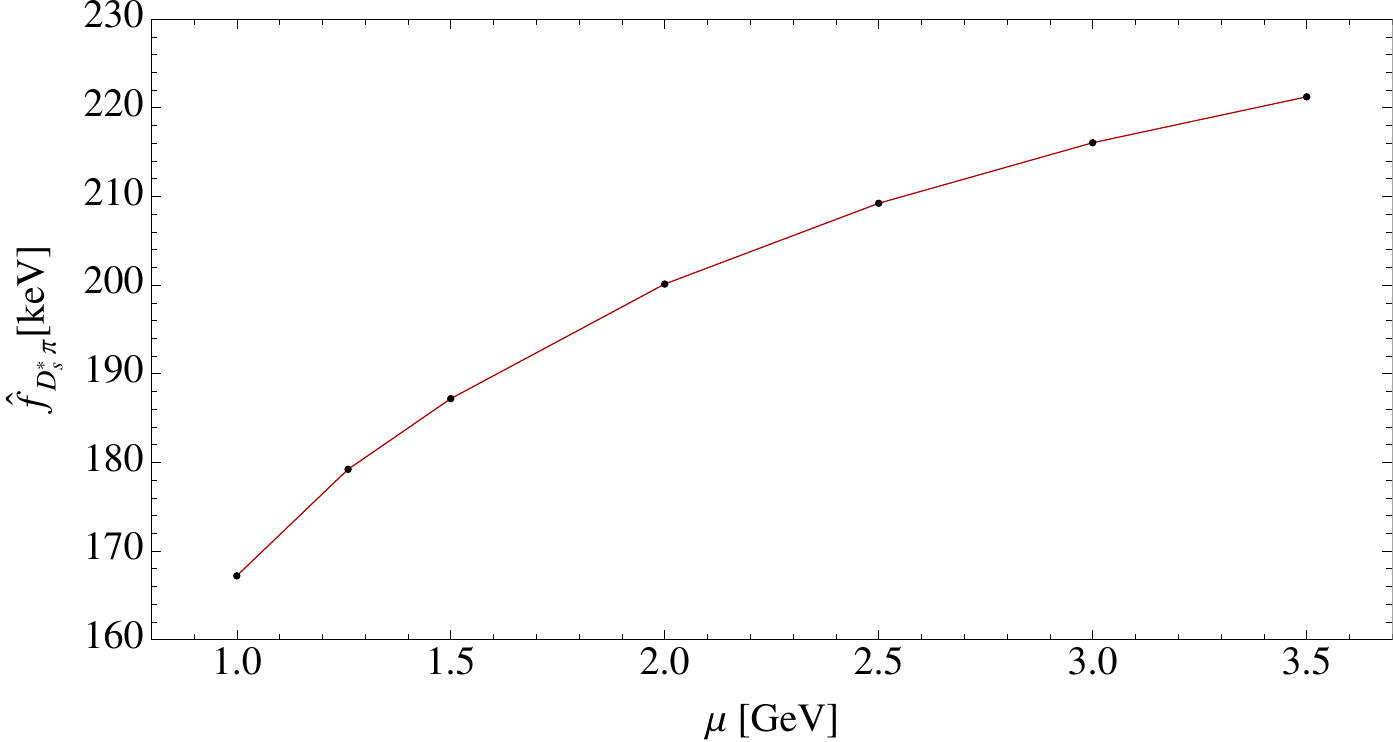}}
\centerline {\hspace*{-3cm} a)\hspace*{6cm} b) }
\caption{
\scriptsize 
{\bf a)} $M_{D^*_s\pi}$ at NLO as function of $\mu$, for the corresponding $\tau$-stability region, for $t_c\simeq 18$ GeV$^2$ and for the QCD parameters in Tables\,\ref{tab:param} and \ref{tab:alfa}; {\bf b)} The same as a) but for the renormalization group invariant coupling $\hat{f}_{D^*_s\pi}$.
}
\label{fig:dstarspi-mu} 
\end{center}
\end{figure}

\begin{figure}[hbt] 
\begin{center}
{\includegraphics[width=6.2cm  ]{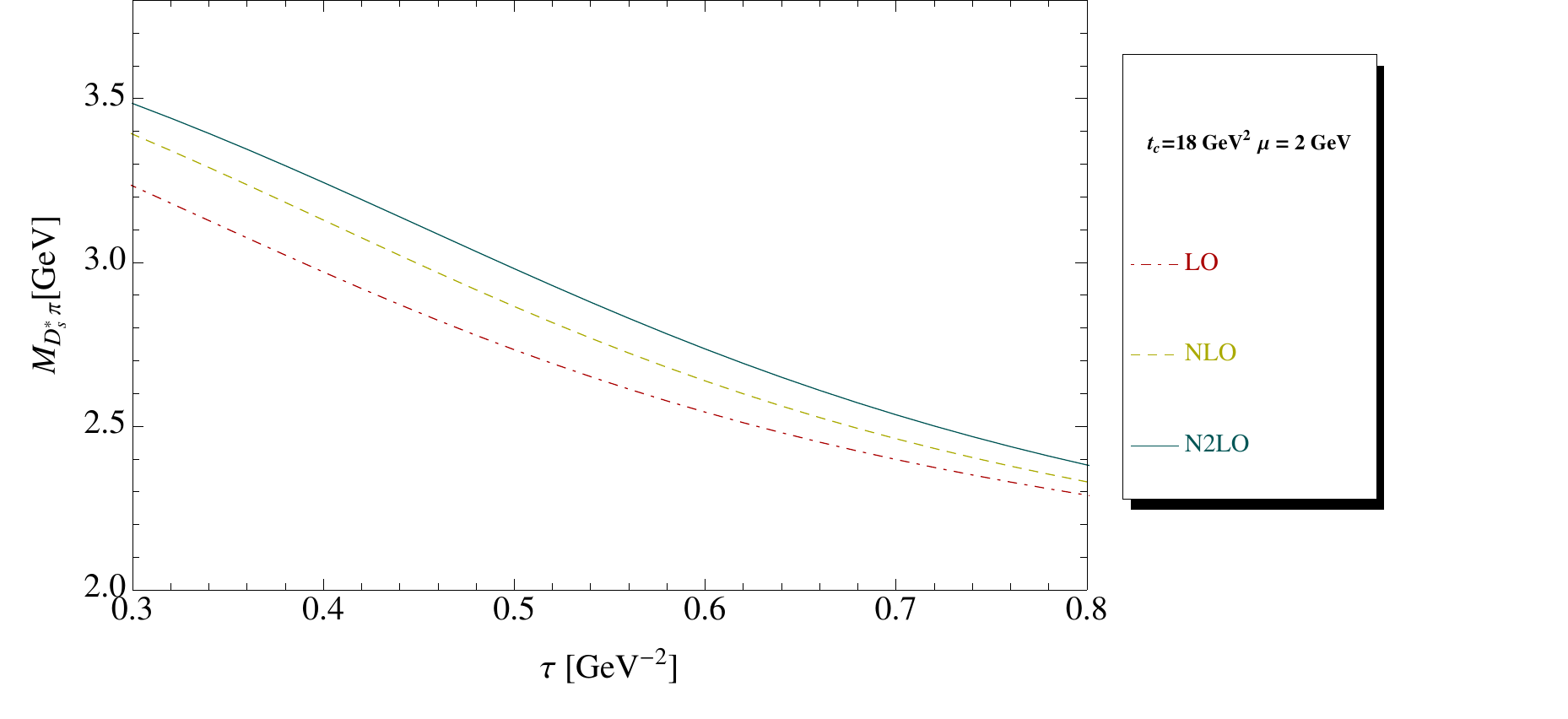}}
{\includegraphics[width=6.2cm  ]{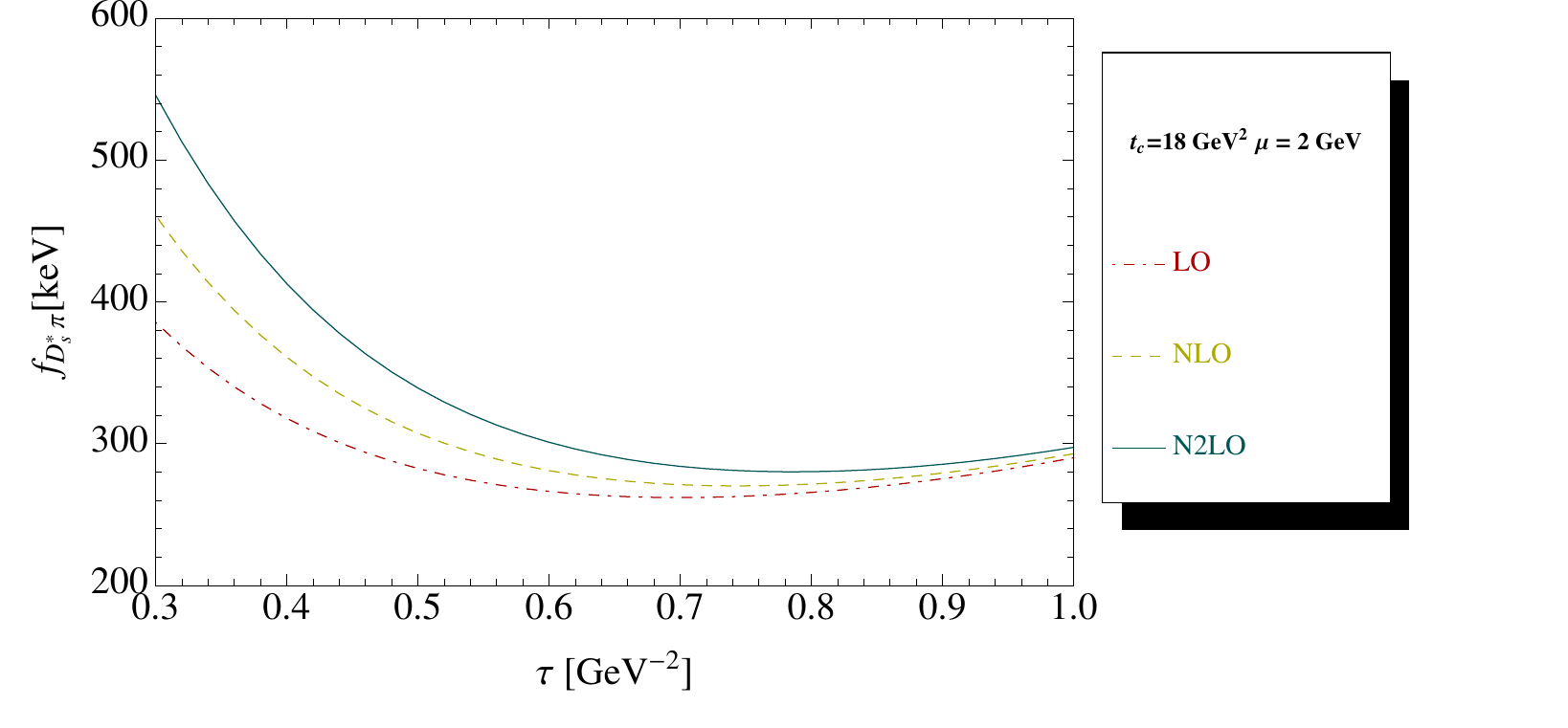}}
\centerline {\hspace*{-3cm} a)\hspace*{6cm} b) }
\caption{
\scriptsize 
{\bf a)} $M_{D^*_s\pi}$  as function of $\tau$ for different truncation of the PT series at a given value of $t_c$=18 GeV$^2$, $\mu=2$ GeV and for the QCD parameters in Tables\,\ref{tab:param} and \ref{tab:alfa}; {\bf b)} The same as a) but for the coupling $f_{D^*_s\pi}$.
}
\label{fig:dstarspi-lo-n2lo} 
\end{center}
\end{figure}

\begin{figure}[hbt] 
\begin{center}
{\includegraphics[width=6.2cm  ]{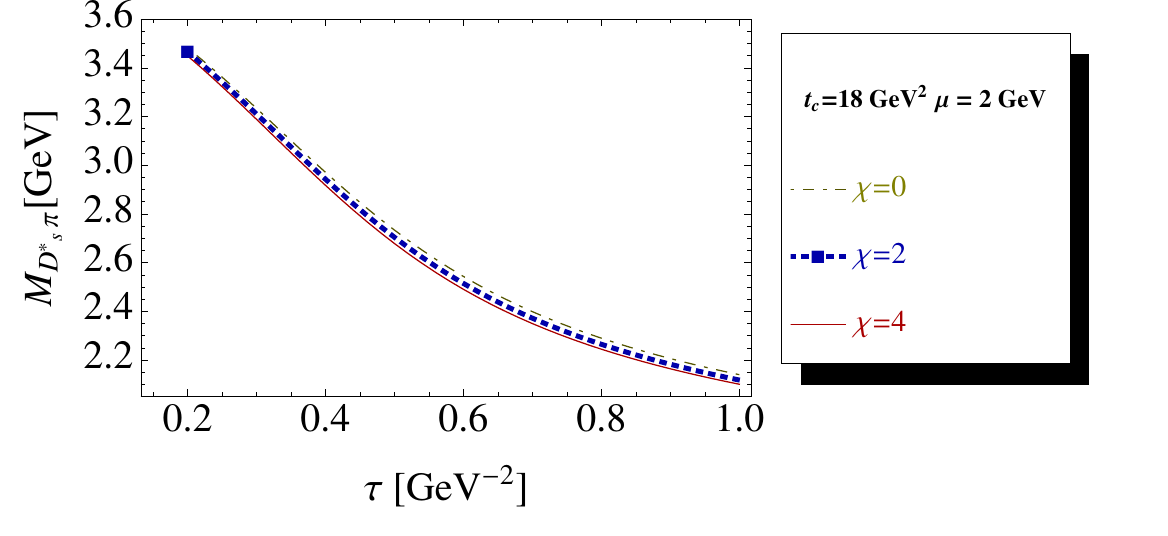}}
{\includegraphics[width=6.2cm  ]{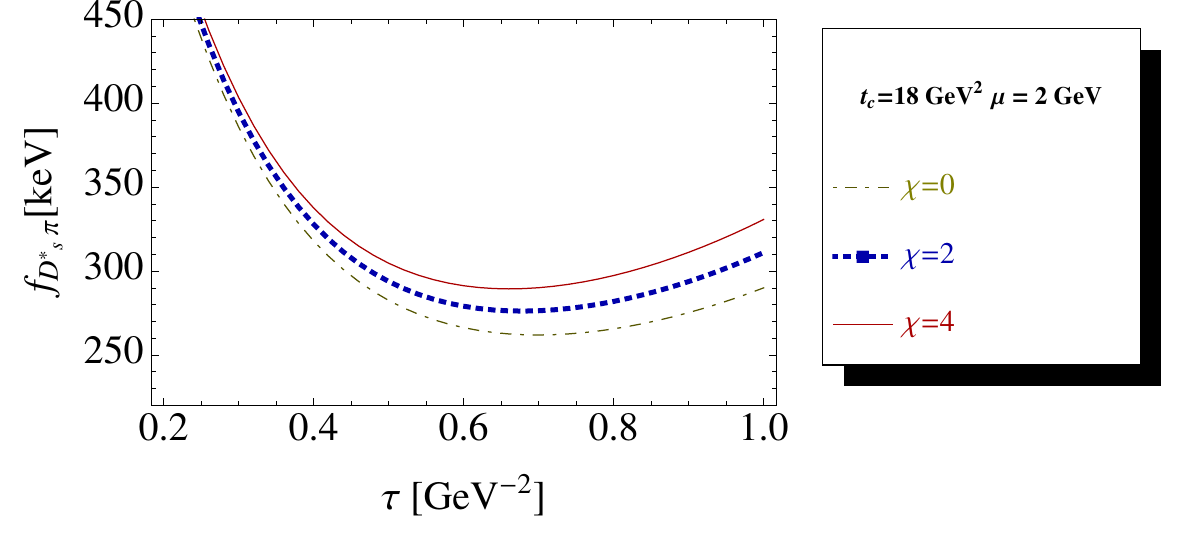}}
\centerline {\hspace*{-3cm} a)\hspace*{6cm} b) }
\caption{
\scriptsize 
{\bf a)} $M_{D^*_s\pi}$  as function of $\tau$, for different values of the $d=7$ condensate contribution ($\chi$ measures the violation of factorization),  at a given value of $t_c$=18 GeV$^2$, $\mu=2$ GeV and for the QCD parameters in Tables\,\ref{tab:param} and \ref{tab:alfa}; {\bf b)} The same as a) but for the coupling $f_{D^*_s\pi}$.
}
\label{fig:dstarspi-d7} 
\end{center}
\end{figure}

\hspace*{0.5cm}
\b The analysis is similar to the previous ones. We show the results of the analysis
in Figs\,\ref{fig:dstarspi-lo} to \ref{fig:dstarspi-d7}. Here, the optimal result is obtained for $\mu\simeq $ 2.5 GeV.
Using the errors quoted in Table\,\ref{tab:errorc}, we deduce at N2LO:
\bea
M_{D^*_s\pi}&\simeq&( 2395\pm 48)~{\rm MeV}~,\nnb\\
\hat f_{D^*_s\pi}&\simeq& ( 215\pm 35 )~{\rm keV}~~~~\Lrar ~~~~f_{D^*_s\pi}\simeq (308\pm 49)~{\rm keV}~,
\eea
where the different sources of errors come from Table\,\ref{tab:errorc}.
\nin
\nin
\nin
\nin
\nin
\subsection{$0^{+}$ scalar $ D_s\pi$ molecule}
\begin{figure}[hbt] 
\begin{center}
{\includegraphics[width=6.2cm  ]{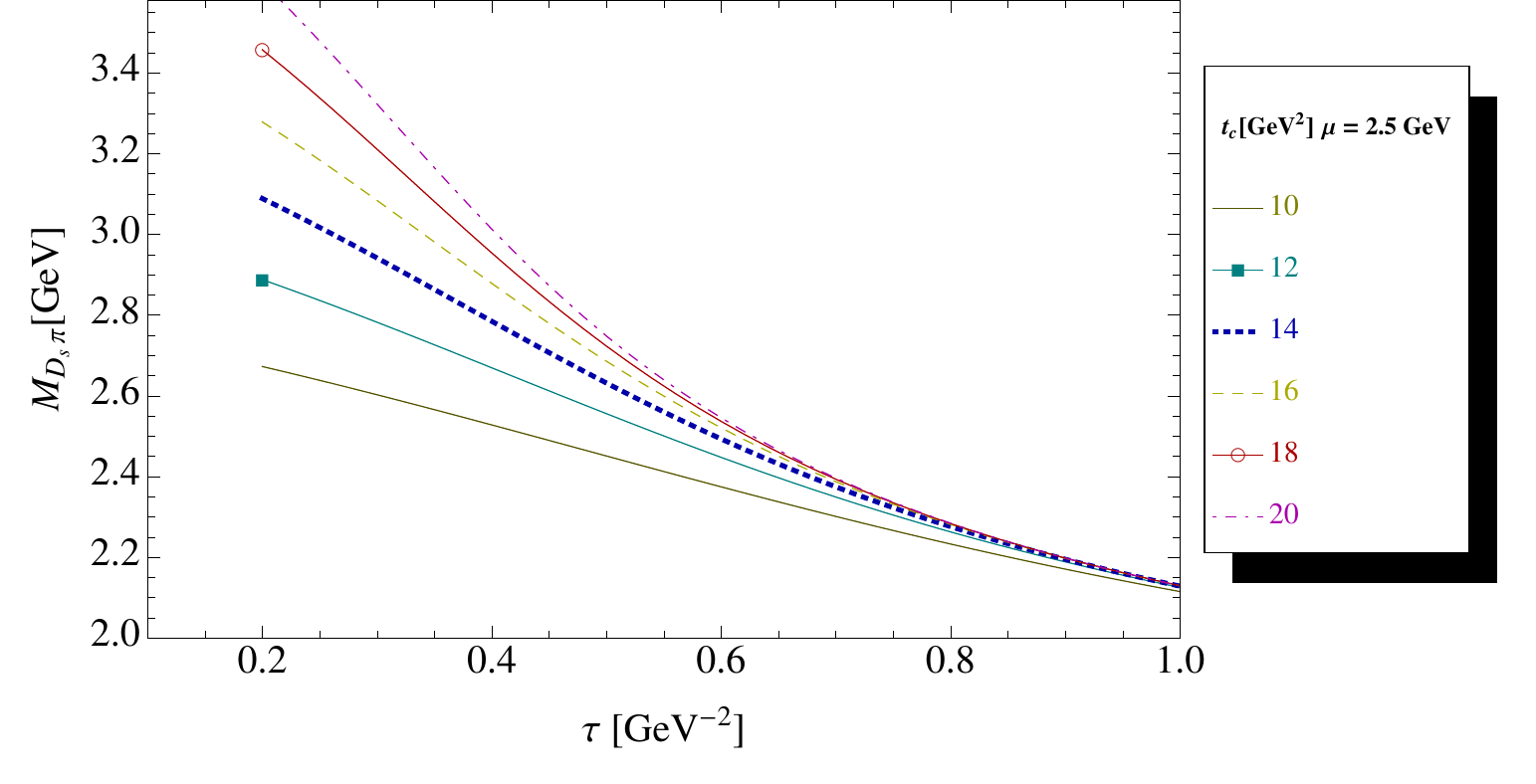}}
{\includegraphics[width=6.2cm  ]{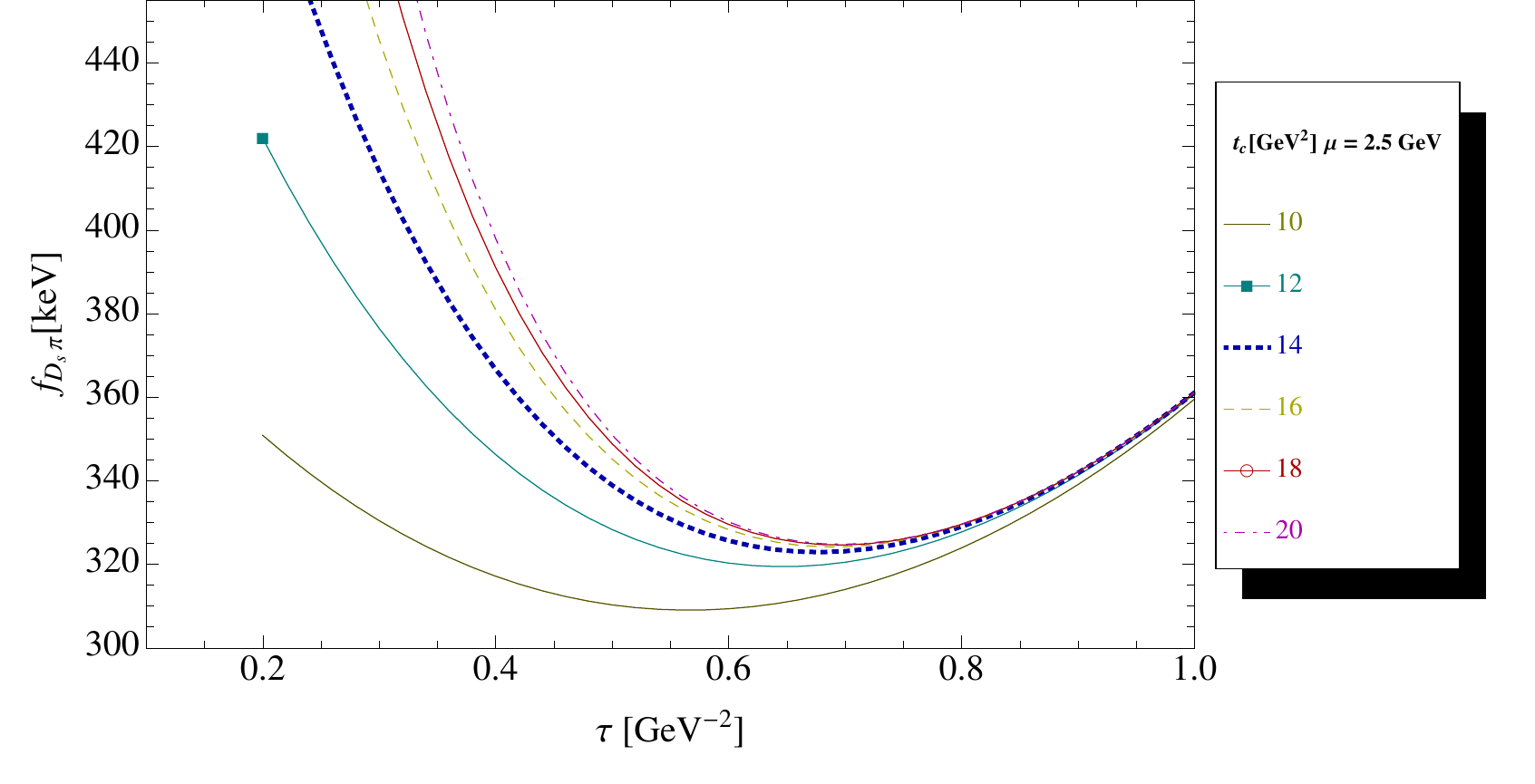}}
\centerline {\hspace*{-3cm} a)\hspace*{6cm} b) }
\caption{
\scriptsize 
{\bf a)} $M_{D_s\pi}$  at LO as function of  $\tau$ and different values of $t_c$. We use $\mu=2.5$ GeV, the mixing parameter $k=0$ and the QCD parameters in Tables\,\ref{tab:param}  and \ref{tab:alfa}; {\bf b)} The same as a) but for the coupling $f_{D_s\pi}$.
}
\label{fig:dspi-lo} 
\end{center}
\end{figure} 
\nin
\begin{figure}[hbt] 
\begin{center}
{\includegraphics[width=6.2cm  ]{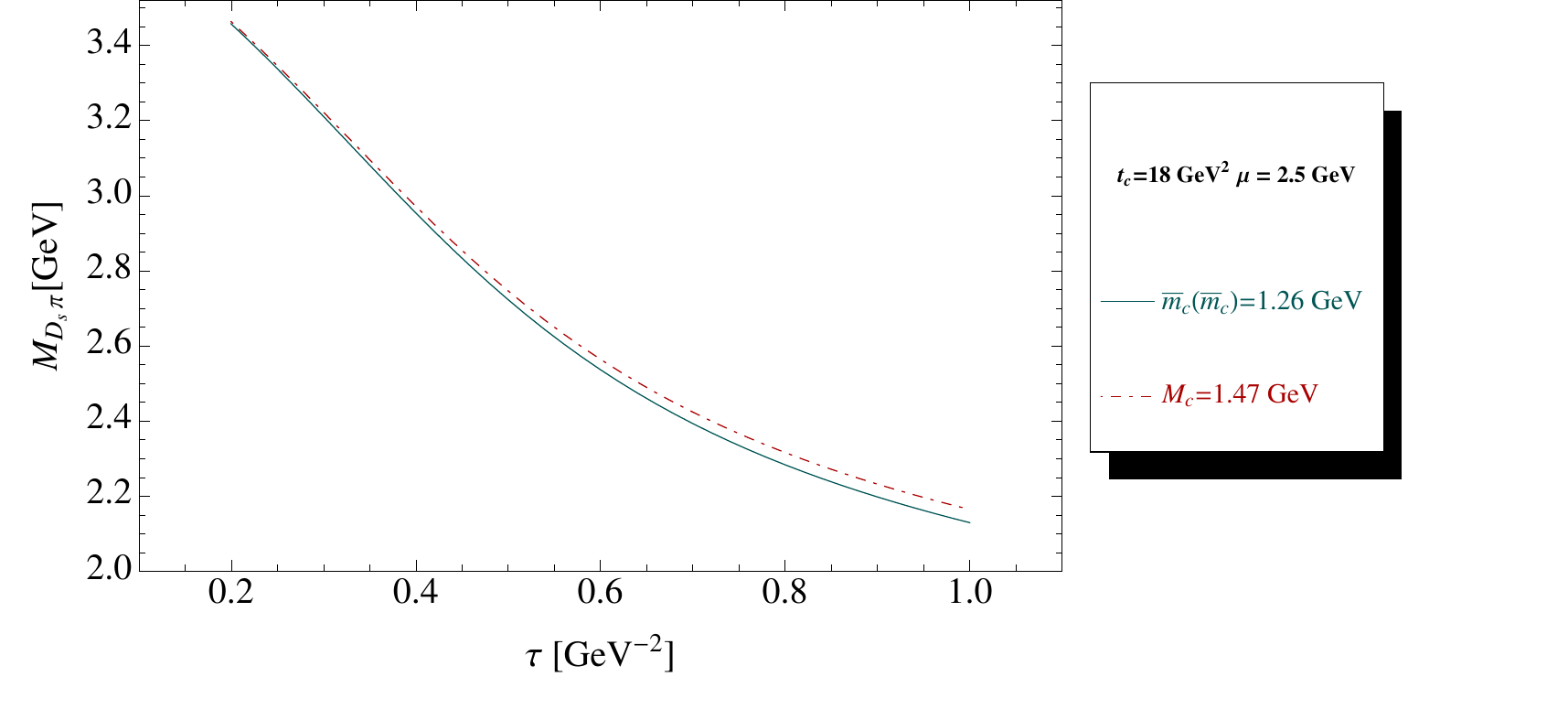}}
{\includegraphics[width=6.2cm  ]{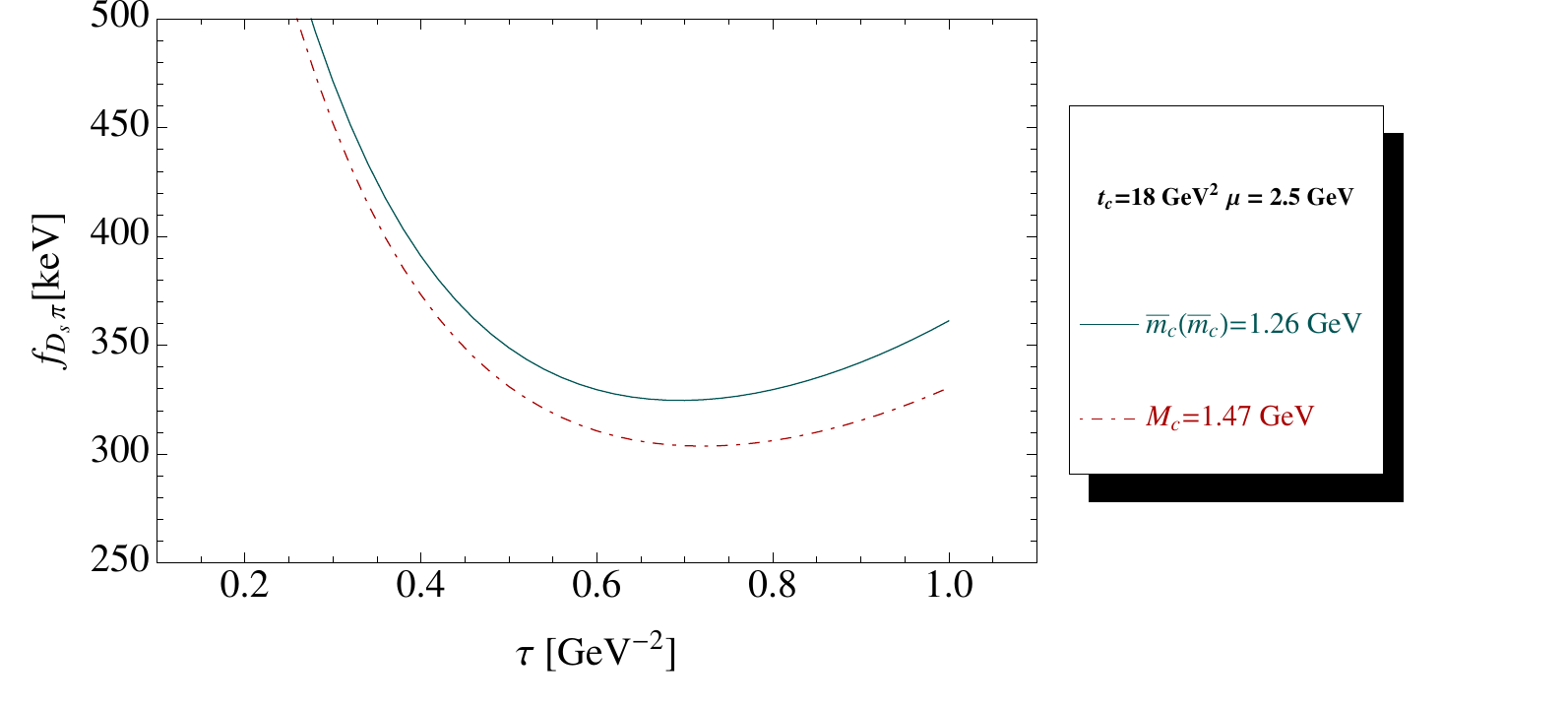}}
\centerline {\hspace*{-3cm} a)\hspace*{6cm} b) }
\caption{
\scriptsize 
{\bf a)} $M_{D_s\pi}$  at LO as function of $\tau$ for a given value of $t_c=18$ GeV$^2$,  $\mu=2.5$ GeV, mixing of currents $k=0$ and for the QCD parameters in Tables\,\ref{tab:param} and \ref{tab:alfa}. The OPE is truncated at $d=6$.  We compare the effect of  the on-shell or pole mass $M_c=1.47$ GeV and of the running mass $\bar m_c(\bar m_c)=1.26$ GeV; {\bf b)} The same as a) but for the coupling $f_{D_s\pi}$.
}
\label{fig:dspimasspole} 
\end{center}
\end{figure} 
\nin
\begin{figure}[hbt] 
\begin{center}
{\includegraphics[width=6.2cm  ]{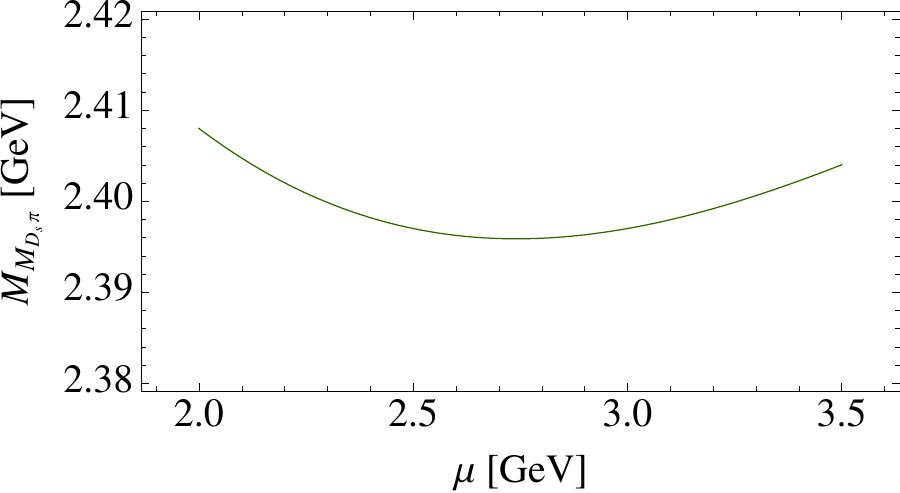}}
{\includegraphics[width=6.2cm  ]{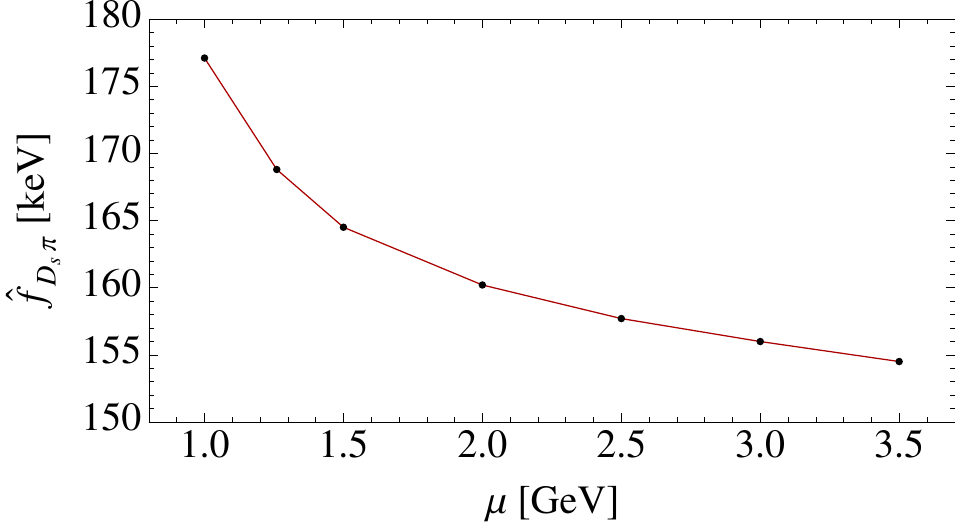}}
\centerline {\hspace*{-3cm} a)\hspace*{6cm} b) }
\caption{
\scriptsize 
{\bf a)} $M_{D_s\pi}$ at LO as function of $\mu$, for the corresponding $\tau$-stability region, for $t_c\simeq 18$ GeV$^2$ and for the QCD parameters in Tables\,\ref{tab:param} and \ref{tab:alfa}; {\bf b)} The same as a) but for the renormalization group invariant coupling $\hat{f}_{D_s\pi}$.
}
\label{fig:dspi-mu} 
\end{center}
\end{figure} 
\nin
\begin{figure}[hbt] 
\begin{center}
{\includegraphics[width=6.2cm  ]{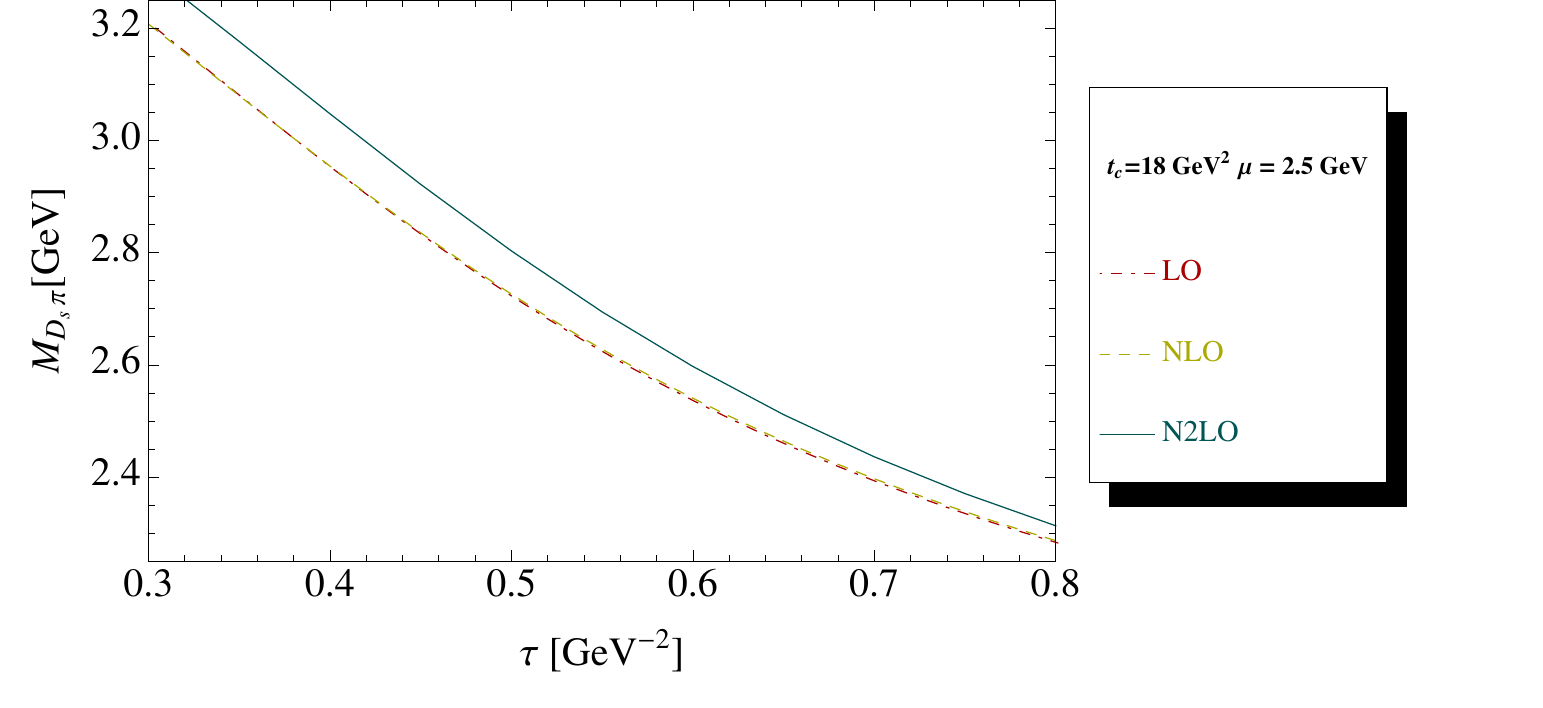}}
{\includegraphics[width=6.2cm  ]{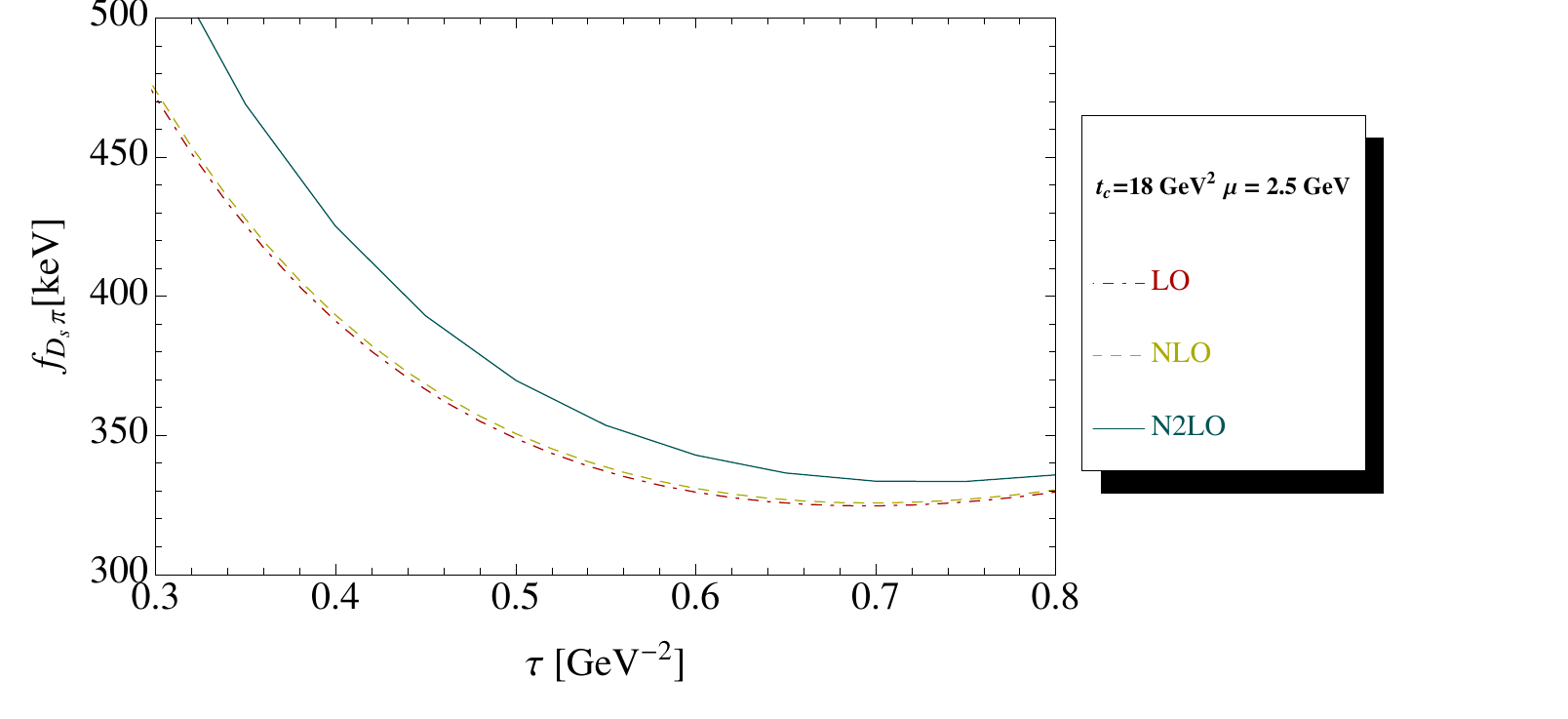}}
\centerline {\hspace*{-3cm} a)\hspace*{6cm} b) }
\caption{
\scriptsize 
{\bf a)} $M_{D_s\pi}$  as function of $\tau$ for different truncation of the PT series at a given value of $t_c$=18 GeV$^2$, $\mu=2.5$ GeV and for the QCD parameters in Tables\,\ref{tab:param} and \ref{tab:alfa}; {\bf b)} The same as a) but for the coupling $f_{D_s\pi}$.
}
\label{fig:dspi-lo-n2lo} 
\end{center}
\end{figure} 
\nin
\begin{figure}[hbt] 
\begin{center}
{\includegraphics[width=6.2cm  ]{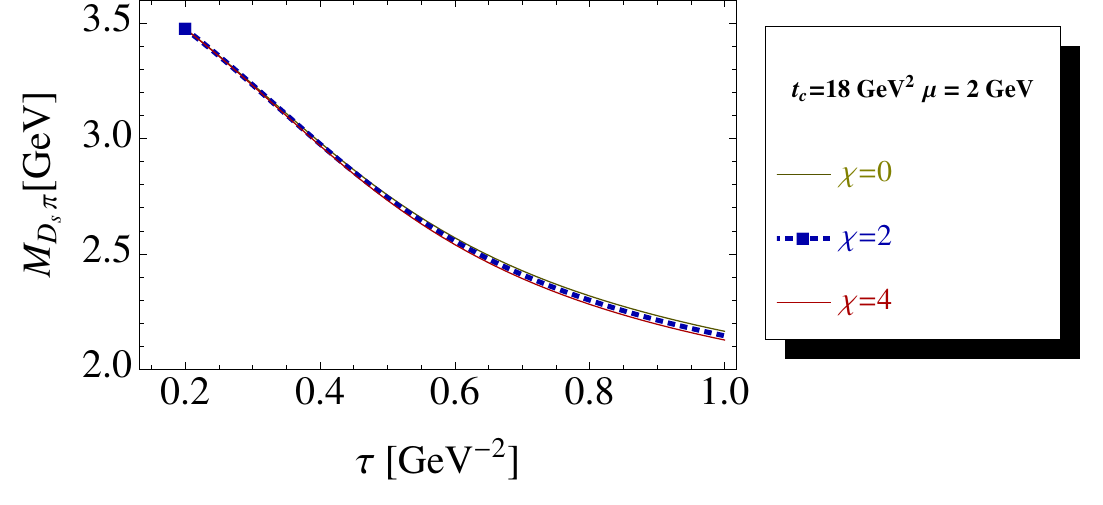}}
{\includegraphics[width=6.2cm  ]{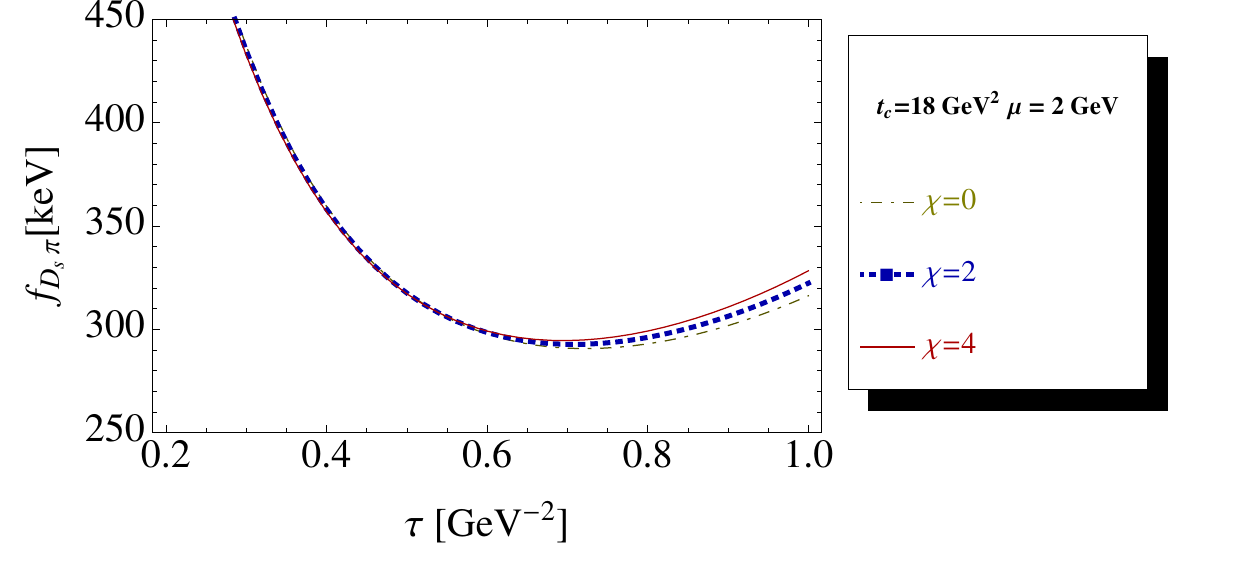}}
\centerline {\hspace*{-3cm} a)\hspace*{6cm} b) }
\caption{
\scriptsize 
{\bf a)} $M_{D_s\pi}$  as function of $\tau$, for different values of the $d=7$ condensate contribution ($\chi$ measures the violation of factorization),  at a given value of $t_c$=18 GeV$^2$, $\mu=2$ GeV and for the QCD parameters in Tables\,\ref{tab:param} and \ref{tab:alfa}; {\bf b)} The same as a) but for the coupling $f_{D_s\pi}$.
}
\label{fig:dspi-d7} 
\end{center}
\end{figure} 
\nin
\hspace*{0.5cm}
\b The analysis is similar to the previous ones. We show the results of the analysis
in Figs\,\ref{fig:dspi-lo} to \ref{fig:dspi-d7}. Using the error quoted in Table\,\ref{tab:error}, we deduce at N2LO and for $\mu=2.5$ GeV:
\bea
M_{D_s\pi}&\simeq&( 2404\pm 37)~{\rm MeV}~,\nnb\\
\hat f_{D_s\pi}&\simeq& ( 160\pm 22 )~{\rm keV}~~~~\Lrar ~~~~f_{D_s\pi}\simeq (331\pm 46)~{\rm keV}~,
\eea
where the different sources of errors come from Table\,\ref{tab:errorc}.
\subsection{$0^{+}$ scalar four-quark state}
\begin{figure}[hbt] 
\begin{center}
{\includegraphics[width=6.2cm  ]{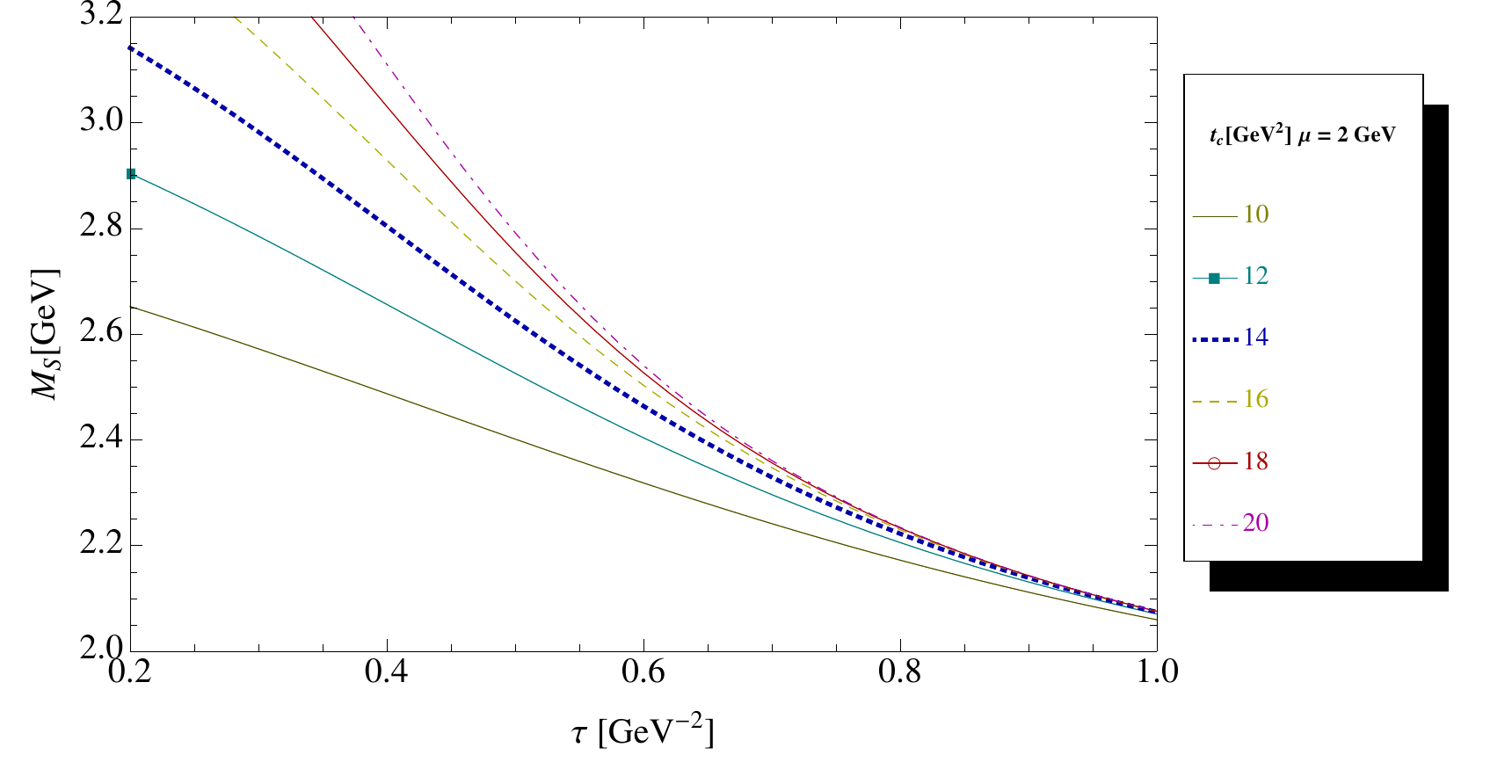}}
{\includegraphics[width=6.2cm  ]{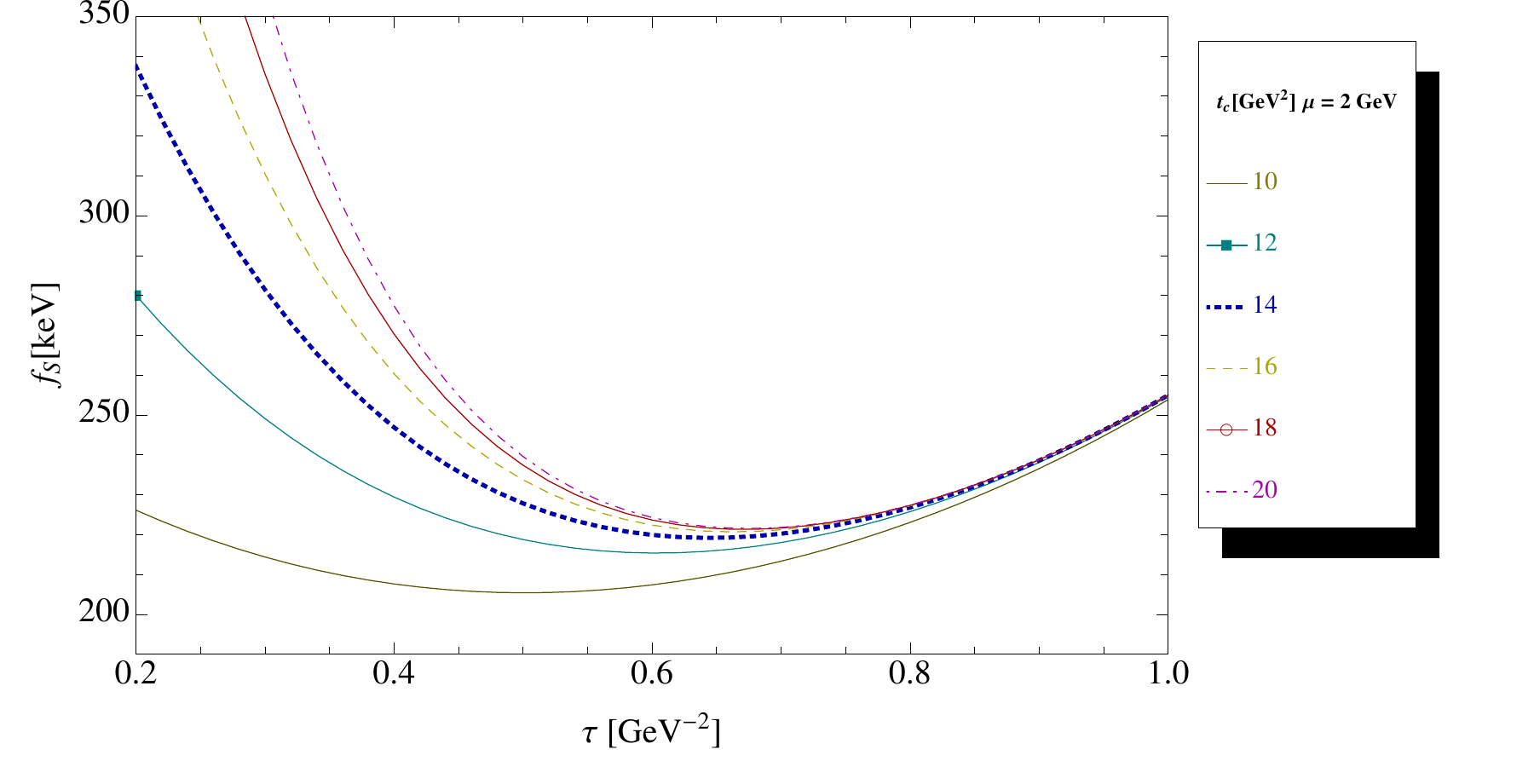}}
\centerline {\hspace*{-3cm} a)\hspace*{6cm} b) }
\caption{
\scriptsize 
{\bf a)} $M_{S_c}$  at LO as function of $\tau$ for  $\mu=2$ GeV, $t_c$=18 GeV$^2$, mixing currents $k=0$  and for the QCD parameters in Tables\,\ref{tab:param}  and \ref{tab:alfa};  {\bf b)} The same as a) but for the coupling $f_{S_c}$.
}
\label{fig:sc-lo} 
\end{center}
\end{figure}

\hspace*{0.5cm}\b The analysis is very similar to the previous one. In  Fig.\,\ref{fig:sc-lo}, we show the $\tau$-behaviour of the mass and coupling at lowest order (LO) for different values of $t_c$, for $\mu=2$ GeV and for the input parameters in Tables\,\ref{tab:param} and \ref{tab:alfa}. 
One can see a $\tau$-stability of about 0.6 GeV$^{-2}$ starting from $t_c=12$ GeV$^2$ while $t_c$-stability is reached from $t_c=18$ GeV$^2$. 
\nin
\begin{figure}[hbt] 
\begin{center}
{\includegraphics[width=6.2cm  ]{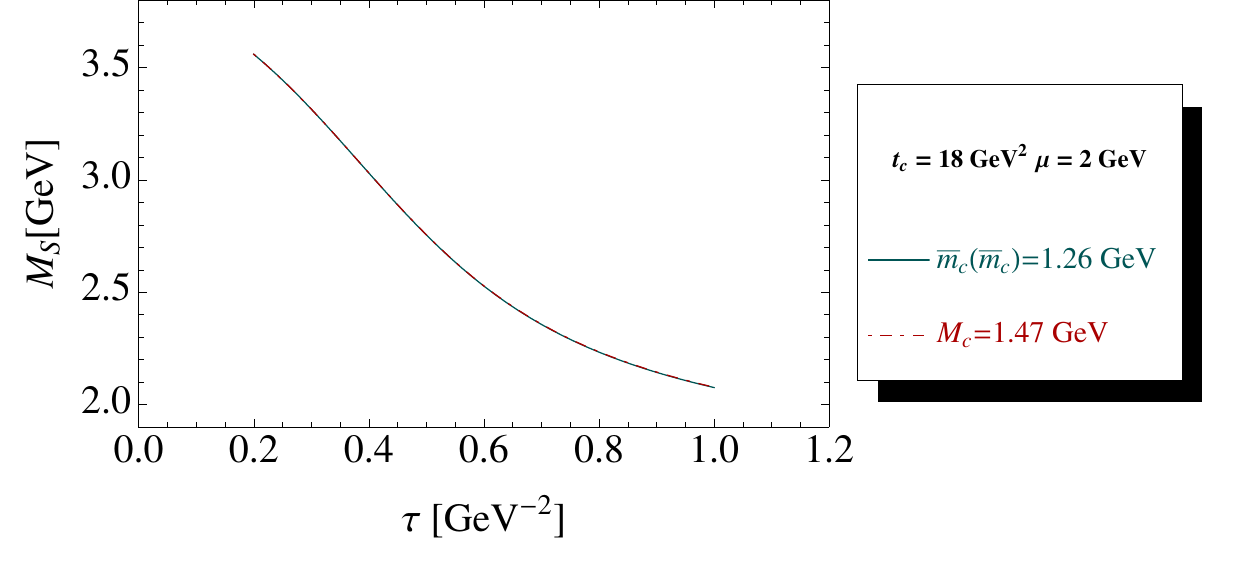}}
{\includegraphics[width=6.2cm  ]{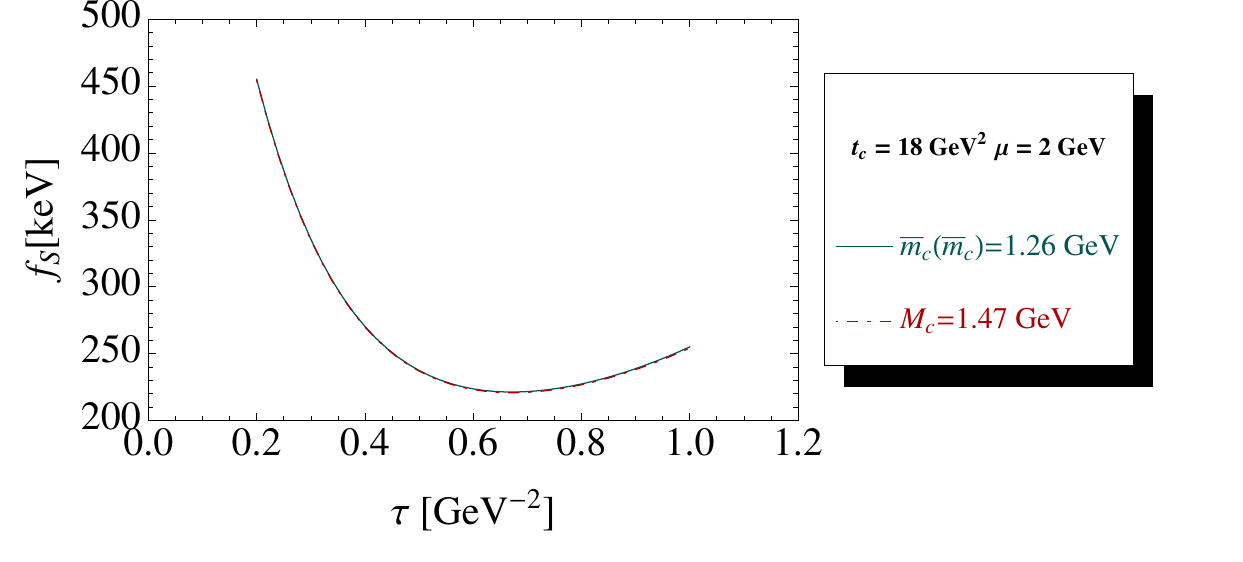}}
\centerline {\hspace*{-3cm} a)\hspace*{6cm} b) }
\caption{
\scriptsize 
{\bf a)} $M_{S_c}$  at LO as function of $\tau$ for a given value of $t_c=18$ GeV$^2$,  $\mu=2$ GeV, mixing of currents $k=0$ and for the QCD parameters in Tables\,\ref{tab:param}  and \ref{tab:alfa};  The OPE is truncated at $d=6$.  We compare the effect of  the on-shell or pole mass $M_c=1.47$ GeV and of the running mass $\bar m_c(\bar m_c)=1.26$ GeV; {\bf b)} The same as a) but for the coupling $f_{S_c}$.
}
\label{fig:scmasspole} 
\end{center}
\end{figure} 
\nin
\begin{figure}[hbt] 
\begin{center}
{\includegraphics[width=6.2cm  ]{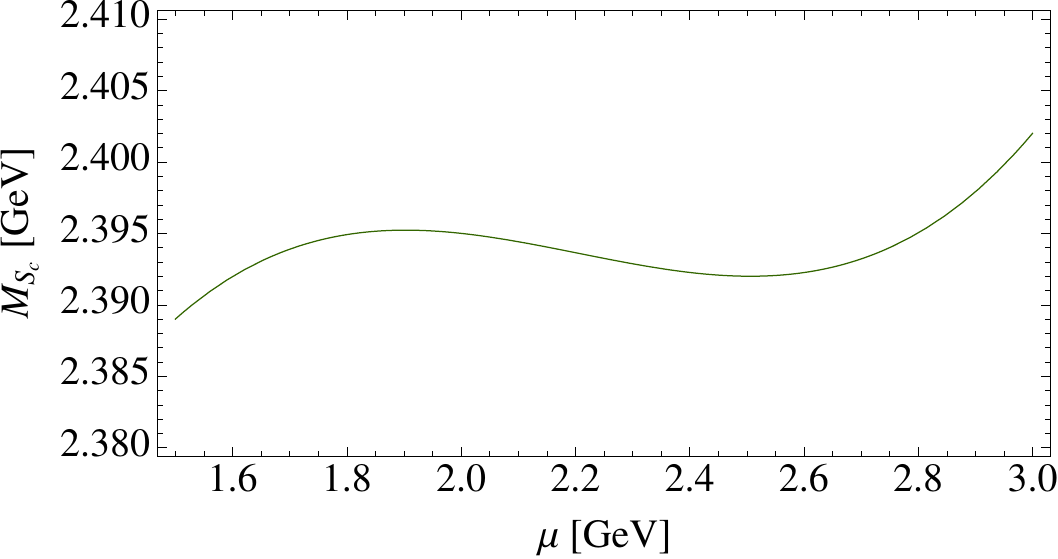}}
{\includegraphics[width=6.2cm  ]{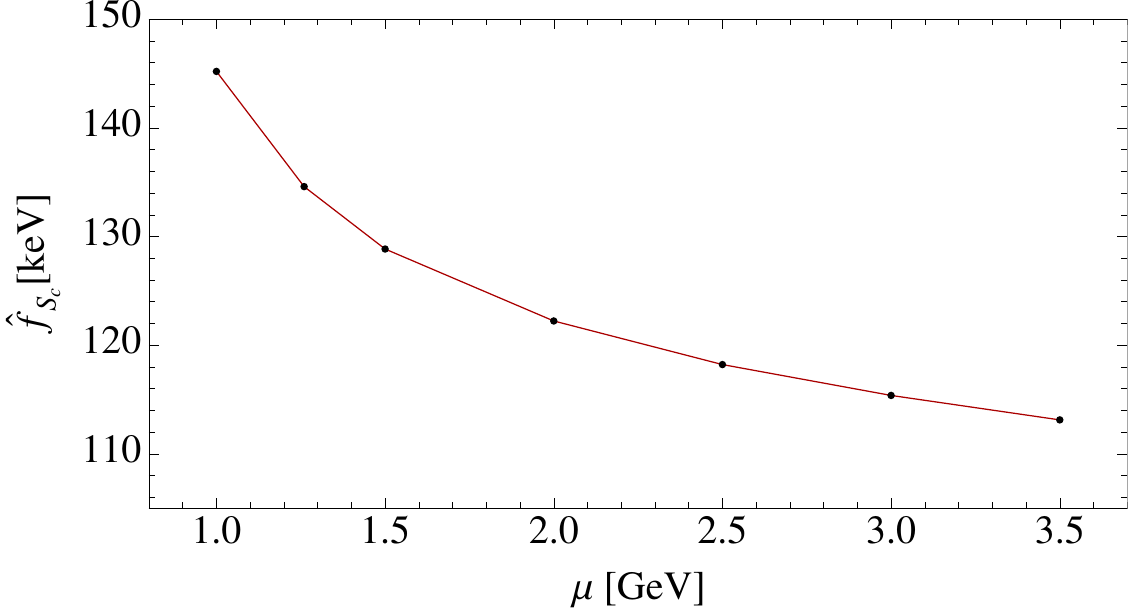}}
\centerline {\hspace*{-3cm} a)\hspace*{6cm} b) }
\caption{
\scriptsize 
{\bf a)} $M_{S_c}$ at NLO as function of $\mu$, for the corresponding $\tau$-stability region, for $t_c\simeq 18$ GeV$^2$ and for the QCD parameters in Tables\,\ref{tab:param}  and \ref{tab:alfa};  {\bf b)} The same as a) but for the renormalization group invariant coupling $\hat{f}_{S_c}$.
}
\label{fig:sc-mu} 
\end{center}
\end{figure} 
\nin
\begin{figure}[hbt] 
\begin{center}
{\includegraphics[width=6.2cm  ]{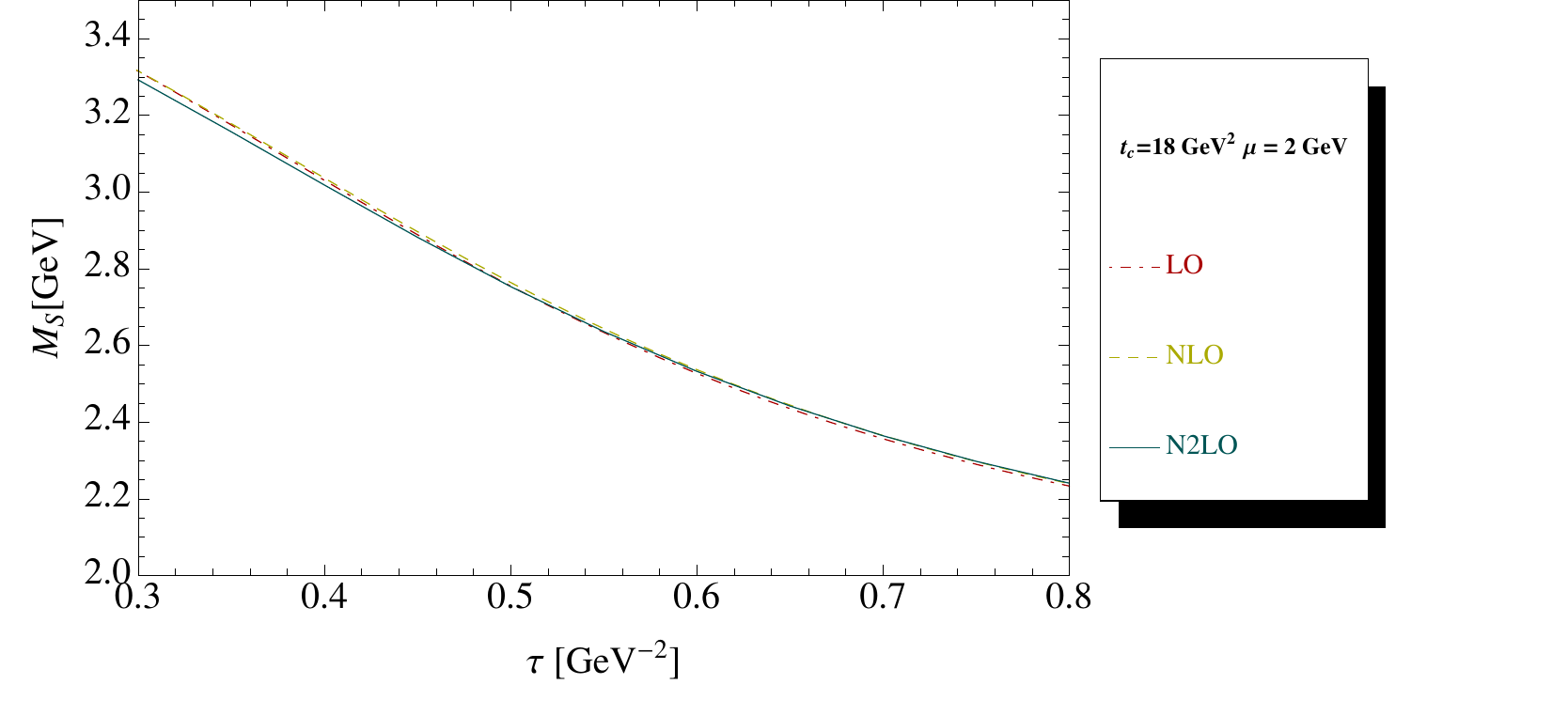}}
{\includegraphics[width=6.2cm  ]{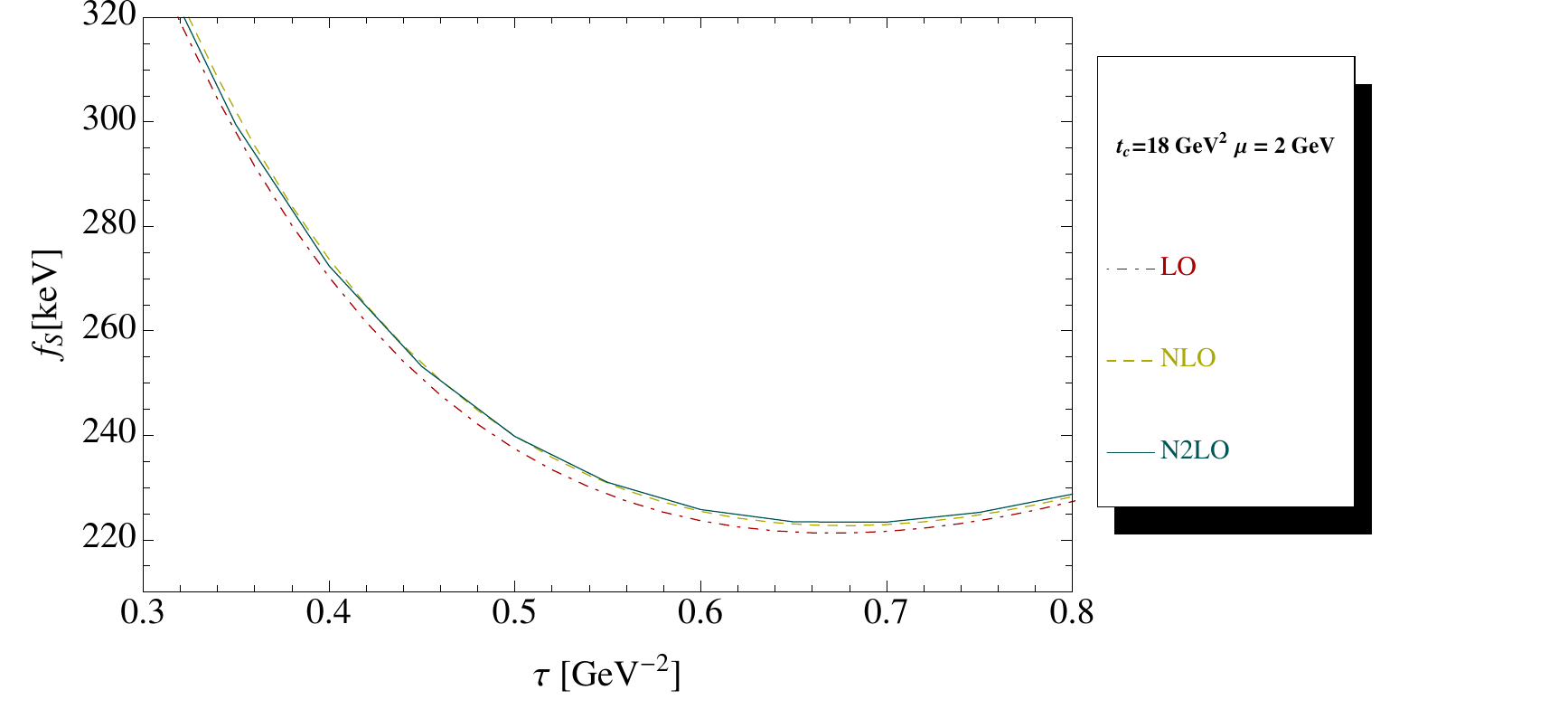}}
\centerline {\hspace*{-3cm} a)\hspace*{6cm} b) }
\caption{
\scriptsize 
{\bf a)} $M_{S_c}$  as function of $\tau$ for different truncation of the PT series at a given value of $t_c$=18 GeV$^2$, $\mu=2$ GeV and for the QCD parameters in Tables\,\ref{tab:param}  and \ref{tab:alfa};  {\bf b)} The same as a) but for the coupling $f_{S_c}$.
}
\label{fig:sc-lo-n2lo} 
\end{center}
\end{figure} 
\nin
\subsection{Error induced by the truncation of the OPE}
\begin{figure}[hbt] 
\begin{center}
{\includegraphics[width=6.2cm  ]{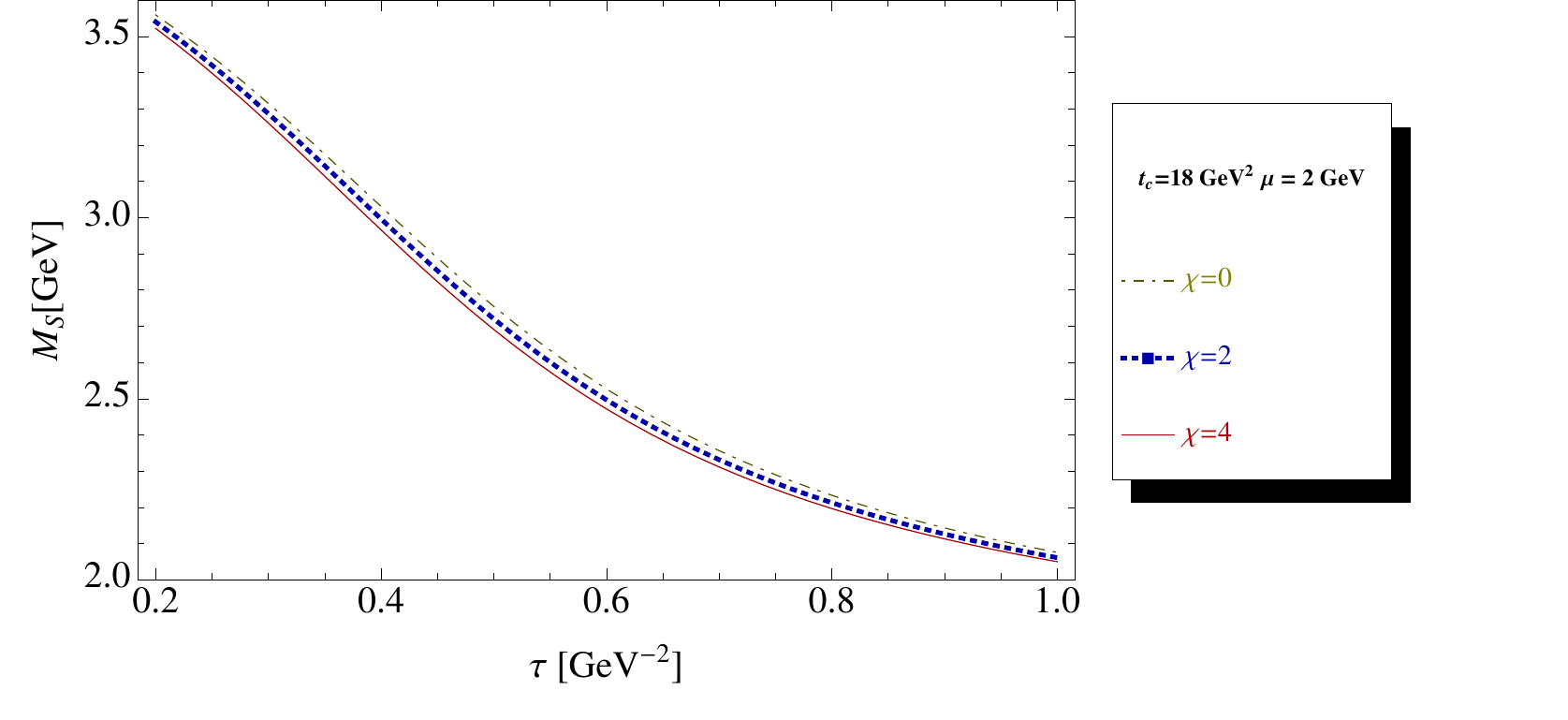}}
{\includegraphics[width=6.2cm  ]{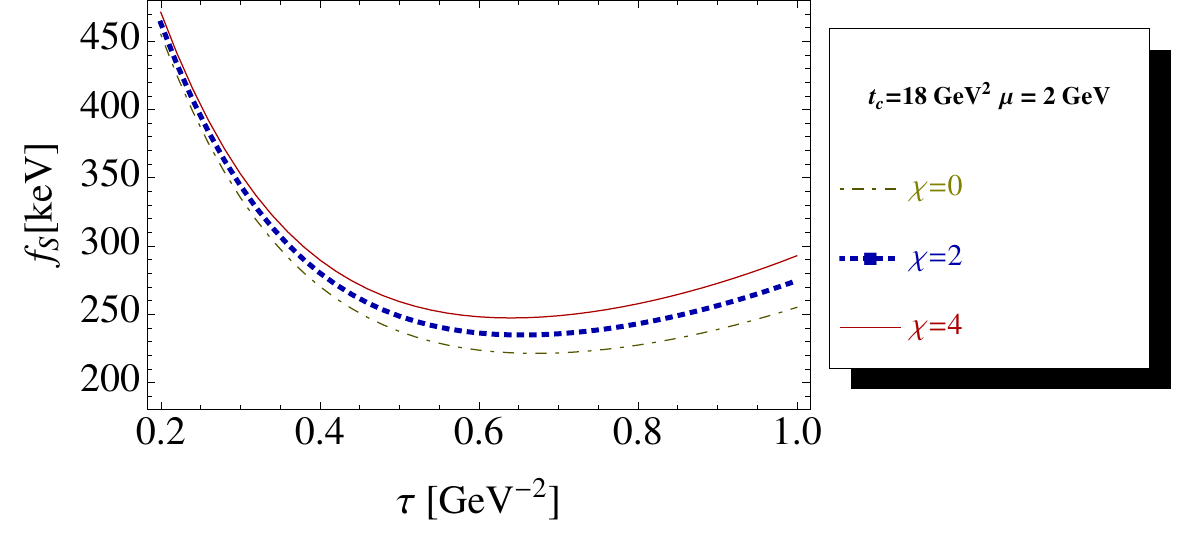}}
\centerline {\hspace*{-3cm} a)\hspace*{6cm} b) }
\caption{
\scriptsize 
{\bf a)} $M_{S_c}$  as function of $\tau$, for different values of the $d=7$ condensate contribution ($\chi$ measures the violation of factorization),  at a given value of $t_c$=18 GeV$^2$, $\mu=2$ GeV and for the QCD parameters in Tables\,\ref{tab:param} and \ref{tab:alfa}; {\bf b)} The same as a) but for the coupling $f_{S_c}$.
}
\label{fig:sc-d7} 
\end{center}
\end{figure}

We show in Fig.\,\ref{fig:sc-d7} the effect of a class of $d=7$ condensates for different values of the factorization violation parameter $\chi$.
Our estimate of the error induced by the truncation of the OPE corresponds to the choice $\chi=4$. 
\nin
At lowest order, one obtains for $\mu$=2 GeV:
\beq
M_{S_c}^{LO}\simeq( 2381\sim 2386)~{\rm MeV}~~~~{\rm and}~~~~
f_{S_c}^{LO}\simeq (215\sim 221)~{\rm keV}~.
\eeq

\b The analysis of the $\mu$-behaviour in Fig.\,\ref{fig:sc-mu} indicates a $\mu$-stability for 
$\mu=2~{\rm GeV}.$

\b The effects of the truncation of the PT series are shown in Fig.\,\ref{fig:sc-lo-n2lo}, where one can notice that the PT corrections are small both for the coupling and
for the mass.

\b From the previous analysis, we deduce the final result including N2LO PT perturbative $\oplus$ $d\leq 6$ dimension contributions. We  take $t_c \simeq (12\sim 18)$ GeV$^2$ and the optimal choice $\mu$=2 GeV:
\bea
M_{S_c}&\simeq&( 2395\pm 68)~{\rm MeV}~,\nnb\\
\hat f_{S_c}&\simeq& (122\pm 26)~{\rm keV}~~~~\Lrar ~~~~f_{S_c}\simeq (221\pm 47 )~{\rm keV}~,
\eea
where the different sources of errors come from Table\,\ref{tab:errorc}.
\subsection{$1^{+}$ axial-vector four-quark state}
\begin{figure}[hbt] 
\begin{center}
{\includegraphics[width=6.2cm  ]{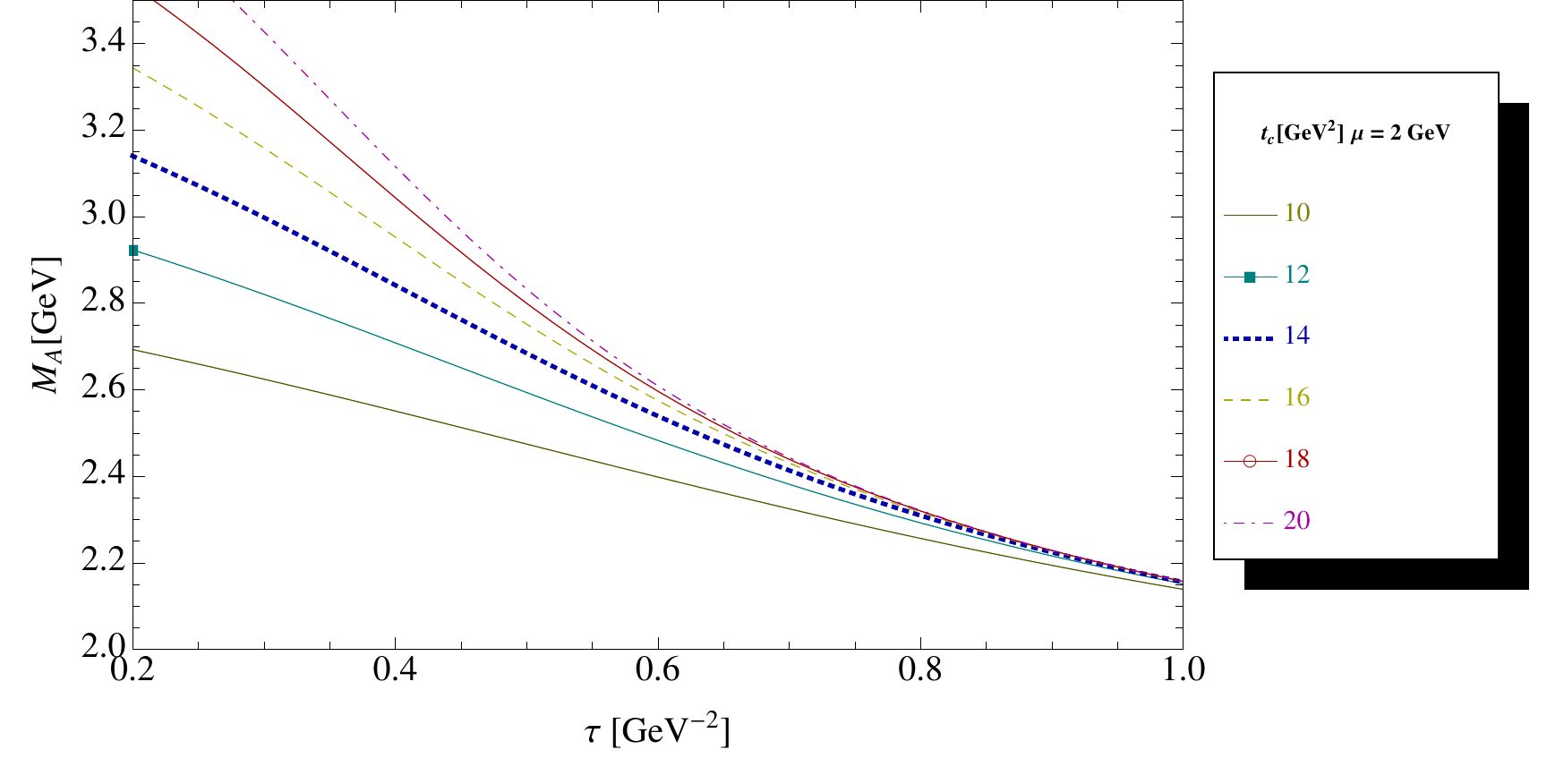}}
{\includegraphics[width=6.2cm  ]{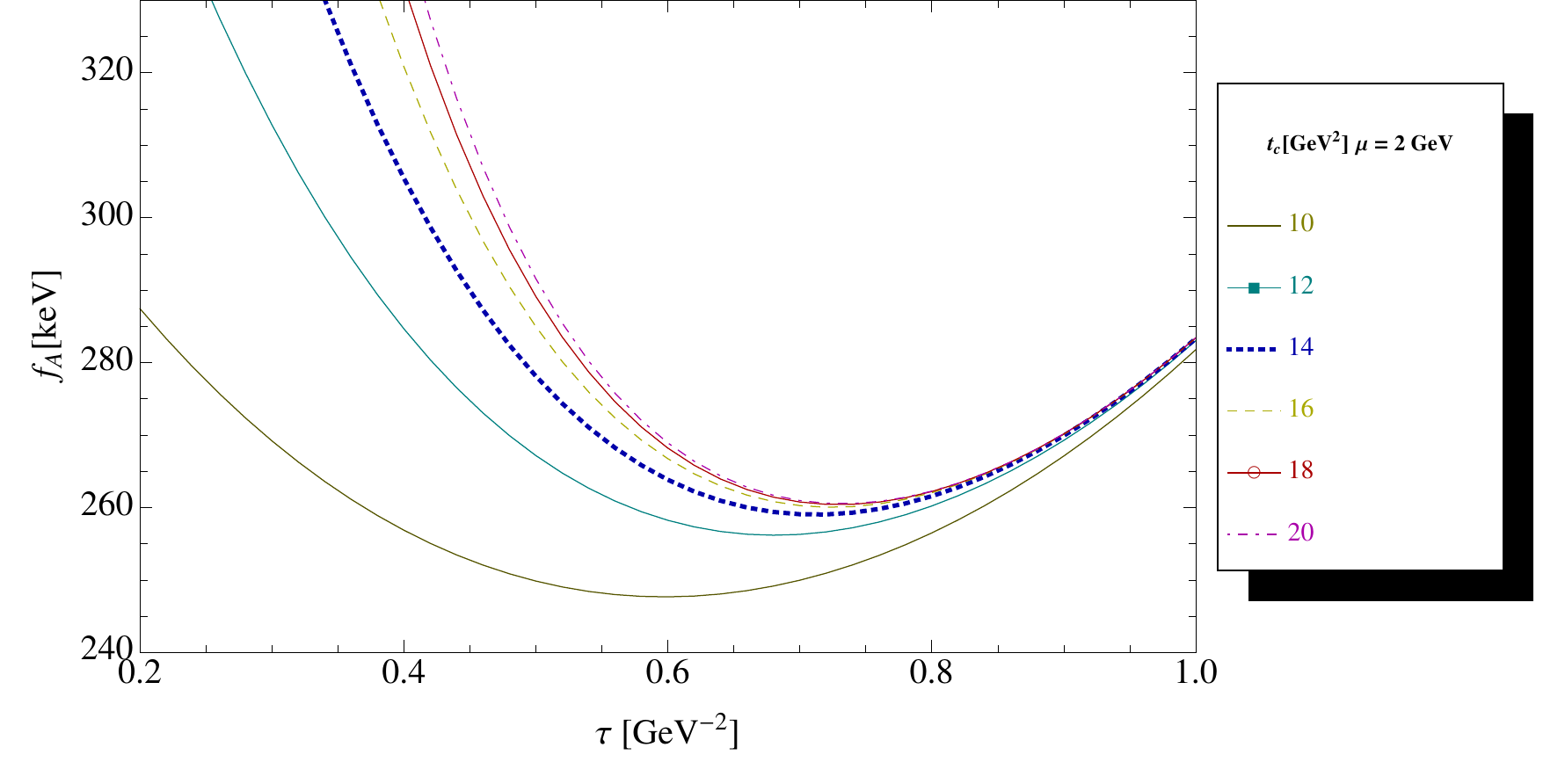}}
\centerline {\hspace*{-3cm} a)\hspace*{6cm} b) }
\caption{
\scriptsize 
{\bf a)} $M_{A_c}$  at LO as function of $\tau$ for  $\mu=2$ GeV, $t_c$=18 GeV$^2$, mixing currents $k=0$  and for the QCD parameters in Tables\,\ref{tab:param} and \ref{tab:alfa}; {\bf b)} The same as a) but for the coupling $f_{A_c}$.
}
\label{fig:ac-lo} 
\end{center}
\end{figure} 
\nin
\begin{figure}[hbt] 
\begin{center}
{\includegraphics[width=6.2cm  ]{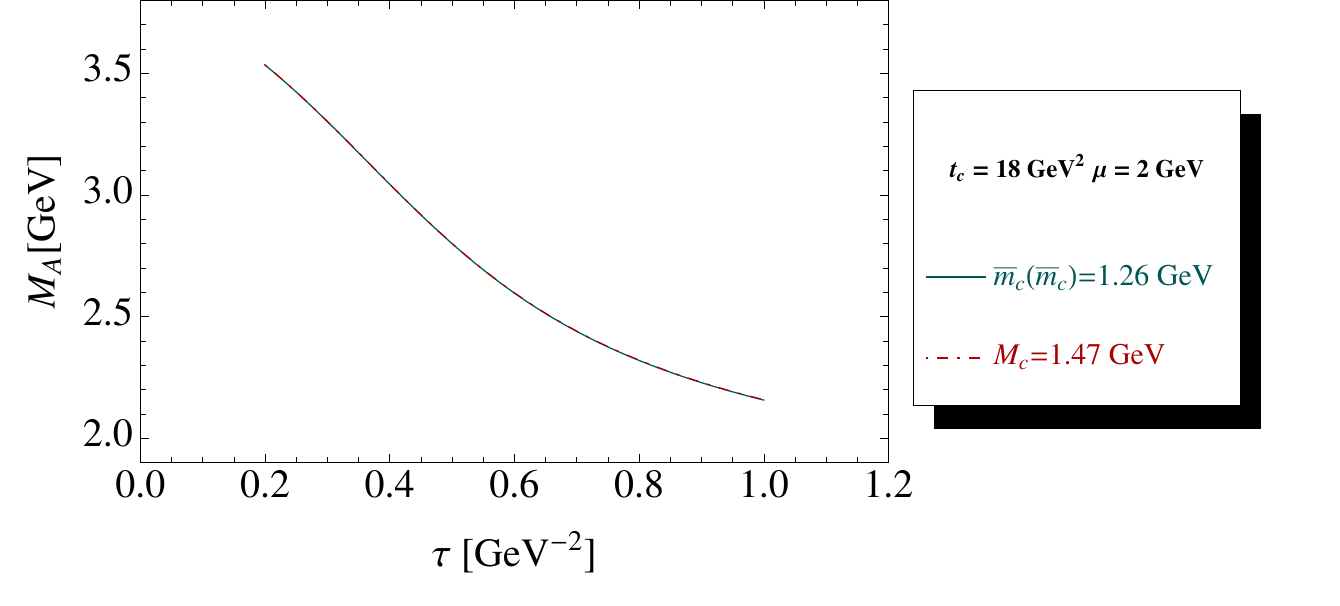}}
{\includegraphics[width=6.2cm  ]{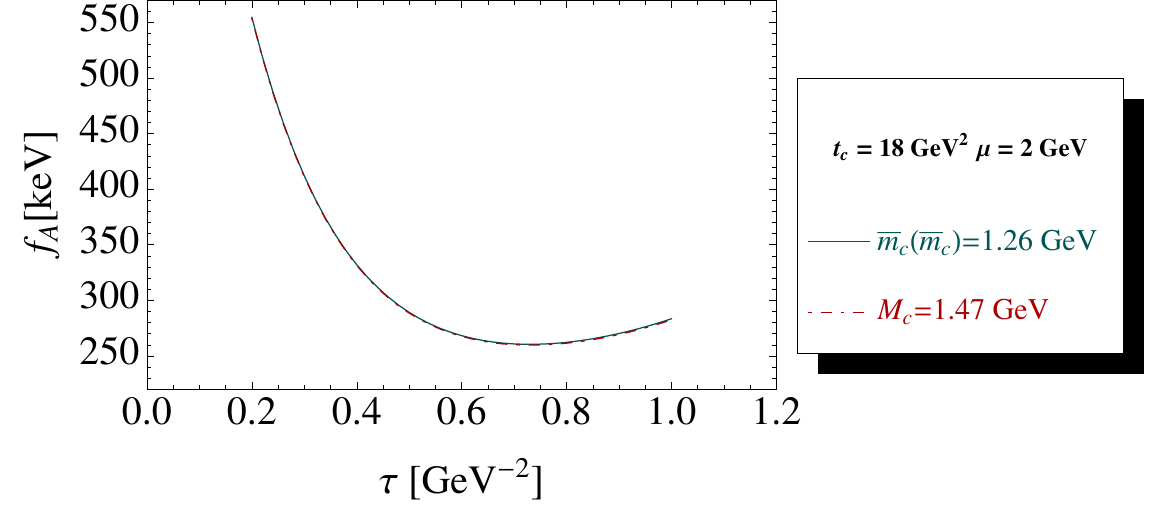}}
\centerline {\hspace*{-3cm} a)\hspace*{6cm} b) }
\caption{
\scriptsize 
{\bf a)} $M_{A_c}$  at LO as function of $\tau$ for a given value of $t_c=18$ GeV$^2$,  $\mu=2$ GeV, mixing of currents $k=0$ and for the QCD parameters in Tables\,\ref{tab:param} and \ref{tab:alfa}. The OPE is truncated at $d=6$.  We compare the effect of  the on-shell or pole mass $M_c=1.47$ GeV and of the running mass $\bar m_c(\bar m_c)=1.26$ GeV; {\bf b)} The same as a) but for the coupling $f_{A_c}$.
}
\label{fig:acmasspole} 
\end{center}
\end{figure} 
\nin
\begin{figure}[hbt] 
\begin{center}
{\includegraphics[width=6.2cm  ]{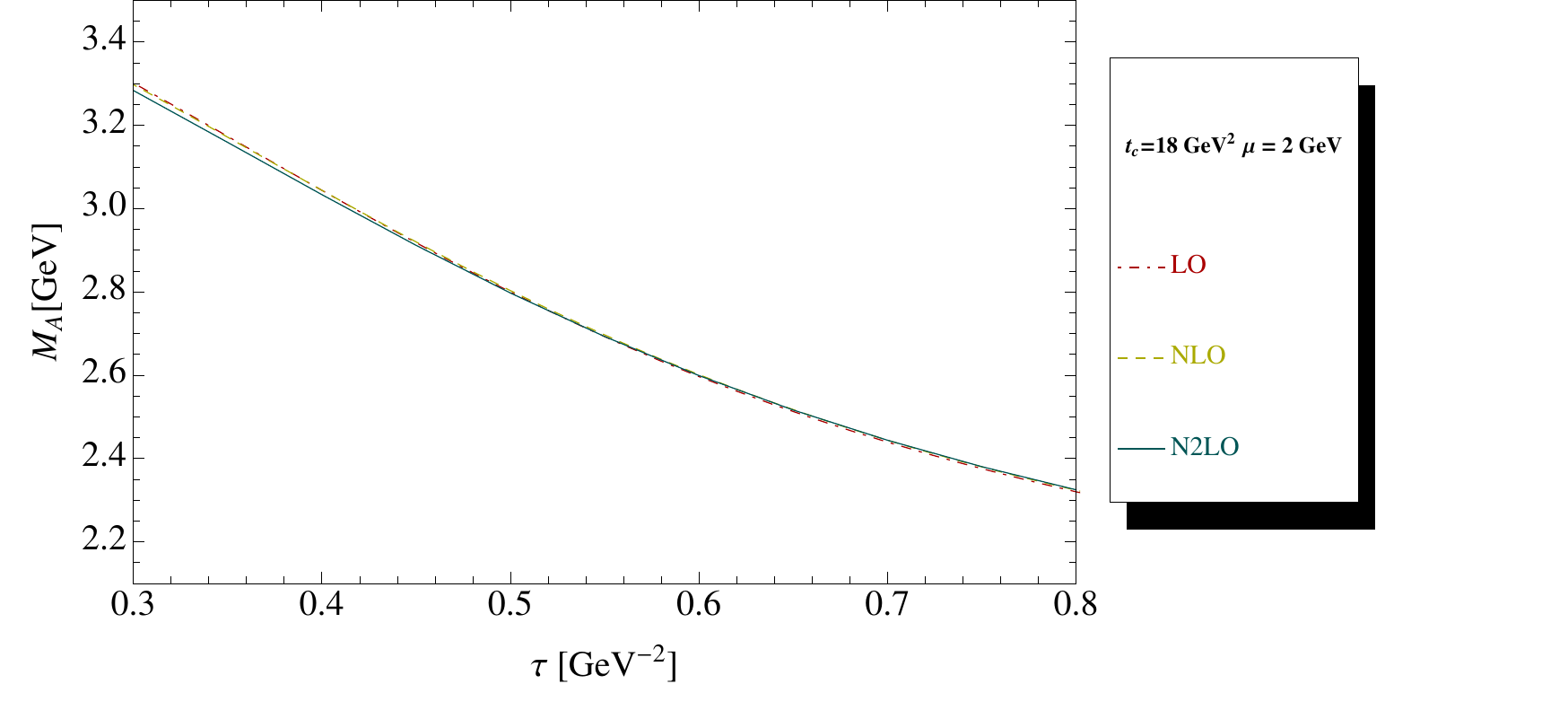}}
{\includegraphics[width=6.2cm  ]{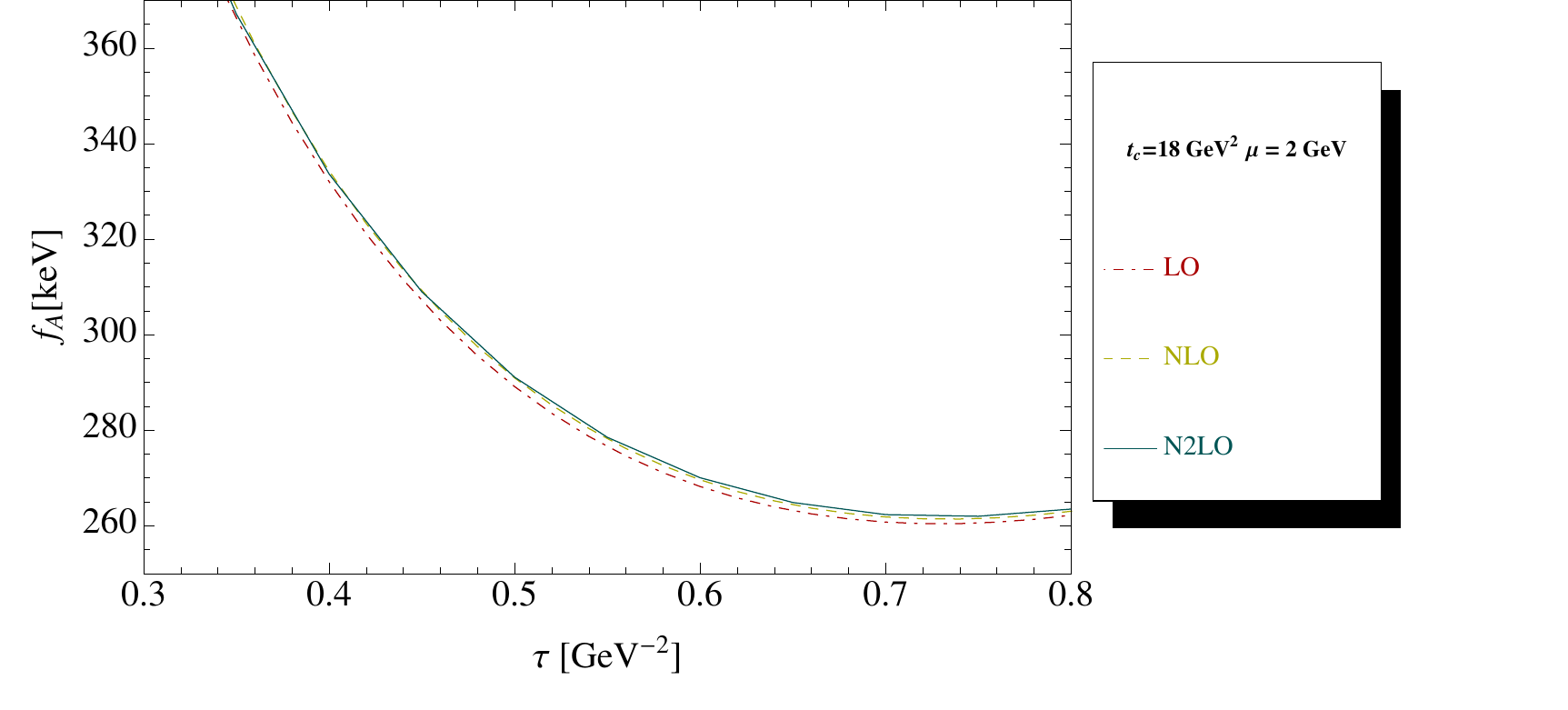}}
\centerline {\hspace*{-3cm} a)\hspace*{6cm} b) }
\caption{
\scriptsize 
{\bf a)} $M_{A_c}$  as function of $\tau$ for different truncation of the PT series at a given value of $t_c$=18 GeV$^2$, $\mu=2$ GeV and for the QCD parameters in Tables\,\ref{tab:param} and \ref{tab:alfa}. {\bf b)} The same as a) but for the coupling $f_{A_c}$.
}
\label{fig:ac-lo-n2lo} 
\end{center}
\end{figure} 
\nin
\begin{figure}[hbt] 
\begin{center}
{\includegraphics[width=6.2cm  ]{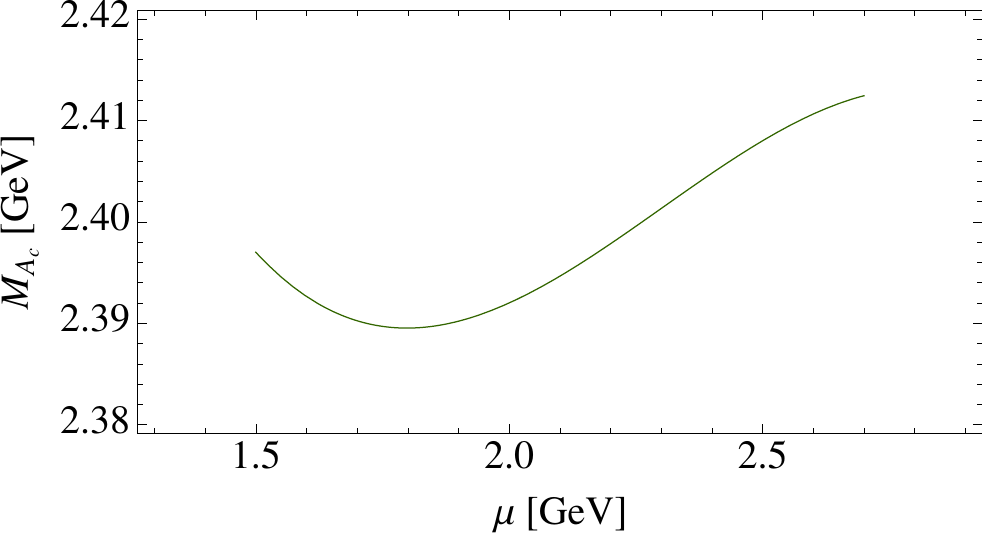}}
{\includegraphics[width=6.2cm  ]{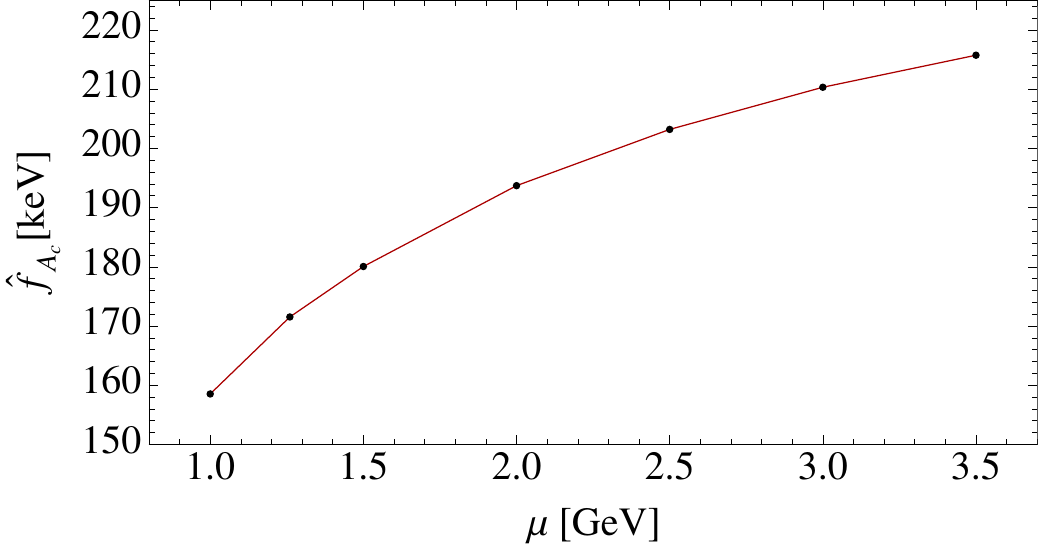}}
\centerline {\hspace*{-3cm} a)\hspace*{6cm} b) }
\caption{
\scriptsize 
{\bf a)} $M_{A_c}$ at NLO as function of $\mu$, for the corresponding $\tau$-stability region, for $t_c\simeq 18$ GeV$^2$ and for the QCD parameters in Tables\,\ref{tab:param} and \ref{tab:alfa}. {\bf b)} The same as a) but for the renormalization group invariant coupling $\hat{f}_{A_c}$.
}
\label{fig:ac-mu} 
\end{center}
\end{figure} 
\nin
\begin{figure}[hbt] 
\begin{center}
{\includegraphics[width=6.2cm  ]{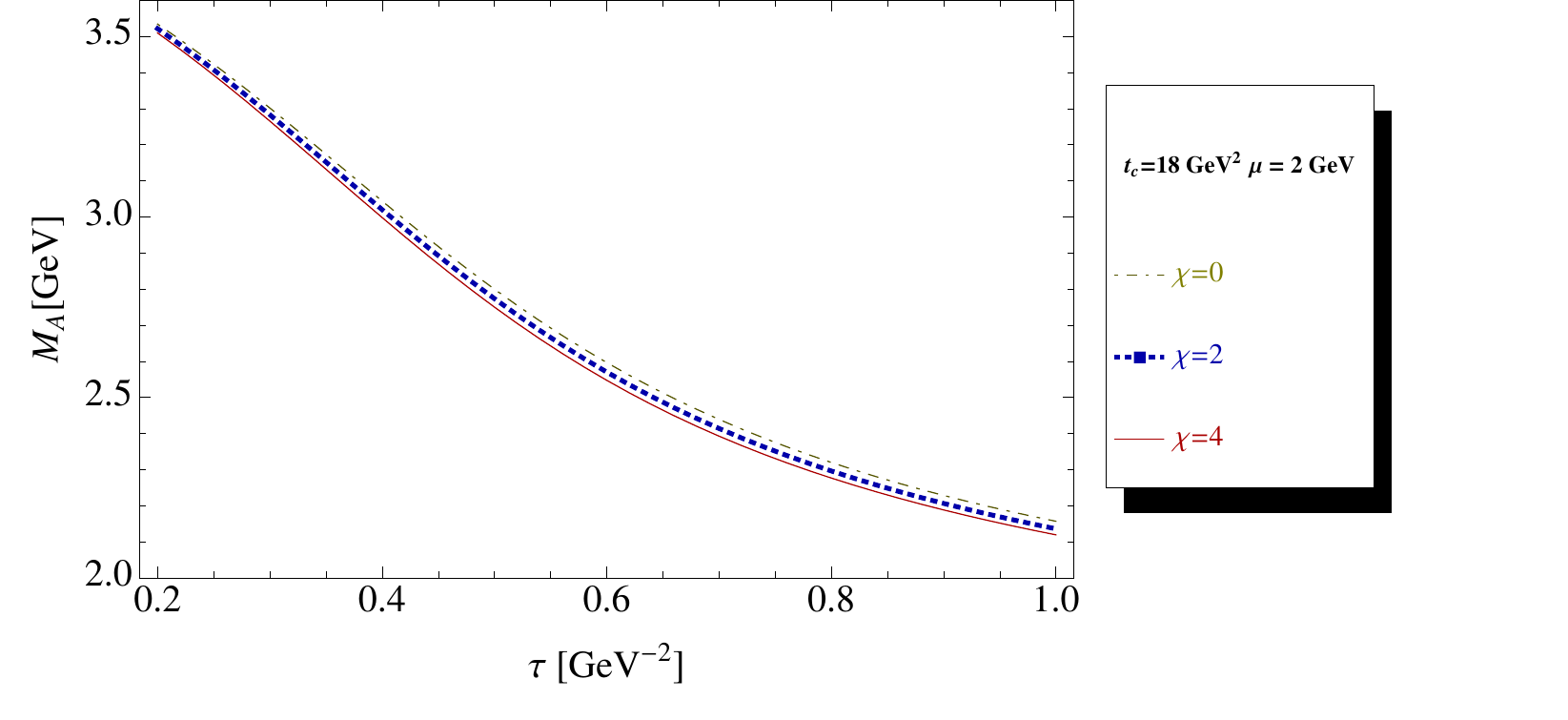}}
{\includegraphics[width=6.2cm  ]{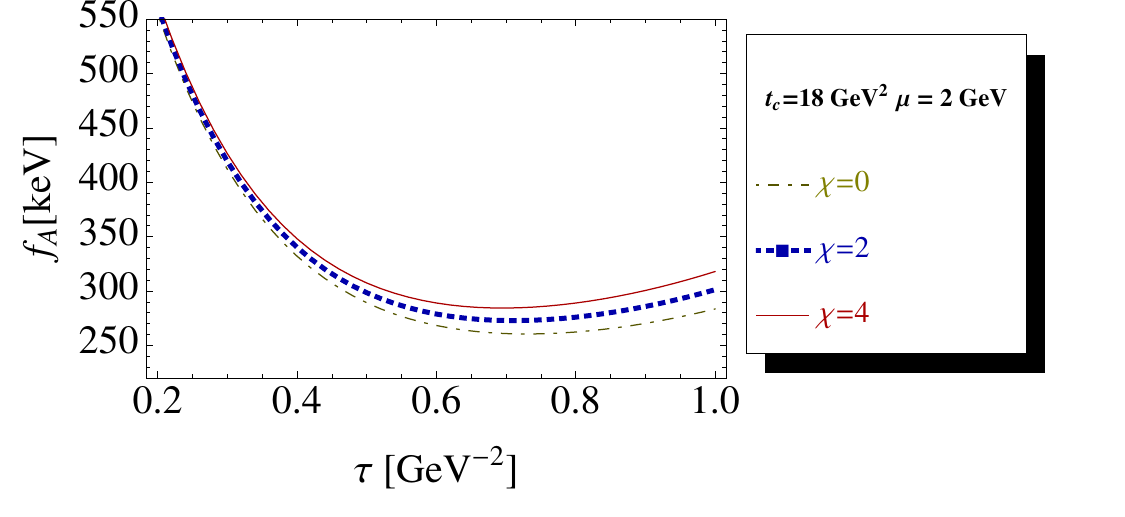}}
\centerline {\hspace*{-3cm} a)\hspace*{6cm} b) }
\caption{
\scriptsize 
{\bf a)} $M_{A_c}$  as function of $\tau$, for different values of the $d=7$ condensate contribution ($\chi$ measures the violation of factorization),  at a given value of $t_c$=18 GeV$^2$, $\mu=2$ GeV and for the QCD parameters in Tables\,\ref{tab:param} and \ref{tab:alfa}; {\bf b)} The same as a) but for the coupling $f_{A_c}$.
}
\label{fig:ac-d7} 
\end{center}
\end{figure} 
\nin
We repeat the previous analysis for the $1^{+}$ axial-vector four-quark state. 
In  Fig.\,\ref{fig:ac-lo}, we show the $\tau$-behaviour of the mass and coupling at lowest order (LO) for different values of $t_c$, for $\mu=2$ GeV and for the input parameters in Tables\,\ref{tab:param} and \ref{tab:alfa}. 
One can see a $\tau$-stability of about 0.6 GeV$^{-2}$ starting from $t_c=12$ GeV$^2$ while $t_c$-stability is reached from $t_c=18$ GeV$^2$. At lowest order, one obtains for $\mu$=2 GeV:
\beq
M_{A_c}^{LO}\simeq( 2387\sim 2401)~{\rm MeV}~~~~{\rm and}~~~~
f_{A_c}^{LO}\simeq (256\sim 261)~{\rm keV}~.
\eeq

\b The effect on the choice of mass (running or pole) is shown in Fig.\,\ref{fig:acmasspole}.

\b The effects of the truncation of the PT series are shown in Fig.\,\ref{fig:ac-lo-n2lo} where
one can notice that the PT corrections are small both for the coupling and
for the mass.

\b The analysis of the $\mu$-behaviour in Fig.\,\ref{fig:ac-mu} indicates a $\mu$-stability for 
$
\mu=2~{\rm GeV}$. 

\b We show in Fig.\,\ref{fig:ac-d7} the effect of a class of $d=7$ condensates for different values of the factorization violation parameter $\chi$.
Our estimate of the error induced by the truncation of the OPE corresponds to the choice $\chi=4$.

\b From the previous analysis, we deduce the final result including N2LO PT perturbative $\oplus$ $d\leq 6$ dimension contributions:
\bea
M_{A_c}&\simeq&( 2400\pm 47)~{\rm MeV}~,\nnb\\
\hat f_{A_c}&\simeq& (192\pm 41)~{\rm keV}~~~~\Lrar ~~~~f_{A_c}\simeq (260\pm 55)~{\rm keV}~,
\eea
where we  take $t_c \simeq (12\sim 18)$ GeV$^2$ and the optimal choice $\mu$=2 GeV. The different sources of errors come from Table\,\ref{tab:errorc}.
{\scriptsize
\begin{table}[hbt]
 \tbl{Exotic hadron masses and couplings from LSR within stability at N2LO}  
    {\small
 {\begin{tabular}{@{}lllll@{}} \toprule
&\\
\hline
\hline
\bf Nature&\bf$J^{P}$&\bf Mass [MeV] & \bf $\hat f_X$ [keV] & \bf $ f_X(4.5)$ [keV]  \\
\hline
{\bf  $b$-quark channel}&\\
{\it Molecule} &&&\\
$B^*K$&$1^{+}$&$5186\pm 13$ &$4.48\pm 1.45$&$8.02\pm 2.60$ \\
$BK$&$0^{+}$&$5195\pm 15$&$2.57\pm 0.75$&$8.26\pm 2.40$\\
$ B^*_s\pi$&$1^{+}$&$5200\pm 18$&$5.61\pm 0.87$&$10.23\pm 1.59$\\
$ B_s\pi$&$0^{+}$&$5199\pm 24$&$3.15\pm 0.70 $&$10.5\pm 2.30$\\
{\it Four-quark $(su)( \overline{bd})$} &&&\\
$A_b$&$1^{+}$&$5186\pm 16$&$5.05\pm 1.32$&$9.04\pm 2.37$\\
$S_b$&$0^{+}$&$5196\pm 17$&$2.98\pm 0.70$&$9.99\pm 2.36$\\
{\bf  $c$-quark channel}&\\
{\it Molecule}\\
$ D^*K$&$1^{+}$&$2395\pm 48$ &$155\pm 36$&$226\pm 52$ \\
$ DK$&$0^{+}$&$2402\pm 42$&$139\pm 26$&$254\pm 48$\\
$ D^*_s\pi$&$1^{+}$&$2395\pm 48$ &$215\pm 35$&$308\pm 49$ \\
$ D_s\pi$&$0^{+}$&$2404 \pm 37$&$160 \pm 22$&$331 \pm 46$\\

{\it Four-quark $(su)( \overline{cd})$} &&&\\
$A_c$&$1^{+}$&$2400\pm 47$&$192\pm 41$&$260\pm 55$\\
$S_c$&$0^{+}$&$2395\pm 68$&$122\pm 26$ &$221\pm 47$\\
\hline\hline
\end{tabular}}
\label{tab:result}
}
\end{table}
} 
\section{$X(5568)$ as a four-quark state-molecule  mixing}
As one can see from Table\,\ref{tab:result}, our predictions do not support the assignements for the $X(5568)$ being a pure $BK$ molecule or four-quark state\,\cite{TURC,NIELSEN,STEELE,WANG,CHINESE,ALI}. Assuming, for instance, that it can result from the mixing of the $BK$ molecule and four-quark state, we consider the two-component mixing:
\bea
|X\ra &=& |B\bar K\ra \cos\theta + |(bu)( \bar d \bar s)\ra\sin\theta\nnb\\
|X_{\perp}\ra &=& -|B\bar K\ra \sin\theta + |(bu)( \bar d \bar s)\ra\cos\theta~.
\eea
Using our result: $M_{BK}\simeq M_{(bu)( \bar d \bar s)}\simeq 5196$ MeV,
one can deduce for reproducing $M_X=5568$ MeV:
\beq
 \sin2\theta \simeq 0.15~,
 \eeq
 and its orthogonal component:
$
 M_{X_\perp}\simeq 4791~{\rm MeV}~,
$
which is much below the $\bar B K$ threshold. This result may go in line with the one in\,\cite{ESPOSITO}. 
\section{Summary and conclusions}
\hspace*{0.5cm} \b We have studied the masses and couplings of the heavy-light exotic states built from one heavy quark  and the three light quarks u,d,s using Laplace sum rule (LSR) within stability (minimum or inflexion point) criteria where N2LO perturbative 
$\oplus$ dimension $d\leq 6$ condensates contributions have been included. 

\b Our analysis including higher order PT corrections has given a more meaning on the input 
value and definition of the heavy quark mass used. We have shown that in this channel the PT corrections are small which (a posteriori) validates the LO  $\oplus$ the running mass results often used in the current literature.

\b We have compared numerically our  QCD expressions with the one in the literature. To our surprise there are disagreement among these though numerically these differences affect only slightly the predictions given by different authors.

\b Comparing our predictions with other authors, we found that the existing results  are obtained in the region of $\tau$ outside the stability region where they are both sensitive to the change of $\tau$ (sum rule variable) and of $t_c$ (continuum threshold) which renders the predictions unreliable. 

\b Our results within stability criteria are summarized in Table\,\ref{tab:result} where one can notice that, contrary to previous claims\,\cite{TURC,NIELSEN,STEELE,WANG,CHINESE,ALI,SWANSON,GUO,CHEN}, we do not predict a mass of a pure exotic $BK,~B^*K,~B_s\pi$ molecule and/or four-quark $(bu)( \bar d \bar s)$ state around or above the $D0$ candidate $X(5568)$\,\cite{D0} which is not confirmed by LHCb\,\cite{LHCb}.  Our results do not also favour mass values derived from coupled channel analysis and from some other approaches (see e.g \cite{OSET} and references quoted therein).

\b We have extended the analysis to the charm sector where the results are also summarized in Table\,\ref{tab:result}.

\b We notice from Table\,\ref{tab:result} that the molecule and four-quark state are almost degenerated. The same feature appears for the $0^{+}$ scalar and $1^{+}$ axial-vector states. 

\b From our analysis, one may suggest  experimentalists to scan the regions $(2327\sim 2444)$\,MeV and $(5173\sim 5226)$\,MeV for detecting these unmixed $(cuds)$ and $(buds)$ exotic hadrons (if any) via eventually their radiative  or $\pi+hadrons$ decays. In the charm sector, the $D^*_{s0}(2317)$ seen by BABAR\,\cite{BABAR} in the $D_s\pi$ invariant mass, expected to be an isoscalar-scalar state with a width less than 3.8 MeV \,\cite{PDG}  could be a good candidate for one of such states.

\b If the $X(5568)$ is experimentally confirmed, one can, for instance, interpret it, within our results in Table\,\ref{tab:result}, by a state resulting from a  mixing of a $BK$ molecule with a scalar four-quark  $(ds)(\overline {bu})$ state
with a mixing angle:  $\sin 2\theta\simeq 0.15~.$  This result may go in line with the one in\,\cite{ESPOSITO} . 


\begin{thebibliography}{999}
\bibitem{D0}V.M. Abazov et al.,  the $D0$ collaboration, arXiv:1602.07588v2 [hep-ex] (2016).
\bibitem{LHCb} J. van Tilburg, the $LHCb$ collaboration, LHCb-CONF-2016-004, 51\`eme Rencontre de Moriond, La Thuile, Italy (2016).

\bibitem{SVZa} M.A. Shifman, A.I. Vainshtein and V.I. Zakharov,
{\it Nucl. Phys.} {\bf B147} (1979) 385. 
\bibitem{SVZb}M.A. Shifman, A.I. Vainshtein and V.I. Zakharov,
{\it Nucl. Phys.} {\bf B147} (1979) 448.

\bibitem{SNB4a} S. Narison, {\it Nucl. Part. Phys. Proc.} {\bf 207-208} (2010) 315.

\bibitem{SNB4b} S. Narison, {\it Nucl. Part. Phys. Proc.} {\bf 258-259} (2015) 189.

\bibitem{SNB1} S. 
Narison, {\it QCD as a theory of hadrons,
Cambridge Monogr. Part. Phys. Nucl. Phys. Cosmol.} {\bf 17} (2002) 1
[hep-ph/0205006].

\bibitem{SNB2}S. Narison, {\it QCD
spectral sum rules ,  World Sci. Lect. Notes Phys.} {\bf 26}
(1989) 1.

\bibitem{SNB3a}S. Narison, {\it Phys. Rept.} {\bf 84} (1982) 263.
\bibitem{SNB3b}S. Narison, {\it Acta Phys. Pol.} {\bf B26} (1995) 687.
\bibitem{SNB3c} S. Narison, hep-ph/9510270 (1995).

\bibitem{RRY} L.J. Reinders, H. Rubinstein and S. Yazaki, {\it Phys. Rept. }
{\bf 127}  (1985) 1. 

\bibitem{CK}E. de Rafael, hep-ph/9802448.

\bibitem{TURC}S. S. Agaev, K. Azizi and H. Sundu,  arXiv:1603.02708.

\bibitem{TURC1}S. S. Agaev, K. Azizi and H. Sundu,  {\it Phys. Rev.} {\bf D93} (2016) $n^o$ 7, 074024.

\bibitem{NIELSEN} C.M. Zanetti, M. Nielsen and K.P. Khemchandani, arXiv:1602.09041v1 [hep-ph].

\bibitem{STEELE} W. Chen et al., arXiv:1602.08916v1 [hep-ph].

\bibitem{WANG} Z.-G. Wang, arXiv:1602.08711v1 [hep-ph].


\bibitem{BELLa}J.S. Bell and R.A. Bertlmann, {\it Nucl. Phys.} {\bf B227} (1983) 435.

\bibitem{BELLb}R.A. Bertlmann, {\it Acta Phys. Austriaca} {\bf 53} (1981) 305.  

\bibitem{SNRAF}S. Narison, E. de Rafael, {\it Phys. Lett.} {\bf B103} (1981) 57.

\bibitem{DRSR}S. Narison, {\it Phys. Lett.}  {\bf B210} (1988)  238. 

\bibitem{SNFORM1} S. Narison, {\it Phys. Lett.} {\bf B337} (1994) 166.

\bibitem{SNGh3} S. Narison, {\it Phys. Lett.} {\bf B322}  (1994) 327.

\bibitem{SNGh1} S. Narison, {\it Phys. Lett.} {\bf B387}  (1996) 162. 

\bibitem{SNGh5} S. Narison, {\it Phys. Lett.} {\bf B358} (1995) 113.
 
 \bibitem{SNmassb} S. Narison, {\it Phys.Lett.} {\bf B466} (1999) 345.
 
 \bibitem{SNhl}S. Narison, {\it Phys. Lett.} {\bf B605} (2005) 319.
 
  \bibitem{SNmassa} S. Narison, {\it  Phys.Rev.} {\bf D74} (2006) 034013.

\bibitem{SNFORM2} S. Narison, {\it Phys. Lett.}  {\bf B668} (2008) 308.

 \bibitem{HBARYON1} R.M. Albuquerque, S. Narison, {\it Phys. Lett.} {\bf B694} (2010) 217.

\bibitem{HBARYON2}R.M. Albuquerque, S. Narison, M. Nielsen, {\it Phys. Lett.} {\bf B684} (2010) 236.

\bibitem{NAVARRA}S. Narison, F. Navarra and M. Nielsen, {\it Phys. Rev.} {\bf D83} (2011) 016004.

\bibitem{SNFB12a} S. Narison,  {\it Phys. Lett.}  {\bf B718} (2013)  1321.
\bibitem{SNFB12b}S. Narison,  {\it Nucl. Part. Phys. Proc. } {\bf 234}  (2013) 187.
\bibitem{PERISa}E. de Rafael, {\it Nucl. Part. Phys. Proc.} {\bf 96} (2001) 316.
 \bibitem{PERISb}S. Peris, B. Phily and E. de Rafael, {\it Phys. Rev. Lett.} {\bf 86} (2001) 14.
 
 \bibitem{STEVENSON}P.M. Stevenson, {\it Nucl.Phys.} {\bf B868} (2013) 38.


\bibitem{CNZ1} K. Chetyrkin, S. Narison and V.I. Zakharov, {\it Nucl. Phys.} 
{\bf B550} (1999)  353.
\bibitem{CNZ2}S. Narison and V.I. Zakharov, {\it  Phys. Lett.} {\bf B522} (2001) 266.
\bibitem{ZAK1} V.I. Zakharov, {\it Nucl. Phys. Proc. Suppl.}, {\bf 164} (2007) 240.
\bibitem{ZAK2}S. Narison,  {\it Nucl. Phys. Proc. Suppl.} {\bf 164} 
 (2007) 225. 
 \bibitem{ZAK3}S. Narison and V.I. Zakharov, {\it Phys. Lett.} {\bf B679} (2009) 355.


\bibitem{FENOSOA1}F. Fanomezana, S. Narison and A. Rabemananjara, {\it Nucl. Part. Phys. Proc.} {\bf 258-259} (2015) 152.

\bibitem{FENOSOA2}F. Fanomezana, S. Narison and A. Rabemananjara, {\it Nucl. Part. Phys. Proc.} {\bf 258-259} (2015) 156.

\bibitem{PICH} A. Pich and E. de Rafael, {\it Phys. Lett.} {\bf B158} (1985) 477.
\bibitem{SNPIVO} S. Narison and A. Pivovarov, {\it Phys. Lett.} {\bf B327}(1994) 341.

 \bibitem{FNR}E.G. Floratos, S. Narison and E. de Rafael, {\it Nucl. Phys.} {\bf B155} (1979) 155.
 \bibitem{BECCHI}C. Becchi et al.,
 {\it Z. Phys.} {\bf C8} (1981)  335.
 \bibitem{BROAD}D.J. Broadhurst, {\it Phys. Lett.}  {\bf B101} (1981)  423.
 \bibitem{RVS}K.G. Chetyrkin,  J.H. K\"uhn and M. Steinhauser, {\it Eur. Phys. J.} {\bf C21} (2001) 319.

\bibitem{TAR}R. Tarrach, {\it Nucl. Phys.} {\bf B183} (1981) 384.

\bibitem{COQUEa} R. Coquereaux, {\it Annals of Physics} {\bf 125} (1980) 401.
\bibitem{COQUEb}P. Binetruy and T. S\"ucker, {\it Nucl. Phys.} {\bf B178} (1981) 293.

\bibitem{SNPOLEa} S. Narison, {\it  Phys. Lett.} {\bf B197} (1987) 405.

\bibitem{SNPOLEb}S. Narison, {\it  Phys. Lett.} {\bf B216} (1989) 191.

\bibitem{BROAD2a} N. Gray et al., 
{\it Z. Phys.} {\bf  C48} (1990) 673.

\bibitem{AVDEEV}L.V. Avdeev and M. Yu. Kalmykov, {\it Nucl. Phys.} {\bf B502}
(1997) 419.

\bibitem{BROAD2b}J. Fleischer et al.,
{\it Nucl. Phys.} {\bf B539} (1999) 671.

\bibitem{CHET2a}K.G. Chetyrkin and M. Steinhauser, {\it Nucl. Phys.} {\bf B573}
(2000) 617.

\bibitem{CHET2b}K. Melnikov and T. van Ritbergen, {\it Phys. Lett.} {\bf B482} (2000)99.



\bibitem{SNTAU}S. Narison, {\it Phys. Lett.} {\bf B673} (2009) 30.

\bibitem{BNPa}E. Braaten, S. Narison and A. Pich, {\it Nucl. Phys.} {\bf B373} (1992) 581.

\bibitem{BNPb}S. Narison and A. Pich, {\it Phys. Lett.} {\bf  B211} (1988) 183.

\bibitem{BETHKE} For a recent review, see e.g:  S. Bethke, {\it Nucl. Phys. Proc. Suppl.} {\bf 234} (2013) 229.

\bibitem{PDG} K.A. Olive et al. (Particle Data Group), {\it Chin. Phys. } {\bf C38} (2014) 090001.

\bibitem{SNH10a}S. Narison,  {\it Phys. Lett.} {\bf B693} (2010)  559; Erratum ibid 705 (2011) 544.
\bibitem{SNH10b}S. Narison,  {\it Phys. Lett.} {\bf B706} (2011)  412.
\bibitem{SNH10c}S. Narison,  {\it Phys. Lett.} {\bf B707} (2012)  259. 

\bibitem{IOFFEa} B.L. Ioffe and K.N. Zyablyuk, {\it Eur. Phys. J.} {\bf  C27}
(2003)  229.
\bibitem{IOFFEb} B.L. Ioffe, {\it Prog. Part. Nucl. Phys.} {\bf 56} (2006) 232.

 \bibitem{SNmass98a}H.G. Dosch and S. Narison,  {\it Phys. Lett.}  {\bf B417} (1998) 173.
\bibitem{SNmass98b} S. Narison,  {\it Phys. Lett.} {\bf B216} (1989) 191.
 
 \bibitem{SNLIGHT}S. Narison,  {\it Phys. Lett.}  {\bf B738} (2014) 346.
 
 \bibitem{JAMI2a}Y. Chung et al., {\it Z. Phys.} {\bf C25} (1984)  151.
\bibitem{JAMI2b}  H.G. Dosch, {\it Non-Perturbative Methods (Montpellier 1985)} ed. S. Narison, WSC. 
\bibitem{JAMI2c}H.G. Dosch, M. Jamin and S. Narison, {\it Phys. Lett.} {\bf B220} (1989)  251.

\bibitem{HEIDa}B.L. Ioffe, {\it Nucl. Phys.} {\bf B188} (1981)  317.

\bibitem{HEIDb}B.L. Ioffe, {\it Nucl. Phys.} {\bf B191} (1981) 591.

\bibitem{HEIDc}A.A.Ovchinnikov and A.A.Pivovarov,
{\it Yad.\ Fiz.}  {\bf 48} (1988) 1135.

\bibitem{LNT}G. Launer, S. Narison and R. Tarrach, {\it  Z. Phys.} {\bf C26}
(1984) 433.

\bibitem{SNIa}S. Narison, {\it Phys. Lett.} {\bf B300} (1993) 293.

\bibitem{SNIb}S. Narison, {\it Phys. Lett.} {\bf B361} (1995) 121.

\bibitem{YNDU}F.J. Yndurain, {\it Phys. Rept.} {\bf 320} (1999) 287.

\bibitem{BELLc}R.A. Bertlmann and H. Neufeld, {\it Z. Phys.} {\bf C27} (1985)  437.


\bibitem{SNG1} S. Narison, {\it Phys. Lett.} {\bf B361} (1995) 121.

\bibitem{SNG2}S. Narison,  {\it Phys. Lett.} {\bf B624} (2005) 223.

\bibitem{SNGH}S. Narison, {\it Phys. Lett.} {\bf B387} (1996) 162.

\bibitem{PICHTAU}A. Pich and A. Rodriguez-S\'anchez, arXiv:1605.06830 [hep-ph] (2016).

\bibitem{DAVIER} M. Davier et al., {\it Eur. Phys. J.} {\bf C56} (2008) 305.

\bibitem{BOITOa} D. Boito et al, {\it Phys. Rev.} {\bf D95} (2015) {\bf n$^0$3}, 034003.
 
\bibitem{BOITOb} D. Boito et al, {\it Nucl. Part. Phys. Proc.} {\bf  270-272} (2016) 103.
\bibitem{ROSNERb}J. Rosner and S. Stone,   arXiv:1509.02220 [hep-ph] (2015).
 \bibitem{LATT13}S. Aoki et al., FLAG working group, {\it Eur. Phys. J.} {\bf C74} (2014) 2890 .
  \bibitem{BALIa} G. S. Bali, C. Bauer and A. Pineda,  {\it Phys. Rev. Lett.} {\bf 113} (2014) 092001.

 \bibitem{BALIb}   T. Lee, {\it Phys. Rev.} {\bf D82}(2010)114021.
 
\bibitem{FESRa} R.A. Bertlmann, G. Launer and E. de Rafael, {\it Nucl. Phys.} {\bf B250} (1985) 61.

\bibitem{FESRb}R.A. Bertlmann et al., {\it Z.\ Phys.}  {\bf C39} (1988) 231.


\bibitem{SNFB15a}S. Narison, {\it Int. J. Mod. Phys.}  {\bf A30} (2015) {\bf n$^0$ 20}, 1550116.
\bibitem{SNFB15b}S. Narison, {\it Nucl. Part. Phys. Proc.} {\bf 270-272} (2016) 143. 

\bibitem{CHINESE}W. Wang and R.-L. Zhu, arXiv: 1602.08806 [hep-ph] (2016).

\bibitem{ALI} A. Ali et al., arXiv: 1604.01731 [hep-ph] (2016).

\bibitem{SWANSON}T.J. Burns and E.S. Swanson, arXiv: 1603.04366 [hep-ph] (2016).

\bibitem{GUO}F.R. Guo, U.G. Meissner and B.S. Zou, arXiv: 1603.06316 [hep-ph] (2016).

\bibitem{CHEN}X. Chen, J. Ping, arXiv: 1604.05651 [hep-ph] (2016). 

\bibitem{OSET}M. Albaladejo et al., arXiv: 1603.09230 [hep-ph] (2016).

\bibitem{BABAR}B. Aubert et al., {\it the BABAR collaboration}, {\it Phys. Rev. Lett.} {\bf 90} (2003) 242001. 

\bibitem{ESPOSITO} A. Esposito, A. Pilloni and A.D. Polosa, {\it Phys. Lett.} {\bf B758} (2016) 292.


\end{thebibliography}
\end{document}